\documentclass[dsc,numbers]{coppe}
\usepackage[utf8]{inputenc}
\usepackage{amsmath,amssymb}
\usepackage{hyperref}
\usepackage{subfigure}
\usepackage{multirow}
\usepackage[portuguese,linesnumbered,vlined,ruled]{algorithm2e}
\usepackage{longtable}
\usepackage{lscape}
\usepackage{float}

\makelosymbols
\makeloabbreviations

\begin{document}
  \title{Relatório Técnico: Controle Distribuído de Tráfego Baseado em Veículos Conectados e Comunicações Veiculares Centradas em Interesses}
  \foreigntitle{Distributed Traffic Control Based on Connected Vehicles}
  \author{Fabrício}{Barros Gonçalves}
  \advisor{Prof.}{Felipe}{Maia Galvão França}{Ph.D.}
  \advisor{Prof.}{Claudio Luis}{de Amorim}{Ph.D.}

  \examiner{Prof.}{Felipe Maia Galvão França}{Ph.D.}
  \examiner{Prof.}{Cláudio Luis de Amorim}{Ph.D.}
  \examiner{Prof.}{Luis Felipe Magalhães de Moraes}{Ph.D.}
  \examiner{Prof.}{Valmir Carneiro Barbosa}{Ph.D.}
  \examiner{Prof.}{Paulo Cezar Martins Ribeiro}{Ph.D.}
  \examiner{Prof.}{Fábio Protti}{D.Sc.}
  \examiner{Prof.}{Diego Moreira de Araujo Carvalho}{D.Sc.}
  \department{PESC}
  \date{06}{2017}

  \keyword{RAdNet-VE}
  \keyword{HRAdNet-VE}
  \keyword{Agentes}
  \keyword{Sistemas Multiagentes}
  \keyword{Controle Inteligente de Tráfego}
  \keyword{Planejamento e Orientação de Rotas}

  \maketitle

   \frontmatter
   \begin{abstract}

Por mais que sistemas avançados de gerenciamento de tráfego consigam lidar com o problema da heterogeneidade dos fluxos de tráfego das vias que incidem nas interseções de uma rede viária, estes têm o seu desempenho comprometido, quando o volume de tráfego da rede viária não é distribuído de maneira uniforme. Para tratar este problema, um sistema avançado de informações ao motorista deve ter total ciência do estado de operação de um sistema avançado de gerenciamento de tráfego. No entanto, este requisito não pode ser completamente satisfeito, devido a existência de lacunas existentes no estado da arte de sistemas avançados de gerenciamento de tráfego e sistemas avançados de informações ao motorista, em específico, a cooperação entre estes dois tipos de sistemas. Por isto, este trabalho propõe um sistema de controle distribuído de tráfego, onde agentes embutidos em veículos conectados, sinalizações semafóricas, elementos urbanos e um centro de controle de tráfego interagem uns com os outros, a fim de promover uma maior fluidez do tráfego veicular. Para tanto, os agentes dependem fortemente de uma rede veicular heterogênea. Por isto, este trabalho também propõe uma rede veicular heterogênea cujo protocolo de comunicação é capaz de satisfazer os requisitos de comunicação de aplicações de serviços de sistemas inteligentes de transporte. De acordo com os resultados obtidos por meio de simulações, o sistema de controle distribuído de tráfego foi capaz de maximizar o fluxo de veículos e a velocidade média dos veículos, e minimizar o tempo de espera, número de paradas, tempo de viagem, consumo de combustível e emissões (CO, CO$_2$, HC, NOx e PMx). 

\end{abstract}

   \begin{foreignabstract}

Although advanced traffic management systems can deal with the heterogeneous traffic flows approaching of intersections, their performances are compromised, when the traffic volume is not distributed uniformly. To evenly distribute the traffic flow, an advanced driver information system should be aware of the traffic control operations. However, such requirement can not ultimately be satisfied due to the gaps in state of the art in advanced traffic management systems. Therefore, this study proposes a distributed traffic control system, in which agents embedded in connected vehicles, traffic signals, urban elements and a traffic control center interact with each other to provide a greater traffic fluidity. Therefore, the agents depend strongly on a heterogeneous vehicular network. In this sense, this study also proposes a heterogeneous vehicular network whose communication protocol can satisfy the communication requirements of intelligent transportation systems service applications. According to the results obtained from simulations, the distributed traffic control system was able to maximize the flow of vehicles and the mean speed of the vehicles, and minimize the wait time, travel time, fuel consume and emissions (CO, CO$_2$, HC, NOx and PMx).

\end{foreignabstract}
  \tableofcontents
  \listoffigures
  \listoftables
%   \printlosymbols
%   \printloabbreviations
   \chapter{Lista de Siglas e Abreviaturas}

\begin{table}[!h]
\small\begin{tabular}{lp{11cm}}
\textbf{3GPP}       & Third Generation Partnership Project                                            \\
\textbf{AODV}       & Ad Hoc On-Demand Distance Vector                                                \\
\textbf{CCN}        & Content-Centric Network                                                         \\
\textbf{D2D}        & Device-to-Device                                                                \\
\textbf{DSR}        & Dynamic Source Routing                                                          \\
\textbf{eNodeB}     & Evolved Node B                                                                  \\
\textbf{GPS}        & Global Positioning System                                                       \\
\textbf{GPSR}       & Greedy Perimeter Stateless Routing                                              \\
\textbf{GSM}        & Global System for Mobile Communications                                         \\
\textbf{HRAdNet-VE} & \textbf{H}eterogeneous Inte\textbf{R}est-Centric Mobile \textbf{Ad} Hoc \textbf{Net}work for \textbf{V}ehicular \textbf{E}nvironments \\
\textbf{HRVEP}      & HRAdNet-VE Protocol                                                             \\
\textbf{ICN}        & Information-Centric Network                                                     \\
\textbf{LTE}        & Long Term Evolution                                                             \\
\textbf{MFU}        & Most Frequently Used                                                            \\
\textbf{MIMO}       & Multiple Input Multiple Output                                                  \\
\textbf{OBU}        & On-Board Unit                                                                   \\
\textbf{RadNet}     & Inte\textbf{R}est-Centric Mobile \textbf{Ad} Hoc \textbf{Net}work                                          \\
\textbf{RadNet-VE}  & Inte\textbf{R}est-Centric Mobile \textbf{Ad} Hoc \textbf{Net}work for \textbf{V}ehicular \textbf{E}nvironments               \\
\textbf{RCEMC}  & Roteamento Baseado em Caminho Espacialmente mais Curto             \\
\textbf{RCTMC}  & Roteamento Baseado em Caminho Temporalmente mais Curto               \\
\textbf{ROOV}  & Roteamento Orientado à Ondas Verdes               \\
\textbf{RSU}        & Roadside Unit                                                                   \\
\textbf{RVEP}       & RadNet-VE Protocol                                                              \\
\textbf{SER}        & Scheduling by Edge Reversal                                                     \\
\textbf{SMER}       & Scheduling by Multiple Edge Reversal                                            \\
\textbf{UE}         & User Equipment                                                                 \\
\textbf{UMTS}       & Universal Mobile Telecommunication System                                       \\
\textbf{URI}        & Universal Resource Identifier                                                   \\
\textbf{VANET}      & Vehicular Ad Hoc Networks                                                        \\
\textbf{WAVE}       & Wireless Access Vehicular Environment                                           \\
\textbf{WSA}        & WAVE Service Advertisement                                                      \\
\textbf{WSM}        & WAVE Short Message 
\end{tabular}
\end{table}

  \mainmatter
  \chapter{Introdução}

Por mais que sistemas avançados de gerenciamento de tráfego \cite{Bazzan:2014} consigam lidar com o problema da heterogeneidade dos fluxos de tráfego das vias  de entradas das interseções de uma rede viária, estes têm o seu desempenho comprometido, quando o volume de tráfego da rede viária não é distribuído de maneira uniforme. Como consequência, os intervalos de indicação da luz verde, nas vias pouco utilizadas, são desperdiçados. Por isto, um segundo problema precisa ser enfrentado, que é o problema de planejamento e orientação de rotas. Para tratar este problema, um sistema avançado de informações ao motorista \cite{Bazzan:2014} deve ter total ciência do estado de operação de um sistema avançado de gerenciamento de tráfego. No entanto, este requisito não pode ser completamente satisfeito, devido a existência de lacunas existentes no estado da arte de sistemas avançados de gerenciamento de tráfego, sistemas avançados de informações ao motorista e redes veiculares. 

\section{Motivação}\label{sec:motivacao}

Após todo o levantamento bibliográfico acerca dos métodos de controle tráfego (\cite{Zheng:2011,Bazzan:2009,DeOliveira:2006,Cheng:2006,Migdalas:1995,Kerner:2012,Xie:2011a,Xie:2011b,Gerhenson:2005,Einhorn:2012,Slager:2010,Helbing:2009,Lammer:2008}) e métodos de planejamento e orientação de rotas (\cite{Kraus:2008,Leontiadis:2011,Schunemman:2009,Wedde:2013,Claes:2011}, percebeu-se que tais métodos podem ser complementares uns aos outros. Métodos como os utilizados em sistemas multiagentes baseados em alocação de recursos podem tirar proveito dos tempos das fases das interseções, ainda que estes possam ser alterados, de acordo com alguma política de adaptação de intervalos de indicações de luzes verdes em função dos fluxos de tráfegos em vias de entrada de interseções. Dessa forma, um mecanismo de planejamento e orientação de rotas pode distribuir o volume de tráfego sobre uma rede viária controlada por sinalizações semafóricas, alocando espaços nas vias para os veículos e fazendo com que eles trafeguem em vias participantes de rotas ótimas, sem que estes precisem parar constantemente. Neste caso, a cooperação entre sistemas avançados de gerenciamento de tráfego e sistemas avançados de informações ao motorista permite que o planejamento de rotas tenha ciência das programações das sinalizações semafóricas ao longo do tempo. Para tanto, é fundamental que o método de controle de tráfego seja adequado para a geração de uma agenda global de intervalos de indicações de luzes verdes, de modo que esta sirva de base para construção de uma solução para o problema de planejamento e orientação de rotas.

Os métodos de controle de tráfego baseados em otimização \cite{Zheng:2011,DeOliveira:2006,Cheng:2006,Migdalas:1995} necessitam de uma infraestrutura computacional centralizada e de alto custo, que é utilizada para o processamento das otimizações de fases  de todas as interseções pertencentes ao sistema de controle de tráfego. A disponibilização de uma agenda de intervalos de indicações de luzes verde por meio de uma infraestrutura computacional centralizada gera um grande custo de comunicação e, leitura e escrita, à medida que uma grande quantidade de veículos solicita cálculos de rotas ótimas. Neste caso, estrutura centralizada pode se tornar um gargalo e, com isto, pode influenciar no tempo de resposta das requisições, fazendo com que o planejamento e orientação de rotas trabalhe com dados defasados. Vale ressaltar que, enquanto os veículos esperam por rotas ótimas, eles se movem ao longo das vias da rede viária. Logo, se a latência de resposta de uma requisição para o cálculo de uma rota ótima é alta, os motoristas dos veículos podem não ser informados a tempo de realizar alguma manobra, podendo, então, ser necessário iniciar um processo de recálculo de rota ótima. 

Os métodos de controle de tráfego baseados em adaptação \cite{Bazzan:2009, Kerner:2012,Xie:2011a,Xie:2011b,Gerhenson:2005,Einhorn:2012,Slager:2010,Helbing:2009,Lammer:2008} são capazes de coletar os dados por meio de detectores de tráfego instalados nas vias e, de tempos em tempos, realizar uma computação local, a fim de calcular os novos tamanhos de intervalos de indicações de luzes verdes das sinalizações semafóricas. Neste sentido, as decisões de ajustar as fases das sinalizações semafóricas são tomadas localmente. Por meio das decisões de todas as sinalizações semafóricas pertencentes ao sistema de controle de tráfego, a otimização do fluxo de tráfego é realizada de maneira descentralizada. Esses métodos são categorizados da seguinte forma: baseados em aprendizado por reforço e baseados em auto-organização.

Devido à incapacidade de se adaptarem facilmente às flutuações frequentes dos fluxos de tráfego, que é ocasionada pela necessidade de uma estabilidade do volume de tráfego para a construção de uma base de conhecimento acerca dos padrões macroscópicos dos fluxos de tráfego das vias, métodos de controle de tráfego baseados em adaptação por meio de aprendizado por reforço tendem a gerar agendas de tempos similares as que podem ser geradas por sinalizações semafóricas de tempos pré-fixados. Assim, qualquer mudança no padrão macroscópico de tráfego, que resulte em um padrão não existente na base de conhecimento de uma via, pode gerar um gargalo em uma parte específica da rede viária. Ainda que os métodos de controle de tráfego baseados em adaptação por meio de aprendizado por reforço possam criar um sistema totalmente descentralizado de controle de tráfego com baixo custo de operação, devido à baixa adaptabilidade em relação aos fluxos de tráfego com flutuações frequentes, tais métodos não são adequados para a construção de sistemas avançados de gerenciamento de tráfego, onde uma das expectativas é cooperar com sistemas avançados de informações ao motorista, no que tange o fornecimento de planejamento e orientação de rotas.

Por outro lado, embora existam métodos de baseados em adaptação por auto-organização, estes também apresentam dificuldades, quando se deseja utilizá-los na construção de sistemas avançados de controle de tráfego com intuito de cooperar com sistemas avançados de informações ao motorista, compartilhando agendas de intervalos de indicações de luzes verdes das sinalizações semafóricas, de modo que um mecanismo de planejamento de rotas possa utiliza-las para alocar espaços nas vias de entradas das interseções. Nestes métodos, o controle das sinalizações semafóricas é programado para identificar as chegadas de pelotões de veículos nas interseções. Sendo assim, as agendas de intervalos de indicações de luzes verdes das sinalizações semafóricas precisam ser atualizadas a cada detecção de um pelotão de veículos. Isto faz com que o aproveitamento da agenda de tempos das programações das sinalizações semafóricas por parte de um mecanismo de planejamento e orientação de rotas seja comprometido. Por causa de uma grande variação de estado de tais agendas, um mecanismo de planejamento e orientação de rotas pode calcular uma rota ótima com base em cópias de agendas de intervalos de indicações de luzes verdes desatualizadas. Como consequência, isto pode levar a um desbalanceamento na distribuição do volume de tráfego sobre as vias de uma rede viária. Por fim, as sinalizações semafóricas tendem a sofrer com o alto \textit{overhead} de comunicação e, de leitura e escrita, à medida que recebem e processam requisições de novos cálculos de rotas ótimas e, em seguida, registram os veículos nas agendas de intervalos de indicações de luzes verdes, geradas com intuito de alocar espaços nas vias de entrada em cada uma das interseções pertencentes as rotas ótimas calculadas.

Devido à alta mobilidade dos veículos e a topologia dinâmica das redes \textit{ad hoc} veiculares, é difícil fornecer serviços de sistemas inteligentes de transporte, usando uma rede veicular baseada em uma única tecnologia de acesso à comunicação sem fio, tal como a de rádios de comunicação dedicada de curta distância. Atualmente, as tecnologias de acesso à comunicação sem fio disponíveis para ambientes veiculares são aquelas baseadas em rádios de comunicação de curta distância (IEEE 802.11 \cite{IEEE80211:2012}, and IEEE 802.11p \cite{IEEE80211p:2010}) e aquelas baseadas em redes celulares (GSM, UMTS, and LTE) \cite{Zheng:2015}\cite{Sommer:2015}. No entanto, essas tecnologias têm suas limitações, quando são usadas em ambientes veiculares. No que diz respeito às tecnologias de acesso à comunicação sem fio baseadas em comunicações dedicadas de curta distância, essas foram projetadas para fornecer comunicações sem fio sem necessitar de uma infraestrutura pervasiva em ambientes como rodovias, estradas e ruas. Por outro lado, embora redes celulares possam fornecer uma ampla cobertura geográfica, ela não pode fornecer trocas de informações de tempo real em áreas locais de maneira eficiente. Por isso, integrar redes baseadas em diferentes tecnologias de acesso à comunicação sem fio é fundamental para o desenvolvimento de aplicações de serviços de sistemas inteligentes de transporte. De acordo com \citet{Zheng:2015}, redes veiculares heterogêneas podem ser uma excelente plataforma para satisfazer os requisitos de comunicação de aplicações de serviços de sistemas inteligentes de transporte.

Para esse fim, métodos de comunicação propostos para facilitar a comunicação entre nós em redes \textit{ad hoc} veiculares podem ser adotados em redes veiculares heterogêneas, pois eles são projetados para lidar com características relacionadas às redes \textit{ad hoc} veiculares. Tais características são as seguintes: conectividade intermitente, topologias altamente dinâmicas e mudanças constantes de densidade. Apesar disto, muitos destes métodos não adequados somente para comunicações infraestrutura-infraestrutura e inadequados para comunicações veículo-a-veículo e veículo-infraestrutura, pois eles são projetados para redes centradas em IP. 

Nestas redes, nós origem precisam conhecer os endereços dos nós destino, a fim de estabelecer comunicação fim-a-fim e descobrir rotas. Em ambientes veiculares altamente dinâmicos, os nós produzem um alto custo de mensagens, a fim de encontrar rotas e atualizar suas tabelas de roteamento. Além disto, a comunicação entre nós é intermitente, pois a topologia de rede é altamente dinâmica. Por estas razões, \citet{Bai:2010} e \citet{Amadeo:2016} têm argumentado por uma troca de paradigma em redes veiculares.

Neste sentido, alguns pesquisadores têm identificado as redes centradas em informação (ICNs - \textit{Information-Centric Networks}) \cite{Ahlgren:2012} como um paradigma chave, pois elas oferecem uma solução atrativa para ambientes móveis e altamente dinâmicos tais como as VANETs. Entre os modelos arquiteturais encontrados na literatura de ICN, as redes centradas em conteúdos (CCNs - \textit{Content-Centric Networks}) têm ganhado proeminência em trabalhos sobre redes veiculares \cite{Arnould:2011,Amadeo:2012a,Amadeo:2013,Wang:2012a,Wang:2012b}. Embora as CCNs sejam mais promissoras que os modelos centrados em IP em ambientes veiculares, existem algumas limitações quanto a adoção destas em projetos de VANETs. Por exemplo, a inundação da rede com pacotes de interesses cuja causa provém das políticas de encaminhamento de pacotes de interesses. Nestas, tais pacotes são encaminhados para todos os vizinhos de um nó, à medida que este os recebe. Isto possibilita o surgimento de broadcast storms. Além disto, o modelo da CCN usa estruturas de dados semelhantes às tabelas de roteamento e adota algoritmos similares ao AODV \cite{Perkins:1994}, DSR \cite{Jhonson:1996}  e GPSR \cite{Karp:2000}. Tais algoritmos são vulneráveis em ambientes veiculares altamente dinâmicos devido a intermitência de caminho \cite{Yu:2013}. Por fim, embora existam estudos promissores no campo de redes veiculares, estes têm somente focado em cenários relacionados a serviços populares de dados compartilháveis \cite{Yu:2013}. Consequentemente, cenários em que aplicações precisam trocar um grande montante de dados sensíveis a atraso não têm sidos estudados. Aplicações com estas características fazem uso de serviços de dados sem cache \cite{Yu:2013}, tais como: controle de sinalizações semafóricas por meio da cooperação de veículos conectados, controle cooperativo e adaptativo de cruzeiro, entre outros.

\section{Objetivos}

O objetivo principal desta tese é tratar os problemas de controle de tráfego e, planejamento e orientação de rotas em sistemas inteligentes de transporte. Para alcançar este objetivo, esta tese propõe um controle distribuído de tráfego baseado em veículos conectados, uma vez que sistemas inteligentes de transporte têm se tornado cada vez mais dependentes de ambientes de redes veiculares. Esta proposta, por sua vez, define os seguintes objetivos específicos:

\begin{enumerate}
	\item Projetar um novo modelo de rede centrada em informação, que leve em consideração os requisitos de comunicação de aplicações de serviços de sistemas inteligentes de transporte;

	\item Criar um protocolo de comunicação para ambientes de redes veiculares heterogêneas, a fim de 
satisfazer os requisitos de comunicação de aplicações de serviços de sistemas inteligentes de transporte, que são: baixa latência de comunicação entre nós, altas taxas de entrega de dados, escopos de comunicação bem definidos, escalabilidade e baixo custo de comunicação de estruturas de grupos \cite{Willke:2009}\cite{Zheng:2015}.  

    \item Controlar interseções de uma rede viária por meio de sinalizações semafóricas, utilizando uma estratégia de controle distribuído, que faz uso de cooperações entre veículos conectados e sinalizações semafóricas em um ambiente veicular, a fim de detectar flutuações de tráfego nas vias de entrada das interseções e, com base nestas flutuações, realizar os ajustes dos tamanhos dos intervalos de luzes de maneira ótima;

    \item Controlar interseções compartilhadas entre sistemas coordenados de sinalizações semafóricas, utilizando uma estratégia de controle distribuído baseada na cooperação entre as sinalizações semafóricas pertencentes à infraestrutura de controle de controle de tráfego de uma rede viária, a fim de fornecer acesso ininterrupto a pelotões de veículos em interseções participantes de corredores, onde as sinalizações semafóricas têm seus intervalos de luzes verdes sincronizados uns com os outros;

    \item Utilizar as configurações de controle de interseções, que são produzidas pelas estratégias de controle distribuído, a fim de gerar agendas de intervalos de luzes verdes, que forneçam dados sobre os períodos em que as vias de entrada das interseções recebem luzes verdes ao longo do tempo;

    \item Compartilhar as configurações de controle de interseções entre as sinalizações semafóricas, de modo que estas possam ser utilizadas para gerar agendas de intervalos de luzes verdes localmente, a fim de que cada sinalização semafórica do sistema de controle de tráfego conheça os períodos em que as vias de entrada das interseções de uma rede viária recebem luzes verdes ao longo do tempo;

    \item Construir uma estratégia de planejamento e orientação de rotas que tire proveito dos períodos em que as vias de entrada das interseções de uma rede viária recebem luzes verdes, a fim de calcular rotas ótimas e, com base nestas, alocar espaços nas vias, a fim de distribuir uniformemente o volume de tráfego sobre as vias de uma rede viária;
    
\end{enumerate}

\section{Principais Contribuições}

Esta tese contribui diretamente em três áreas distintas de pesquisa, a saber: redes veiculares, controle de tráfego e, planejamento e orientação de rotas.

As contribuições na área de redes veiculares são as seguintes:

\begin{itemize}
   	\item Proposta de uma nova ICN cuja troca de mensagens é baseada somente em interesses definidos por aplicações para VANETs. Esta proposta recebeu o nome de RAdNet-VE \cite{Goncalves:2016b}, que é uma extensão da RAdNet \cite{Dutra:2012} para ambientes veiculares;
     \item Demonstração da viabilidade da RAdNet-VE como uma VANET, utilizando simulações de ambientes de redes veiculares com nós equipados com interfaces de acesso à comunicação baseadas nos padrões IEEE 802.11n e IEEE 802.11p;
     \item Proposta de uma nova ICN cuja troca de mensagens é baseada somente em interesses definidos por aplicações de serviços de sistemas inteligentes de transporte, operando em ambientes de redes veiculares heterogêneas. Esta proposta recebeu o nome de HRAdNet-VE, que é uma extensão da RAdNet-VE  para ambientes veiculares heterogêneas;
     \item Demonstração da viabilidade da HRAdNet-VE como uma rede veicular heterogênea, utilizando simulações de ambientes de redes veiculares com nós equipados com interfaces de acesso à comunicação baseadas nos padrões IEEE 802.11n, IEEE 802.11p e LTE;
\end{itemize}

As contribuições na área de controle de tráfego são as seguintes:
  
\begin{itemize}
 	 \item Extensões das estratégias de controle distribuído de tráfego propostas por \citet{Paiva:2012}.
     \item Propostas de estratégias de controle de tráfego distribuído tolerante às ausências de funcionamento de sinalizações semafóricas. Tal mecanismo faz uso de veículos conectados para controlar interseções isoladas e cooperar em operações de sistemas coordenados de sinalizações semafóricas;
     \item Avaliações experimentais das estratégias de controle de tráfego distribuído tolerante às ausências de funcionamento de sinalizações semafóricas, operando sobre um ambiente de rede veicular heterogênea;
 \end{itemize}

Por fim, as contribuições na área de planejamento e orientação são as seguintes:
    
 \begin{itemize}
 	\item Extensões da estratégia de planejamento e orientação de rotas orientados à ondas verdes proposta por \citet{Faria:2013};
 	\item Proposta de um mecanismo de geração e compartilhamento de agendas de intervalos de indicações de luzes verdes, utilizando dados relativos ao estado do controle de interseções, sendo estas isoladas ou participantes de sistemas coordenados de sinalizações semafóricas;
     \item Proposta de um mecanismo de alocação de espaços em vias baseado nos tamanhos de veículos e comprimento de vias, de modo que este possa ser utilizado para prever e evitar congestionamentos, à medida que o volume de tráfego de uma rede viária é distribuído ao longo do tempo.
 \end{itemize}

\section{Organização do Trabalho}

Os estudos, propostas e análises realizados durante este trabalho foram organizados como segue. O Capítulo 2 apresenta o referencial teórico. O Capítulo 3 apresenta duas propostas de redes veiculares centradas em interesses. O Capítulo 4 apresenta as definições dos agentes utilizados nos sistemas multiagentes propostos nesta tese. O Capítulo 5 apresenta uma proposta de sistema multiagente para controle inteligente de tráfego baseado em sinalizações semafóricas inteligentes e veículos conectados. O Capítulo 6 apresenta a proposta de um sistema multiagente de planejamento e orientação inteligentes de rota baseado nos interesses de motoristas. O Capítulo 7 apresenta as avaliações experimentais das propostas apresentadas nos Capítulos 3, 5 e 6. Por fim, o Capítulo 7 apresenta as conclusões e trabalhos futuros acerca das propostas apresentadas nesta tese.

  \chapter{Referencial Teórico}

Para a elaboração da proposta desta tese, foi necessário realizar um estudo aprofundado acerca das teorias e conceitos que pudessem fundamentar este trabalho. Para tanto, foi realizado um levantamento bibliográfico sobre redes \textit{ad hoc} veiculares, sistemas inteligentes de transporte e fundamentos de engenharia de tráfego. Para fins mais práticos, também foram realizados levantamentos bibliográficos a respeito dos seguintes assuntos: agentes e sistemas multiagentes, escalonamento por reversão múltipla de arestas e escalonamento em sistemas flexíveis de manufatura. 

A apresentação de cada um dos levantamentos bibliográficos para a composição deste capítulo é dada, de acordo com a disposição das seções que seguem abaixo.

\section{Redes \textit{Ad Hoc} Veiculares}

Esta seção tem como objetivo apresentar um levantamento bibliográfico acerca do referencial teórico em torno de redes \textit{ad hoc} veiculares. Portanto, este referencial teórico aborda os seguintes assuntos: definição e características de redes \textit{ad hoc} veiculares; modelos de comunicação; tecnologias de acesso à comunicação sem fio; taxonomia de aplicações para redes \textit{ad hoc} veiculares; requisitos de comunicação de aplicações para redes ad hoc veiculares; e técnicas de roteamento em redes ad hoc veiculares. Cada um destes assuntos é apresentado conforme as subseções a seguir.

\subsection{Definição e Características}

Redes \textit{ad hoc} veiculares são compostas de veículos equipados com dispositivos de comunicação sem fio, que fazem com que os veículos atuem como nós móveis, a fim de realizar algum tipo de computação, e roteadores cujo intuito é rotear e ou encaminhar mensagens de controle e dados ao longo das redes formadas espontaneamente pelos nós móveis. Similares às redes \textit{ad hoc} móveis, redes ad hoc veiculares são formadas por nós móveis que se comunicam por meio de dispositivos de comunicação sem fio de curto alcance, além de serem auto-organizáveis, autogerenciáveis e de largura de banda reduzida. Mesmo assim, redes \textit{ad hoc} veiculares possuem características que as diferem das redes ad hoc móveis \cite{Li:2007}, que são:

\begin{itemize}
  \item \textbf{Topologia altamente dinâmica}: devido à alta velocidade dos veículos, a topologia das redes \textit{ad hoc} veiculares está sempre mudando. 
  
  \item \textbf{Desconexões frequentes da rede}: devido à topologia de rede ser altamente dinâmica, o estado das conexões entre veículos em uma rede \textit{ad hoc} veicular pode mudar com grande frequência. 
  
  \item \textbf{Disponibilidade de energia e recursos}: uma característica comum dos nós em uma rede \textit{ad hoc} veicular é a disponibilidade de energia e recursos (processamento e armazenamento), pois os veículos possuem mecanismos que permitem a auto recarga de suas baterias e espaços para acomodar dispositivos capazes de armazenar e processar dados. 
  
  \item \textbf{Comunicação geográfica}: comparada às outras redes, que usam comunicação \textit{unicast} e \textit{multicast} e os nós são identificados por meio de um identificador único ou um identificador de grupo, as redes \textit{ad hoc} veiculares muitas vezes usam comunicação geográfica para endereçar áreas, onde pacotes de rede precisam ser encaminhados. Para tanto, redes \textit{ad hoc} veiculares usam comunicação \textit{geocast}, que é basicamente um \textit{multicast} baseado em localização. O objetivo da comunicação \textit{geocast} é entregar um pacote originado a partir de um nó para todos os outros nós em uma região geográfica específica.
  
  \item \textbf{Modelagem de mobilidade e predição}: devido à alta mobilidade dos nós e o dinamismo da topologia, o uso de modelos de mobilidade e de predição contribui significantemente para o projeto de protocolos de rede \textit{ad hoc} veiculares, pois é possível prever a posição futura de um nó. A mobilidade dos veículos é limitada pelos mapas viários de rodovias, estradas e ruas, e também pela velocidade com que os mesmos percorrem estas vias.
  
  \item \textbf{Vários ambientes de comunicação}: redes ad hoc veiculares operam em dois ambientes típicos de comunicação, que são rodovias ou estradas, e ruas. Em cenários de tráfego em rodovias ou estradas, o ambiente é relativamente simples e constante, diferente dos cenários de tráfego em ruas nas cidades. As ruas em uma cidade são muitas vezes separadas por obstáculos e, por isto, nem sempre é possível estabelecer ou manter a comunicação entre os nós.
  
  \item \textbf{Restrições severas na latência da rede}: em algumas aplicações de redes \textit{ad hoc} veiculares, a rede não necessita de altas taxas de transferências. No entanto, essas aplicações exigem que as redes \textit{ad hoc} veiculares tenham uma baixa latência, pois elas atuam em cenários onde uma informação deve ser propagada dentro de um curto espaço de tempo. 
  
  \item \textbf{Interações com sensores on-board}: os nós são equipados com sensores \textit{onboard}, a fim de fornecer informações  que possam ser utilizadas para estabelecer a comunicação entre nós e também para roteamento e ou encaminhamento de mensagens. Uma vez que dispositivos GPS (\textit{Global Positioning System})  estão cada vez mais presentes em veículos, eles podem ser utilizados para fornecer informações de localização, de modo que estas possam ser utilizadas para comunicação \textit{geocast}.

\end{itemize}

Após a apresentação da definição e características das redes \textit{ad hoc} veiculares, segue, na próxima subseção, uma descrição acerca dos modelos de comunicação encontrados em ambientes de redes \textit{ad hoc} veiculares.

\subsection{Modelos de Comunicação}

\begin{figure}[t]
	\centering
    \includegraphics[scale=0.6]{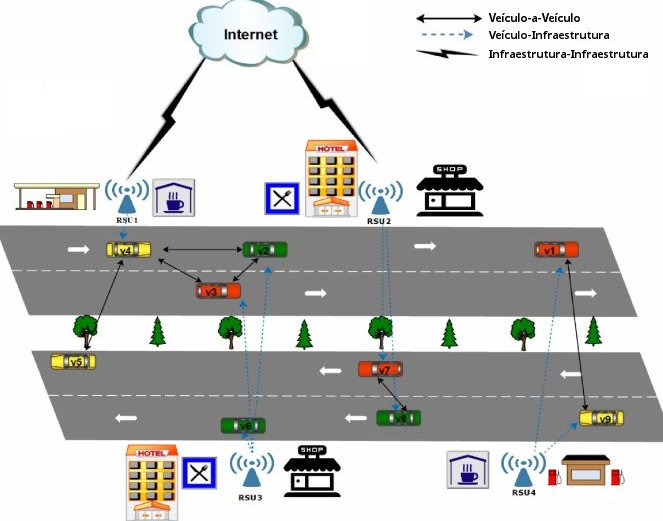}
    \caption{Modelos de comunicação em redes \textit{ad hoc} veiculares.}
    \label{fig:modelocomunicao}
    
\end{figure}

Segundo \citet{Aslam:2012}, as redes \textit{ad hoc} veiculares baseiam-se em três modelos de comunicação: comunicação veículo-a-veículo, comunicação veículo-infraestrutura e comunicação infraestrutura-infraestrutura. Veja Figura \ref{fig:modelocomunicao}.

\begin{figure}[t]

\center
\subfigure[]{
	\includegraphics[width=6.5cm]{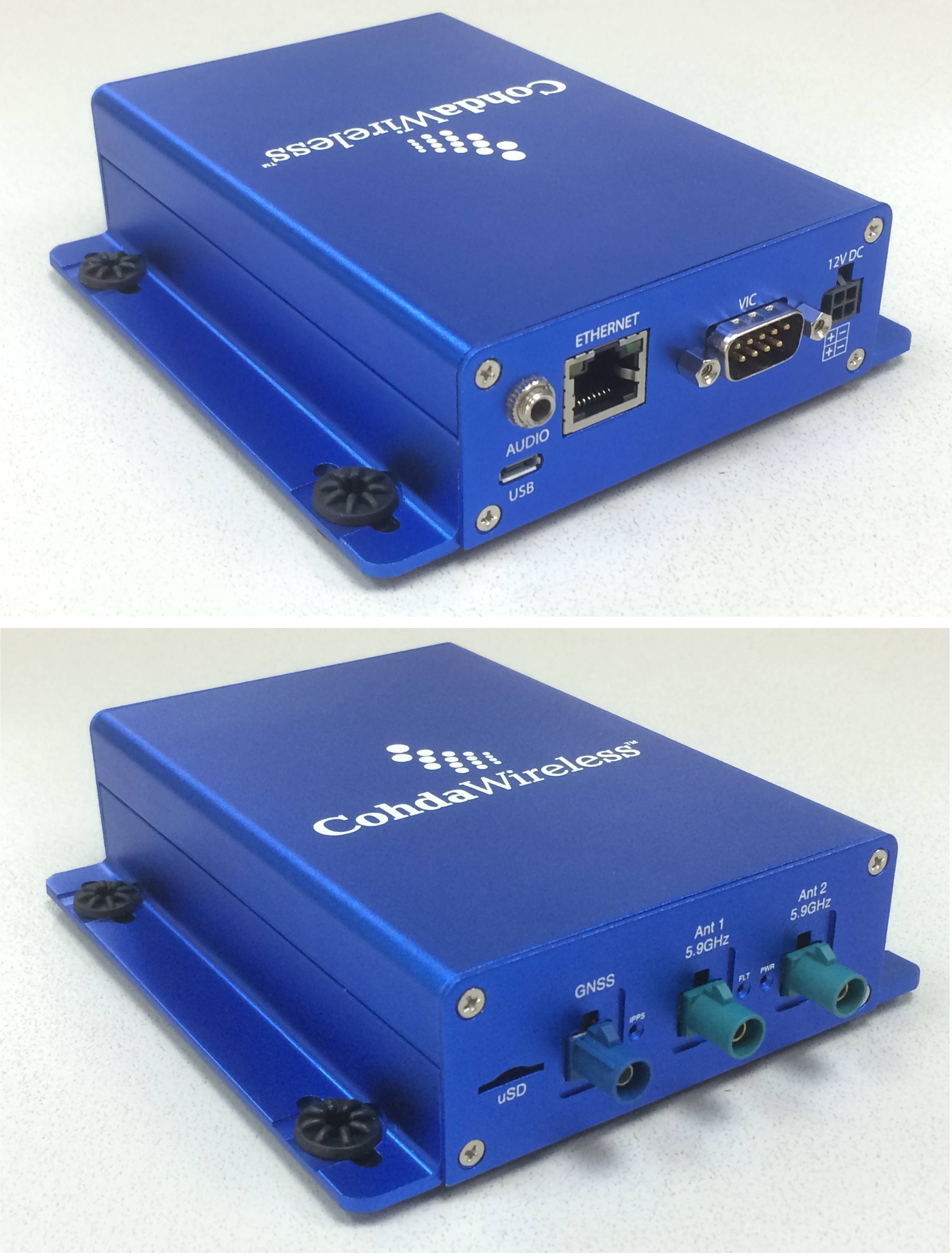}
    \label{fig:obu}
}
\qquad
\subfigure[]{
	\includegraphics[width=6.5cm]{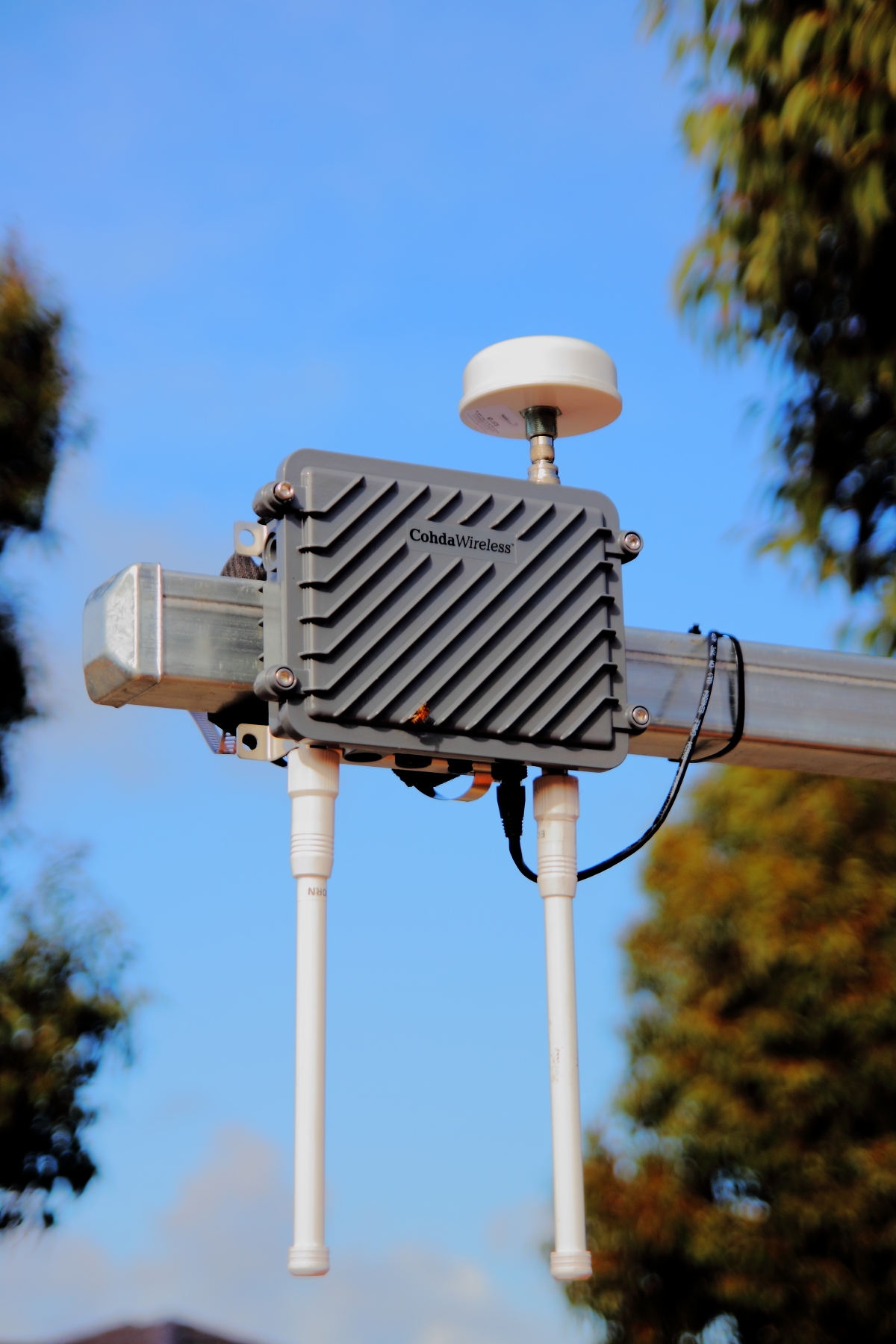}
    \label{fig:rsu}
}
\caption{Exemplos de unidades de bordo e unidades de acostamento: (a) Unidade de Bordo; (b) Unidade de Acostamento. Fonte: \citet{Cohda:2016}}

\end{figure}

No modelo de comunicação veículo-a-veículo, \textit{broadcast/multicast multi-hop} são utilizados para transmitir dados sobre um grupo de receptores. Para que isto ocorra, veículos devem ser equipados com algum tipo de interface de comunicação sem fio ou unidade on-board (\textit{On-Board Unit} - OBU) capazes de formar uma rede \textit{ad hoc} veicular. Além disto, os veículos também devem ser equipados com dispositivos que permitam obter dados detalhados de posição dos mesmos, tais como receptores GPS. A Figura \ref{fig:obu} mostra um exemplo de unidade \textit{on-board}. Em comunicação veículo-a-veículo, existem dois tipos de encaminhamento de mensagens, são eles: \textit{broadcasting} ingênuo e \textit{broadcasting} inteligente. Em \textit{broadcasting} ingênuo, veículos enviam mensagens de \textit{broadcast} periodicamente e em intervalos regulares. As limitações do método de \textit{broadcasting} ingênuo é  que um grande número de mensagens de \textit{broadcast} é  gerado, aumentando o risco de colisão de mensagens, resultando, então, em baixas taxas de entrega de mensagens, além do aumento na latência de comunicação. O \textit{broadcasting} inteligente trata os problemas inerentes ao \textit{broadcasting} ingênuo, limitando o número de mensagens de \textit{broadcast}. Por exemplo, quando um veículo que detectou um evento em uma via recebe uma mensagem de pelo menos um dos veículos atrás dele, o mesmo assume que pelo menos um dos veículos de trás receberam uma mensagem de \textit{broadcast}. Isto faz com o veículo em questão cesse o envio de mensagens de \textit{broadcast}, pois, em \textit{broadcasting} inteligente, assume-se que o veículo de trás tem a responsabilidade de encaminhar a mensagem de \textit{broadcast} para o restante dos veículos. Se um veículo recebe a mesma mensagem de mais de uma fonte, o mesmo repassa somente a primeira mensagem recebida.

No modelo de comunicação veículo-infraestrutura, a comunicação é feita a partir do envio de mensagens de \textit{broadcast} para nós que estão a um salto de distância. Dessa forma, unidades de acostamento (\textit{Roadside Unities} - RSUs) enviam mensagens de \textit{broadcast} para veículos equipados com dispositivos de comunicação sem fio. Para tanto, unidades de acostamento podem ser posicionadas a cada quilômetro ou menos, fazendo com que altas taxas de transferência de dados sejam mantidas em condições de tráfego pesado de veículos. A Figura \ref{fig:rsu} mostra um exemplo de unidade de acostamento. Dependendo da aplicação, a comunicação veículo-infraestrutura pode ser disponibilizada de duas maneiras, que são: comunicação veículo-infraestrutura esparsa e comunicação veículo-infraestrutura ubíqua \cite{Sichitiu:2008}. Em comunicação veículo-infraestrutura esparsa, é possível fornecer serviços de comunicação por meio de \textit{hot spots}. Dessa forma, é possível disponibilizar uma infraestrutura de comunicação de maneira gradual, fazendo com que não seja necessário um investimento significativo, antes que um benefício concreto seja obtido. A comunicação veículo-infraestrutura ubíqua tem como objetivo, fornecer serviços de comunicação por toda uma rede viária. No entanto, isto pode exigir grandes investimentos para fornecer a cobertura total da rede viária.

No modelo de comunicação infraestrutura-infraestrutura, as unidades de acostamento podem ser conectadas às unidades de acostamento localizadas em suas adjacências. A conexão entre elas pode ser por meio de cabos ou sem fio. Quando duas unidades de acostamento estão conectadas uma à outra, é dito que a comunicação entre elas é direta \cite{Aslam:2012}. No entanto, ao longo de uma rodovia, podem existir unidades de acostamento que não possuem qualquer conexão com unidades de acostamento adjacentes. Logo, para que estas unidades possam se comunicar umas com as outras, elas usam os veículos que estejam passando por elas como meio de estabelecer comunicações indiretas umas com as outras \cite{Aslam:2012}.

\subsection{Tecnologias de Acesso à Comunicação Sem Fio}

Em um ambiente de rede veicular, a escolha da tecnologia de acesso à comunicação sem fio tem um papel preponderante no que diz respeito à satisfação dos requisitos de comunicação das aplicações para redes \textit{ad hoc} veiculares, pois a tecnologia influencia diretamente no modo como os nós devem se comunicar. Logo, tecnologias de acesso à comunicação sem fio baseadas em padrões como IEEE 802.11 \cite{IEEE80211:2012}, IEEE 802.11p \cite{IEEE80211p:2010}, IEEE 802.16e \cite{IEEE80216:2012} e 3GPP (\textit{Third Generation Partnership Project}) formam a base de qualquer pilha de comunicação em redes que operam em ambientes veiculares.

O padrão IEEE 802.11 é um padrão de comunicação que define as operações de redes locais sem fio. No início da pesquisa de rádios de curto alcance para comunicação interveicular, este padrão foi amplamente utilizado no desenvolvimento de protótipos de redes \textit{ad hoc} veiculares, devido à grande disponibilidade de dispositivos de baixo custo baseados no padrão IEEE 802.11a no mercado \cite{Sommer:2015}. As redes formadas por dispositivos baseados no padrão IEEE 802.11 conseguem fornecer altas taxas de transferência de dados dentro de um alcance relativamente curto \cite{Hossain:2010}. No entanto, em cenários onde veículos trafegam em alta velocidade, o curto alcance de comunicação desses dispositivos faz com que aconteçam desconexões constantes. Para evitar isto, muitos pontos de acesso precisam ser disponibilizados ao longo de rodovias, estradas e ruas. Porém, isto incorre em altos custos de implantação. Atualmente, o padrão IEEE 802.11 tem sido utilizado por meio da disponibilização de pontos de acesso em interseções de vias de redes viárias com intuito de fornecer acesso à Internet, capturar dados de tráfego ou obter dados que possam auxiliar no roteamento de mensagens ao longo de uma rede \textit{ad hoc} formada pelos veículos nas vias.

O padrão IEEE 802.11p é um padrão novo de comunicação sem fio, sendo ele pertencente à família de padrões IEEE 802.11. Este padrão tem como objetivo fornecer acesso à comunicação sem fio em ambientes exclusivamente veiculares. O IEEE 802.11p foi criado a partir de melhorias feitas no padrão IEEE 802.11a, de modo que este último pudesse ser utilizado em aplicações de sistemas inteligentes de transporte. De acordo com \citet{Menouar:2006}, tais melhorias advêm dos conceitos de segurança veicular ativa e dos requisitos de comunicação das categorias de aplicações para redes \textit{ad hoc} veiculares. Com dispositivos de comunicação sem fio baseados em IEEE 802.11p, nós de rede são capazes de transmitir dados com taxas de transferência variando entre 6 a 27 Mbps, desde que os mesmos estejam a uma distância máxima de 1000m uns dos outros. Atualmente, dispositivos de comunicação baseados no padrão em discussão já podem ser encontrados no mercado, tais como os fornecidos pela Qualcomm e Codah Wireless. Isto tem possibilitado o desenvolvimento de uma gama de aplicações de sistemas inteligentes de transporte \cite{Hossain:2010}.

O padrão IEEE 802.11e é o padrão de acesso à comunicação sem fio de banda larga móvel do WiMax e permite a convergência de redes de banda larga móveis e fixas. Este padrão tem como características: altas taxas de transferência de dados e inclusão de técnicas MIMO (Multiple Input Multiple Output) com antenas, usando tecnologia de diversidade espacial em conjunto com esquemas de subcanalização, codificação e modulação, que permitem alcançar taxas de transferência de dados entre 28 e 63 Mbps por setor. A mobilidade do IEEE 802.16e suporta velocidades superiores a 160 km/h, além de fornecer esquemas otimizados de \textit{handover} com latências menores que 50ms, a fim de satisfazer requisitos de tempo real. Além disto, o padrão adota diferentes possibilidades de faixas de frequência entre 1,25 a 20 MHz. Isto permite que o padrão se adapte às diferentes realidades mundiais de alocação de frequências.

O padrão 3GPP é uma família de padrões relacionados às tecnologias de acesso à comunicação sem fio para redes celulares \cite{Sommer:2015}, tais como: GSM (\textit{Global System for Mobile Communication}), UMTS (\textit{Universal Mobile Telecommunication System}) e LTE (\textit{Long Term Evolution}). Tradicionalmente, redes celulares baseadas no padrão GSM são chamadas de redes celulares de segunda geração (2G), pois elas sucederam os sistemas celulares analógicos. Uma vez que o padrão UMTS substituiu o GSM, redes celulares baseadas no padrão UMTS são chamadas de redes celulares de terceira geração (3G). Por fim, redes celulares baseadas no padrão LTE têm sido chamadas de redes celulares de quarta ou quinta geração (4/5G). Embora cada um dos padrões para redes celulares represente uma evolução tecnológica, o conceito principal em torno das redes celulares é o mesmo. Sendo assim, uma ampla área de geográfica é coberta por uma rede de estações base, onde cada uma serve uma parte desta área, que é uma célula. Nesta rede, um dispositivo se conecta a uma estação base que serve a célula em que ele se encontra. O dispositivo sempre muda para a estação base mais apropriada, à medida que ele se move em direção aos limites de uma célula. Devido as suas características, o padrão LTE tem sido visto como o padrão tecnológico proeminente, no que tange o fornecimento tanto de comunicação veículo-infraestrutura quanto comunicação infraestrutura-infraestrutura, quando se trata da disponibilização de redes veiculares, pois redes celulares LTE são capazes de transferir dados com taxas de transferência de dados entre 50 e 100 Mbps \cite{Zheng:2015}. Além disto, tal rede é capaz de suportar nós que se movem até 350 km/h \cite{Zheng:2015}. Por fim, as redes LTE são capazes de fornecer alta capacidade com ampla cobertura. Neste caso, o padrão LTE pode suportar até 1200 veículos por célula em ambientes rurais com um delay abaixo de 55ms.

\subsection{Taxonomia de Aplicações para Redes \textit{Ad Hoc} Veiculares}

Aplicações de redes ad hoc veiculares tentam resolver diversos problemas cujos requisitos de comunicação são muito específicos. Por esta razão, \citet{Willke:2009} classificaram as aplicações de redes ad hoc veiculares em quatro categorias, a saber:

\begin{itemize}
	\item \textbf{Serviços de informações gerais:} aplicações desta categoria exigem pouco custo de comunicação e uma alta taxa de entrega de informações. Além disto, nós equipados com dispositivos GPS e interfaces de comunicação em redes sem fio usam um banco de dados local para registrar encontros com outros nós ao longo das vias, onde estes trafegam. Com isto, é possível que veículos submetam consultas aos bancos de dados dos nós, a fim de obter informações a respeito de serviços oferecidos ao longo das vias, condições das vias e volume de tráfego. Além disto, os nós podem ser utilizados para disseminar informações para os diversos nós da rede, tais como localização e, metadados de serviços de notícias, publicidade e entretenimento.
	\item \textbf{Serviços de informações de segurança:} nesta categoria, as aplicações são propensas a ser sensíveis à latência do que à capacidade de transferência de dados da rede \textit{ad hoc} veicular, pois mensagens de aviso de emergência ou acidente devem ser disseminadas em um curto espaço de tempo. Sendo assim, por meio destas mensagens é possível identificar a localização e o estado de movimento de veículos com comportamento anormal, causado por um acidente, avaria mecânica ou outro tipo de falha. Para tanto, veículos devem ser equipados com GPS e dispositivos de comunicação sem fio de curto alcance (DSRC). Com isto, mensagens de aviso podem ser transmitidas por toda uma região e, dessa forma, notificar os nós a respeito de acidentes, avarias, anomalias de tráfego e condições de superfície das vias. 
	\item \textbf{Controle de movimento individual:} aplicações de controle de movimento individual utilizam comunicação veículo-a-veículo para trocar dados a respeito da posição, velocidade, aceleração, comportamento e estado dos atuadores dos veículos (aceleradores e freios). Nestas aplicações, a latência da rede \textit{ad hoc} veicular deve ser baixa, pois os dados mencionados anteriormente podem ser utilizados para controlar a aceleração e a frenagem de veículos ou, ainda, gerar avisos de colisão ou propor alternativas para prevenção de acidentes.
	\item \textbf{Controle de movimento de grupo:} nesta categoria, as aplicações estão envolvidas na coordenação de movimento de grupo entre veículos. Para tanto, os veículos se comunicam com sua vizinhança de veículos e usam técnicas de controle distribuído, a fim de se deslocarem, com pequenas proximidades uns dos outros, sem colidir. Para regular o movimento dos grupos de veículos, podem ser utilizados modelos de regulação de movimentos. Dentre estes, se destacam o planejamento de grupo com regulação individual, regulação de movimento baseada em líder e regulação de movimento baseada em líder virtual. Em planejamento de grupo com regulação individual, comunicação veículo-a-veículo pode ser usada para otimizar os planos de trajetória de veículos, interseções de vias e outros recursos. Em regulação de movimento baseada em líder, um veículo transmite a referência de movimento e comandos para o grupo. Cada veículo combina esta informação do líder com dados de movimentos de outros veículos próximos a ele, a fim de determinar seu próprio curso de ação. Por fim, em regulação de movimento baseada em líder virtual, para coordenar o movimento de veículos, diretivas comuns devem ser transmitidas entre os veículos. Estas diretivas podem ser definidas a partir de um processo de consenso distribuído. Aplicações de líder virtual podem usar comunicação veículo-a-veículo para realizar manobras de grupo relativamente complexas e manter uma distância segura entre os veículos. 

\end{itemize}

Com base na taxonomia de aplicações de redes \textit{ad hoc} veiculares apresentada acima, é possível identificar que cada categoria de aplicação estabelece seus próprios requisitos de comunicação. Neste sentido, a próxima subseção apresenta os requisitos de comunicação para cada uma das categorias de aplicação para redes \textit{ad hoc} veiculares. Além disto, a próxima subseção também apresenta mecanismos e abordagens que podem ajudar no projeto de um protocolo de comunicação para redes \textit{ad hoc} veiculares, de modo que este venha a satisfazer tais requisitos. 

\subsection{Requisitos de Comunicação de Aplicações para Redes \textit{Ad Hoc} Veiculares} 

Segundo \citet{Willke:2009}, tais requisitos são: latência de entrega de mensagens, confiabilidade de entrega de mensagens, escala,  escopo de comunicação e estrutura de grupo de comunicação. Por isto, é necessário conhecer a importância de cada um destes requisitos. 

A latência e a confiabilidade de entrega de mensagens são medidas críticas de desempenho. A escala é importante, uma vez que a comunicação pode acontecer de um para muitos ou de muitos para muitos, podendo envolver um grande número de nós. O escopo de comunicação tem um efeito significativo sobre como as mensagens são roteadas ou encaminhada e, de como a rede é organizada, podendo, assim, influenciar diretamente na escalabilidade da aplicação. Por fim, estrutura de grupo se refere à capacidade de os nós poderem estabelecer relacionamentos persistentes ou de simplesmente poderem se comunicar com outros nós. 
	
Com isso, é possível salientar que aplicações pertencentes à categoria de serviços de informações gerais podem tolerar atrasos na entrega de mensagens e ainda operar corretamente. Além disto, estas aplicações também podem tolerar falhas de comunicação intermitente, tais como perda de respostas a uma consulta e quadros de mídia. Outro ponto importante relacionado às aplicações de serviços de informações gerais é a escala, pois este tipo de aplicação necessita transmitir mensagens em áreas grandes. Isto faz com que o escopo de comunicação também seja grande. Por não lidar com controle de movimento de veículos, aplicações de serviços de informações gerais não mantêm estruturas de grupo entre veículos. 
	
Embora tenham características semelhantes às de serviços de informações gerais, aplicações de serviços de informações de segurança têm como requisito fundamental duras restrições de tempo real e podem apresentar falhas, estas oriundas de atrasos infrequentes de entrega de mensagens. Para tanto, estas aplicações demandam de tempos de processamento e latência menores que 40 ms e frequência de repetição de envio de mensagens seja de 50 Hz. Além disto, é necessário que as aplicações garantam altas taxas de entrega de informação. No que diz respeito à escala, escopo de comunicação e estrutura em grupo, aplicações de serviços e informações de segurança são semelhantes às aplicações de serviços de informações gerais. 
	
Como as aplicações de serviços de informações de segurança, aplicações de controle de movimento individual têm como requisito fundamental duras restrições de tempo real. Apesar disto, ainda pode haver falhas, por causa de atrasos infrequentes de entrega de mensagens. Além disto, estas aplicações fazem uso de dados de vizinhanças de veículos para garantir a segurança de condutores e manter uma distância otimizada entre veículos. Portanto, em casos de atraso de entrega de mensagens, as aplicações passam a operar às cegas, no que diz respeito aos possíveis perigos ao longo das vias. Diferentemente de aplicações de serviços de informações gerais e de serviços de informações de segurança, aplicações de controle de movimento individual operam em redes ad hoc veiculares de média escala, a fim de fornecer comunicação entre veículos que estejam próximos uns dos outros. Por isto, o escopo de comunicação é menor, quando comparado ao escopo de comunicação dos dois tipos de aplicações cujos requisitos de comunicação foram discutidos imediatamente acima. No que tange a estrutura de grupos, aplicações de controle de movimento individual não envolvem grupos persistentes e, por isto, os relacionamentos entre veículos são apenas transientes. 
	
Em aplicações de controle de movimento de grupo, o requisito de latência de entrega de mensagens varia de um modelo de regulação de movimento para outro. Neste caso, no modelo de planejamento de grupo com regulação individual, aplicações podem tolerar o atraso da entrega de mensagens, sem que haja falha, porque as restrições tempo real não são tão rígidas quantos as dos demais modelos. No que tange o requisito de confiabilidade de entrega de mensagens, as aplicações de movimento de grupo devem ter a capacidade de determinar se as mensagens enviadas para seus destinatários são efetivamente recebidas pelos veículos que devem recebê-las. Assim, os veículos podem realizar alguma ação, caso as entregas das mensagens não sejam confirmadas dentro de um tempo estipulado. A escala de aplicações de controle de movimento de grupo é na maioria das vezes média, semelhante às aplicações de controle de movimento individual. Sendo assim, o escopo de comunicação dessas aplicações é limitado às vizinhanças de veículos ou pequenas regiões. Diferentemente das outras categorias de aplicações de redes ad hoc veiculares, que têm estrutura de grupo transiente, aplicações de controle de movimento em grupo possuem estrutura de grupo persistente, pois nelas são envolvidos relacionamentos persistentes entre veículos específicos. Estes veículos, por sua vez, podem compartilhar missões ou destinos.
	
Para finalizar a apresentação do referencial teórico acerca de redes \textit{ad hoc} veiculares, a próxima seção apresentará as principais técnicas de roteamento adotadas em redes \textit{ad hoc} veiculares.  

\subsection{Técnicas de Roteamento em Redes \textit{Ad Hoc} Veiculares}

Devido à natureza dinâmica dos nós de uma rede \textit{ad hoc} veicular, encontrar e manter rotas é uma tarefa difícil. O problema de roteamento em redes \textit{ad hoc} veiculares tem sido amplamente estudado e, por isto, muitos protocolos de roteamento têm sido propostos. De acordo com [38], os protocolos de roteamento podem ser agrupados em cindo categorias: roteamento \textit{ad hoc}, roteamento baseado em localização geográfica, roteamento baseado em \textit{cluster}, roteamento \textit{broadcast} e roteamento \textit{geocast}.

Como mencionado anteriormente, redes \textit{ad hoc} veiculares e redes \textit{ad hoc} móveis compartilham os mesmos princípios, ou seja, elas não contam com uma infraestrutura fixa de comunicação e possuem muitas similaridades, tais como: auto-organização, autogerenciamento, largura de banda baixa e alcance curto de comunicação. Por este motivo, muitos protocolos de roteamento \textit{ad hoc} têm sido aplicados em redes \textit{ad hoc} veiculares, pois eles foram projetados para redes ad hoc móveis de propósitos gerais. Além disto, estes protocolos não mantêm rotas, a menos que seja necessário e, por isto, podem reduzir o overhead de mensagens de controle, especialmente em cenários com um número pequeno de nós. No entanto, isto não é uma realidade em ambientes veiculares e, por isto, protocolos de roteamento ad hoc sofrem com o dinamismo topológico das redes ad hoc veiculares, uma vez que este é causado pela alta mobilidade dos nós de rede. Por isto, em cenários de redes ad hoc veiculares, as técnicas de roteamento ad hoc são incapazes de encontrar rapidamente, manter e atualizar rotas longas em uma rede ad hoc veicular. Logo, o roteamento ad hoc pode apresentar grandes perdas de pacotes, devido às falhas de rotas, uma vez que é quase impossível finalizar um \textit{handshake} de três vias, quando uma conexão TCP (\textit{Transmission Control Protocol}) está sendo estabelecida. 

O movimento dos nós em redes \textit{ad hoc} veiculares é frequentemente restrito em movimentos bidirecionais, sendo estes limitados ao longo de rodovias, estradas e ruas. Por isto, estratégias de roteamento baseadas em dados de localização geográfica, obtidos por meio de mapas de ruas, modelos de tráfego ou até de sistemas de navegação a bordo dos veículos, fazem sentido. Embora os nós, em uma rede ad hoc veicular, possam usar dados de localização geográfica em decisões de roteamento, ainda existem alguns desafios que precisam ser superados. Em cenários de cidades, o encaminhamento de mensagens é frequentemente limitado, devido à dificuldade em manter conexões entre os nós. Tal dificuldade é causada por obstáculos presentes neste tipo de cenário, tais como prédios e árvores. Além disto, a construção de uma topologia para roteamento pode degradar o desempenho de roteamento, ou seja, as mensagens poderão percorrer longos caminhos com altos \textit{delays} de comunicação fim-a-fim. No que tange a mobilidade, esta pode induzir o roteamento a loops. Por fim, mensagens podem ser encaminhadas para direções erradas, causando \textit{delays} desnecessários de comunicação fim-a-fim ou, até mesmo, partições de rede.

Em roteamento baseado em \textit{clusters}, uma rede virtual é criada por meio de agrupamento de nós, a fim de fornecer escalabilidade na comunicação. Com base nisto, cada cluster possui um líder, que é responsável pela coordenação intra e inter \textit{cluster}. Os nós dentro de um \textit{cluster} se comunicam diretamente uns com os outros. A criação de uma infraestrutura de rede virtual é crucial para a escalabilidade de protocolos de acesso ao meio, protocolos de roteamento e infraestrutura de segurança. O agrupamento estável dos nós é a chave para criar esta infraestrutura. Embora existam propostas de protocolos de roteamento baseados em \textit{clusters} para redes ad hoc móveis, estes não são adequados, no que diz respeito às redes \textit{ad hoc} veiculares, devido ao comportamento dos motoristas, restrições na mobilidade e às altas velocidades dos nós. Por isto, as técnicas de redes \textit{ad hoc} móveis para formação de clusters tornam-se instáveis, quando aplicadas em redes \textit{ad hoc} veiculares. 

O roteamento broadcast é frequentemente utilizado em aplicações de redes \textit{ad hoc} veiculares cujo objetivo é compartilhar dados entre os veículos. Tais dados podem estar relacionados ao tráfego, condições climáticas, emergências, condições das vias, entre outros. O roteamento \textit{broadcast} também é usado em protocolos de roteamento unicast, de modo que uma rota eficiente até um destino seja encontrada. A maneira mais simples de implementar um roteamento \textit{broadcast} é o \textit{flooding}. Nesta estratégia, cada nó, ao receber mensagens, as retransmitem para todos os nós vizinhos, a menos que não as tenha recebido. O \textit{flooding} funciona bem para um número limitado de nós. No entanto, quando o número de nós aumenta, o desempenho diminui drasticamente. Além disto, a demanda por largura de banda para a transmissão de uma mensagem pode aumentar exponencialmente. Como cada nó recebe e transmite a mesma mensagem ao mesmo tempo, isto causa contenções, colisões de pacotes, \textit{broadcast storms} e alto consumo da largura de banda.

Por fim, o roteamento \textit{geocast} é basicamente um roteamento \textit{multicast} baseado em localização. O objetivo deste tipo de roteamento é entregar mensagens de uma origem para todos os nós de uma região geográfica específica, que é chamada de zona de relevância. Devido a esta característica, muitas aplicações de rede ad hoc veiculares podem se beneficiar do roteamento geocast. Por exemplo, se um veículo detecta que se envolveu em um acidente por meio de sensores veiculares de acionamento de \textit{airbags}, ele pode informar instantaneamente o acidente aos veículos dentro da zona relevância. 

Após a apresentação do referencial teórico sobre redes \textit{ad hoc} veiculares, o texto desta tese avança para a próxima seção, cujo objetivo é apresentar um estudo teórico a respeito de sistemas de transporte e sistemas inteligentes de transporte.

\section{Sistemas Inteligentes de Transporte}

Esta seção tem como objetivo apresentar o referencial teórico a respeito de sistemas inteligentes de transporte. Este referencial teórico tem como base a terceira seção do Capítulo 1 do livro de \citet{Bazzan:2014}.

Sistemas inteligentes de transporte podem ser vistos como sistemas em que tecnologias de informação e comunicação são aplicadas em áreas relacionadas à rede de transporte (por exemplo, infraestrutura, veículos e usuários de vias, gerenciamento de tráfego e mobilidade e a integração dinâmica entre todos estes). Sistemas inteligentes de transporte também podem ser vistos como um termo geral para a aplicação integrada de tecnologias de comunicação, controle e processamento de informações no sistema de transporte. 

A informação está no centro de sistemas inteligentes de transporte. Por isto, muitas ferramentas de sistemas inteligentes de transporte são dedicadas à coleta, processamento, integração e fornecimento de informações. A informação não é importante somente para operadores e autoridades de transporte, mas também fornecedores de transporte público e comercial, e usuários de vias. O objetivo disto é permitir a tomada de decisões de maneira mais inteligente, a fim de tornar os sistemas de transporte mais seguros e eficientes. 

Desde a década de 90, vários elementos, que hoje são conhecidos como sistemas inteligentes de transporte, foram definidos com ênfase em vigilância de tráfego, controle, otimização, e simulação de sistemas de tráfego e transporte. O surgimento de sistemas inteligentes de transporte é frequentemente creditado aos avanços na computação e na comunicação. À medida que os custos dos dispositivos de computação vêm diminuindo desde a década de 90, tem sido possível embutir microprocessadores e mais inteligência em sistemas de transporte. Dessa forma, técnicas e assuntos relacionados à computação estão se tornando comuns em sistemas de transporte, tais como: computação ubíqua, internet das coisas e computação nas nuvens. Isto permite que indivíduos (motoristas, passageiros, entre outros) tenham acesso contínuo e ubíquo à informação, que é entregue tanto pelo setor público quanto pelos próprios usuários das vias (por exemplo, plataformas colaborativas como o Waze). Isto constitui uma clara mudança de paradigma, pois, a uma década atrás, o principal fornecedor de informações de tráfego e transporte era o setor público (autoridade de tráfego). Esta mudança no paradigma, por sua vez, significa que as autoridades de tráfego têm perdido cada vez mais o controle sobre a rede de transporte, porque os usuários das vias estão cada vez mais informados. Consequentemente, estes usuários podem se adaptar continuamente às mudanças das condições de tráfego, a fim de atingir os seus objetivos de mobilidade. 

Sistemas inteligentes de transporte envolvem cinco áreas ou sistemas, a saber: sistemas avançados de gerenciamento de tráfego, sistemas avançados de informações ao motorista, sistemas avançados de assistência à direção, sistemas avançados de transporte público, e operação de veículos comerciais. Os sistemas avançados de gerenciamento de tráfego visam as tecnologias de gerenciamento relacionadas a dispositivos de controle de tráfego, situações de emergências, monitoramento de emissões e comunicações entre várias partes do sistema, tais como dispositivos de monitoramento de tráfego, controladores de sinalização semafóricas e outros dispositivos relacionados à segurança. Os sistemas avançados de informações ao motorista têm como objetivo fornecer informações aos usuários das vias e a outros participantes do sistema de transporte em ambientes como os de rodovias, estradas e ruas. Tais informações são, em muitos casos, coletadas e processadas por sistemas avançados de gerenciamento de tráfego e, em seguida, são transmitidas para aplicações que fazem uso destas. Enquanto os sistemas avançados de gerenciamento de tráfego referem-se primariamente à infraestrutura, os sistemas avançados de informações ao motorista são direcionados aos usuários do sistema. Os sistemas avançados de assistência à direção têm como objetivo aplicar tecnologias avançadas em veículos e vias, a fim de reduzir acidentes e melhorar a segurança do tráfego. Estes sistemas incluem controle e aviso anti-colisão, assistência à direção, controle automático lateral ou longitudinal, entre outros. Os sistemas avançados de transporte público aplicam as tecnologias de sistemas avançados de gerenciamento de tráfego, sistemas avançados de informações ao motorista e sistemas avançados de assistência à direção no transporte público, a fim de melhorar a qualidade do serviço e aumentar a eficiência por meio de monitoramento automático de veículos e bilhetagem eletrônica. Por fim, operação de veículos comerciais funciona como sistemas avançados de transporte público, mas aplicado a operações de veículos comerciais (monitoramento automático de veículos, gerenciamento de frota, agendamento e pagamento eletrônicos).

Com o fim da apresentação do referencial teórico acerca de sistemas de transporte e sistemas inteligentes de transporte, a apresentação do referencial teórico desta tese avança para um outro assunto, que são os fundamentos de engenharia de tráfego. Portanto, a próxima seção realiza a apresentação destes.

\section{Fundamentos de Engenharia de Tráfego}

Esta seção tem como objetivo apresentar o levantamento bibliográfico do referencial teórico acerca da estrutura de rede de tráfego, assim como, o referencial teórico relacionado aos parâmetros microscópicos e macroscópicos de tráfego, diagrama fundamental do fluxo de tráfego e controle de tráfego por meio de sinalização semafórica. Com relação à seção sobre estrutura de rede de tráfego, o texto tem como base primeira seção do sétimo dois do livro de \citet{Bazzan:2014}. No que tange o referencial teórico de parâmetros microscópico e macroscópico de tráfego, o texto tem como base a primeira seção do segundo capítulo do livro de \citet{Chowdhury:2003}. O referencial teórico a respeito do diagrama fundamental do fluxo de tráfego teve como base o texto da segunda seção do capítulo seis do livro de \citet{Garber:2009}. Por fim, o referencial teórico acerca do controle de tráfego por meio de sinalizações semafórica teve como base documentações do \citet{DENATRAN:2014} e da quarta seção do segundo capítulo do livro de \citet{Chowdhury:2003}.

\subsection{Estrutura de Rede de Tráfego}

Redes são úteis para descrever sistemas de transporte a partir do ponto de vista dos componentes físicos. Elas são compostas por nós (vértices) e conexões (arestas). Os nós representam estações, interseções, entre outros. Em sistemas de transporte, existem normalmente mais de uma maneira de viajar entre dois nós, ou seja, as redes são redundantes. As conexões representam rodovias, estradas, ruas, linhas de trens, linhas de ônibus, entre outros. As conexões conduzem fluxos (veículos em uma direção ou em duas direções) e têm uma determinada capacidade. Existem várias maneiras de representar o custo de atravessar uma conexão. Neste sentido, este pode ser representado por uma função relacionada à capacidade da conexão e a ocupação desta. Os custos podem ser explícitos (pedágios), implícitos (tempo de viagem) ou representações mais abstratas.

\begin{figure}[t]
	\centering
    \includegraphics[scale=0.3]{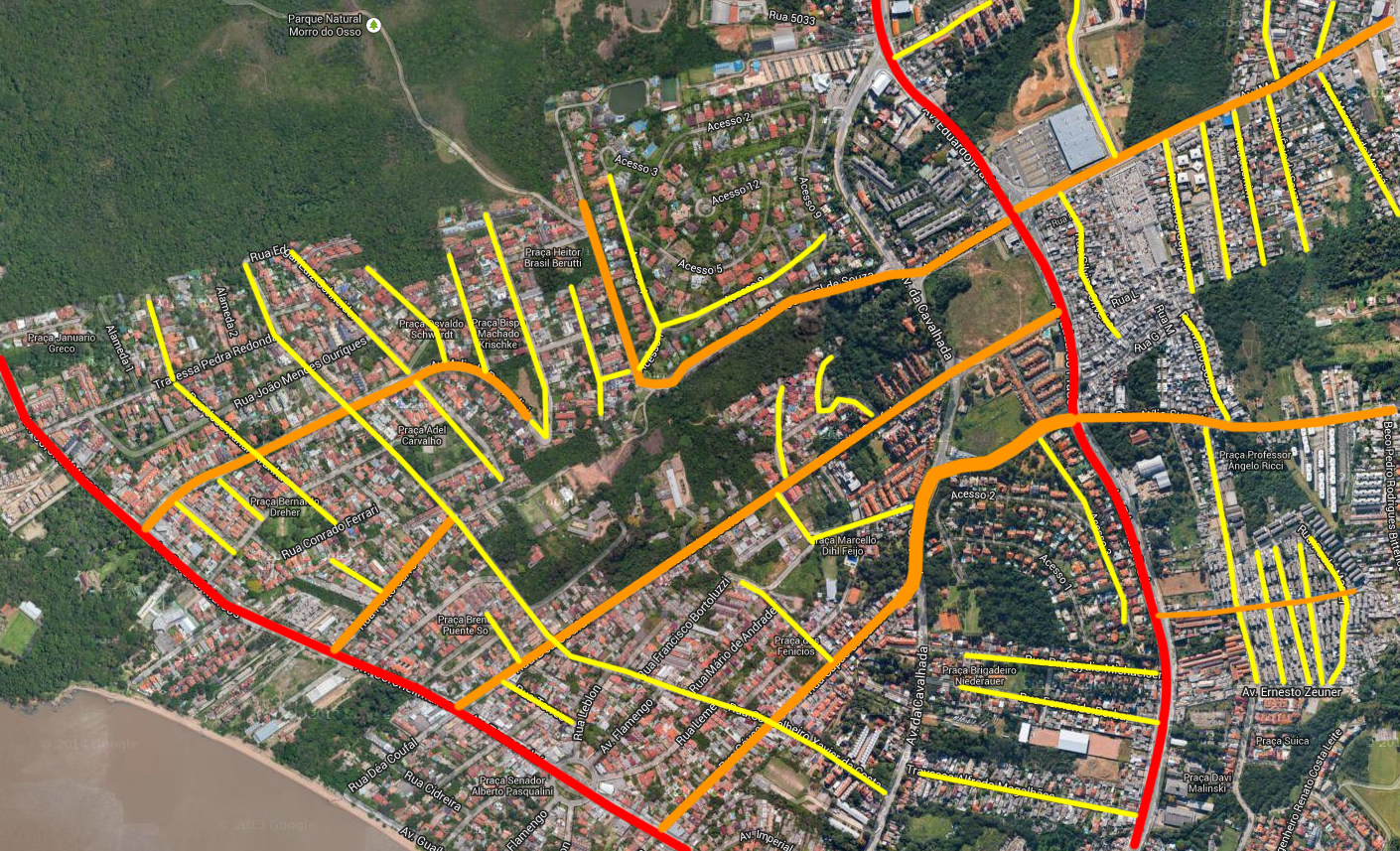}
    \caption{Classificação hierárquica das vias urbanas.}
    \label{fig:hierarquiavias}
\end{figure}

\begin{figure}[t]
	\centering
    \includegraphics[scale=0.5]{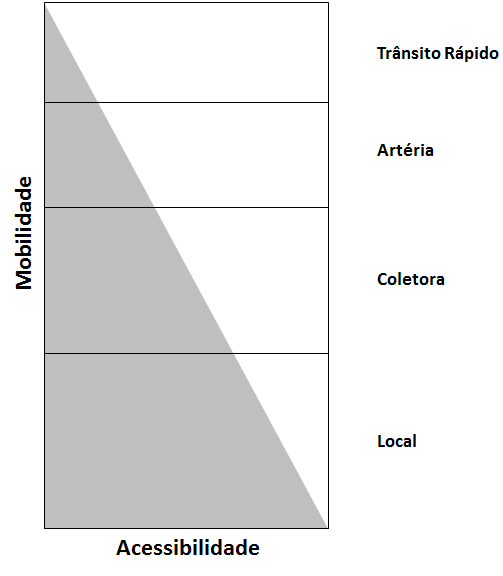}
    \caption{Relação entre acessibilidade e mobilidade dos diferentes tipos de vias.}
    \label{fig:relacaovias}
\end{figure}

Cada autoridade de tráfego tem nomes particulares para os vários tipos de conexões, mas uma hierarquia amplamente aceita é baseada em suas capacidades e funções. A respeito da capacidade, as vias são organizadas, de modo que o fluxo de tráfego escoe de uma via de menor capacidade para uma via de maior capacidade. Neste sentido, vias locais escoam o fluxo de tráfego para vias coletoras, que escoam o seu fluxo de tráfego para vias arteriais, que, por fim, escoam o seu fluxo para vias de trânsito rápido. A vias locais são aquelas caracterizadas por interseções em nível não semaforizadas, destinada apenas ao acesso local ou a áreas restritas \cite{CTB:2016}. As vias coletoras são aquelas destinadas a coletar e distribuir o trânsito que tenha necessidade de entrar ou sair das vias de trânsito rápido ou arteriais, possibilitando o trânsito dentro das regiões da cidade \cite{CTB:2016}. A vias arteriais são aquelas caracterizadas por interseções em nível, geralmente controlada por semáforo, com acessibilidade aos lotes lindeiros e às vias secundárias e locais, possibilitando o trânsito entre as regiões da cidade \cite{CTB:2016}. Por fim, as vias de trânsito rápido são aquelas caracterizadas por acessos especiais com trânsito livre, sem interseções em nível, sem acessibilidade direta aos lotes lindeiros e sem travessia de pedestres em nível \cite{CTB:2016}. Com base nestas definições de vias, é possível perceber uma relação entre os níveis de acessibilidade e mobilidade. Uma via local fornece alta acessibilidade, mas possui baixa mobilidade. Diferentemente, uma via de trânsito rápido tem acesso muito limitado, mas possui alta mobilidade. A Figura \ref{fig:relacaovias}, apresenta graficamente a relação entre acessibilidade e mobilidade dos diferentes tipos de vias.

A tarefa de descrever fluxos de tráfego envolve a medida e a análise de parâmetros macroscópicos ou microscópicos. Os parâmetros macroscópicos podem ser o volume, a velocidade e a densidades dos fluxos de tráfego. Os parâmetros microscópicos podem ser a velocidade individual dos veículos, intervalo de tempo entre veículos (\textit{headway}) e espaçamento entre veículos (\textit{gap}). Detalhes sobre os parâmetros microscópicos e macroscópicos serão apresentados na próxima seção. No entanto, antes de detalhar cada um destes parâmetros, é importante mencionar que o tipo de via influencia diretamente os valores obtidos a partir destes parâmetros. 

\subsection{Parâmetros Macroscópicos e Microscópicos de Tráfego}

Como mencionado anteriormente, descrever fluxos de tráfego envolve uma série de parâmetros que podem ser macroscópicos ou microscópicos. Neste sentido, três parâmetros básicos são utilizados para descrever fluxos de tráfego. Tais parâmetros são: fluxo, velocidade e densidade.  
 
O fluxo ($q$) é um parâmetro macroscópico de tráfego, que é o número de veículos passando em um determinado ponto de uma via durante um determinado período de tempo, que geralmente é uma hora. O fluxo é frequentemente expressado em veículos por hora. Um parâmetro importante derivado do fluxo é o valor de fluxo máximo, que é frequentemente referenciado como a capacidade ($q_m$) de uma via. O fluxo pode ser determinado por:
\begin{equation}
	q = \frac{n  \times \times 3600}{T} 
\end{equation}, 
onde $n$ é o número de veículos passando em um determinado ponto de uma via em $T$ segundos.

A velocidade claramente é um parâmetro microscópico de cada veículo. Assim, um valor de velocidade média ($\overline{u}$) pode ser associado a um fluxo de tráfego, fazendo com que a velocidade se torne um parâmetro macroscópico na perspectiva de um fluxo de tráfego. A velocidade média horária ($\overline{u}_t$) é a média aritmética das velocidades dos veículos passando em um determinado ponto de uma via durante um período de tempo. A velocidade média horária é frequentemente expressada em milhas por hora (nos Estados Unidos), quilômetros por hora, metros por segundo ou pés por segundo. A velocidade média horária é calculada a partir de 
\begin{equation}
	\overline{u}_t = \frac{1}{n} \sum^{n}_{i=1}u_i
\end{equation},
onde $n$ é o número de veículos passando em um determinado ponto de uma via, e $u_i$ é a velocidade do $i$-ézimo veículo. A velocidade média espacial ($\overline{u}_s$)é a  média harmônica das velocidades dos veículos passando em um determinado ponto de uma via durante um período de tempo. Ela é obtida a partir da divisão da distância total viajada por dois ou mais veículos pelo tempo gasto por estes veículos para viajar esta distância. Esta é a velocidade que está envolvida nos relacionamentos entre fluxo e densidade. A velocidade média espacial é calculada a partir de 
\begin{equation}
	\overline{u}_s = \frac{nL}{\sum^{n}_{i=1}t_i}
\end{equation}, 
onde $n$ é o número de veículos passando em um determinado ponto de uma via, $t_i$ é o tempo que o $i$-ézimo veículo leva para viajar através de seção de via, e $L$ é o tamanho da seção de via. A velocidade média horária é sempre maior que a velocidade média espacial. A diferença entre estas velocidades tende a diminuir, à medida que os valores absolutos de velocidades aumentam. 
 
A densidade ($k$) é um parâmetro macroscópico de tráfego, que pode ser definido como o número de veículos presentes sobre uma unidade de tamanho de uma via em um determinado instante no tempo. A densidade é tipicamente expressada em veículos por milha (nos Estados Unidos) ou veículos por quilômetro. Existem dois parâmetros importantes derivados a partir da densidade: a densidade de congestionamento ($k_j$) e a densidade ótima ($k_o$). A densidade de congestionamento ocorre sob condições extremas de congestionamento, quando o fluxo e a velocidade do tráfego se aproximam de zero. A densidade ótima ocorre sob condições de fluxo máximo. 
 
A densidade está também relacionada a dois parâmetros microscópicos de tráfego, que são: o intervalo de tempo entre veículos (\textit{headway}) e o espaçamento entre veículos (gap). O intervalo de tempo entre veículos ($h$) é definido como a diferença de tempo entre o momento em que a frente de um veículo chega em um determinado ponto de uma via e o momento em que a frente do próximo veículo atinge o mesmo ponto da via. Este parâmetro é tipicamente expresso em segundos. O espaçamento entre veículos ($d$) é definido como a distância entre a frente de um veículo e a frente do veículo imediatamente atrás. Este parâmetro é tipicamente expresso em metros. 

Por fim, é importante ressaltar que, a medição de cada um destes parâmetros é influenciada por fatores como características das vias, características dos veículos, características dos motoristas e fatores ambientais, tais como condições climáticas de uma região geográfica.

\subsection{Diagrama Fundamental de Fluxo de Tráfego}

A teoria de fluxo de tráfego tem como base o relacionamento entre os parâmetros macroscópicos de fluxo de tráfego apresentados na seção anterior, que são: fluxo ($q$), densidade ($k$) e velocidade média espacial $\overline{u}_s$. Esta relação se dá a partir de 
\begin{equation}
	q = k \times \overline{u}
    \label{eq:relacionamento}
\end{equation}. 

Outros relacionamentos que existem entre os parâmetros macroscópicos de fluxo de tráfego podem ser obtidos de acordo com as Equações \ref{eq:relvelocidademedia},\ref{eq:relespacamento1}, \ref{eq:reldensidade}, \ref{eq:relespacamento2} e \ref{eq:relintervalo}.
\begin{equation}
	\overline{u}_s = q \times \overline{d}
    \label{eq:relvelocidademedia}
\end{equation}, 
onde $\overline{d}$ é o espaçamento entre veículos, que é dado por
\begin{equation}
	\overline{d} = \frac{1}{k}
    \label{eq:relespacamento1}
\end{equation} 
\begin{equation}
	\overline{k} = q \times \overline{t}
    \label{eq:reldensidade}
\end{equation}, 
onde $\overline{t}$ é a média de tempo para atravessar uma seção de via, que é dado por 
\begin{equation}
	\overline{t} = \frac{n}{\sum^{n}_{i=1}t_i}
\end{equation}, 
onde $n$ é o número de veículos passando em uma seção de via, e $t_i$ é o tempo que o $i$-ézimo veículo leva para viajar através de uma seção de via.
\begin{equation}
	\overline{d} = \overline{u}_s \times \overline{h}
    \label{eq:relespacamento2}
\end{equation},
onde $\overline{h}$ é a média de intervalos de tempo entre veículos, que pode ser obtida a partir de 
\begin{equation}
	\overline{h} = \overline{t} \times \overline{d}
    \label{eq:relintervalo}
\end{equation}.

\begin{figure}[t]
	\centering
    \includegraphics[scale=0.5]{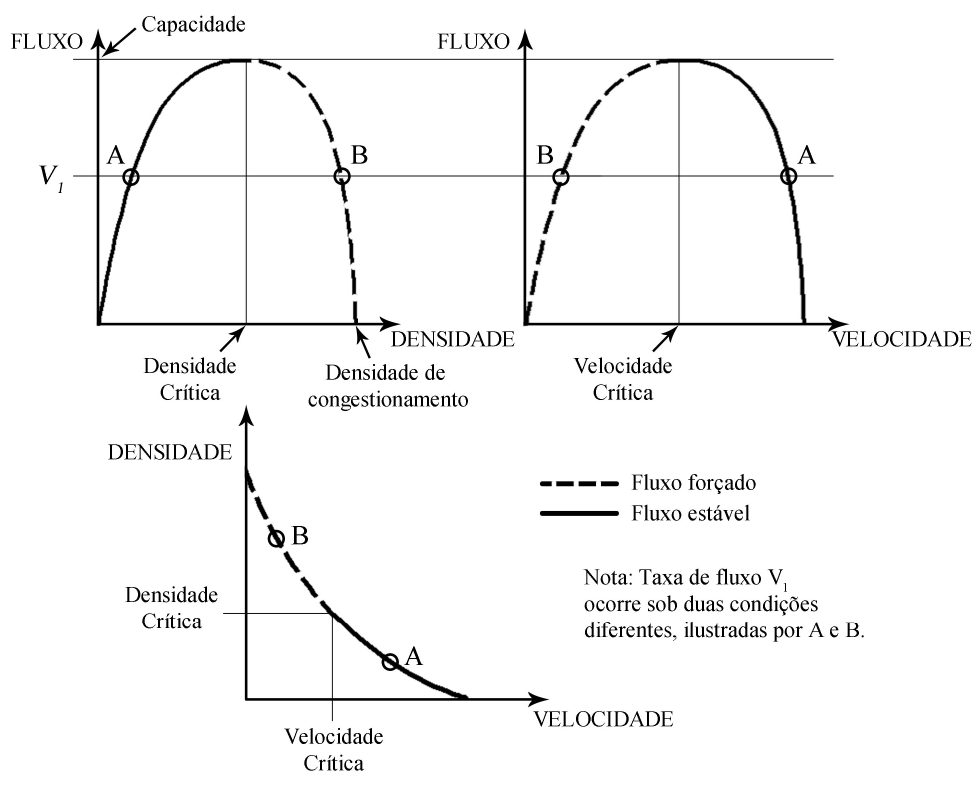}
    \caption{Diagrama fundamental de fluxo de tráfego. As partes tracejadas dos gráficos representam o regime de congestionamento. Os pontos de transição em cada curva são a densidade crítica e velocidade crítica. Fonte: \cite{Paiva:2012}.} 
    \label{fig:digramafluxotrafego}
\end{figure}

Com base no relacionamento entre a densidade e o fluxo (veja Figura \ref{fig:digramafluxotrafego}), as seguintes hipóteses podem ser feitas acerca deste relacionamento:

\begin{enumerate}
	\item Quando a densidade é zero, o fluxo também é zero, pois não existem veículos na via;
    \item Quando a densidade cresce, o fluxo também cresce, pois a quantidade de veículos trafegando na via também aumenta;
    \item Quando a densidade atingir o seu valor máximo (densidade de engarrafamento, $k_j$), o fluxo deve ser zero, uma vez que os veículos se encontram parados na via;
    \item Baseando-se em 2 e 3, conclui-se que, à medida que a densidade cresce, o fluxo inicialmente cresce até atingir um valor máximo. Aumentos adicionais na densidade levam a redução do fluxo, que poderá chegar a zero, caso a densidade atinja o seu valor máximo.

\end{enumerate}

O diagrama fundamental de fluxo de tráfego é usado para caracterizar diferentes regimes de fluxo. Um regime de fluxo livre é um regime em que cada veículo viaja na velocidade desejada do motorista. Este fluxo ocorre quando poucos veículos estão na via e existem faixas suficientes para permitir ultrapassagens sem qualquer atraso. Além disto, a velocidade neste regime de tráfego é denotada por $\overline{u}_f$. Com o aumento na densidade de veículos, os motoristas começam a ter dificuldades de manter suas velocidades desejadas, devido a uma redução de velocidade causado por veículos mais lentos a frente de seus veículos. Isto resulta em uma diminuição contínua da velocidade média. Por isto, este regime é chamado de parcialmente restrito, pois nem todos os veículos conseguem se mover na velocidade desejada por seus motoristas e, tão pouco, realizar ultrapassagens. Quando as ultrapassagens não são possíveis, os veículos começam a viajar em pelotões cuja velocidade é determinada pelos veículos à frente destes. Neste caso, é dito que o regime de fluxo é restrito. A transição do regime de tráfego parcialmente restrito para o restrito é claramente observada no relacionamento entre velocidade e fluxo (ver Figura \ref{fig:digramafluxotrafego}). Esta transição ocorre no ponto de máximo da curva em que o diagrama fundamental apresenta o relacionamento entre o fluxo e a velocidade. A velocidade neste ponto é denotada por $\overline{u}_o$, uma vez que a via está sob condições de fluxo máximo.

O relacionamento densidade-velocidade reflete o comportamento dos motoristas, pois estes ajustam as velocidades de seus veículos a partir da percepção da proximidade de outros veículos e seus conceitos sobre segurança em determinadas condições de tráfego. Dessa forma, a densidade é criada a partir do ajuste da velocidade do veículo de cada um dos motoristas na via. 

\subsection{Controle de Tráfego por meio de Sinalização Semafórica}\label{sec:controle_trafego}

Quando se aborda o problema de controle de tráfego, é comum o uso de sinalizações semafóricas como mecanismo de controle de tráfego. De acordo com \abbrev{DENATRAN}{Departamento Nacional de Trânsito} \cite{DENATRAN:2014}, a sinalização semafórica é um subsistema da sinalização viária que é composto de indicações luminosas acionadas alternada ou intermitentemente por meio de um sistema eletromecânico ou eletrônico (controlador). Além disto, este tipo de sinalização tem como objetivo transmitir diferentes mensagens aos usuários das vias, regulando o direito de passagem ou advertindo sobre condições especiais nas vias \cite{DENATRAN:2014}. Em situações específicas, tais como o uso de dispositivos de detecção do tráfego, equipamentos de fiscalização não metrológico e centrais de controle podem ser associados a sinalização semafórica \cite{DENATRAN:2014}. Por fim, a operação da sinalização semafórica deve ser contínua e criteriosamente avaliada quanto a sua real necessidade e adequação de sua programação \cite{DENATRAN:2014}.

Toda sinalização semafórica tem seu funcionamento baseado em um ciclo composto de três indicações: SIGA, ATENÇÃO e PARE. Cada uma destas três indicações é representada por uma luz de cor específica. A luz verde representa a indicação SIGA. Nesta indicação, veículos podem atravessar a interseção. A luz amarela representa a indicação ATENÇÃO. Nesta indicação, os motoristas devem reduzir a velocidade de seus veículos, de modo que possam se preparar para uma parada antes da faixa de contenção pintada sobre as vias de entrada de uma interseção. Por fim, a luz vermelha representa a indicação PARE. Além disto, cada uma destas indicações tem um tempo de duração, que ajuda a definir o início e o fim de uma indicação ao longo do tempo. Dessa forma, se as luzes das sinalizações semafóricas estão verdes ou amarelas para as vias de entrada de uma interseção, então, as luzes vermelhas das sinalizações semafóricas sobre as vias com movimentos conflitantes aos das vias entrada devem estar acesas. Esta regra de segurança tem como objetivo evitar que veículos atrevessem uma interseção ao mesmo tempo. Além desta regra de segurança, também existem regras de igualdade cujo intuito é distribuir de maneira justa os tempos das fases das sinalizações semafóricas que controlam as vias de entrada de uma interseção. Portanto, para todas estas vias, deve ser fornecido um tempo mínimo de duração para a indicação SIGA. No que tange a indicação ATENÇÃO, esta tem um tempo fixo de duração.

\begin{figure}[t]
	\centering
    \subfigure[]{
    	\includegraphics[width=5cm]{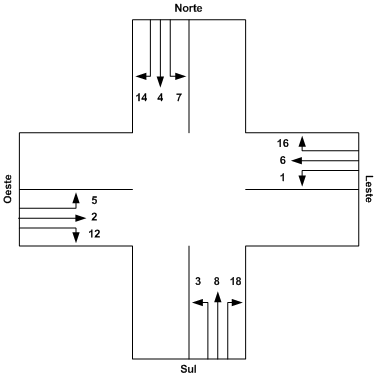}
        \label{fig:intersecao_numerada}
    }
    \quad
    \subfigure[]{
    	\includegraphics[width=5cm]{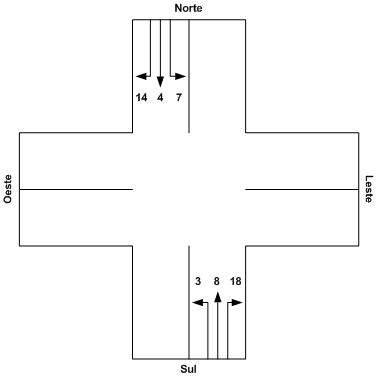}
        \label{fig:grupo_norte_sul}
    }
    \quad
    \subfigure[]{
    	\includegraphics[width=5cm]{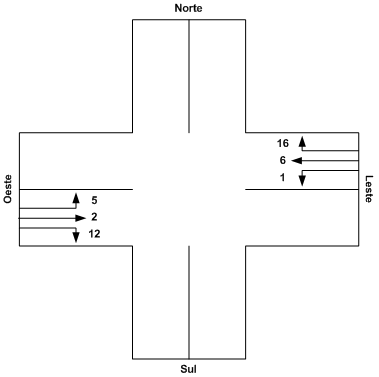}
        \label{fig:grupo_leste_oeste}
    }
    \caption{Exemplo de uma interseção com doze vias de entrada com movimentos e grupos de movimentos identificados: (a) Enumeração de movimentos em uma interseção sinalizada; (b) Grupo de movimentos norte-sul-norte; (c) Grupo de movimentos leste-oeste-leste. Fonte: \citet{Kyte:2014}.}
\end{figure}

Existem termos que são comumente utilizados em controle de tráfego por meio de sinalizações semafóricas, são eles: fase, intervalo, ciclo, tamanho do ciclo \cite{Chowdhury:2003}, movimento e grupos de movimento \cite{Kyte:2014}. A fase é o tempo em que um conjunto de movimentos de tráfego recebe o direito de atravessar a interseção. O intervalo é o tempo em que as indicações de uma sinalização semafórica se mantêm constantes. O ciclo é uma sequência completa de indicações de uma sinalização semafórica até que uma nova sequência seja iniciada. O tamanho do ciclo é o tempo que uma sinalização semafórica leva para exibir uma sequência completa de indicações. O intervalo é o tempo em que as indicações das sinalizações semafóricas se mantêm constantes. Um ciclo inclui vários intervalos para as indicações de uma sinalização semafórica, ou seja, intervalos para indicações das luzes verde, amarela e vermelha. O movimento é definido pela direção do fluxo de tráfego e a manobra que cada veículo precisa realizar em uma interseção. Em uma interseção sinalizada, os movimentos são identificados por meio de números, como pode ser visto na Figura \ref{fig:intersecao_numerada}. Neste caso, movimentos que viram a esquerda são enumerados com números ímpares (veja Figura \ref{fig:intersecao_numerada}). Para movimentos que atravessam a interseção, utiliza-se uma enumeração com números pares (veja Figura \ref{fig:intersecao_numerada}). No que diz respeito aos movimentos que viram à direita, utiliza-se o mesmo número do movimento compatível que atravessa a interseção ou este número antecedido pelo dígito 1 (veja Figura \ref{fig:intersecao_numerada}). Os movimentos são classificados, de acordo com as restrições impostas a eles, a saber: movimento sem oposição, movimento protegido, movimento permitido e movimento proibido. O movimento sem oposição é aquele que nenhum outro movimento se opõe a ele. O movimento protegido é aquele que pode ter um movimento oposto a ele, mas a indicação da sinalização dá a este movimento o direito de passagem. O movimento permitido é aquele que é permitido trafegar através de uma interseção, mas deve dar direito de passagem, se um movimento oposto de maior prioridade existir. Por fim, o movimento não permitido é aquele que um fluxo de veículos é totalmente proibido de trafegar ou tem tráfego proibido em certos períodos do dia. Os grupos de movimentos são classificados em compatíveis ou conflitantes. Em geral, movimentos norte-sul/sul-norte conflitam com movimentos leste-oeste/oeste-leste (veja Figura \ref{fig:intersecao_numerada}). Os movimentos norte-sul e sul-norte são parte de um grupo chamado de grupo de concorrência, como pode ser visto na Figura \ref{fig:grupo_norte_sul}, pois estes movimentos podem trafegar concorrentemente pela interseção. O mesmo se aplica os movimentos Leste-Oeste e Oeste-Leste, como pode ser visto na Figura \ref{fig:grupo_leste_oeste}. Por fim, de acordo com o plano de fases da sinalização semafórica e restrições nos movimentos, um movimento no grupo de concorrência Norte-Sul-Norte pode ser servido ao mesmo tempo que qualquer movimento do mesmo grupo. Este conceito se aplica aos movimentos do grupo Leste-Oeste-Leste.  

As sinalizações semafóricas podem operar em quatro diferentes modos, a saber: pré-temporizado, semiatuado, totalmente atuado e baseado em computadores \cite{Chowdhury:2003}. No modo de operação pré-temporizado, o tamanho do ciclo, intervalos e fases são pré-definidos ou fixos e, por isto, não são sensíveis às variações do volume de tráfego. O modo de operação semiatuado é utilizado em interseções onde uma via com volume maior e outra com volume menor são claramente identificadas. Neste caso, os detectores de tráfego são utilizados somente na via de menor volume. Desta forma, a sinalização semafórica da via com maior volume de tráfego mantém a luz verde acesa até o momento em que veículos sejam detectados na via de menor volume. Quando isto acontece, após a sinalização semafórica da via com maior volume ter permanecido um tempo mínimo com a luz verde acesa, a sinalização semafórica da via com menor volume pode ter sua luz verde acesa. No modo de operação totalmente atuado, é necessário que todas as vias tenham detectores de tráfego. Com isto, o tempo em que a luz verde permanece acesa nas sinalizações semafóricas pode ser alocado de acordo com o volume de tráfego observado em cada via de entrada da interseção. Por fim, o modo de operação baseado em computador refere-se ao uso de um computador para ligar a operação de um grupo de interseções sinalizadas na forma de um sistema coordenado. Assim, o computador seleciona ou computa planos ótimos de sinalizações semafóricas coordenadas para todo o sistema, baseando-se nas informações de tráfego fornecidas pelos detectores de tráfego instalados nas vias. 

Uma sinalização semafórica pode ser responsável pelo controle de uma interseção isolada, que é uma interseção que opera a parte de um sistema coordenado de sinalizações semafóricas. Um sistema coordenado de sinalizações semafóricas é um grupo de sinalizações semafóricas que têm seus intervalos indicações de luzes verdes sincronizados, de modo que os veículos não precisem parar em cada uma interseção controlada por uma das sinalizações semafóricas deste sistema. Sistemas coordenados de sinalizações semafóricas são utilizados, quando as sinalizações semafóricas estão relativamente próximas umas das outras e formam um corredor. Nesse tipo de sistema, as sinalizações semafóricas devem ter o mesmo tamanho de ciclo. Para sincronizar os intervalos de verde, sistemas coordenados de sinalizações semafóricas fazem uso de um termo chamado de \textit{offset}. O \textit{offset} é a diferença de tempo entre o início do intervalo de luz verde entre duas interseções vizinhas. 

Ao findar toda essa discussão em torno dos fundamentos de engenharia de tráfego, o texto segue para próxima seção, a fim de apresentar o estudo que diz respeito ao assunto de agentes e sistemas multiagentes.

\section{Agentes e Sistemas Multiagentes}

Esta seção tem como objetivo apresentar o levantamento bibliográfico acerca de agentes e sistemas multiagentes. Os textos compreendidos nas subseções seguintes tem como base o capítulo de livro escrito por \cite{Garcia:2005}.

Apesar de não existir uma definição universalmente aceita acerca de agentes, será adotada uma definição mais genérica baseada em \citet{Ferber:1999}. Um agente é uma entidade real ou virtual, capaz de agir em um ambiente, de se comunicar com outros agentes, que é movida por um conjunto de inclinações (sejam objetivos individuais a atingir ou uma função de satisfação a otimizar). Além disto, um agente possui recursos próprios, além de ser capaz de perceber o seu ambiente (de modo limitado) e possuir (eventualmente) uma visão parcial deste sistema. Outra característica importante de um agente é que ele possui autonomia e oferece serviços. Com base nisto, sistemas multiagentes são as atividades de um conjunto de agentes autônomos em um universo multiagente. Uma vez que são autônomos, os agentes devem coordenar dinamicamente suas atividades e cooperar uns com outros para que o objetivo do sistema seja alcançado. 

Com base nessa definição, as características-chave que efetivam a caracterização de um agente são: autonomia de decisão, autonomia de execução, competência para decidir e a existência de uma agenda própria \cite{Huhms:1998}. A autonomia de decisão está relacionada à capacidade de analisar uma situação, gerar alternativas de atuação e escolher a situação que melhor atende seus objetivos. Em certos casos, o agente não conhece o cenário de atuação, mas tem capacidade de escolher uma experiência prévia semelhante e, em seguida, adaptar a solução ao novo cenário. A autonomia de execução é a capacidade de operar no ambiente sem intervenção de outro agente (geralmente humanos). A competência para decidir é a capacidade de configurar sua atuação sem intervenção externa. Por fim, a existência de uma agenda própria, que é a capacidade de criar uma agenda de objetivos que concretizem suas metas. 

Continuando a apresentação do referencial teórico desta tese, a próxima seção apresenta um estudo sobre escalonamento por reversão de arestas.

\section{Escalonamento por Reversão de Arestas}

Esta seção tem como objetivo apresentar o levantamento bibliográfico a respeito dos algoritmos de escalonamento por reversão de arestas conhecidos como SER \cite{Barbosa:1989,Barbosa:1996} e SMER \cite{Barbosa:2001}. A apresentação dos detalhes de cada um destes algoritmos é feita de acordo com a disposição das próximas subseções.

\subsection{SER}

Escalonamento por Reversão de Arestas ou \textit{Scheduling by Edge Reversal} (SER) é um poderoso algoritmo paralelo e distribuído desenvolvido por \citet{Barbosa:1989}, a fim de controlar a operação concorrente entre os elementos de sistemas de vizinhança restrita de topologia genérica. O SER tem como base um conjunto de processos que executam à medida que acessam recursos atômicos compartilhados. Sistemas de recursos compartilhados são representados por meio de um grafo orientado finito $G = (V, E)$, onde os processos são os vértices de $G$ e uma aresta orientada existe entre quaisquer dois nós, desde que estes compartilhem um recurso. 

\begin{figure}[t]
	\centering
    \subfigure[]{
    	\includegraphics[width=5cm]{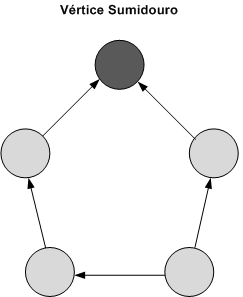}
        \label{fig:orientacaoaciclica}
    }
    \qquad
    \subfigure[]{
    	\includegraphics[width=5cm]{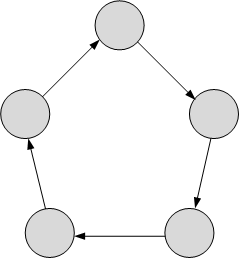}
        \label{fig:grafodeadlock}
    }
    \caption{Estados de um grafo direcionado no contexto do SER. (a) Grafo com orientação acíclica; (b) Grafo em \textit{deadlock}.} 
\end{figure}

O SER inicia sua execução a partir de qualquer orientação acíclica $\omega$ em $G$, devendo existir pelo menos um vértice sumidouro, ou seja, um vértice que tenha todas as arestas entre seus vizinhos direcionadas para ele. A Figura \ref{fig:orientacaoaciclica} apresenta um exemplo de grafo com uma orientação acíclica contendo um vértice sumidouro. A partir de uma orientação acíclica inicial, pode-se garantir que não haverá \textit{deadlock} ou \textit{starvation} durante a execução do algoritmo. A Figura \ref{fig:grafodeadlock} apresenta um grafo em estado de \textit{deadlock}, uma vez que as orientações das arestas entre os vértices deste grafo formam um ciclo. Quando um vértice se torna um sumidouro, este tem a permissão de operar sobre os recursos compartilhados entre ele e seus vizinhos. Uma vez que um vértice sumidouro tenha finalizado a sua operação, ele reverte a orientação de todas as suas arestas, enviando mensagens para todos os seus vizinhos e, em seguida, se torna um vértice fonte. Este processo dá início a uma nova orientação acíclica $\omega '$. A partir desta nova orientação acíclica, um novo conjunto de vértices sumidouros é formado e assim por diante. 

\begin{figure}[t]
	\centering
    \includegraphics[scale=1.2]{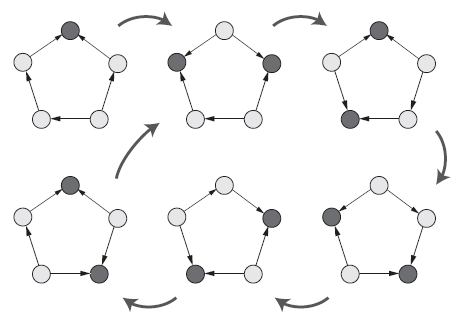}
    \label{fig:periodoser}
    \caption{Exemplo de um período, onde $p=5$, $m=2$ e $\gamma(G) = 2/5$. Adaptado de \citet{Barbosa:1996}.}
\end{figure}

Como o número das orientações acíclicas é finito, eventualmente um conjunto de orientações se repetirá ao longo do tempo. Isto define um período de tamanho $p$ de orientações. Dentro de um período, cada vértice opera exatamente o mesmo número constante de vezes, que é denotado por $m$. Assim, a concorrência de um período é definida pelo coeficiente $\gamma(G) = m/p$. Quanto mais alto é o coeficiente de concorrência de um período, menor será a ociosidade dos vértices durante a evolução de um período. Segundo \citet{Barbosa:1996}, para qualquer sistema conectado, o coeficiente de concorrência de um período deve respeitar a seguinte regra: $1/n \leq m/p \leq 1/2$, tal que $n = |N|$.

Embora o SER garanta que todos os nós possam operar com a mesma frequência sobre os recursos compartilhados em um sistema distribuído, esta condição é indesejável, quando os vértices precisam operar com diferentes frequências sobre tais recursos. Para tanto, \citet{Barbosa:2001} desenvolveram uma generalização do SER, que é apresentada na próxima seção.

\subsection{SMER}

\begin{figure}[t]
	\centering
    \includegraphics{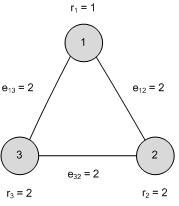}
    \caption{Exemplo de um multigrafo no contexto do SMER. Adaptado de \citet{Paiva:2012}.}
    \label{fig:multigrafoSMER}
\end{figure}

O Escalonamento por Reversão Múltipla de Arestas ou \textit{Scheduling by Multiple Edge Reversal} (SMER) é uma generalização do SER em que taxas de acesso pré-determinadas para os recursos atômicos compartilhados são impostas aos processos em um sistema distribuído de compartilhamento de recursos representado por uma transformação do grafo $G$ em um multigrafo $M = (V", E")$, onde $N" = N$ e, para cada ($v_i, v_j$) $\in E$, existe $e_{i,j}$ arestas em $E"$. Diferente do SER, múltiplas arestas $e_{i,j} \in E$ podem existir entre quaisquer dois vértices $i$ e $j$ ($i, j \in V"$) nas dinâmicas do SMER, denotando um conjunto de arestas cujo nome é arco \cite{Barbosa:2001}. A Figura \ref{fig:multigrafoSMER} apresenta um exemplo de multigrafo, apresentando três vértices, os arcos entre os vértices e as reversibilidades de cada vértice.

Além disto, a cada vértice $i$ é associado um parâmetro $r_i$ cujo nome é reversibilidade. A reversibilidade é o número de arestas que devem ser revertidas por um vértice $i$ em direção a cada um de seus vértices vizinhos, após este findar uma operação que esteja acessando os recursos atômicos compartilhados. No SMER, um vértice $i$ é chamado de sumidouro, se este possuir, em cada arco $e_{i,j}$ com seus vizinhos, um número $r_i$ de arestas direcionadas para si. Diferente do SER, um vértice pode operar mais de uma vez consecutivamente durante a execução do algoritmo SMER. Para evitar que dois vértices executem sobre os mesmos recursos compartilhados, \cite{Barbosa:2001} introduziram uma restrição sobre os arcos que conectam os vértices do multigrafo $M$. Esta restrição consistem na seguinte regra: um arco $e_{i,j}$ deve ter um número de arestas que seja maior ou igual a maior reversibilidade entre $i$ e $j$ (máximo\{$r_i$, $r_j$\}), e menor ou igual a $r_i + r_j - 1$. Além disto, para garantir que o número de arestas de um arco $e_{i, j}$ seja mínimo, \citet{Barbosa:2001} propuseram a Equação \ref{eq:numarestas}.
\begin{equation}
	e_{i, j} = r_i + r_j - mdc(r_i, r_j)
    \label{eq:numarestas}
\end{equation}, onde $mdc(r_i, r_j)$ é o máximo divisor comum entre as reversibilidades dos vértices $i$ e $j$. Por fim, para qualquer par de vértices $i$ e $j$ em $M$, a razão entre o número de vezes que $i$ tem prioridade sobre $j$ é dada por $r_j/r_i$, que é a taxa relativa de execução entre dois vértices. Assim, o SER passa a ser um caso particular do SMER, em que todos os vértices possuem a mesma reversibilidade, uma vez que a razão entre as reversibilidades de cada par de vértices é igual a um, fazendo vigorar a justiça desta forma.

\begin{figure}[t]
	\centering
    \includegraphics[scale=0.7]{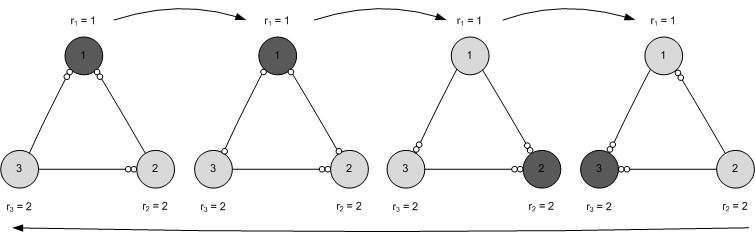}
    \caption{Exemplo de execução do SMER sobre um multigrafo com três vértices. Adaptado de \citet{Paiva:2012}}
    \label{fig:dinamica_smer}
\end{figure}

A Figura \ref{fig:dinamica_smer} exibe um exemplo de execução do SMER sobre um multigrafo com três vértices, em que o vértice um tem uma frequência de execução duas vezes maior que os seus vizinhos. As reversibilidades dos vértices são calculadas com base nas demandas $d_i$ de cada vértice $i$, que são: $d_1 = 2$, $d_2 = 1$ e $d_3 = 1$. Além disto, o cálculo das reversibilidades de cada um dos vértices teve como base a Equação \ref{eq:reversibilidade}.
\begin{equation}
	r_i = \frac{mmc(d_1, d_2, ..., d_n)}{d_i}
    \label{eq:reversibilidade}
\end{equation}
Com base nas demandas, primeiramente calcula-se o mínimo múltiplo comum entre $d_1$, $d_2$ e $d_3$ ($mmc(d_1, d_2, d_3) = 2)$. Em seguida, calcula-se as reversibilidades como se segue:
$$ r_1 = \frac{(d_1, d_2, d_3)}{d_1} = \frac{2}{2} = 1$$
$$ r_2 = \frac{(d_1, d_2, d_3)}{d_2} = \frac{2}{1} = 2$$
$$ r_3 = \frac{(d_1, d_2, d_3)}{d_3} = \frac{2}{1} = 2$$
 Por fim, utiliza-se a Equação \ref{eq:numarestas} para calcular o número de arestas de um arco $e_{i,j}$ compreendido entre cada par de vértices $i, j$. O cálculo do número de arestas de cada arco é dado como segue:
$$ e_{1,2} = r_1 + r_2 - mdc(r_1, r_2) = 1 + 2 - 1 = 2$$
$$ e_{1,3} = r_1 + r_3 - mdc(r_1, r_3) = 1 + 2 - 1 = 2$$
$$ e_{2,3} = r_2 + r_3 - mdc(r_2, r_3) = 2 + 2 - 2 = 2$$.

Como pode ser visto na Figura \ref{fig:dinamica_smer}, o SMER inicia sua execução a partir de um estado inicial em que o vértice um é o sumidouro. Após operar sobre os recursos compartilhados entre os vértices dois e três, o vértice um reverte $r_1$ arestas para os seus vizinhos. Uma vez que $r_1$ é igual a um, o vértice um inicia novamente a operação sobre os recursos compartilhados com seus vizinhos, pois ainda restaram uma aresta em cada um dos arcos que interligam o vértice um aos vértices dois e três. Após o término desta segunda operação, o vértice um reverte novamente $r_1$ arestas para seus vizinhos. Neste momento, o vértice um se torna um vértice fonte e o vértice dois se torna um sumidouro, permitindo este operar sobre os recursos compartilhados com os vértices um e três. Ao findar a sua operação, o vértice dois reverte $r_2$ arestas para seus vizinhos, tornando-se um vértice fonte, pois não existem arestas direcionadas para ele. Após isto, o vértice três passa a ser o vértice sumidouro, permitindo-lhe operar sobre os recursos compartilhados com os vértices um e dois. Após o fim de sua operação, o vértice três reverte $r_3$ arestas, finalizando o período $\omega$ e dando início ao período $\omega '$.

\begin{figure}[t]
	\centering
    \includegraphics{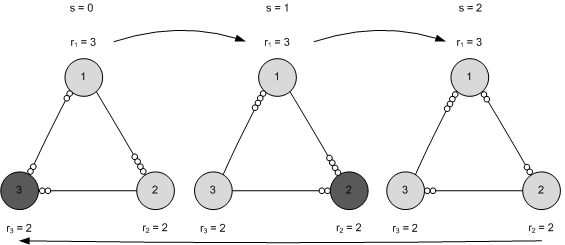}
    \caption{Exemplo de uma orientação inicial de um multigrafo que leva a uma situação de \textit{deadlock}. Adaptado de \citet{Paiva:2012}.}
    \label{fig:smer_deadlock}
\end{figure}

Tanto para o SER quanto para o SMER, determinar uma orientação inicial acíclica é fundamental para execução destes algoritmos distribuídos. No entanto, este estado inicial não pode levar o grafo a uma orientação que leve a uma situação de \textit{deadlock}. Em específico, no SMER, existe uma maneira de saber se um estado $s \geq 0$ levará o grafo a um estado de situação de \textit{deadlock}. Para tanto, considere o conjunto $K$ de ciclos simples não-dirigidos em $G$. Para $k \in K$, tem-se que $v_i \in k$, significando que o vértice $i$ pertence ao ciclo não-dirigido $k$, e ($v_i, v_j$) $\in k$, indicando que a aresta entre os vértices $i$ e $j$ faz parte do ciclo não-dirigido $k$. Além disto, também considere a soma das reversibilidades dos vértices em $k$, que é dada por 
\begin{equation}
	\rho(k) = \sum_{v_i \in k}r_{n_i}
    \label{eq:soma_reversibilidades}
\end{equation}, e o número de arestas orientadas do multigrafo em $k$ na direção transversal cuja maioria das arestas está orientada, que é dado por
\begin{equation}
	\sigma_s(k) = max \left\lbrace \sum_{(v_i, v_j) \in k^+}e_{s}^{ij}, \sum_{v_i \in k}e_{s}^{ij} \right\rbrace
    \label{eq:soma_arestas}
\end{equation}, onde $k^+$ possui arestas orientadas em sentido horário e $k^-$ possui arestas orientadas em sentido anti-horário. Com base nas Equações \ref{eq:soma_reversibilidades} e \ref{eq:soma_arestas}, é possível verificar se o multigrafo entrará em um estado de \textit{deadlock}, comparando $\sigma(k)$ com $\rho(k)$. Se $\sigma(k) < \rho(k)$, então é garantido que o multigrafo não entrará em um estado de \textit{deadlock}. Como pode ser visto na Figura \ref{fig:smer_deadlock}, o SMER inicia sua execução a partir de uma orientação inicial acíclica que posteriormente leva o grafo a um estado de \textit{deadlock}. No multigrafo, a soma das reversibilidades dos três vértices é 
$$\rho(k) = \sum_{v_i \in k}r_{n_i} = r_{1} + r_{2} + r{3} = 3 + 2 + 2 = 7$$. A soma das arestas orientadas no sentido horário é:
$$ \sigma_s(k^+) = \sum_{(v_i, v_j) \in k^+}e_{s}^{ij} = e^{31}_0 + e^{12}_0 + e^{23}_0 = 2 + 4 + 2 = 8$$. A soma das arestas orientadas no sentido anti-horário é:
$$ \sigma_s(k^-) = \sum_{(v_i, v_j) \in k^-}e_{s}^{ij} = e^{13}_0 + e^{32}_0 + e^{21}_0 = 2 + 0 + 0 = 2$$. Substituindo esses resultados na Inequação \ref{eq:inequacao_deadlock}, identifica-se que o multigrafo entrará em um estado de \textit{deadlock}, pois $\sigma(k) = 8$ e $\rho(k) = 7$. Estes resultados, por sua vez, não satisfazem a Inequação \ref{eq:inequacao_deadlock}.

\begin{equation}
	max \left\lbrace \sum_{(v_i, v_j) \in k^+}e_{s}^{ij}, \sum_{v_i \in k}e_{s}^{ij} \right\rbrace < \rho(k)
    \label{eq:inequacao_deadlock}
\end{equation}

Embora seja simples detectar \textit{deadlocks} em grafos de baixa complexidade, esta tarefa se torna difícil em grafos complexos e gera um grande \textit{overhead} de comunicação durante a detecção destes estados \cite{Paiva:2012}. Para tratar este problema, \citet{Santos:2012} propôs um método baseado em mudanças de reversibilidade. Este método permite que os vértices de um multigrafo mudem suas taxas relativas de execução durante a execução do algoritmo SMER.

O método de \citet{Santos:2012} consiste em iniciar o multigrafo em uma condição específica, de maneira que todos os vértices tenham a mesma frequência de operação. Logo, isto faz com que o algoritmo SMER inicie sua execução se comportando como o SER. Desta forma, é possível assegurar que o estado inicial do multigrafo não levará o algoritmo SMER entre em \textit{deadlock}. A partir disto, à medida que os vértices em $M$ se tornam sumidouros, estes passam a ter o direito de mudar as suas reversibilidades $r_i$ com intuito de ajustar as suas frequências de operação de acordo com uma demanda $d_i$ ao longo do tempo. Esta mudança de frequência só é possível, quando os vértices sumidouros têm conhecimento das reversibilidades de seus vizinhos. Com isto, é possível recalcular a quantidade de arestas presentes em cada arco $e_{ij}$ que interliga o vértice sumidouro aos seus vizinhos. Este cálculo é realizado, usando a Equação \ref{eq:numarestas}. Uma vez calculado o novo número de arestas para os arcos entre os vértices sumidouros e seus vizinhos, os vizinhos dos sumidouros não sofrem qualquer mudança, mas o sistema distribuído de compartilhamento de recursos automaticamente sofre um redimensionamento de seu período com base nas novas reversibilidades \cite{Paiva:2012}. Por fim, o mecanismo de mudança de reversibilidade garante a ausência de \textit{deadlocks}, ainda que as reversibilidades tenham seus valores aumentados ou diminuídos em função da demanda dos vértices. Esta garantia se baseia na regra definida pela Inequação \ref{eq:inequacao_deadlock}, pois aumentos ou diminuições nas reversibilidades dos vértices sumidouros são reproduzidos em ambos os lados da inequação.

\begin{figure}[t]
	\centering
    \includegraphics[scale=0.7]{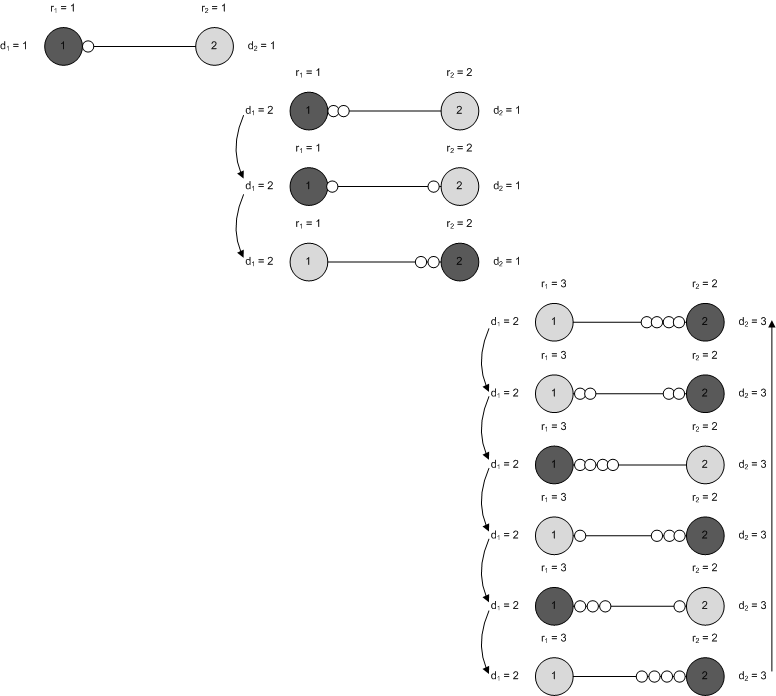}
    \caption{Exemplo de uma execução do algoritmo SMER com o mecanismo de mudança de reversibilidade. Adaptado de \cite{Santos:2012}}
    \label{fig:mudanca_reversibilidade}
\end{figure}

A Figura \ref{fig:mudanca_reversibilidade} mostra um exemplo de uma execução do algoritmo SMER, utilizando o mecanismo de mudança de reversibilidade, tendo como base um multigrafo com dois vértices. Neste exemplo, ambos os vértices iniciam com demandas e reversibilidades iguais a um. Além disto, o estado inicial do multigrafo tem o vértice um como sumidouro. O vértice um tem sua demanda aumentada para dois, fazendo com que sua reversibilidade se mantenha em um e a reversibilidade do vértice dois aumente para dois. Estes valores de reversibilidade foram obtidos a partir do uso da Equação \ref{eq:reversibilidade} da seguinte forma:
$$r_1 = \frac{mmc(d_1,d_2)}{d_1} = \frac{2}{2} = 1$$
$$r_2 = \frac{mmc(d_1,d_2)}{d_2} = \frac{2}{1} = 2$$. 
A partir das novas reversibilidades, o número de arestas no arco entre os vértices um e dois aumentou para dois. Este valor foi obtido, empregando Equação \ref{eq:numarestas}.
$$e_{12} = r_1 + r_2 - mdc(r_1, r_2) = 2 + 1 - mdc(2, 1) = 2 + 1 - 1 = 2$$. Após finalizar sua operação, o vértice um reverte $r_1$ arestas para o vértice dois. Como o número de arestas orientadas para o vértice um ainda é igual ao valor de sua reversibilidade, o vértice um inicia uma nova operação. Após o fim desta operação, o vértice um reverte novamente $r_1$ arestas para o vértice dois, se tornando um vértice fonte. Uma vez que o vértice dois tem o número de arestas orientadas para ele sendo igual a sua reversibilidade, o vértice dois se torna um sumidouro. Neste momento, o vértice dois realiza uma mudança de reversibilidade em função de sua demanda $d_2$, que é igual a três. Esta mudança de reversibilidade se deu da seguinte forma:
$$r_1 = \frac{mmc(d_1,d_2)}{d_1} = \frac{6}{2} = 3$$
$$r_2 = \frac{mmc(d_1,d_2)}{d_2} = \frac{6}{3} = 2$$.
Com estes novos valores de reversibilidade, o resultado do cálculo do número de arestas para o arco $e_{12}$ foi o seguinte:
$$e_{12} = r_1 + r_2 - mdc(r_1, r_2) = 3 + 2 - mdc(3, 2) = 3 + 2 - 1 = 4$$.
Dessa forma, o arco $e_{ij}$ passou a ter quatro arestas. Após isto, o período se torna estável e o algoritmo SMER continua sua execução normalmente.

Para finalizar a apresentação do referencial teórico desta tese, a próxima seção apresenta um breve estudo sobre escalonamento em sistemas flexíveis de manufatura.

\section{Escalonamento em Sistemas Flexíveis de Manufatura \textit{Job-Shop}}

Esta seção tem como objetivo apresentar o referencial teórico sobre escalonamento em sistemas flexíveis de manufatura. Nas próximas subseções, portanto, são apresentados os seguintes assuntos: sistemas flexíveis de manufatura, escalonamento em sistemas flexíveis de manufatura do tipo \textit{job-shop} e heurísticas para geração de regras de despacho.

\subsection{Sistemas Flexíveis de Manufatura}

Sistemas Flexíveis de Manufatura podem ser definidos como sistemas de fabricação formados por estações de trabalho que compartilham sistemas de transporte e controle, objetivando a produção de um determinado espectro de produtos, sem que haja a necessidade de interromper o processo de produção para uma reconfiguração.

Segundo \citet{Browne:1984}, Sistemas Flexíveis de Manufatura podem ser divididos em duas categorias: flexibilidade de máquina e flexibilidade de roteamento. A primeira categoria consiste na capacidade que o sistema tem de mudar, de modo que novos tipos de produtos sejam produzidos, bem como na capacidade de alterar a ordem das operações em uma parte deste sistema. A segunda categoria consiste na capacidade que sistema tem de usar várias máquinas para executar as mesmas operações em uma parte deste sistema, bem como a capacidade de absorver grandes mudanças em termos de volume, capacidade e capabilidade de processo. Dentro desta classificação, o problema de orientação de rotas pode ser tratado como um problema de Sistemas Flexíveis de Manufatura com flexibilidade de roteamento. Em Sistemas Flexíveis de Manufatura, existem diferentes processos com uma variedade de produtos que podem ou não seguir a mesma rota. Quando todos os grupos de produtos percorrem uma mesma rota para chegar às estações de trabalho, o Sistema Flexível de Manufatura é chamado de sistema de produção contínua ou \textit{Flow Shop}. Diferente disto, quando cada tipo de produto tem sua própria rota, o Sistema Flexível de Manufatura é chamado de sistema de produção descontínua ou \textit{Job Shop}.

\subsection{Escalonamento em \textit{Job-Shops}}

O problema de escalonamento em sistemas flexíveis de manufatura do tipo \textit{job-shop} (Problema de Escalonamento \textit{Job-Shop}) pode ser descrito como um conjunto de $J$ de $n$ \textit{jobs} ($j_1, j_2, j_3, ..., j_n$), que é processado em um conjunto $M$ de $m$ máquinas ($m_1, m_2, m_3, ..., m_r$). O processamento do \textit{job} $j_i$ na máquina $m_r$ é chamado de operação $o_{ir}$. A operação $o_{ir}$ requer o uso exclusivo da máquina $m_r$ durante um período ininterrupto de tempo, que é o seu tempo de processamento $p_{ir}$. Uma agenda é um conjunto $C$ de tempos de conclusão $c_{ir}$ para cada operação. O tempo gasto para completar todos os $jobs$ é chamado de \textit{makespan} $L$. A solução ou a otimização para este tipo de problema tem como objetivo determinar uma agenda que minimize $L$.

\begin{figure}[t]
	\centering
    \includegraphics[scale=0.73]{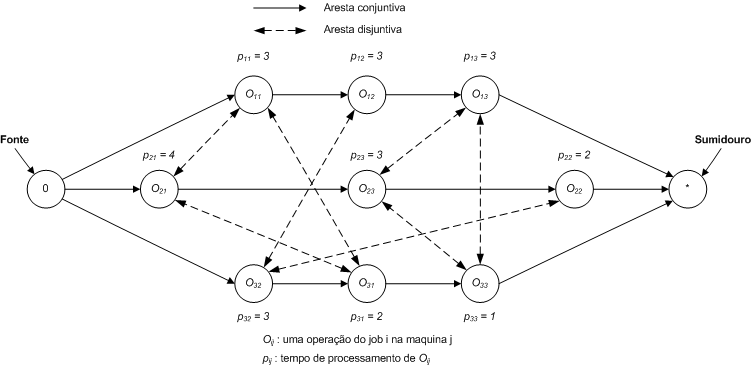}
    \caption{Grafo disjunto de um Problema \textit{Job-Shop} 3 $\times$ 3. Adaptado de \citet{Yamada:1996}.}
    \label{fig:jobshop}
\end{figure}

\begin{figure}[t]
	\centering
    \includegraphics[scale=0.6]{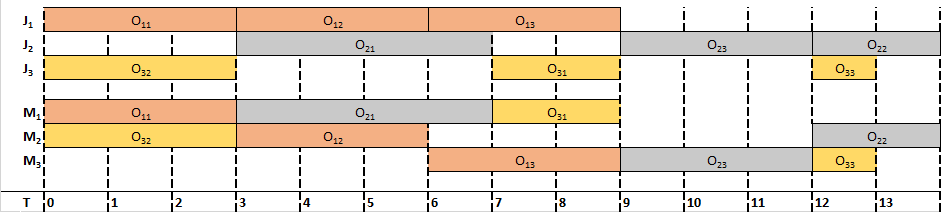}
    \caption{Mapeamento de um Problema \textit{Job-Shop} 3 $\times$ 3 em um Grafo de Gantt}
    \label{fig:jobshop_gantt}
\end{figure}

O Problema de Escalonamento \textit{Job-Shop} pode ser formalmente descrito por um grafo disjunto $G = $($V, C \cup D$), onde $V$ é o conjunto de vértices, $C$ é o conjunto de arcos conjuntivos e $D$ é o conjunto de arcos disjuntivos. A Figura \ref{fig:jobshop} ilustra um grafo com dez operações (em três \textit{jobs}) e três máquinas. Os vértices de $G$ correspondem às operações (O vértice origem ($0$) e o sumidouro ($*$) são as operações de início e fim, respectivamente). Os arcos conjuntivos de $C$ representam a sequências de máquinas das operações. Os arcos disjuntivos de $D$ representam os pares de operações, que devem ser realizadas na mesma máquina. Para melhor ilustrar o mapeamento das operações nas máquinas, a Figura \ref{fig:jobshop_gantt} apresenta o funcionamento o escalonamento de operações de um Sistema Flexível de Manufatura do tipo \textit{Job-Shop} por meio de um gráfico de Gantt.

O escalonamento de um \textit{job-shop} pode ser visto como uma definição de uma ordenação entre todas as operações que devem ser processadas na mesma máquina, fixando precedências entre estas operações. No modelo de grafo disjunto, isto é feito, tornando os arcos não direcionados (disjuntivos) em direcionados. Uma seleção é um conjunto de arcos direcionados selecionados com base nos arcos disjuntivos. Por definição, uma seleção é completa, se todas as disjunções são selecionadas. Além disto, uma seleção é consistente, se o grafo direcionado resultante é acíclico. Dessa forma, uma seleção consistente define uma família de agendas e cada agenda pertence exatamente a uma determinada família. Por fim, o \textit{makespan} ótimo para uma seleção consistente é o maior caminho no grafo direcionado, que é gerado a partir do grafo disjunto. Portanto, o problema consiste em encontrar uma seleção consistente que minimize o tamanho de um caminho mais longo no grafo direcionado gerado com base grafo disjunto.

\subsection{Heurísticas para Geração de Regras de Despacho}

O Problema de Escalonamento \textit{Job-Shop} está entre os problemas de otimização combinatória mais difíceis. Ele não somente é um problema do tipo NP-completo, mas, entre os problemas desta classe, ele está entre os piores. De acordo com a literatura, é possível resolver problemas do cacheiro viajante gerados aleatoriamente com um número de cidades entre 300 e 400 ou problemas de cobertura de conjunto com centenas de restrições e milhares de variáveis, mas não é possível escalonar de maneira ótima dez \textit{jobs} em dez máquinas. Uma vez que o escalonamento \textit{job-shop} é um problema prático muito importante, é natural olhar para métodos de aproximação, a fim de produzir o escalonamento de \textit{jobs} em um tempo aceitável.

A maioria dos métodos heurísticos de escalonamento \textit{job-shop} descritos na literatura são baseados em regras de despacho. Tais regras de despacho são baseadas em critérios, dentre eles: tempo de processamento e data de entrega. 

As regras de despacho baseadas em tempo de processamento têm como características a escolha de \textit{jobs} com o ciclo de processamento menor ou maior e também com o menor ou maior número de tarefas ainda não concluídas. Tais regras são as seguintes:

\begin{itemize}
	\item \textbf{\textit{Shortest Processing Time} (SPT):} escolhe a operação mais rápida, em outras palavras, com o menor tempo de processamento;
    \item \textbf{\textit{Largest Imminent Operation Time} (LI):} escolhe a operação com o maior tempo de processamento;
    \item \textbf{\textit{Most Work Remaining} (MWR):} seleciona o produto com a maior soma de operações ainda não efetuadas. 
    \item \textbf{\textit{Least Work Remaining} (LWR):} seleciona o produto com a menor soma de operações ainda não efetuadas. 
    
\end{itemize}

As regras de despacho baseadas na data de entrega são as seguintes:

\begin{itemize}
	\item \textbf{\textit{Earliest Due Date} (EDD):} seleciona a operação cujo produto tem a data de entrega mais cedo;
    \item \textbf{\textit{Minimum Slack Time} (MST):} seleciona a operação que tem a menor folga. A folga é a diferença entre o tempo que falta para a data de entrega e tempo necessário para executar as operações restantes;
    \item \textbf{\textit{Earliest Operation Due Date} (ODD):} seleciona a operação que o tempo de término mais próximo;
    \item \textbf{\textit{Operation Slack Time} (OST):} seleciona a operação que tem a menor folga.
\end{itemize}

Outras regras de despacho podem ser encontradas na literatura de escalonamento job-shop. No entanto, elas não serão apresentadas aqui, uma vez que esta tese usa somente duas das regras de despacho citadas acima, que são: SPT e EDD.

Aqui, se dá o fim da apresentação do referencial teórico desta tese. No entanto, existem algumas considerações que precisam ser feitas, de modo que fique claro a relação de todo o referencial com o restante desta tese. Portanto, a próxima seção tem esta finalidade.

\section{Considerações Finais}\label{sec:consideracoes_finais_cap_2}

Este capítulo apresentou o assuntos pertinentes ao referencial teórico desta tese, que são: redes \textit{ad hoc} veiculares; sistemas inteligentes de transporte; fundamentos de engenharia de tráfego; agentes e sistemas multiagentes; escalonamento por reversão de arestas; e escalonamento em sistemas flexíveis de manufatura. Cada um destes assuntos teve como objetivo o fornecimento de um embasamento conceitual e teórico acerca das temáticas e trabalhos relacionados que envolvem os problemas atacados por esta tese, que são: controle de tráfego por meio de sinalização semafórica, planejamento e orientação de rotas e comunicações heterogêneas em ambientes veiculares.

As soluções para os problemas de controle de tráfego por meio de sinalizações semafóricas e planejamento e orientação de rotas para motoristas dependem de comunicação com baixa latência e alta confiabilidade, sendo esta escalável e de baixo custo de mensagens. Para tanto, esta tese se apropriou dos conceitos e características das redes \textit{ad hoc} veiculares, levando em consideração as tecnologias de acesso à comunicação sem fio e técnicas de roteamento, a fim de propor uma rede veicular, que, por meio de seu protocolo de comunicação, satisfaça os requisitos de comunicação das categorias de aplicações para redes \textit{ad hoc} veiculares \cite{Willke:2009,Zheng:2015}.

Tendo em vista que esta tese contribui também para área de sistemas inteligentes de transporte, foi necessário compreender os conceitos básicos e a importância de tais tipos de sistemas inteligentes. Este breve estudo contribuiu para compreensão do termo sistemas inteligentes de transporte, assim como, as suas principais áreas de aplicação. Assim, esta tese apropriou-se dos conceitos relacionados aos sistemas avançados de gerenciamento de tráfego e sistemas avançados de informações ao motorista, a fim de propor tanto um mecanismo de controle de tráfego quanto um mecanismo para planejamento e orientação de rotas. Este último mecanismo, por sua vez, tira proveito das agendas de intervalos de indicações de luzes verdes geradas com base nas configurações de controle das sinalizações semafóricas, a fim de orientar o motorista, no que diz respeito a escolha da melhor rota para o destino desejado por este \cite{Rillings:1991}.

Devido aos estudos relacionados aos sistemas inteligentes de transporte, tornou-se imprescindível um estudo a respeito dos fundamentos de engenharia de tráfego. Este estudo forneceu um conhecimento sobre a estrutura de uma rede de tráfego, que foi de grande importância durante a confecção de mapas viários durante a criação de cenários de uso para execução de experimentos. Além disto, o estudo também forneceu um conhecimento sobre os parâmetros macroscópicos e microscópicos. Alguns destes parâmetros foram utilizados como medidas de desempenho durante a execução dos experimentos de simulação presentes nesta tese. 

Devido a natureza distribuída e a cooperação entre as entidades de mundo real, em específico, veículos e sinalizações semafóricas, inerentes aos mecanismos de controle de tráfego e orientação de rotas fez-se necessário a busca de um paradigma de sistemas de computação que pudesse servir como base teórica para a especificação destas entidades de mundo real na forma de entidades computacionais, que não somente realizam algum tipo de tarefa, mas também cooperam umas com as outras. Sendo assim, esta tese apropriou-se dos conceitos e características relacionados ao paradigma de agentes e sistemas multiagentes, para especificar duas aplicações de sistemas inteligentes de transporte como sistemas multiagentes, uma vez que este tipo de sistema está relacionado à área de Inteligência Artificial Distribuída.

A fim de atacar o problema de controle ótimo de tráfego por meio de sinalizações semafóricas, esta tese lança a mão sobre as teorias e algoritmos relacionados à técnica de escalonamento por reversão múltipla de arestas. O escalonamento por reversão múltipla de arestas tem um papel relevante nesta tese, pois ele não somente foi utilizado para controlar o acesso às interseções, mas também foi utilizado para geração e compartilhamento das agendas dos tempos das fases das sinalizações semafóricas. Além disto, esta técnica também é utilizada por esta tese, no que diz ao controle de interseções compartilhadas entre corredores contendo redes de sinalizações semafóricas, que, por sua vez, precisam ter os seus tempos dos inícios de seus intervalos de indicação de luzes verdes sincronizados, a fim de facilitar o escoamento de pelotões de veículos que precisam trafegar ao longo dos seguimentos de via componentes destes corredores. Tal técnica também é utilizada em um dos trabalhos relacionados a esta tese. 

Por fim, para atacar o problema de orientação de rotas, foi necessário buscar um modelo que pudesse não somente fornecer um arcabouço teórico, mas também um arcabouço prático, que pudesse fornecer subsídios para o desenvolvimento de um procedimento heurístico, cujo intuito não somente é calcular rotas ótimas, mas também alocar espaços de uso por parte dos veículos nas vias que os conduzem ao longo de um conjunto de interseções de uma rede viária. O problema de orientação de rotas em uma rede viária é similar ao problema de planejamento de rotas de veículos guiados automaticamente em sistemas flexíveis de manufatura do tipo \textit{job-shop} \cite{Perez:2010}. Por este motivo, esta tese faz uso das teorias e heurísticas de escalonamento em sistemas flexíveis de manufatura do tipo \textit{job-shop}.

Após o levantamento bibliográfico acerca do referencial teórico desta tese, a pesquisa avançou na direção do levantamento do estado da arte em métodos de controle de tráfego por meio de sinalizações semafóricas, orientação de rotas e redes \textit{ad hoc} veiculares centradas em informação. Portanto, o próximo capítulo descreve o levantamento bibliográfico a respeito do estado da arte de cada um destes tópicos.

  \chapter{Estado da Arte}

Este capítulo tem como objetivo apresentar o estado da arte acerca dos seguintes assuntos: controle de tráfego, orientação de rotas e redes ad hoc veiculares centradas em informação. O estado da arte de cada um destes assuntos é apresentado como seguem as próximas seções. 
Por fim, são apresentadas as considerações finais acerca do que foi apresentado neste capítulo, destacando lacunas que precisam ser preenchidas em cada um dos assuntos supracitados.

\section{Controle de Tráfego por meio de Sinalizações Semafóricas}

Tendo em vista os possíveis modos de operação de sinalizações semafóricas, o modo de operação baseado em computadores merece bastante atenção, uma vez que permite o emprego de métodos computacionais, permitindo, assim, o projeto e o desenvolvimento de sistemas avançados de gerenciamento de tráfego. Analisando a bibliografia de controle de sinalização semafórica por meio de computadores, é possível destacar trabalhos em duas linhas metodológicas:  métodos baseados em otimização \cite{Zheng:2011,DeOliveira:2006,Cheng:2006,Migdalas:1995} e métodos baseados em adaptação \cite{Bazzan:2009, Kerner:2012,Xie:2011a,Xie:2011b,Gerhenson:2005,Einhorn:2012,Slager:2010,Helbing:2009,Lammer:2008}.

Os métodos baseados em otimização tentam alcançar uma solução ótima que satisfaça as restrições impostas a um espaço de busca, que é formado por planos de temporização e fases de sinalizações semafóricas. Para tanto, segundo a literatura, o problema de controle ótimo de tráfego poder ser modelado como problema de programação linear mista, game-playing, problema de satisfação de restrições distribuídas ou problema de programação linear de dois níveis. Apesar disto, todos os métodos baseados em otimização compartilham as mesmas deficiências. Tais métodos são computacionalmente caros e, por isto, eles demandam uma grande quantidade de tempo para convergir soluções ótimas. Vale ressaltar, que o tempo para que estes métodos convirjam soluções ótimas é influenciado pelo tamanho da rede semafórica e a configuração de cada uma das interseções presentes em tal rede. Portanto, o sistema de controle de tráfego requer uma estrutura computacional dedicada compatível com o tamanho e a configuração da rede semafórica. Além disto, os métodos de otimização são cegos diante de situações anormais de tráfego, como, por exemplo, muitos veículos chegando ou saindo de um determinado lugar (estádio, shopping, centros econômicos, entre outros) ao mesmo tempo. Tendo em vista que o volume de tráfego muda constantemente, métodos baseados em adaptação tendem a ser melhores que os métodos baseados em otimização.

Neste sentido, \citet{Gerhenson:2005} propõem que o controle ótimo de tráfego por meio de sinalizações semafóricas não é necessariamente um problema de otimização, uma vez que o volume de tráfego muda constantemente. Portanto, o controle ótimo de tráfego por meio de sinalizações semafóricas deve ser tratado como um problema de adaptação. Dentre os métodos baseados em adaptação, destacam-se os que se baseiam em aprendizado por reforço e os que se baseiam em auto-organização. Embora os métodos baseados em adaptação por meio do aprendizado por reforço possam ajustar os tamanhos de ciclos, intervalos e fases das sinalizações semafóricas, eles necessitam de um grau de estabilidade por parte dos volumes de tráfego, a fim de possibilitar a construção de uma base de conhecimento acerca dos padrões macroscópicos do tráfego. Sendo assim, tais métodos não funcionam bem em redes viárias cujas vias apresentam constante variação no volume de tráfego. No que tange os métodos baseados em adaptação por meio de auto-organização, eles são mais promissores que os métodos baseados em adaptação por meio de aprendizado por reforço, uma vez que eles conseguem ajustar os tamanhos de ciclos, intervalos e fases em tempo real. Para realizar o controle ótimo de sinalizações semafóricas, os métodos baseados em adaptação por meio de auto-organização levam em consideração a formação de comboios de veículos ao longo das vias que participam das interseções sinalizadas, uma vez que veículos normalmente trafegam juntos e aproximadamente na mesma velocidade. Desta forma, um grupo de veículos tende a se comportar como um comboio coerente, devido às limitações para ultrapassagem. Neste sentido, os comboios de veículos não são importantes somente para as sinalizações semafóricas pertencentes às vias em que eles estejam trafegando, mas também para as sinalizações semafóricas pertencentes às interseções vizinhas, de modo que elas possam operar de maneira síncrona e coordenada umas com as outras. A consequência disto é que os métodos baseados em adaptação por meio de auto-organização conseguem compactar os veículos em comboios, à medida que estes trafegam ao longo da rede viária controlada por sinalizações semafóricas. Além disto, eles conseguem maximizar a velocidade média destes comboios, aumentando, a taxa de vazão de veículos do sistema de controle de tráfego. 

Apesar de os métodos baseados em adaptação por meio de auto-organização serem os mais promissores dentre os métodos apresentados até aqui, eles compartilham uma deficiência com os demais métodos, ou seja, todos os métodos são sensíveis às escolhas de rotas por parte dos motoristas dos veículos. Em outras palavras, tanto métodos baseados em otimização quanto métodos baseados em adaptação podem ter seus resultados comprometidos à medida que haja um desbalanceamento severo na densidade das vias. Neste caso, as vias podem ter um aumento severo de densidade, devido às rotas escolhidas pelos motoristas dos veículos ou uso de sistemas de navegação baseados em mapas estáticos, que, por não terem ciência de rotas alternativas, acabam escolhendo rotas que levam ao aumento da concentração de veículos em determinadas vias, enquanto outras ficam subutilizadas ou até mesmo ociosas. Ainda que tais métodos possam aumentar o tamanho dos ciclos e ajustar os tamanhos dos intervalos de luzes verde das sinalizações semafóricas, estas operações não devem ultrapassar  os limites recomendados de tempo, que são definidos pelas especificações técnicas para o funcionamento de sinalizações semafóricas, uma vez que valores muito altos de tamanhos de ciclo implicam em tempos de espera elevados \citet{DENATRAN:2014}. Sendo assim, é necessário que o fluxo de veículos seja distribuído, a fim de evitar congestionamentos.

\section{Planejamento e Orientação de Rotas}

Para dissipar rapidamente os fluxos de veículos e evitar congestionamentos, veículos precisam ser uniformemente distribuídos ao longo de rotas alternativas. Para tanto, é preciso que um mecanismo tenha conhecimento das escolhas de rotas dos veículos que trafegam ao longo de uma rede viária, de modo que os motoristas possam ser orientados a seguir um caminho que os levem aos destinos pretendidos por seus motoristas, sem que um congestionamento seja causado. Um importante componente em sistemas avançados de informações ao motorista, que é responsável por isto, é o mecanismo de planejamento e orientação de rota. Embora existam soluções comercialmente disponíveis para sistemas avançados de informações ao motorista com orientação de rotas, que conseguem utilizar informações dinâmicas sobre o volume tráfego (tais como: \abbrev{TMC}{Traffic Message Channel} \cite{TMC:2016}, TomTom \cite{TomTom:2016}, Waze \cite{Waze:2016}, entre outros), elas não são capazes de antecipar congestionamentos. Tais soluções reagem somente após os congestionamentos terem ocorrido e, com isso, acabam propagando os efeitos destes sobre uma parte substancial da rede viária. Isto se deve aos altos intervalos de atualização do estado do tráfego, que variam entre 2 a 30 minutos \cite{Wedde:2013}. 

Para tratar as deficiências descritas acima, novas técnicas para orientação de rotas têm sido propostas. Analisando a bibliografia pertinente ao tema, é possível destacar duas categorias, a saber: técnicas de orientação de rotas usando redes \textit{ad hoc} veiculares \cite{Kraus:2008,Leontiadis:2011,Schunemman:2009} e técnicas de orientação de rotas baseadas em sistemas multiagentes \cite{Wedde:2013,Claes:2011}. 

Nas técnicas de orientação de rotas usando redes ad hoc veiculares, informações detalhadas a respeito das condições de tráfego são propagadas por meio de redes \textit{ad hoc} veiculares. Com base neste tipo de rede, alguns trabalhos merecem destaque devido aos seus resultados, tais como: \citet{Kraus:2008}, \citet{Leontiadis:2011}, \citet{Schunemman:2009}, \citet{Lakas:2010}. Em \citet{Kraus:2008} e \citet{Leontiadis:2011}, os autores usaram em redes \textit{ad hoc} veiculares uma técnica de redes \textit{ad hoc} móveis conhecida como \textit{gossiping} para propagar informação de tráfego. Em \citet{Kraus:2008}, os resultados das simulações mostraram que somente entre 20\% a 30\% dos veículos tiveram seus tempos de viagem similares a aqueles cujas informações de tráfego foram obtidas a partir de uma fonte centralizada de informações de tráfego. Apesar disto, esta proposta apresentou problemas de escalabilidade. Á medida que o número de nós realizando \textit{gossiping} aumenta, também aumenta o tempo de viagem dos veículos que tentaram seguir a mesma rota alternativa. Percebe-se, então, que os nós não têm conhecimento do estado global acerca do volume de tráfego presente na rede viária por onde eles trafegam. Em \citet{Leontiadis:2011}, os resultados mostraram que 64\% dos veículos que realizam \textit{gossiping} tiveram uma redução nos seus tempos de viagem. No entanto, devido à falta de coordenação, os veículos podem congestionar rotas alternativas. Em \citet{Schunemman:2009}, cada veículo é capaz de informar sua velocidade média aos outros veículos em sua vizinhança e sua rota de viagem com informações a respeito das condições do tráfego. Com esta abordagem, foi possível reduzir o tempo de viagem em 50\%, utilizando uma taxa de penetração de veículos com capacidade de comunicação igual a 80\%. No entanto, esta abordagem faz uso intensivo de comunicação veículo-a-veículo. Em \citet{Lakas:2010}, cada veículo é capaz de requisitar informações sobre rotas, utilizando comunicação veicular. Com base nestas informações, os veículos são capazes de selecionar as rotas menos congestionadas. Observando os resultados obtidos em \citet{Lakas:2010}, percebe-se que a proposta ajuda a reduzir congestionamentos.  Analisando essas técnicas de orientação de rotas, usando redes \textit{ad hoc} veiculares, a conclusão a respeito é que redes \textit{ad hoc} veiculares facilitam a propagação de informações de tempo real acerca das condições de tráfego. No entanto, é importante frisar que aplicações para redes \textit{ad hoc} veiculares, cujo objetivo é propagar, manter e processar informações de tempo real a respeito das condições de tráfego de uma rede viária, possuem requisitos de comunicação específicos, conforme descrito em \citet{Willke:2009}. Neste sentido, é necessária uma atenção especial, no que tange a arquitetura de rede e o protocolo de comunicação adotados \cite{Shen:2014}. Por fim, para que técnicas de orientação de rotas possam distribuir uniformemente os fluxos de tráfego sobre uma rede viária, são necessárias decisões coordenadas entre os motoristas. Dentro desta perspectiva, técnicas de orientação de rotas baseadas em sistemas multiagentes são mais adequados para o problema de orientação de rotas e demandam coordenação entre as entidades constituintes do tráfego (tais como: veículos, sinalizações semafóricas e dispositivos de controle).

Com base nisto, é possível destacar as principais técnicas de sistemas multiagentes para o problema de orientação de rotas, a saber: abordagens bioinspiradas \cite{Nartz:2010,Claes:2011,Wedde:2013} abordagens baseadas em alocação de recursos \cite{Vasirani:2011,Vasirani:2012,Desai:2013,Shen:2014}. Em algumas abordagens bioinspiradas, os veículos são modelados como formigas, que depositam uma substância química chamada feromônio para marcar o caminho mais curto entre o formigueiro e a fonte de comida. A partir desta técnica, é possível identificar quais vias estão sendo mais utilizadas para escoar o tráfego, uma vez que o feromônio identifica a frequência com que uma determinada via está sendo usada. Dessa forma, o mecanismo de orientação de rotas selecionará uma rota com a menor quantidade possível de feromônio. Sempre que um veículo passa por uma interseção, o processo de busca pela melhor rota se repete, adotando a interseção como formigueiro \cite{Nartz:2010,Claes:2011}. Embora \citet{Wedde:2013} tenham proposto uma abordagem bioinspirada baseada em enxame de abelhas, tal abordagem é similar à abordagem proposta por \citet{Claes:2011}. No entanto, o algoritmo de roteamento proposto por \citet{Wedde:2013} é um algoritmo probabilístico, enquanto o algoritmo de roteamento proposto por \citet{Claes:2011} é um algoritmo proativo e utiliza um mecanismo de reserva com intuito de evitar ciclos. Além disto, ambas as abordagens foram projetadas, levando em consideração uma rede \textit{ad hoc} veicular cujo modelo de comunicação adotado é o veículo-infraestrutura. No entanto, os autores dessas abordagens não detalharam as configurações de rede adotadas durante os experimentos de simulação. Por fim, os resultados de ambos os trabalhos mostram ganhos nos tempos de viagem, quando comparados com motoristas normais e aqueles motoristas que fazem uso de sistemas de navegação baseados em serviços de TMC. No que tange as abordagens baseadas em alocação de recursos, estas consistem a partir de agentes de infraestrutura, que reservam espaços nas interseções para agentes motoristas, ou puramente de agentes veículos, que trocam preferências de rotas, de modo que este possam estabelecer negociações virtuais sucessivas e, com isto, tomar cooperativamente a melhor decisão para alocação de rotas para cada agente. Dessa forma, agentes motoristas escolhem as rotas baseados no custo da reserva do espaço e ou preferências pessoais de tempos de viagem. Em \citet{Vasirani:2011,Vasirani:2012}, os gerenciadores de interseções concorrem pelo fornecimento de reservas. Os agentes motoristas participam na alocação da capacidade da rede viária por meio de um mecanismo combinatório baseado em leilão. A avaliação experimental mostrou 30\% na redução do atraso para os motoristas que submeteram ofertas de alto valor. Em \citet{Shen:2014}, um centro de gerenciamento de tráfego detecta eventos que demandam o roteamento de veículos. Uma vez detectados tais eventos, agentes de infraestrutura são notificados. Tais agentes de infraestrutura são sinalizações semafóricas. Em seguida, as sinalizações semafóricas enviam alertas de re-roteamento para o primeiro veículo de cada uma das vias de entrada das interseções em que estão localizadas. Após isto, os veículos checam, se existe a necessidade de requerer uma nova alocação de rota. Se verdadeiro, os veículos requisitam às sinalizações semafóricas uma nova alocação de rota. Ao receber estas requisições, as sinalizações semafóricas calculam os caminhos ótimos para cada requisição recebida, levando em consideração os dados contidos nesta. Após o cálculo ótimo de rota, as sinalizações semafóricas enviam a rota para o veículo correspondente. Com base nas novas rotas recebidas, os motoristas seguem os caminhos que lhes são sugeridos, até que recebam um novo alerta de alguma sinalização semafórica. Em \citet{Shen:2014}, os resultados mostraram uma redução no tempo médio de viagem de 51.5\%, em um cenário baseado em grade manhattan, e 11.9\% em um cenário baseado em um mapa realístico. Em \citet{Desai:2013}, agentes veículos tomam decisões cooperativa de alocação de rotas, à medida que se aproximam das interseções nas vias em que trafegam. Cada agente veículo troca de maneira autônoma informações de preferência de rota calculada para chegar em uma alocação inicial de rotas. A alocação é melhorada usando um número de sucessivas negociações virtuais. Em \citet{Desai:2013}, os resultados mostraram que quando a abordagem é comparada com um algoritmo de caminho mínimo para melhorias no tempo de viagem, a abordagem oferece um ganho entre 21\% a 43\%, quando a demanda de tráfego está abaixo da capacidade da rede viária, e ganho entre 13\% a 17\%, quando a demanda de tráfego excede a capacidade da rede viária. Analisando as abordagens de sistemas multiagentes baseados em alocações recursos, \citet{Vasirani:2011,Vasirani:2012} e \citet{Shen:2014} o modelo de comunicação adotado foi o veículo-infraestrutura. Já, em \citet{Desai:2013}, o modelo de comunicação adotado foi o veículo-a-veículo. Embora essas abordagens tenham adotado modelos de comunicação diferentes, seus autores, assim como os das abordagens de sistemas multiagentes bioinspirados, também não detalharam modelos de pilhas de protocolo, mecanismos de endereçamento de nós e algoritmos de encaminhamento ou roteamento de mensagens adotados.

\section{Redes Ad Hoc Veiculares Centradas em Informação}

Esta seção apresenta o estado da arte em redes \textit{ad hoc} veiculares centradadas em informação, que foi apresentado em \citet{Goncalves:2016b}.

Muitos métodos têm sido propostos para facilitar a comunicação entre os nós de redes \textit{ad hoc} veiculares, mas o uso de modelos de pilhas de protocolos e esquemas de endereçamento, ambos projetados para redes centradas em IP (Internet Protocol), os tornam inadequados para ambientes veiculares altamente dinâmicos \cite{Yu:2013}. Por isto, Bai e Krishnamachari \cite{Bai:2010} têm argumentado a respeito de uma troca de paradigma de comunicação em redes veiculares. Para este fim, alguns pesquisadores têm identificado as redes centradas em informação (\textit{Information-Centric Network} -- ICN)\cite{Ahlgren:2012} como um paradigma chave, pois elas oferecem uma solução atrativa para ambientes móveis e altamente dinâmicos tais como as redes \textit{ad hoc} veiculares. Entre os modelos arquiteturais encontrados na literatura de ICN, as redes centradas em conteúdos (\textit{Content-Centric Networks} -- CCN) têm ganhado proeminência em trabalhos sobre redes veiculares  \cite{Arnould:2011,Amadeo:2012a,Amadeo:2013,Wang:2012a,Wang:2012b}.

Em \citet{Arnould:2011} aplicaram o modelo de redes centradas em conteúdos para disseminar informações críticas em uma VANET híbrida. Tal estudo propôs um mecanismo de entrega de dados chamado pacote de evento, que não requer o envio prévio de um pacote de interesse para que a entrega de dados aconteça. Desta forma, o \textit{publisher} detecta eventos críticos, usando sensores embutidos em um veículo e, em seguida, transmite pacotes de evento contendo informações relacionadas a eventos sensíveis a atraso, tais como informações de acidentes, alertas de segurança e avisos de colisões. Porém, os autores modificaram a arquitetura original da CCN, para controlar a disseminação de pacotes de evento de acordo com largura de banda demandada por estes pacotes. Este aumento de complexidade pode causar falhas na entrega de dados em ambientes de alta demanda.
    
Em \citet{Amadeo:2012a,Amadeo:2013} propuseram um framework para redes \textit{ad hoc} veiculares que tem como base a arquitetura de uma rede centrada em conteúdos. De acordo com \citet{Amadeo:2012b}, redes \textit{ad hoc} veiculares baseadas em redes centradas em conteúdos têm desempenho melhor que as redes \textit{ad hoc} veiculares baseadas em IP, quando se trata da transmissão de dados e do balanceamento de carga dos veículos na rede, além de sofrerem menos com a degradação de desempenho à medida que o volume de dados aumenta. Além disto, os autores dividiram os pacotes de interesse em dois subtipos, que são: interesse básico (B-Int) e interesse avançado (A-Int). Em ambos os trabalhos, B-Int foi usado quando um consumidor desejava descobrir fornecedores de conteúdo e requisitar o primeiro segmento de dados do conteúdo descoberto, enquanto o A-Int foi usado para requisitar os segmentos de dados restantes a partir dos fornecedores descobertos após o envio de B-Int. Os autores também introduziram uma nova estrutura de dados chamada de \textit{Content Provider Table} (CPT) para substituir a \textit{Forwarding Information Base} (FIB). A CPT foi usada para armazenar informações a respeito de fornecedores que já tenham sido descobertos e associa o endereço MAC destes nós com o conteúdo. Com esta mudança, os autores descartaram a principal premissa de uma rede centrada em conteúdo, que é a independência do conteúdo em relação a sua localização física. Tal ruptura conceitual pode levar ao comprometimento do suporte à mobilidade dos nós \cite{Prates:2014}.
    
Em \citet{Wang:2012a} apresentaram uma arquitetura de rede centrada em conteúdo, mas não consideraram os tipos de aplicações possíveis ou eventos que poderiam afetar significantemente sua arquitetura. Além disto, a arquitetura proposta não pode ser usada eficientemente em aplicações baseadas fortemente na comunicação veículo-a-veículo. Em \citet{Wang:2012b} também propuseram um mecanismo de disseminação de pacotes para reduzir a latência da entrega de conteúdo. Este mecanismo utilizou temporizadores para coordenar o envio de pacotes entre os nós da rede. No entanto, Wang et al. não definiram limites para a disseminação de interesses em uma área geográfica, tornando persistente o problema de inundação de pacotes de interesses \cite{Prates:2014}. Por fim, a avaliação feita pelos autores não considerou como a mobilidade dos nós impactaria o mecanismo proposto.

\section{Considerações Finais}

Após todo o levantamento bibliográfico acerca dos métodos de controle de tráfego e métodos de orientação de rotas, percebeu-se que tais métodos podem ser complementares uns aos outros. Sendo mais específico, métodos como os utilizados em sistemas multiagentes baseados em alocação de recursos podem tirar proveito dos tempos das fases das sinalizações semafóricas, ainda que estes possam ser alterados, de acordo com alguma política de adaptação de tamanhos de intervalos de luzes verdes em função dos fluxos de tráfegos em vias que incidem nas interseções. Dessa forma, um mecanismo de orientação de rotas pode distribuir o volume de tráfego sobre uma rede viária controlada por sinalizações semafóricas, fazendo com que veículos trafeguem em vias participantes de rotas ótimas, sem que estes precisem realizar paradas desnecessárias. Neste caso, a cooperação entre sistemas avançados de gerenciamento de tráfego e sistemas avançados de informações ao motorista permite que a orientação de rotas tenha ciência das programações das sinalizações semafóricas ao longo do tempo. 

No entanto, os métodos de controle de tráfego por meio de sinalização semafórica baseados em operação por computadores apresentados acima não são adequados para fornecer agendas de tempos a partir das programações das sinalizações semafóricas, de modo que estas possam ser consultadas por veículos à medida que estes se movem ao longo de uma rede viária. Embora os métodos de controle de tráfego por meio de sinalização semafórica baseados em operação por computadores não tenham levado em consideração o uso da comunicação entre veículos e sinalizações semafóricas, isto não é o fator determinante que indica a deficiência destes, quando se deseja criar uma cooperação entre um sistema avançado de gerenciamento de controle de tráfego e um sistema avançado de informações ao motorista. O uso de redes \textit{ad hoc} veiculares em conjunto com tais métodos de controle de tráfego por meio de sinalização semafórica não contribui necessariamente para que eles possam ser aproveitados em uma cooperação entre as duas categorias de aplicações de sistemas inteligentes de transporte, uma vez que o único benefício alcançado seria a substituição da coleta de dados macroscópicos de tráfego por meio de detectores de tráfego instalados nas vias por coleta de dados feita pelas sinalizações semafóricas à medida que estas recebem dados sobre os veículos que se aproximam de uma interseção. Sendo assim, as deficiências dos métodos de controle de tráfego por meio de sinalização semafórica baseados em operações por computadores consistem de suas próprias concepções.

No que tange os métodos de controle de tráfego baseados em otimização, estes necessitam de uma estrutura de computação centralizada e de alto custo para processamento das otimizações de fases de todo o sistema de controle de tráfego. Além disto, ainda que uma agenda de intervalos de indicações de luzes verdes fosse disponibilizada para consulta, esta seria utilizada na manutenção da alocação e desalocação de espaços nas vias de entrada das interseções, que seriam utilizados pelos veículos, à medida que um mecanismo de planejamento e orientação de rotas fornecesse rotas ótimas, de acordo com a necessidade de distribuição dos veículos sobre a rede viária. Isto gera um grande \textit{overhead} de comunicação e, de leitura e escrita de dados. Além disso, qualquer necessidade de ajuste no tamanho do intervalo de indicação de luz verde de uma única interseção, faz com que o sistema de controle de tráfego tenha que ajustar os tamanhos dos intervalos de indicação de luzes verdes de todas as sinalizações semafóricas pertencentes ao sistema de controle de tráfego e, em seguida, notificar os veículos sobre o evento de modificação das agendas de intervalos de indicações de luzes verdes. Como consequência, todos os veículos requisitarão novos cálculos de rotas ótimas, com o intuito de novamente tirar proveito dos novos tamanhos de intervalos de indicações de luzes verdes disponíveis no sistema de controle de tráfego. Neste instante, a estrutura centralizada pode se tornar um gargalo e, com isto, pode influenciar no tempo de resposta das requisições, fazendo com que o planejamento e orientação de rotas trabalhe com dados defasados, no que diz respeito das alocações dos espaços nas vias de entrada das interseções. Vale ressaltar que, enquanto os veículos esperam por rotas ótimas, eles se movem ao longo das vias da rede viária. Logo, se a latência de resposta de uma requisição para o cálculo de uma rota ótima é alta, os motoristas dos veículos podem não ser informados a tempo de realizar alguma manobra, podendo, então, ser necessário iniciar um processo de recálculo de rota ótima. 

Com relação aos métodos baseados em adaptação, estes se dividem em duas subcategorias:  métodos baseados em aprendizado por reforço e métodos baseados em auto-organização. Nestas duas categorias, as sinalizações semafóricas são capazes de coletar os dados por meio de detectores de tráfego instalados nas vias e, de tempos em tempos, realizar uma computação local, a fim de calcular os novos tamanhos de intervalos de indicações de luzes verdes das sinalizações semafóricas. Neste sentido, as decisões de ajustar as fases das sinalizações semafóricas são tomadas localmente. Por meio das decisões de todas as sinalizações semafóricas pertencentes ao sistema de controle de tráfego, a otimização do fluxo de tráfego é realizada de maneira descentralizada. Com base nessas sinalizações semafóricas, é possível construir uma agenda de intervalos de indicações de luzes verdes de maneira centralizada ou descentralizada. 

Para manter as agendas de intervalos de indicações de luzes verdes de maneira centralizada, é necessária a disponibilidade de uma infraestrutura computacional como a que é fornecida aos métodos baseados em otimização. Dessa forma, os métodos baseados em adaptação são afetados pelos mesmos problemas relacionados à infraestrutura computacional adotada pelos métodos baseados em otimização. Para contornar este problema, as agendas de intervalos de indicações de luzes verdes podem ser mantidas de maneira descentralizada. Desta forma, sinalizações semafóricas podem manter sua própria agenda de intervalos de indicações de luzes verdes ou, até mesmo, uma cópia da agenda de intervalos de indicações de luzes verdes de cada uma das sinalizações semafóricas do sistema de controle de tráfego. Assim, as requisições por cálculos de rotas ótimas devem se dá por meio da comunicação entre veículos e as sinalizações semafóricas. Se cada sinalização semafórica mantém sua própria agenda de intervalos de indicações de luzes verdes, ela precisa consultar outras sinalizações semafóricas para calcular uma rota ótima e, em seguida, devolvê-la como resposta de uma requisição enviada por um veículo. Neste cenário, o tempo de obtenção de uma rota é influenciado pelo tamanho da rede de sinalizações semafóricas. Assim, uma rede de sinalizações semafóricas de larga escala pode fazer com que o tempo de resposta não seja adequado e, por isto, os veículos podem receber informações equivocadas a respeito de rotas ótimas. Neste caso, essa estratégia de mantença das agendas de intervalos de indicações de luzes verdes pode levar aos mesmos problemas de latência apresentados na discussão relacionada aos métodos de controle de tráfego baseados em otimização. 

Uma opção a essa estratégia é fazer com que cada sinalização semafórica tenha uma cópia das agendas de intervalos de indicações de luzes verdes das demais sinalizações semafóricas do sistema de controle de tráfego. Neste sentido, a computação para calcular rotas ótimas é feita localmente nas sinalizações semafóricas que venham a receber requisições enviadas por veículos. Independente da estratégia adotada para a mantença de agendas de intervalos de indicações de luzes verdes, os veículos que trafegam ao longo de uma rede viária controlada pelo sistema de controle de tráfego precisam ser notificados sobre as mudanças destas agendas, de modo que eles possam requisitar novos cálculos de rotas ótimas. No entanto, se a estratégia adotada é a que cada sinalização semafórica mantenha somente a sua agenda de intervalos de indicações de luzes verdes, ela pode inviabilizar o planejamento e orientação de rotas, pois um grande \textit{overhead} de comunicação e, de leitura e escrita será demandado, uma vez que os veículos reagirão ao recebimento das notificações de mudanças nas agendas de intervalos de indicações de luzes verdes das sinalizações semafóricas, requisitando novos cálculos de rotas ótimas. Em um cenário como este, à medida que os veículos requisitam novos cálculos de rotas ótimas, uma execução do algoritmo de cálculo de rotas ótimas é realizada sobre toda a rede de sinalizações semafóricas. Tendo em vista o grande número de mensagens geradas durante a execução do algoritmo de cálculo de rotas ótimas, os tempos de respostas para cada uma das requisições enviadas pelos veículos podem ser afetados drasticamente, uma vez que uma instância de tal algoritmo é criada e executada, à medida que uma sinalização semafórica recebe requisições por cálculos de rotas ótimas enviadas pelos veículos. Consequentemente, toda a rede de sinalizações semafóricas é inundada por mensagens desse algoritmo. No caso do uso de uma estratégia de mantença de agendas de intervalos de indicações de luzes verdes, em que cada sinalização semafórica mantém sua própria agenda de tempos e cópias das de outras sinalizações semafóricas do sistema de controle de tráfego, todo o \textit{overhead} de comunicação é retirado, pois este tem suas instâncias executadas localmente nas sinalizações semafóricas. 

Tendo em vista que o planejamento e orientação de rotas auxilia na distribuição equilibrada do volume de tráfego ao longo de uma rede viária, as sinalizações semafóricas precisam manter um grau de estabilidade no que diz respeitos aos seus tempos de fases e, também, precisam adaptar-se a qualquer alteração no fluxo de tráfego. Com base nisto, as sinalizações semafóricas precisam atualizar suas agendas de intervalos de indicações de luzes verdes, conforme as modificações de seus planos de temporização, realizadas após a leitura e processamento dos dados relacionados aos fluxos de tráfego incidentes nas interseções controladas por elas.  Devido à incapacidade de se adaptarem facilmente às flutuações frequentes dos fluxos de tráfego, que é ocasionada pela necessidade de uma estabilidade do volume de tráfego para a construção de uma base de conhecimento acerca dos padrões macroscópicos dos fluxos de tráfego das vias, métodos de controle de tráfego baseados em adaptação por meio de aprendizado por reforço tendem a gerar agendas de tempos similares as que podem ser geradas por sinalizações semafóricas de tempos pré-fixados. Assim, qualquer mudança no padrão macroscópico de tráfego, que resulte em um padrão não existente na base de conhecimento de uma via, pode gerar um gargalo em uma parte específica da rede viária. Ainda que os métodos de controle de tráfego baseados em adaptação por meio de aprendizado por reforço possam criar um sistema totalmente descentralizado de controle de tráfego com baixo \textit{overhead} de operação, devido à baixa adaptabilidade em relação aos fluxos de tráfego com flutuações frequentes, tais métodos não são adequados para a construção de sistemas avançados de gerenciamento de tráfego, onde uma das expectativas é cooperar com sistemas avançados de informações ao motorista, no que tange o fornecimento de planejamento e orientação de rotas. 

Por outro lado, embora existam métodos de baseados em adaptação por auto-organização, estes também apresentam dificuldades, quando se deseja utilizá-los na construção de sistemas avançados de controle de tráfego com intuito de cooperar com sistemas avançados de informações ao motorista, compartilhando agendas de intervalos de indicações de luzes verdes das sinalizações semafóricas, de modo que um mecanismo de orientação de rotas possa utiliza-las para alocar espaços de uso nas vias de entradas das interseções. Nestes métodos, o controle das sinalizações semafóricas é programado para identificar as chegadas de pelotões de veículos nas interseções. Sendo assim, as agendas de intervalos de indicações de luzes verdes das sinalizações semafóricas precisam ser atualizadas a cada detecção de um pelotão de veículos. Isto faz com que o aproveitamento da agenda de tempos das programações das sinalizações semafóricas por parte de um mecanismo de planejamento e orientação de rotas seja comprometido. Por causa de uma grande variação de estado de tais agendas, um mecanismo de planejamento e orientação de rotas pode calcular uma rota ótima com base em cópias de agendas de intervalos de indicações de luzes verdes desatualizadas. Como consequência, isto pode levar a um desbalanceamento na distribuição do volume de tráfego sobre as vias de uma rede viária. Por fim, as sinalizações semafóricas tendem a sofrer com o alto \textit{overhead} de comunicação e, de leitura e escrita, à medida que recebem e processam requisições de novos cálculos de rotas ótimas e, em seguida, registram os veículos nas agendas de intervalos de indicações de luzes verdes, geradas com intuito de alocar espaços nas vias de entrada em cada uma das interseções pertencentes as rotas ótimas calculadas.

Embora as redes centradas em conteúdos sejam mais promissoras que os modelos centrados em IP em ambientes veiculares, existem algumas limitações quanto a adoção destas em projetos de aplicações para redes ad hoc veiculares, tais como a inundação da rede com pacotes de interesses cuja causa provém das políticas de encaminhamento de pacotes de interesses. Nestas, tais pacotes são encaminhados para todos os vizinhos de um nó, à medida que este os recebe. Isto possibilita o surgimento de \textit{broadcast storms}. Além disto, o modelo de uma rede centrada em conteúdos usa estruturas de dados semelhantes às tabelas de roteamento e adota algoritmos similares ao AODV \cite{Perkins:1994}, DSR \cite{Jhonson:1996}  e GPSR \cite{Karp:2000}. Tais algoritmos são vulneráveis em ambientes veiculares altamente dinâmicos devido a intermitência de caminho \cite{Yu:2013}. Por fim, embora existam estudos promissores no campo de redes veiculares, estes têm somente focado em cenários relacionados a serviços populares de dados compartilháveis \cite{Yu:2013}. Consequentemente, cenários em que aplicações precisam trocar um grande montante de dados sensíveis a atraso não têm sidos estudados.

Após o levantamento bibliográfico referente ao estado da arte de cada um dos assuntos relacionados a esta tese, o trabalho de pesquisa avançou na direção de trabalhos que pudessem fornecer subsídios para a estruturação da proposta de solução para os problemas anteriormente relatados na Seção \ref{sec:motivacao}. Para apresentar os trabalhos que contribuíram para a construção desta tese, o próximo capítulo apresentará os trabalhos relacionados a esta tese.

  \chapter{Trabalhos Relacionados}\label{cap:trabalhos_relacionados}

Este capítulo apresenta os trabalhos relacionados, que foram relevantes ao desenvolvimento desta tese. Portanto na Seção \ref{sec:dutra}, é apresentada de maneira breve os detalhes da rede \textit{ad hoc} móvel centrada em interesses desenvolvida por \citet{Dutra:2012,Dutra:2010}. Posteriormente, na Seção \ref{sec:lucio}, são apresentadas as abordagens desenvolvidas por \citet{Paiva:2012} para controlar e coordenar sinalizações semafóricas por meio de um sistema distribuído de controle de tráfego. Por fim, a Seção \ref{sec:flavio} apresenta o mecanismo desenvolvido por \citet{Faria:2013} cujo objetivo é rotear veículos ao longo de uma rede viária urbana, utilizando os dados obtidos a partir dos planos de temporização das sinalizações semafóricas, com o intuito de calcular rotas ótimas para os veículos.

\section{Rede \textit{Ad Hoc} Móvel Centrada em Interesses}\label{sec:dutra}

Esta seção tem como objetivo apresentar brevemente o trabalho desenvolvido por \citet{Dutra:2012,Dutra:2010}. Primeiramente, serão apresentadas as principais características da rede \textit{ad hoc} móvel centrada em interesses. Em seguida, são apresentada as principais estruturas de dados relacionadas ao trabalho de \citet{Dutra:2012,Dutra:2010}, que são:  o prefixo ativo e o cabeçalho de mensagens de rede. Por fim, é apresentado o protocolo de comunicação, que é executado pelos nós de rede à medida que recebem mensagens uns dos outros. 

\subsection{Principais Características}

A Rede \textit{Ad Hoc} Móvel Centrada em Interesses ou \textit{Inte\textbf{R}est-Centric \textbf{Ad}-Hoc \textbf{Net}work} (RAdNet) \cite{Dutra:2012,Dutra:2010} é uma rede \textit{ad hoc} móvel centrada em informação baseada no modelo Publicador/Subscritor (\textit{Publisher/Subscriber}) \cite{Buschmann:1996}. Dessa forma, na RAdNet, um nó publicador envia uma mensagem de rede com algum interesse para todos os nós da rede e, em seguida, os nós subscritores deste interesse específico recebem a mensagem. Este modelo de comunicação da RAdNet difere das abordagens tradicionais, uma vez que tais abordagens se baseiam em comunicação centrada em IP. No modelo de comunicação centrada em IP, os nós de origem precisam conhecer os endereços dos nós de destino, a fim de estabelecer comunicação fim-a-fim e descobrir rotas até tais nós. 

Segundo \citet{Dutra:2012,Dutra:2010}, o modelo Publicador/Subscritor foi adotado pela RAdNet com o intuito de tratar o problema das entradas e saídas frequentes de nós em uma rede \textit{ad hoc} móvel, pois, à medida que um nó se move, ele pode entrar ou sair dos alcances de transmissão dos nós próximos a ele. Devido à adoção desse modelo, a instabilidade de rede, causada por estas entradas e saídas frequentes do alcance de transmissão de nós vizinhos, não impacta a RAdNet, pois os nós não precisam conhecer a topologia de rede para enviar mensagens, fazendo com que o protocolo de comunicação da RAdNet difira da maioria dos protocolos de roteamento para redes \textit{ad hoc} móveis (por exemplo, DSDV \cite{Perkins:1994}, AODV \cite{Perkins:1999}, DSR \cite{Johnson:1996} e GPSR \cite{Karp:2000}). Além disto, na RAdNet, os nós não necessitam de um endereço para identificá-los unicamente, pois a comunicação é totalmente baseada em interesses.

Um interesse é qualquer termo (ou sequência de caracteres) que tem algum significado para aplicações, permitindo que a informação seja focada na aplicação e não em dispositivos \cite{Dutra:2012,Dutra:2010}. Segundo \cite{Dutra:2012,Dutra:2010}, o uso do interesse reduz o custo de mensagens, assim como, a necessidade de tabelas de roteamento, pois as mensagens com interesses não precisam de um destino único ou predeterminado. O encaminhamento de mensagens ocorre com base em um conjunto de campos probabilisticamente escolhidos, como em um protocolo \textit{gossip}, reduzindo a complexidade das decisões de encaminhamento \cite{Dutra:2012,Dutra:2010}. Na RAdNet, os nós não precisam determinar qual vizinho deve receber mensagens. Antes de encaminhar mensagens, cada nó toma sua própria decisão de encaminhamento de mensagem, usando critérios definidos em um filtro de casamento de dados.

\subsection{Estruturas de Dados da RAdNet}

\begin{figure}[t]
	\centering
    \includegraphics[scale=0.87]{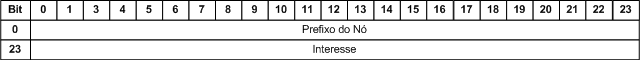}
    \caption{Especificação do Prefixo Ativo. Adaptado de \citet{Dutra:2012,Dutra:2010}.}
    \label{fig:prefixo_ativo_radnet}
\end{figure}

Na RAdNet, cada nó define seu Prefixo Ativo, que é uma estrutura de dados simples, implementada na camada de rede de cada nó. Esta, por sua vez, é dividida em dois campos, como pode ser visto na Figura \ref{fig:prefixo_ativo_radnet}. 
O primeiro campo é o Prefixo do Nó, que é composto por campos escolhidos probabilisticamente. O Prefixo do Nó é formado por $n$ campos com tamanho de $m$ bits, tal que os $n \times m$ bits fornecem identificação de nó para propósitos de endereçamento e filtragem de mensagens por meio de combinação de dados, a fim de encaminhar mensagens. Além disto, cada nó gera sua própria sequência de $n$ campos, onde, para cada campo, o nó associa um valor de $m$ bits usando uma variável aleatória com alguma distribuição de probabilidade. Os campos do Prefixo do Nó são utilizados para duas funções: encaminhamento probabilístico de mensagens e endereçamento. No que diz respeito ao encaminhamento probabilístico de mensagens, cada nó tem uma determinada probabilidade de encaminhar mensagens, de acordo com a construção da sequência de valores dos campos. No que tange o endereçamento, os nós são identificados por meio de uma sequência de valores dos campos do Prefixo do Nó. Por fim, o segundo campo é o Interesse, cuja função é manter o interesse de uma aplicação que esteja sendo executa pelo nó. Portanto, vale ressaltar que todo nó possui um mecanismo de registro de interesses. Tal mecanismo é utilizado sempre pelas aplicações, quando estas iniciam suas execuções, a fim de registrar os seus interesses na camada de rede nós que as mantêm.

O cabeçalho de mensagens da RAdNet é formado por sete campos, a saber: versão, limite de saltos, identificador da mensagem, prefixo do destino, prefixo da origem e o interesse. O campo versão contém a versão do protocolo de comunicação da RAdNet. O campo limite de saltos contém o número de saltos dados por uma mensagem, à medida que esta é encaminhada pelos nós ao longo da rede. O campo tamanho do cabeçalho contém um número inteiro que remente ao comprimento do cabeçalho de mensagens. O campo prefixo de destino contém identificador do destinatário da mensagem, podendo ser um valor correspondendo ao prefixo de um nó ou nulo. O campo prefixo de origem contém o prefixo do nó que envia a mensagem. Por fim, o campo interesse contém um interesse, que é enviado por um nó publicador para os nós subscritores deste.

\begin{figure}[t]
	\centering
    \includegraphics[scale=0.87]{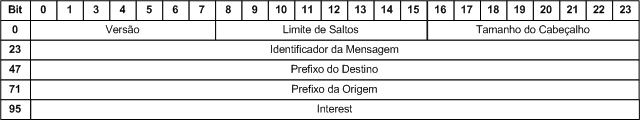}
    \caption{Especificação do cabeçalho de mensagens da RAdNet. Adaptado de \citet{Dutra:2012,Dutra:2010}.}
\end{figure}

\subsection{Protocolo de Comunicação da RAdNet} \label{subsec:rp}

Na RAdNet, cada nó executa um protocolo de comunicação, que é o \textit{RAdNet Protocol} (RP). Este protocolo permite que os nós encaminhem mensagens, de acordo com o resultado do casamento de dados entre seus prefixos ativos e os valores de campos específicos dos cabeçalhos das mensagens recebidas. Estes campos específicos são o prefixo de origem e o interesse. O Algoritmo \ref{alg:protocoloradnet} descreve os processos envolvidos no encaminhamento de uma mensagem da RAdNet.

\begin{algorithm}[!t]
  \SetKwData{Msg}{Msg}
  \SetAlgoLined
  	\footnotesize\Entrada{msg$_j$}
  	
  	\eSe{$msg_j.id \in tabelaIds_{i}[msg_{j}.prefixoOrigem]$} {
	  \textbf{Descartar} msg$_j$\;
  	}{
	  \textbf{Inserir} msg$_j$.id \textbf{em} tabelaIds$_{i}$[msg$_{j}$.prefixoOrigem]\;
	  
      msg$_j$.limiteSaltos := msg$_j$.limiteSaltos - 1\;
	  casInteresses := msg$_{j}$.interesse $\in$ tabelaInteresses$_{i}$\;
	  casPrefixos := $|$prefixo$_i \cap$ msg$_j$.prefixoOrigem$| > 0$  $\vee$ msg$_j$.prefixoDestino =  \textbf{nul}\;
	  
	  \Se{casInteresses = \textbf{verdadeiro}}{
	    \Se{$msg_{j}.prefixoDestino$ = \textbf{nulo} $\vee msg_{j}.prefixoDestino = prefix_i$}{
	      \textbf{Enviar } uma cópia $msg_j$ \textbf{para} aplicação\;
	    }
	  }
	  
	  \eSe{casPrefixos = \textbf{verdadeiro} $\vee$ casInteresses = \textbf{verdadeiro}}{    
	      \eSe{msg$_j$.limiteSaltos $>$ 0} {
		  \textbf{Enviar } msg$_j$ \textbf{para} \textbf{todos} $vizinhos_i$\;
	      }{
			\textbf{Descartar} msg$_j$\;
	      }     
	  }{
	    \textbf{Descartar} msg$_j$\;
	  }
	  
  	}
  \caption{Protocolo de comunicação da RAdNet. Obtido de \citet{Dutra:2012,Dutra:2010}.}
  \label{alg:protocoloradnet}
\end{algorithm}

De acordo com o Algoritmo \ref{alg:protocoloradnet}, o receber uma mensagem ($msg_j$) de qualquer vizinho $j$, ele checa se o nó $i$  já recebeu esta mensagem. Caso já tenha recebido, o nó $i$ descarta $msg_j$. Caso contrário, o algoritmo registra o valor do identificador da mensagem e o prefixo de origem da mensagem na tabela de identificadores ($tabelaIds_i$) do nó $i$. Em seguida, o algoritmo reduz em um o limite de saltos de $msg_j$. Após isto, o algoritmo realiza a filtragem de mensagens. Primeiramente, ele checa se a tabela de interesses do nó $i$ ($tabelaInteresses_i$) possui uma entrada igual ao interesse da mensagem recebida ($msg_j.interesse$). Após esta checagem, ele também checa se o prefixo do nó $i$ ($prefixo_i$) e o prefixo de origem da mensagem recebida ($msg_j.prefixoDestino$) têm um ou mais pares de campos com o mesmo valor ou, se o valor de $msg_j.prefixoDestino$ é igual a nulo. Caso o resultado da primeira checagem dê verdadeiro, o algoritmo checa se o nó $i$ é o destino de $msg_j$. Se verdadeiro, o algoritmo cria uma cópia de $msg_j$ e, em seguida, a encaminha para aplicação associada ao interesse da mensagem recebida. Finalmente, se o resultado da primeira checagem (linha 6) ou o resultado da segunda checagem (linha 7) for verdadeiro e o limite de saltos da mensagem ($msg_j.limiteSaltos$) for maior que zero, o nó $i$ encaminha $msg_i$ para todos os seus vizinhos. Senão, $msg_j$ é descartada.

\section{Controle Distribuído de Tráfego com Fluxos Heterogêneos}\label{sec:lucio}

A seguir, serão apresentados sucintamente os detalhes do trabalho de \citet{Paiva:2012}. Primeiramente, é apresentado as dinâmicas de funcionamento de um controlador inteligente de sinalização semafórica. As dinâmicas de funcionamento deste controlador têm como base as dinâmicas do algoritmo de escalonamento por reversão múltipla de arestas (SMER) \cite{Barbosa:2001}. Por fim, em seguida, é apresentada uma abordagem para o problema de coordenação de redes de sinalização semafórica. Assim como no controlador inteligente de sinalizações semafóricas, \citet{Paiva:2012} aplica também as dinâmicas do SMER com o intuito de solucionar tal problema.

\subsection{Controlador Inteligente de Sinalizações Semafóricas}

\begin{figure}[!t]
	\centering
	\subfigure[]{
		\includegraphics[width=10cm]{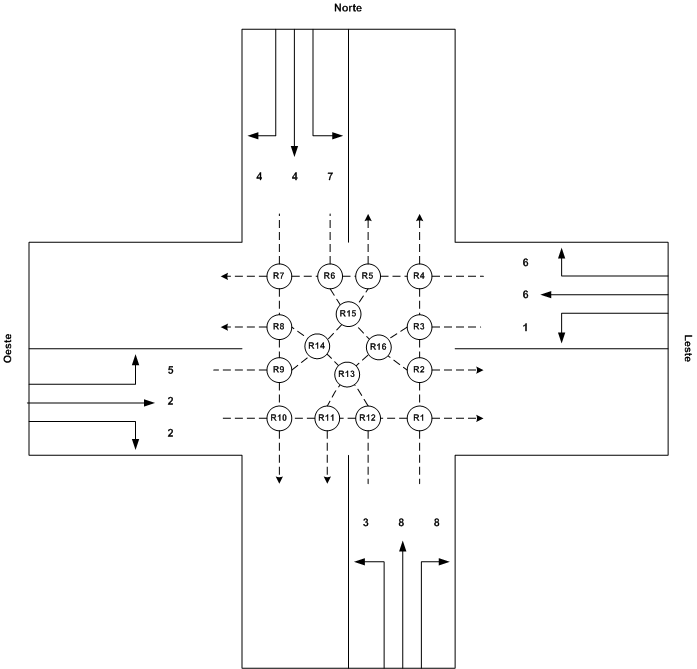}
    	\label{fig:pontos_conflito}
	}	
    \quad
    \subfigure[]{
		\includegraphics[width=10cm]{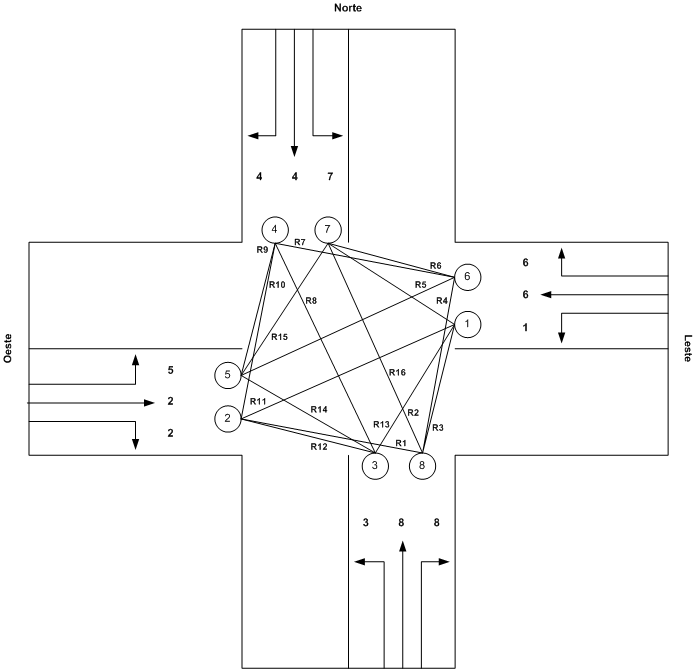}
    	\label{fig:grafo_recursos}
	}

    \caption{Modelagem de uma interseção de vias como um conjunto de recursos compartilhados entre os movimentos das vias de entrada da interseção.(a) pontos de conflito entre os movimentos da conflitantes da interseção; (b) mapeamento dos pontos de conflito na forma de recursos.}
    \label{fig:modelagem_intersecao}

\end{figure}

O controlador proposto por \citet{Paiva:2012} é capaz de responder sensivelmente às flutuações dos fluxos de tráfego que incidem nas interseções por meio de suas vias de entrada. Para tanto, inicialmente, \citet{Paiva:2012} se apropria de uma modelagem proposta por \citet{Soares:2007}, que trata uma interseção de vias como um recurso compartilhado entre fluxos conflitantes de tráfego oriundos das vias de entrada desta interseção. Dessa forma, tais fluxos são considerados processos que necessitam de acesso ao recurso compartilhado, quando estes estão em execução. Por fim, o trabalho de \citet{Paiva:2012} estende o controlador proposto por \citet{Soares:2007}, que consistiu na substituição do algoritmo SER, utilizado por \citet{Soares:2007} em seu trabalho, pelo SMER.

\begin{figure}[!t]
	\centering
	\subfigure[]{
		\includegraphics[width=6.5cm]{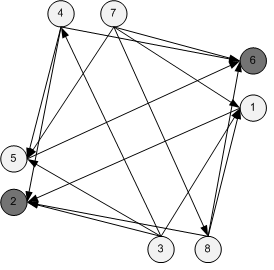}
    	\label{fig:grafo_ser}
	}	
    \quad
    \subfigure[]{
		\includegraphics[width=6.5cm]{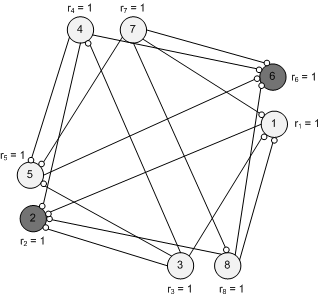}
    	\label{fig:grafo_smer}
	}

    \caption{Grafos criados com base no mapeamento de pontos de conflito na forma de recursos: (a) um exemplo de um grafo criado para o algoritmo SER; (b) um exemplo de um multigrafo para o algoritmo SMER. Ambos os exemplos apresentam grafos com orientações iniciais acíclicas.}
    \label{fig:modelagem_intersecao}

\end{figure}

Assim como em \citet{Soares:2007}, para cada movimento de uma interseção é criado um vértice no grafo (veja \ref{fig:grafo_recursos}), a fim de representar a incidência dos fluxos de tráfegos das vias de entrada da interseção como um processo. Com base nisto, os pontos de conflito entre os movimentos da interseção (veja Figura \ref{fig:pontos_conflito}) são mapeados na forma de recursos (veja \ref{fig:grafo_recursos}), cujos compartilhamentos são representados por arestas, ligando um par de vértices. De acordo com \citet{Soares:2007}, quando um vértice é um sumidouro, os veículos do movimento representado por tal vértice ganham a permissão para acessar a interseção, uma vez que a sinalização semafórica passa a indicar a luz verde. A Figura \ref{fig:grafo_ser} apresenta um exemplo de orientação inicial acíclica, de acordo com a abordagem proposta por \citet{Soares:2007}. Uma vez que \citet{Paiva:2012} usa o SMER, o grafo resultante do mapeamento é um multigrafo, conforme ilustrado na Figura \ref{fig:grafo_smer}. Assim como na Figura \ref{fig:grafo_ser}, a Figura \ref{fig:grafo_smer} apresenta um exemplo de uma orientação inicial acíclica para o multigrafo gerado com base no mapeamento dos pontos de conflito na forma de recursos.

Conforme \citet{Paiva:2012}, no início da operação de sua proposta de controlador inteligente de sinalização semafórica, o controlador ainda não possui informações acerca das demandas das vias de entrada da interseção. Por isto, cada vértice do multigrafo começa com reversibilidade e demanda igual a um. Assim que as sinalizações semafóricas começam a operar, elas começam a extrair dados oriundos dos detectores de tráfego instalados na entrada das vias que estas estão instaladas. Uma vez que as demandas de cada vértice são iniciadas com o valor um, todos os fluxos de tráfego oriundos das vias de entrada das interseções acessam estas com taxas de operação iguais. 

À medida que os sensores acumulam dados sobre os fluxos de tráfego das vias em que eles estão instalados, as sinalizações semafóricas, após um intervalo de tempo, recolhem tais dados, a fim de obter as demandas destas vias. À medida que as sinalizações semafóricas obtêm as demandas de suas vias, os vértices que as representam compartilham as mesmas com os demais vértices do multigrafo.

Uma vez que os vértices têm conhecimento das demandas das vias de entrada, elas passam a ter os dados necessários para mudar suas reversibilidades, para que, em seguida, os intervalos de tempos das indicações das sinalizações semafóricas sejam reconfigurados, a fim de refletir as novas condições de tráfego. As reversibilidades são ajustadas à medida que os vértices se tornam sumidouros, conforme o algoritmo para mudanças de reversibilidades proposto por \citet{Santos:2012}. Para tanto, primeiramente, a sinalização semafórica precisa calcular localmente a sua demanda e as das demais sinalizações da interseção. Assim, a demanda de um vértice $i$ é dada por
\begin{equation}
	d_i = \left\lfloor 100 \times \frac{D_i}{\sum_{j \in V}} + 0,5 \right\rfloor
\end{equation}, onde $D_i$ é a demanda do vértice $i$ e $D_j$ é a demanda de um vértice vizinho $j$. 

Com isto, os ciclos das sinalizações semafóricas podem ser ajustados em função das flutuações dos fluxos de tráfego oriundos das vias de entrada das interseções. Além disto, é também importante ressaltar que este controlador inteligente de sinalizações semafóricas proposto por \citet{Paiva:2012}, foi projetado para operar sobre sinalizações semafóricas de uma interseção isolada. Sendo assim, deve ficar a cargo de um mecanismo inteligente a coordenação de redes de sinalizações semafóricas. Tal mecanismo é proposto por \citet{Paiva:2012} e, por isto, é apresentado na seção seguinte.

\subsection{Coordenador de Redes de Sinalizações Semafóricas}

Uma rede de sinalizações semafóricas é formada com base em um corredor, que é uma sequência de seguimentos de vias, por onde os fluxos de tráfego são escoados. À medida que os fluxos de tráfego são escoados ao longo de um corredor, os veículos atravessam uma série de interseções sinalizadas. Por isto, é desejável que os fluxos de tráfego fluam ao longo de corredores com sinalizações semafóricas, sem que estes sejam interrompidos desnecessariamente, em outras palavras, sem paradas em sinalizações indicando luz vermelha em seus grupos focais.

Neste sentido, segundo a proposta de \citet{Paiva:2012}, a composição das redes de sinalizações semafóricas precisa ser conhecida, ou seja, quais as interseções e suas respectivas sinalizações semafóricas compõem um corredor. Além disto, as sinalizações semafóricas participantes de uma ou mais redes de sinalizações semafóricas, precisam ter ciência de quais redes elas participam. Isto é importante, pois o mecanismo de coordenação de redes de sinalizações semafóricas precisa calcular o \textit{offset} entre sinalizações semafóricas instaladas em interseções vizinhas. Para tanto, o mecanismo precisa de dados a respeito das vias componentes de um corredor, tais como: limite de velocidade, comprimento e sentido do tráfego. Por fim, sugere-se que um sensor seja instalado na via de entrada de cada corredor, a fim de fornecer dados que permitam o cálculo das demandas de cada corredor.

\begin{figure}[!t]
	\centering
    \subfigure[]{
    	\includegraphics[width=15cm]{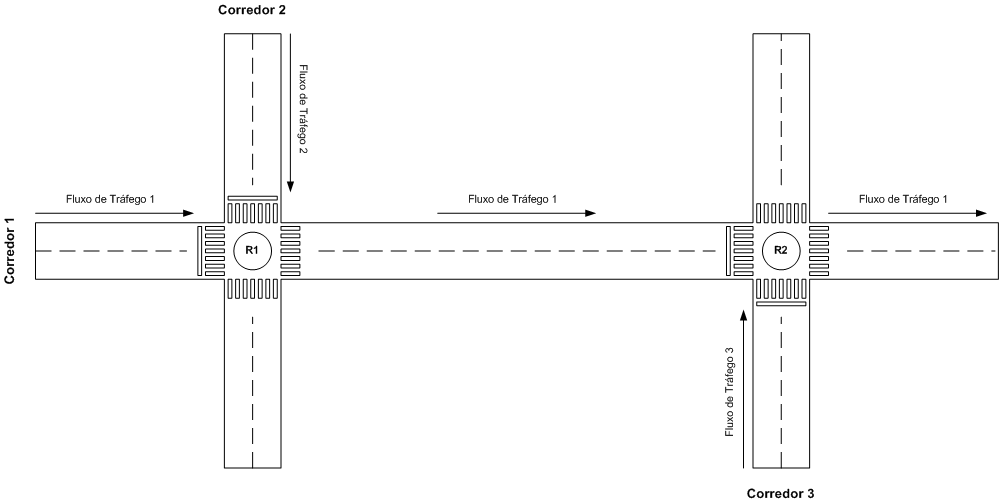}
        \label{fig:corredores_recursos}
    }
    \quad
    \subfigure[]{
    	\includegraphics[width=6.5cm]{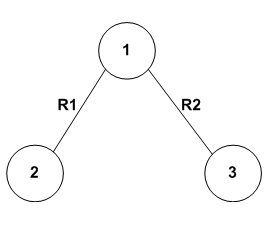}
        \label{fig:grafo_corredores_recursos}
    }
    \quad
    \subfigure[]{
    	\includegraphics[width=6.5cm]{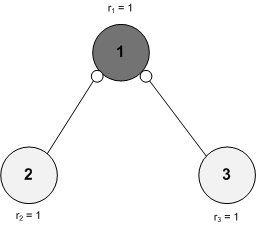}
        \label{fig:grafo_smer_corredor}
    }
    \caption{Controle de coordenação de redes de sinalizações semafóricas: (a) corredores e interseções como recursos compartilhados; (b) grafo criado inicialmente para mapear os corredores, na forma de vértices, e interseções, na forma de arestas, representando o compartilhamento das interseções entre os corredores; (c) multigrafo criado para controlar os corredores de acordo com as demandas de tráfego.}
    \label{fig:corredores_recursos}
\end{figure}

Com isto, o sistema de coordenação de redes semafóricas dá início a sua operação, executando um algoritmo distribuído para prospecção de corredores. O início da execução deste algoritmo se dá com base nas sinalizações semafóricas líderes de suas redes, ou seja, aquelas sinalizações semafóricas instaladas nas vias de entrada de um corredor. Por sua vez, tais sinalizações semafóricas atuam como controladores dos corredores em que elas participam. Segundo \citet{Paiva:2012}, uma sinalização semafórica só pode ser líder de um único corredor. Com base na execução do algoritmo de prospecção de corredores, tais sinalizações semafóricas buscam pontos de interseção entre seus corredores e os demais corredores da rede viária em elas operam. A Figura \ref{fig:corredores_recursos} apresenta um exemplo com três corredores com duas interseções compartilhadas entre eles. Ao encontrar uma interseção, onde dois corredores se cruzam, o algoritmo cria uma aresta entre os vértices que controlam os corredores, que, neste caso, representam as interseções líderes. Esta aresta é criada, a fim de garantir o acesso mutuamente exclusivo por parte dos fluxos de tráfego, que competem um com o outro, para atravessar o ponto de interseção entre seus corredores. Por fim, após o grafo para controle dos corredores ter sido construído (veja Figura \ref{fig:corredores_recursos}), uma sinalização semafórica executa um algoritmo \textit{Alg-Colour}, a fim de gerar um estado inicial acíclico para o grafo (veja Figura \ref{fig:grafo_smer_corredor}). Após isto, este estado inicial passa a ser conhecido por todas as interseções líderes de corredores. 

Em seguida, cada vértice do multigrafo inicia um processo de ajuste dos relógios das sinalizações semafóricas pertencentes ao corredor. Para tanto, o vértice toma a si como ponto de partida, usando o seu relógio como início referencial do tempo de operação da coordenação da rede de sinalização semafórica. Iniciando as sinalizações cujas interseções estão a jusante, ou seja, interseções que receberão o fluxo de veículos da interseção em que a sinalização semafórica líder está instalada, o \textit{offset} de cada uma é dado por

\begin{equation}
	t_d - \frac{d}{v}
\end{equation}, onde $d$ é o distância entre o par de interseções, $v$ é o limite de velocidade do segmento de via e $t_d$ é o tempo que se leva para percorrer o segmento de via. 

Uma vez calculado o tempo de viagem, os \textit{offsets} sinalizações semafóricas das interseções a jusante serão fixados com o valor atual da sinalização semafórica (que, no caso inicial, é a líder e vale zero) mais o valor de $t_d$ \cite{Paiva:2012}. Após isso, a interseção líder se volta para as sinalizações semafóricas instaladas nas interseções imediatamente a montante, ou seja, aquelas interseções que enviam fluxos de veículos para a interseção em que a sinalizações semafórica líder está instalada. Para cada uma, o \textit{offset} é calculado por meio da base referencial local (no caso inicial, zero) menos $t_d$. Dessa forma, a sinalização semafórica líder encerra a primeira geração de \textit{offsets}, ajustando os relógios locais das sinalizações semafóricas das interseções vizinhas. Após isto, cada uma das sinalizações semafóricas que tiveram seus relógios locais ajustados inicia uma nova geração de \textit{offsets}, tendo como base o ajuste de seu relógio local. Uma vez que uma sinalização semafórica já tenha tido o seu relógio local ajustado, esta não participa da geração de \textit{offsets} \cite{Paiva:2012}. Todo esse processo se repete recursivamente, até que todas as sinalizações existentes em corredor tenham sido iniciadas. 

Após o término das gerações de \textit{offsets}, segundo \citet{Paiva:2012}, o sistema coordenado estará pronto para iniciar, tendo como base o estado inicial acíclico gerado anteriormente por uma das sinalizações semafóricas líder de algum corredor, que é representada por um dos vértices do multigrafo. Uma vez que cada vértice representa um corredor, os corredores, cujos vértices são sumidouros, terão seus intervalos de onda verde, a fim de escoar os fluxos de tráfego que os atravessam. Ao final do intervalo de tempo da onda verde, os vértices revertem suas arestas, de modo que os demais vértices do multigrafo se tornem sumidouros. Desta forma, os fluxos de tráfegos dos demais corredores ganham permissão de atravessar as interseções compartilhadas, sem que tenham os seus movimentos cessados temporariamente por uma sinalização semafórica com sua luz vermelha acesa.

De acordo com \citet{Paiva:2012}, assim como no controlador inteligente de sinalizações semafóricas, cada vértice representante de um corredor pode redefinir suas reversibilidades. Para tanto, \citet{Paiva:2012} propôs separar os corredores em grupos de corredores concorrentes. Com base nestes grupos, extrai-se a maior demanda e esta, por sua vez, passa a valer para todos os vértices representantes destes corredores. Com base nisto, \citet{Paiva:2012} aplica o algoritmo de \citet{Santos:2012} para redefinir as reversibilidades dos vértices do multigrafo. 

Por fim, é possível que interseções presentes nos corredores tenham vias de entrada que não fazem parte de outro corredor. Neste caso, as sinalizações semafóricas destas interseções devem possuir dois controladores, um primário e outro secundário. O controlador primário faz com que as sinalizações semafóricas destas interseções operem, contribuindo com a coordenação da rede de sinalizações semafóricas do corredor que elas fazem parte. O controlador secundário faz com que as sinalizações semafóricas operem, como se elas estivessem controlando as vias de entrada de uma interseção isolada. De acordo com \citet{Paiva:2012}, o controlador primário tem prioridade sobre o controlador secundário. Dessa forma, o controlador primário entra em operação, quando as sinalizações semafóricas de uma interseção precisam cooperar com a coordenação da rede de sinalizações semafóricas do corredor, cujo vértice que o representa é um sumidouro. Logo após o vértice representante do corredor deixar de ser um sumidouro, o controlador primário entrega o controle das sinalizações semafóricas para o controlador secundário. 

\section{Roteamento Ecológico de Veículos Orientados a Ondas Verdes}\label{sec:flavio}

Nesta seção, são apresentados brevemente os detalhes acerca do trabalho desenvolvido por \citet{Faria:2013}. Inicialmente, é descrito um algoritmo de cálculo ótimo de rotas que tira proveito dos planos de temporização das sinalizações semafóricas, a fim de determinar a rota temporalmente mais curta entre uma interseção de origem até uma interseção de destino. Após isto, é apresentado um mecanismo de alocação de espaços nas vias de entrada de interseções sinalizadas. Este mecanismo é utilizado para estender o algoritmo de cálculo de rotas ótimas apresentado inicialmente. Por fim, é apresentada uma extensão do modelo \textit{Intelligent Driver Model}. Por meio desta extensão, veículos podem alterar seus comportamentos em função da presença de ondas verdes nas vias em que eles estão trafegando.

\subsection{Predição de Fases de Sinalizações Semafóricas}

Segundo \citet{Faria:2013}, a orientação de rotas pode tirar vantagem dos intervalos de tempo das indicações das sinalizações semafóricas. À medida que um veículo viaja através de uma sequência de segmentos de via cujas sinalizações semafóricas operam em sincronia, uma série de benefícios são obtidos, são eles: tempos mais curto de viagem, paradas menos frequentes, diminuição da formação de filas durante o intervalo de tempo em que as sinalizações semafóricas mantêm suas luzes vermelhas acesas, redução nos níveis de consumo de combustível e nas emissões de gases e material particulado.

Em um ambiente em que todos as sinalizações semafóricas de um sistema de controle de tráfego são pré-temporizadas e as suas temporizações são conhecidas, é possível planejar rotas, a fim de evitar segmentos de via, onde as sinalizações semafóricas estejam indicando luz vermelha. Dessa forma, um veículo, ao entrar em uma interseção, ele pode avaliar quanto tempo será necessário para alcançar a próxima interseção à sua frente. De maneira otimista, o tempo mínimo a trafegar através do segmento de via e alcançar a próxima interseção pode ser obtido com base na divisão do comprimento do seguimento de via pelo limite de velocidade do mesmo. Caso este tempo mínimo esteja dentro do intervalo de verde da próxima sinalização semafórica, o motorista pode manter a velocidade do seu veículo igual a permitida do seguimento de via. Caso contrário, ou seja, a sinalização semafórica esteja indicando luz vermelha, o motorista não terá outra opção senão esperar a próxima indicação de luz verde. Após atravessar a interseção, o veículo deverá seguir por uma das vias de saída desta e, em seguida, alcançar a próxima interseção. Por fim, escolha da próxima interseção a partir do tempo mínimo para alcançá-la.

Com base na temporização das sinalizações semafóricas pré-temporizadas, o trabalho de \citet{Faria:2013} constrói rotas baseadas nos tempos factíveis para atravessar os segmentos de via que ligam pares de interseções. Para tanto, \citet{Paiva:2012} fez uso do algoritmo de caminho mais curto de Dijskstra, ponderando as arestas (segmentos de via ligando um par de interseções) por meio dos valores de tempo mínimo para o alcance de uma próxima interseção. 

Além de sinalizações pré-temporizadas, \citet{Faria:2013} também fez uso de sinalizações semafóricas controladas por meio do controlador proposto por \citet{Paiva:2012}. Neste caso, \citet{Faria:2013} leva em consideração os intervalos de tempo entre as reprogramações das sinalizações semafóricas. Entre uma reprogramação e outra, as sinalizações semafóricas se comportam como se fossem pré-temporizadas, permitindo que o algoritmo para cálculo de rotas ótimas, apresentado nesta seção, seja executa sobre uma rede viária controlado por sistema de controle de tráfego cujas sinalizações semafóricas utilizam o controlador inteligente proposto por \citet{Paiva:2012}.

\subsection{Planejamento Cooperativo de Rotas e Sensível à Demanda}

Uma vez que os veículos são capazes de planejar as suas rotas, pode surgir um desbalanceamento na distribuição do volume de tráfego sobre uma rede viária. Para resolver este problema, \citet{Faria:2013} propõe um mecanismo de alocação de espaço nas vias controladas por sinalizações semafóricas. Assumindo que os níveis de saturação das vias são conhecidos globalmente, o trabalho de \citet{Faria:2013} permite que um veículo determine o atraso que lhe será imposto durante o planejamento de rota e, por meio disto, evite vias que lhe aumentarão o tempo de viagem e o tempo de espera atrás de uma sinalização semafórica com a luz vermelha acesa.

Assim, durante o cálculo de rotas ótimas, o algoritmo reserva espaços nos segmentos de via escolhidos para compor a rota ótima. Esta reserva de espaço é visível para todo o sistema, a fim de permitir que os demais veículos saibam quantos veículos já reservaram espaços nos segmentos de via durante o planejamento de suas rotas. Com base nisto, o algoritmo de cálculo de rotas ótimas de \citet{Faria:2013} determina qual o seguimento de via que terá o maior atraso em função da quantidade de espaços alocados por veículos. Este tempo é adicionado ao tempo mínimo factível para se alcançar a interseção pretendida. Uma vez que a quantidade de espaços alocados varia para cada segmento de via, o algoritmo de cálculo de rotas ótimas de \citet{Faria:2013} uso deste dado para calcular o atraso, que se dá por
\begin{equation}
	t_d = w \times q^{f}_{r}
\end{equation}, onde $w$ é uma parâmetro de regulagem e $q^{f}_{r}$ é a quantidade de veículos que reservaram espaços no seguimento de via. 

Como mencionado anteriormente, a construção de rotas parte de premissas otimista acerca das condições de tráfego da rede viária. Em cenários realísticos, nem sempre é possível que um veículo trafegue na velocidade máxima do seguimento de via, uma vez que os seguimentos de vias são divididos com outros veículos. Por isto, é comum que os veículos encontres veículos mais lento a sua frente. Para tratar este problema, \citet{Faria:2013} propôs que o veículo, ao entrar em cada interseção, recalcule sua rota, tomando como ponto de partida a interseção. 

\subsection{\textit{Intelligent Driver Model} e as Ondas Verdes}

De acordo com \citet{Faria:2013}, o comportamento do motorista ou, até mesmo, de veículos autônomos pode ser ajustado para que as temporização dos sejam aproveitadas ao calcular a aceleração do veículo. Desta fora, veículos podem atravessar várias interseções, aproveitando o intervalo de verde das em que estão trafegando, sem que aconteçam paradas desnecessárias. Para atingir este objetivo, \citet{Faria:2013} estendeu o \textit{Intelligent Driver Model} (IDM) \cite{Treiber:2000}.

No trabalho de \citet{Faria:2013}, os veículos se esforçam para viajar nas ondas verdes sempre que possível, permitindo que eles aproveitem as temporizações das sinalizações semafóricas para se movimentarem, a fim de reduzir as variações de velocidade e aceleração. Neste sentido, os veículos buscam se posicionar o mais próximo possível da cabeça da onda, de modo que haja espaço na cauda da onda, para que outros veículos possam alcançá-la e, em seguida, se juntarem, formando um pelotão. Aquele veículo que entra na onda verde a partir de sua cabeça, ele deve reduzir sua velocidade, a fim de que a onda verde o alcance. Quando a onda verde alcança o veículo, os veículos que se movimentam sobre ela são empurrados para trás, permitindo a acomodação do novo veículo na onda. Por outro lado, se um veículo deseja alcançar uma verde, primeiramente, ele precisa verificar se a calda da onda verde pode ser alcançada. Segundo \citet{Faria:2013}, uma onda verde é alcançável, quando o veículo pode chegar até a sua calda antes de entrar em uma interseção. Dadas a velocidade, aceleração máxima e posicionamento do veículo juntamente com a velocidade e o posicionamento da onda verde, \citet{Faria:2013} descobre o momento em o veículo e a onda se encontrarão de acordo com a Equação \ref{eq:encontro_onda_verde}.

\begin{equation}
	\centering
  	\begin{split}
    	s_{0_v} + v_v t = s_{0_w} + v_w t \\ 
    	s_{0_v} + (v_{0_v} + a_v t) = s_{0_w} + (v_{0_w} + a_w t)
  	\end{split}
    \label{eq:encontro_onda_verde}
\end{equation}

Uma vez que a onda verde se move em uma velocidade constante, sua aceleração máxima $a_w$ é nula. Dessa forma, simplificando a Equação \ref{eq:encontro_onda_verde}, tem-se
\begin{equation}
	a_v t^2 + (v_{0_v} - v_w)t + s_{0_v} - s_{0_w} = 0
    \label{eq:simplificacao}
\end{equation}, onde $t$ é o instante em que o veículo e a onda verde se encontrarão, $a_v$ é a aceleração calculada pelo IDM, $v_{0_v}$ é a velocidade instantânea do veículo, $v_w$ é a velocidade constante da onda verde, $s_{0_v}$ é a posição instantânea da frente do veículo e $s_{0_w}$ é a posição instantânea da cauda da onda verde.

Em \citet{Faria:2013}, a aceleração calculada do IDM é usada no lugar do limite de aceleração do veículo na Equação \ref{eq:simplificacao}, de modo que seja garantido que o veículo responderá ao comportamento do veículo à frente. Isto, segundo \citet{Faria:2013}, garante que não haja colisões entre veículos. 

Com base na Equação \ref{eq:simplificacao}, a posição em que o veículo e a onda verde se encontrarão se dá por
\begin{equation}
	s = a_v t^2 + v_{0_v} t + s_{0_v}
\end{equation}. 
No entanto, segundo \citet{Faria:2013}, é importante verificar se o veículo não excederá o limite de velocidade da via, à medida que este acelera a $a_v m/s^2$ durante $t$ segundos. A velocidade final do veículo durante este trajeto pode ser dada por
\begin{equation}
	v_v = v_{0_v} + a_v t
\end{equation}. Caso esta velocidade seja superior ao limite da via, é necessário calcular o instante em que a velocidade máxima é atingida. Este instante é dado por
\begin{equation}
	t' = \frac{v_r^{max} - v_{0_v}}{a_v}
\end{equation}. A posição do veículo no instante $t'$ é dada por
\begin{equation}
	s'_v = a_v t^2 + v_{0_v} t' + s_{0_v}
\end{equation}, assim como, a posição da cauda da onda verde é dada por 
\begin{equation}
	s'_w = s_{0_w} + v_w t'
\end{equation}. 
Portanto, o tempo em que o veículo viajará na velocidade máxima é dado por 
\begin{equation}
	t'' = \frac{s'_w - s'_v}{v_{r}^{max} - v_w}
\end{equation}. 
Baseado nisto, o instante em que o veículo se encontrará com a cauda da onda verde é dado por $t' + t''$, que corresponde ao tempo em que o veículo levará para chegar até a velocidade máxima da via somado ao tempo em que ele viajará na velocidade constante $v_r^{max}$. A onda verde somente será considerada alcançável, se o encontro entre o veículo e ela acontecer antes que a cauda desta onda atinja a interseção. Isto é dado por meio da Inequação 
\begin{equation}
	t' + t'' < \frac{d_{w,i}}{v_w}
\end{equation}, 
onde $d_{w,i}$ é a distância entre a cauda da onda verde e a interseção à frente, que para onde a onda verde se movimenta. Por fim, uma vez que o veículo esteja dentro da onda verde, o veículo acelerará confortavelmente, se for preciso, para se juntar ao pelotão de veículos formado a partir da cabeça da onda verde e, nela, se manterá.

Na situação em que não houver nenhuma onda verde com a possibilidade de ser alcançada, o veículo terá que esperar pela próxima indicação de luz verde para cruzar a sinalização semafórica à frente. Por isto, o veículo usa dados do plano de temporização da sinalização semafórica à sua frente, a fim de calcular a sua velocidade e sua aceleração. O objetivo disto é obter uma taxa de desaceleração confortável, a fim de minimizar a probabilidade de parada do veículo atrás de uma faixa de retenção de uma via de entrada de uma interseção, enquanto este espera pela próxima indicação de luz verde na sinalização semafórica instalada sobre tal via.

Então, seja $b$ a taxa de frenagem definida na equação do IDM \cite{Treiber:2000} , $a_v^{max}$ o nível máximo de aceleração do veículo e $d_{v,i}$ a distância entre o veículo e a interseção. O tempo mínimo de viagem até a interseção, em condições de tráfego livre, é dado por 
\begin{equation}
	(a_v^{max} - b) t^2 + v_{0_v} t + d_{v,i} = 0
\end{equation}. 
Com base na equação acima, é possível verificar a velocidade final do veículo, à medida que ele mantém a aceleração $a_v^{max}$ por $t$ segundos. Isto se dá por meio de 
\begin{equation}
	v_v = v_{0_v} + a_{v}^{max} t
\end{equation}. 
No entanto, é necessário considerar que o veículo não poderá ultrapassar o limite de velocidade da via em o mesmo está trafegando. O tempo necessário para que o limite de velocidade seja atingido pelo veículo é dado por
\begin{equation}
	t' = \frac{v_r^{max} - v_{0_v}}{a_v^{max} - b}
\end{equation}. 
No instante t', o veículo estará na posição $s_v$, que é dada por
\begin{equation}
	s_v = (a_v^{max} - b) t'^2 + v_{0_v} t' + s_{0_v}
\end{equation}. 
Após atingir a velocidade máxima, o veículo ainda viajará por mais $t''$ segundos, até alcançar a interseção à sua frente. Este tempo é dado por 
\begin{equation}
	t'' = \frac{l_r - s_v}{v_r^{max}}
\end{equation}. Por fim, dado que a via estará livre para que o veículo possa acelerar na sua velocidade máxima e que o limite de velocidade da via será respeitado, o tempo mínimo de viagem até a interseção à frente pode ser calculado por meio da Equação \ref{eq:tempo_minimo_i}.
\begin{equation}
	t_{i}^{min} =\left\{\begin{matrix}
		t & \textrm{, para } v_v \leq v_r^{max} \\ 
		t' + t'' & \textrm{, para } v_v > v_r^{max}
	\end{matrix}\right.
    \label{eq:tempo_minimo_i}
\end{equation}, onde $t_i$ é o tempo em que o intervalo de verde, na sinalização semafórica à frente do veículo, será iniciado.

Segundo \citet{Faria:2013}, para viajar com um comportamento suave, o veículo deve estabelecer uma velocidade média da sua posição até a interseção, que é dada por 
\begin{equation}
	v_{m_v} = \frac{s_{v,i}}{t_i}.
\end{equation}
\begin{equation}
	a = max \left[b_{max}, min \left(\frac{2 v_{m_n} - 2 v_{0_v}}{t_i}, a_f \right)\right].
    \label{eq:aceleracao_confortavel}
\end{equation}
Finalmente, para garantir que a viajem seja realizada com esta velocidade média, o veículo deve estabelecer uma aceleração instantânea de com a Equação \ref{eq:aceleracao_confortavel}.

\section{Considerações Finais}

Este capítulo apresentou os trabalhos relacionados relevantes ao desenvolvimento desta tese, que são: rede \textit{ad hoc} móveis centrada em interesses \cite{Dutra:2012,Dutra:2010}, controle distribuído de tráfego com fluxos heterogêneos \cite{Paiva:2012} e roteamento ecológico de veículos orientados a ondas verdes.

O trabalho de \citet{Dutra:2012,Dutra:2010}, que é a rede \textit{ad hoc} móvel centrada em interesses, contribui relevantemente para este tese, fornecendo um arcabouço teórico para o projeto e desenvolvimento de uma rede \textit{ad hoc} veicular centrada em interesses \cite{Goncalves:2016b}, uma vez que protocolo de comunicação apresentado na Seção \ref{subsec:rp} mostrou-se superior às abordagens tradicionais baseadas em IP para redes \textit{ad hoc} móveis. Embora o trabalho de \cite{Dutra:2012,Dutra:2010} tenha obtido tais resultados, ele não é adequado para um ambiente de comunicação veicular. Esta afirmativa parte da análise do trabalho de \citet{Willke:2009}, que define os requisitos de comunicação das categorias de aplicações para redes \textit{ad hoc} veiculares. Por este motivo, foi necessário estender as estruturas de dados e mecanismos propostos por \citet{Dutra:2012,Dutra:2010}, a fim de satisfazer os requisitos de comunicação das categorias de aplicações para redes \textit{ad hoc} veiculares. 

O trabalho de \citet{Paiva:2012}, que é o controle distribuído de tráfego com fluxos heterogêneos, contribui para esta tese, fornecendo um estudo inicial acerca do uso do algoritmo de SMER, no que tange o controle de sinalizações semafóricas em interseções isoladas e o controle de redes de sinalizações semafóricas coordenadas. Embora as abordagens adotadas por \citet{Paiva:2012} tenham se mostrado eficiente, elas foram construídas sem levar em consideração os modelos de comunicação utilizados em ambientes veiculares. A construção das abordagens de controle de tráfego propostas por \citet{Paiva:2012} se deu por meio de um simulador de tráfego desenvolvido pelo mesmo, que é chamado de MicroLAM. Embora o MicroLAM seja capaz de fornecer um \textit{framework} para implementação de novos componentes para uso em simulações de ambientes de sistemas inteligentes de transporte, existem simuladores já consolidados pela comunidade científica de sistemas de transporte e de redes de computadores, tais como o SUMO (\textit{Simulation of Urban Mobility}) \cite{SUMO:2013} e o Omnet++ \cite{Omnet:2013}. Por meio de bibliotecas de simulação de redes veiculares do Omnet++, tais como Veins \cite{Veins:2013}, é possível não somente construir e simular abordagens como as de \citet{Paiva:2012}, mas também implementar protocolos para redes \textit{ad hoc} veiculares. A biblioteca Veins fornece uma API de comunicação com o SUMO, permitindo que as simulações executadas por meio do do Omnet++ controlem as simulações no SUMO. Outro ponto a respeito do trabalho de \citet{Paiva:2012} é sobre a maneira como são obtidas os dados a respeito dos fluxos de tráfego nas vias de entrada das interseções e corredores. Para obter tais dados, \citet{Paiva:2012} leva em consideração a instalação de sensores de pressão tanto no início das vias de entradas das interseções controladas por sinalizações semafóricas quanto no início de corredores. Devido a isto, as abordagens propostas por \citet{Paiva:2012} levam em consideração somente a contagem de veículos que entram nas vias controladas pelas sinalizações semafóricas. Dessa forma, os ajustes dos intervalos das indicações das sinalizações semafóricas podem ser comprometidos, uma vez que alguns veículos podem ficar retidos na via, quando a sinalização desta estiver com a luz vermelha acesa. Logo, não é possível levar em consideração a quantidade residual de veículos na via. Além disto, fica difícil identificar os tipos de veículos que compõem os fluxos de tráfego. Neste caso, uma via pode ser ocupada por veículos como caminhões ou ônibus e estes serão contabilizados como automóveis. Dessa forma, as leituras dos fluxos de tráfego serão equivocadas, pois, de acordo com \citet{DENATRAN:2014}, caminhões de dois eixos, caminhões de três eixos e ônibus têm fatores de equivalência de tráfego iguais dois, três e dois, respectivamente. Em outras palavras, os fatores de equivalência de tráfego correspondem ao número de veículos que uma determinada classe de veículo corresponde. Portanto, caminhões de dois eixos, caminhões de três eixos e ônibus equivalem a dois, três e dois veículos, respectivamente. Apesar de uma abordagem para controlar corredores e sincronizar as sinalizações semafóricas neste tenha sido proposta por \citet{Paiva:2012}, não fica bem claro o funcionamento desta em seu trabalho, devido à ausência de uma especificação algorítmica. De forma geral, no que tange a descrição dos algoritmos relacionados a cada uma das abordagens propostas por \citet{Paiva:2012}, não há uma especificação das primitivas relacionadas as trocas de mensagens entre as sinalizações semafóricas que controlam interseções isoladas, assim como, as sinalizações semafóricas que compõem o modelo de controle de corredores com sinalizações semafóricas coordenadas. Tendo em vista que a abordagem de \citet{Paiva:2012} é um sistema multiagente, a especificação destas primitivas é de grande importância, pois permitem uma melhor compreensão de como se dá a interação entre os agentes inteligentes, que são as sinalizações semafóricas.  

O trabalho de \citet{Faria:2013}, que é o roteamento ecológico de veículos orientados à ondas verdes, contribui para esta tese, fornecendo um estudo inicial acerca de uma abordagem para o planejamento e orientação de rotas baseados nos intervalos de verde das sinalizações semafóricas. Tal abordagem se baseia no compartilhamento dos planos de temporização das sinalizações semafóricas de uma rede viária controlado por um sistema de controle de tráfego. De acordo com \citet{Faria:2013}, presume-se que os intervalos de verde das sinalizações semafóricas são conhecidos por todo o sistema de planejamento e orientação de rotas. Além disto, por meio destes intervalos de verde bem conhecidos, os veículos alocam espaços nas vias de entrada das interseções que fazem parte de uma rota ótima, que é calculada por meio de um algoritmo de caminho temporalmente mais curto. Por fim, os veículos, à medida que seguem os seus planejamentos de rota, se fazem valer dos intervalos de verde ou da previsão de quando estes irão iniciar, para poderem ajustar suas velocidades, enquanto trafegam de uma interseção a outra, evitando paradas desnecessárias. Tudo isso foi possível, pois \citet{Faria:2013} desenvolveu seu trabalho com base no simulador desenvolvido por \citet{Paiva:2012}, o MicroLAM. Por este motivo, algumas lacunas no trabalho de \cite{Faria:2013} precisam ser preenchidas. Então, por se tratar também de um sistema multiagente, o trabalho de \citet{Faria:2013} não deixa claro como os agentes veículos e sinalizações semafóricas interagem. Além disto, não fica claro como os agentes do tipo sinalizações semafóricas mantêm as alocações de espaços nas vias durante os intervalos de indicação de verde. Outro ponto importante é que não existe uma definição clara a respeito do agente responsável em executar o algoritmo de planejamento e orientação de rotas, além de como este algoritmo é executado pelo sistema de planejamento e orientação de rotas, ou seja, se é executado localmente pelos agentes ou se é executado de maneira distribuída. No que tange o processo de alocação definido juntamente com o algoritmo de planejamento e orientação de rotas, não existe menção sobre como as estruturas de dados responsáveis pela mantença das alocações de espaços nas vias são manutenidas, à medida que os agentes veículos replanejam suas rotas. Ainda sobre o processo de alocação de espaços nas vias, este não leva em consideração os tamanhos dos veículos, assim como, não leva em consideração o tamanho das vias. Por fim, à medida que as sinalizações semafóricas mudam os seus planos de temporização, seja por necessidade de adaptar-se às flutuações dos fluxos de tráfego ou pela cooperação em operações de coordenação de redes de sinalizações semafóricas em corredores controlados, os veículos não são notificados sobre estas mudanças. Isto faz com que os planejamentos de rotas se tornem obsoletos, uma vez que as agendas de tempos, criadas a partir dos planos de temporização das sinalizações semafóricas, são modificadas. A consequência disto é a inconsistência nas alocações de espaços nas vias de entrada das interseções.

Após a apresentação e discussão em torno dos trabalhos relacionados a esta tese, o texto avança para as duas primeiras contribuições desta pesquisa de doutorado. Portanto, o próximo capítulo apresentará os detalhes acerca dos projetos da Rede \textit{Ad Hoc} Veicular Centrada em Interesses \cite{Goncalves:2016a,Goncalves:2016b} e a Rede \textit{Ad Hoc} Veicular Heterogênea Centrada em Interesses.

  \chapter{Redes Veiculares Centradas em Interesses}\label{cap:radnet-ve}

Este capítulo apresenta as duas primeiras contribuições desta tese, que são a Rede \textit{Ad Hoc} Veicular Centrada em Interesses \cite{Goncalves:2016b} e a Rede \textit{Ad Hoc} Veicular Heterogênea Centrada em Interesses. 

Os detalhes acerca destas duas redes veiculares centradas em interesses, assim como, as relações destas com as tecnologias de acesso à comunicação sem fio serão apresentadas nas seções que seguem.

\section{Rede \textit{Ad Hoc} Veicular Centrada em Interesses}

Entender os requisitos de comunicação das categorias de aplicações para redes \textit{ad hoc} veiculares é fundamental para o projetar um protocolo de comunicação eficiente. Neste sentido, inicialmente, esta seção apresenta como os o projeto da Rede \textit{Ad Hoc} Veicular Centrada em Interesses\cite{Goncalves:2016b} ou RAdNet-VE (\textit{Inte\textbf{R}est-Centric Mobile \textbf{Ad} Hoc \textbf{Net}work for \textbf{V}ehicular \textbf{E}nvironments}) trata tais requisitos. Em seguida, é descrito como o projeto da RAdNet-VE estendeu o cabeçalho de mensagens da RAdNet (\textit{Inte\textbf{R}est-Centric Mobile \textbf{Ad} Hoc \textbf{Net}work}). 

\subsection{Tratando os Requisitos de Comunicação das Categorias de Aplicações para Redes \textit{Ad Hoc} Veiculares}

Os projetos de protocolos de comunicação para redes \textit{ad hoc} veiculares devem levar em conta os requisitos de latência, confiabilidade e escala. Além disto, o escopo de comunicação e serviços de comunicação de membros de grupos também devem ser bem definidos, a fim de satisfazer os requisitos de comunicação das categorias de aplicação para redes \textit{ad hoc} veiculares \cite{Willke:2009}. Portanto, a RAdNet-VE herda características da RAdNet \cite{Dutra:2012,Dutra:2010} e, além disto, adota mecanismos e abordagens descritas a seguir.

Quando uma aplicação precisa se comunicar com baixa latência, o protocolo de comunicação deve ser capaz de fornecer comunicação fim-a-fim com baixo \textit{delay} \cite{Willke:2009}. Neste sentido, a RAdNet-VE deve fornecer baixa latência, pois ela não sofre com o dinamismo das redes \textit{ad hoc} veiculares. Na RAdNet-VE, os nós não precisarão de informações a respeito da topologia de rede, pois eles não precisarão encontrar, manter e atualizar rotas para outros nós na rede. Consequentemente, a largura de banda, normalmente ocupada por mensagens de controle, deverá ser liberada. Na RAdNet, o uso de prefixos como um conjunto de valores probabilísticos permite que mensagens sejam encaminhadas, de acordo com a distribuição de probabilidade usada para construir os prefixos. Dessa forma, os nós não precisam determinar o melhor caminho entre a origem e o destino de uma mensagem. O mecanismo adotado pela RAdNet permite que os veículos troquem mensagens por múltiplos caminhos.

As aplicações necessitam de um protocolo, que possa entregar mensagens a um grupo de nós. Este protocolo deve assegurar uma alta probabilidade de entrega de mensagens \cite{Willke:2009}. Como nos nós da RAdNet, os nós da RAdNet-VE devem armazenar o identificador e o prefixo de origem das mensagens, a fim de compará-los com os de mensagens recebidas. Isto permite que mensagens já recebidas pelos nós não sejam encaminhadas. No entanto, isto não é suficiente para assegurar altas taxas de entrega de mensagens na RAdNet-VE, pois ambientes veiculares são caracterizados por serem ambientes com alta densidade de veículos. Assim, quando os veículos na mesma vizinhança recebem uma mensagem, eles a encaminham para todos os nós vizinhos, de acordo com um filtro de casamento de dados da RAdNet, causando um overhead desnecessário. Para evitar isto, o mecanismo de encaminhamento de mensagens da RAdNet deve ser estendido, adicionando campos ao cabeçalho original de mensagem da RAdNet. Estes campos devem armazenar a posição relativa da origem da mensagem e identificador de via. Além disto, devem ser levados em consideração o uso de dispositivos GPS em cada nó da RAdNet-VE e o acesso a bancos de dados de mapas por parte das aplicações para redes \textit{ad hoc} veiculares. 

Portanto, quando um nó receber uma mensagem, ele pode calcular a distância relativa entre ele e a origem da mensagem recebida, além de poder armazenar tanto o prefixo quanto a posição relativa da origem da mensagem recebida. É importante pontuar que, os nós devem armazenar somente os prefixos e posições relativas de seus vizinhos, que são aqueles nós a um salto de distância. Uma vez que os nós passam a conhecer as posições relativas de seus vizinhos, aqueles que são os nós mais distantes da origem de uma mensagem, eles devem encaminhar as mensagens recebidas ou devem atuar passivamente e não encaminhar mensagens, se estes não obedecerem esta restrição. Além disto, o encaminhamento de mensagens deve acontecer entre nós que estejam na mesma via cujo identificador seja igual ao do campo identificador de via da mensagem recebida. A adição destas restrições ao mecanismo de encaminhamento de mensagens deve permitir que as mensagens sejam entregues a muitos nós, usando poucos saltos, assim como, deve fornecer baixa latência de comunicação entre os nós. 

O uso de posições relativas no encaminhamento de mensagens sobre longas distâncias deve permitir que o protocolo de comunicação escale apropriadamente em ambientes veiculares com alta densidade \cite{Willke:2009}. No entanto, algumas aplicações para redes \textit{ad hoc} veiculares precisam propagar mensagens em uma determinada direção. Embora os nós da RAdNet possam encaminhar mensagens sobre longas distâncias, usando muitos saltos, o mecanismo de encaminhamento de mensagens da RAdNet não satisfaz tal requisito, pois ela foi projetada para satisfazer os requisitos de aplicações para rede \textit{ad hoc} móveis. Isto, portanto, torna necessária a adição de um campo direção ao cabeçalho de mensagens da RAdNet. Este campo deve admitir somente os seguintes valores:

\begin{itemize}
  \item \textbf{-1:} mensagens podem somente ser encaminhadas na direção oposta à do nó, à medida que este se move;
  \item \textbf{0:} mensagens podem ser encaminhadas em todas as direções;
  \item \textbf{-1:} mensagens podem somente ser encaminhadas na mesma direção que a do nó.
\end{itemize}

A adição do campo direção no cabeçalho de mensagem da RAdNet permite que o fluxo de encaminhamento de mensagens seja unidirecional ou bidirecional. Portanto, o mecanismo de encaminhamento de mensagens deve ser estendido, adicionando mais uma restrição, a fim de permitir que os nós encaminhem mensagens, de acordo com a direção em que eles se movem. Então, ao receber uma mensagem, o nó deve usar o campo de posição relativa para calcular seu posicionamento na via em relação à origem da mensagem. Uma vez que cada nó possui seu próprio dispositivo GPS, ele deve ser capaz de obter o seu posicionamento atual em relação aos nós que trafegam em uma via. Em outras palavras, os nós devem ser capazes de saber se eles estão atrás ou à frente da origem de uma mensagem. Então, para representar o posicionamento dos nós, adota-se o valor -1, quando o nó estiver atrás da origem da mensagem, ou 1, quando ele estiver à frete desta. Dessa forma, ao receber uma mensagem cuja origem está à frente e o valor do campo direção é igual a -1, o nó deve ser capaz de encaminhar a mensagem para seus vizinhos, senão, ele deve descartar a mensagem. Por outro lado, ao receber uma mensagem de uma origem que está atrás e o valor do campo direção é 1, o nó deve ser capaz de encaminhar esta mensagem para os seus vizinhos, caso contrário, ele deve descartar a mensagem. Além disto, os nós devem encaminhar mensagens, se o identificador de via da mensagem recebida corresponde a via em eles estão operando.

As aplicações relacionadas ao controle de movimento individual ou em grupo operam em um escopo de comunicação bem definido, que pode ser uma vizinhança de veículos ou uma região pequena na rede \cite{Willke:2009}. Portanto, os protocolos de comunicação devem assegurar a entrega seletiva de mensagens, que podem ser baseadas em trajetória, proximidade de veículo ou identificação de veículo. Na RAdNet, os nós não precisam manter ou atualizar rotas, mas eles podem encaminhar mensagens até um número máximo de saltos, de acordo com o resultado do filtro de casamento de dados \cite{Dutra:2012,Dutra:2010}. No entanto, tal comportamento não permite que aplicações de controle de movimentos individuais de veículos ou de grupos de veículos troquem mensagens dentro de um escopo bem definido de comunicação, pois o protocolo de comunicação da RAdNet leva em consideração um único valor de número máximo de saltos para encaminhar mensagens. Por isto, o mecanismo de registro de interesses deve ser estendido, assim como, o mecanismo de encaminhamento de mensagens. Neste sentido, deve se fazer uso do campo identificador de via, a fim de limitar o escopo de comunicação na via em que os nós estejam operando. Além disto, na RAdNet-VE, as aplicações para redes \textit{ad hoc} veiculares devem registrar seus interesses com seus respectivos números máximos de saltos. Com isto, o escopo de comunicação se torna restrito, de acordo com os números máximos de saltos registrados com os interesses e as vias em que os nós estejam operando.

A respeito de serviços de membros de grupos, as aplicações relacionadas ao controle de movimentos individuais ou de grupos precisam de um protocolo que possibilite a mantença de estruturas de grupos persistentes. Uma vez que a RAdNet-VE é uma rede centrada em informações, ela não leva em consideração espaços de endereçamento de grupos ou mecanismos centralizados ou distribuídos para gerenciamento de membros de grupos. Assim como a RAdNet, a RAdNet-VE é baseada no modelo de comunicação \textit{Publisher/Subscriber} \cite{Buschmann:1996}. Portanto, usando o mecanismo de interesses contidos nas mensagens, serviços para membros de grupos devem ser implementados de maneira completamente distribuída.

Com base em de todas essas propostas de mecanismos e abordagens para satisfazer os requisitos de comunicação das categorias de aplicações para redes \textit{ad hoc} veiculares, as seções seguintes descrevem como se deram as extensões das estruturas de dados e mecanismos da RAdNet, tais como: prefixo ativo, cabeçalho de mensagens, mecanismo de encaminhamento de mensagens e mecanismo de registro de interesses.

\subsection{Descrição das Extensões das Estruturas de Dados e Mecanismos da RAdNet}

\begin{figure}[!t]
	\centering
    \subfigure[]{
    	\includegraphics[width=12cm]{prefixo_ativo_radnet}
        \label{fig:prefixo_ativo_radnet2}
    }
    \subfigure[]{
    	\includegraphics[width=15cm]{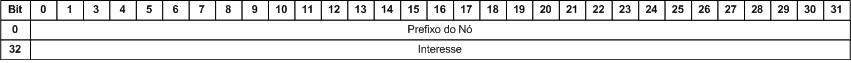}
        \label{fig:prefixo_ativo_radnet_ve}
    }
    \caption{Comparação entre as especificações dos Prefixos Ativos da RAdNet e a RAdNet-VE: (a) Prefixo Ativo da RAdNet; (b) Prefixo Ativo da RAdNet-VE.}
    \label{fig:prefixos_ativos}
\end{figure}

A proposta da RAdNet-VE estende certas estruturas de dados e mecanismos da RAdNet. No que tange às estruturas de dados, a proposta da RAdNet-VE altera os tamanhos dos campos do Prefixo Ativo, assim como, estende e modifica os tamanhos dos campos do cabeçalho de mensagens da RAdNet. No projeto da RAdNet, os campos prefixo do nó e interesse têm tamanhos de 24 bits, como pode ser visto na Figura \ref{fig:prefixo_ativo_radnet2}. Devido às modificações dos tamanhos dos campos do cabeçalho de mensagens da RAdNet, os tamanhos dos campos prefixo do nó e interesse aumentaram de 24 bits para 32 bits, como pode ser visto na Figura \ref{fig:prefixo_ativo_radnet_ve}. 

\begin{figure}[t]
	\centering
    \subfigure[]{
    	\includegraphics[width=11cm]{cabecalho_radnet}
        \label{fig:cabecalho_radnet2}
    }
    \subfigure[]{
    	\includegraphics[width=15cm]{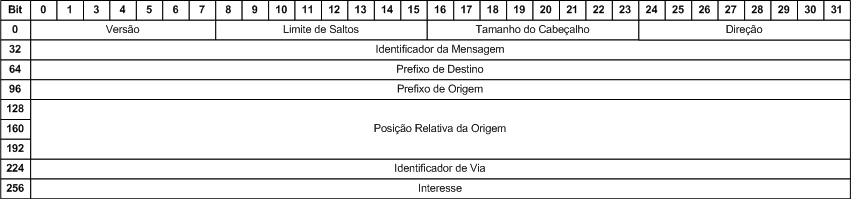}
        \label{fig:cabecalho_radnet_ve}
    }
    \caption{Comparação entre as especificações dos cabeçalhos de mensagens da RAdNet e da RAdNet-VE: (a) cabeçalho de mensagens da RAdNet; (b) cabeçalho de mensagens da RAdNet-VE.}
    \label{fig:cabecalhos_mensagens}
\end{figure}

Acerca das alterações de tamanho de alguns campos do cabeçalho de mensagem da RAdNet e extensões do mesmo, ambas procederam como segue:

\begin{itemize}
	\item Um aumento de 24 bits para 32 bits no tamanho dos seguintes campos do cabeçalho de mensagens da RAdNet: identificador de mensagem, prefixo de destino, prefixo de origem e interesse. A decisão de aumentar os tamanhos destes campos teve como objetivo padronizá-los com tamanhos cujos valores são de base dois. Estas alterações são ilustradas na Figura \ref{fig:cabecalho_radnet_ve}.
    \item A adição de três campos ao cabeçalho de mensagem da RAdNet: posição relativa da origem de mensagem (96 bits), direção de encaminhamento de mensagens (8 bits) e o identificador de via (32 bits). Estes novos campos e seus respectivos tamanhos são exibidos pela Figura \ref{fig:cabecalho_radnet_ve}. Nesta tese, assume-se que todos os nós são equipados com dispositivos GPS e que as aplicações que executam sobre estes nós podem acessar bancos de dados de mapas, tais como Open Street Maps e Google Maps. Além disto, a posição relativa da origem da mensagem deve armazenar três valores de 32 bits, que são: longitude, latitude e altitude;
\end{itemize}

No que diz respeito às extensões dos mecanismos da RAdNet, estas procederam da seguinte forma:

\begin{itemize}
    \item \textbf{Registro dos interesses com seus respectivos números máximos de saltos na camada de rede do nó:} Esta extensão apenas realiza um mapeamento entre um interesse e um número máximo de saltos. É importante salientar que um interesse só pode mapear um e somente um número máximo de saltos. No entanto, isto não impede que as aplicações para redes \textit{ad hoc} veiculares possam usar uma mesma identificação de interesse. Para resolver este problema, recomenda-se especificar os interesses usando a seguinte estrutura: \textit{radnet-ve://user service bundle/user service/application name/interest}, tal que \textit{user bundle} é um agrupamento lógico de serviços de sistemas inteligentes de transporte, \textit{user service} é o tipo de serviço de usuário de sistema inteligente de transporte, \textit{application name} é o nome da aplicação pertencente ao tipo de serviço de usuário de sistemas inteligentes de transporte e, por fim, \textit{interest} é o interesse no qual a aplicação é uma subscritora. Os componentes desta URI se baseiam na especificação da \textit{National ITS Architecture 7.1} do \textit{United States Department of Transportation} \cite{DOT:2016}. Com base nesta estrutura é possível não somente identificar os interesses por aplicações, mas também é possível identificar unicamente as aplicações instaladas nos nós. Portanto, um interesse pode ser identificado de acordo com a seguinte URI, por exemplo: \textit{radnet://avss/safety\_readiness/obstacle\_notifier/obstacle\_on\_road}. Por fim, o mecanismo de registro de interesses deve ser capaz de manter em memória os interesses mais utilizados, à medida que os nós se comunicam uns com os outros. Isto deve permitir que muitos interesses sejam registrados na camada de rede do nó. No entanto, somente os mais utilizados são mantidos em memória, a fim de facilitar a consulta por interesses, nos quais os nós são subscritores. Para tanto, o mecanismo de registro de interesses deve ser implementado, levando em consideração um algoritmo MFU (\textit{Most Frequently Used}).
    \item \textbf{Adição de novas regras de filtragem ao mecanismo de encaminhamento de mensagens:} é importante pontuar que, a adição destas novas regras não afeta as que foram herdadas do protocolo de comunicação da RAdNet.
\end{itemize}

Por fim, com base nas extensões descritas acima, foi possível iniciar o projeto do protocolo de comunicação da RAdNet-VE. Pretende-se com este protocolo, encaminhar mensagens, levando em conta os seguintes dados:
\begin{enumerate}
	\item O identificador da via em que os nós (veículos e unidades de acostamento) estejam operando;
    \item O resultado do casamento de dados entre o prefixo do nó e o prefixo da origem da mensagem;
    \item O resultado do casamento de dados entre os interesses registrados nos nós e aqueles contidos nas mensagens;
    \item A distância entre um nó e a origem da mensagem;
    \item A posicionamento do nó em relação à origem da mensagem;
    \item O número máximo de saltos registrado juntamente com o interesse na camada de rede do nó ou um número máximo de saltos padrão, que é somente usado, se os interesses contidos nas mensagens não existirem no registro de interesses na camada de rede do nó.
\end{enumerate}

A próxima seção tem como objetivo descrever os detalhes do projeto do protocolo de comunicação da RAdNet-VE cujo nome é o seguinte acrônimo: RVEP (\textit{RAdNet-VE Protocol}).

\subsection{Projeto do Protocolo de Comunicação da RAdNet-VE}\label{subsec:radnet_ve}

Na RAdNet-VE, todos os nós devem ser inicializados, antes que eles possam transmitir mensagens. Portanto, durante o processo de inicialização do nó, as seguintes variáveis de controle e estruturas de dados são inicializadas: contador de mensagens enviadas, prefixo do nó, tabela de prefixos de origem e mensagens recebidas, tabela de interesses e números máximos de saltos, tabela de posições relativas de vizinhos dentro da área coberta pelo rádio de comunicação e lista de identificadores de vias. Seguindo com o processo de inicialização, o nó registra os interesses das aplicações com seus respectivos números máximos de saltos. Por fim, a configuração da lista de identificadores de via depende das aplicações instaladas no nó e de como elas lidam com as características das vias, onde o nó esteja operando. Portanto, as aplicações precisam registrar os identificadores de vias na camada de rede do nó. 

A identificação deve seguir a seguinte estrutura: \textit{radnet-ve://user service bundle/user service/application name/roadway identifier/lane}, tal que \textit{roadway identifier} é o identificador único da via em que o nó está operando, que é obtido por meio de um banco de dados de mapas pela aplicação; e \textit{lane} é a faixa da via, caso haja a necessidade de diferenciar as faixas da via, a fim de estabelecer a comunicação entre os nós. Portanto, uma via pode ser identificada de acordo com a seguinte URI, por exemplo: \textit{radnet://avss/safety\_readiness/obstacle\_notifier/Rod.+Gov.+Mario+Covas}. Caso seja necessário identificar as faixas de uma via com duas faixas, estas podem ser identificadas de acordo com as seguintes URIs, por exemplo: \textit{radnet://avss/safety\_readiness/obstacle\_notifier/Rod.+Gov.+Mario+Covas/0}, para a faixa da direita, e \textit{radnet://avss/safety\_readiness/obstacle\_notifier/Rod.+Gov.+Mario+Covas/1}, para a faixa da esquerda.

Se uma aplicação instalada em uma unidade de acostamento é responsável em notificar veículos sobre obstáculos em uma rodovia, a lista de identificadores pode conter uma entrada. Embora a via tenha mais de uma faixa, pretende-se notificar todos os veículos, independente das faixas em que eles estejam trafegando. Um outro exemplo pode ser uma aplicação responsável em controlar as sinalizações semafóricas de uma interseção complexa, semelhante a apresentada na Figura \ref{fig:modelagem_intersecao} desta tese. Neste caso, cada sinalização semafórica da interseção é responsável em executar uma instância diferente da aplicação. Dessa forma, a lista de identificadores de vias de cada sinalização semafórica poderá ter mais de duas entradas, sendo que uma delas está relacionada a via de entrada, onde a sinalização semafórica está instalada. As demais vias, neste caso, estão relacionadas às vias de saída, de acordo com os movimentos permitidos na interseção a partir da via de entrada. Um último exemplo pode ser as aplicações instaladas nos veículos. Devido à mobilidade dos veículos, estas aplicações constantemente atualizam a lista de identificadores de vias, à medida que trafegam em diferentes vias ao longo de uma determinada rota. 

Para o nó enviar uma mensagem, o protocolo de comunicação deve receber quatro itens, que são: prefixo de destino, interesse, direção, identificador de via. Em \cite{Goncalves:2016b}, eram levados em consideração somente os três itens. A inclusão de mais um item tem como objetivo tirar a responsabilidade da camada de rede, no que diz respeito à inclusão automática do identificador de via, quando a mensagem é construída. Antes de enviar a mensagem para seus vizinhos, o nó constrói a mensagem da seguinte maneira: 

\begin{enumerate}
	\item Configura o campo versão com a versão atual do protocolo;
    \item Configura o campo limite de saltos com o valor zero, de modo que este valor incremente, à medida que a mensagem for encaminhada por outros nós;
    \item Configurar o campo tamanho do cabeçalho com o valor inteiro correspondente;
    \item Configurar o campo identificador da mensagem com o valor atual do contador de mensagens enviadas;
    \item Configurar o campo prefixo de destino com o dado de entrada correspondente;
    \item Configurar o campo prefixo de origem com o prefixo do nó;
    \item Configurar o campo interesse com o dado de entrada correspondente;
    \item Configurar o campo posição com o dado obtido por meio do dispositivo GPS instalado no nó;
    \item Configurar o campo direção com o dado de entrada correspondente;
    \item Configurar o campo identificador de via com o dado de entrada correspondente.
\end{enumerate}

\begin{algorithm}[!t]
  \SetKwData{Msg}{Msg}
  \SetAlgoLined
  	\footnotesize\Entrada{msg$_j$}
  	\eSe{$msg_j.id \in tabelaIds_{i}[msg_{j}.prfxOrg]$} {
	  \textbf{Descartar} msg$_j$\;
  	}{
	  \textbf{Inserir} msg$_j$.id \textbf{em} tabelaIds$_{i}$[msg$_{j}$.prfxOrg]\;
	  msg$_j$.limSaltos := msg$_j$.limSaltos + 1\;
      \Se{msg$_j$.limSaltos = 1}{
      	\textbf{Inserir} msg$_j$.posicao \textbf{em} tabelaPos$_{i}$[msg$_{j}$.prfxOrg]\;
      }
	  \eSe{msg$_j$.idVia $\in$ lstIdVias$_i$}{
      	pos$_i$ := calcPos(posicao$_i$, msg$_j$.posicao, msg$_j$.idVia)\;
        prfxEnc := prfx$_i$\;
        dist := calcDist(posicao$_i$, msg$_j$.posicao, msg$_j$.idVia)\;
        \ParaCada{posicao $\in$ tabelaPos$_i$ } {
          \Se{pos$_i$ = msg$_j$.direcao}{
              distViz := calcDist(position, msg$_j$.position, msg$_j$.roadId)\;
              \Se{distViz $>$ dist $\wedge$ distViz $\leq$ diametroRadio/2}{
                  encPrfx := prefixo do vizinho\;
                  dist := distViz\;
              }
          }
        }
        casInt := msg$_{j}$.interesse $\in$ tabelaInt$_{i}$\;
        casPrfx := $|$prefix$_i \cap$ msg$_j$.prfxOrg$| > 0$\;
        \Se{casInt = \textbf{verdadeiro} $\wedge$ (pos = msg$_j$.direcao $\vee$ msg$_j$.direcao = 0)}{
          \Se{msg$_{j}$.prfxDest = \textbf{nulo} $\vee$ msg$_{j}$.prfxDest = prefix$_i$}{
            \textbf{Enviar } uma cópia de msg$_j$ \textbf{para} aplicação\;
          }
        }
        \eSe{casPrfx = \textbf{verdadeiro} $\vee$ casInt = \textbf{verdadeiro}}{
          encMsg := msg$_{j}$.prfxDest = \textbf{nulo} $\vee$ msg$_{j}$.prfxDest $\neq$ prefixo$_i$\;
          posNo := encPrfx = prefix0$_i$ $\wedge$ (pos = msg$_j$.direcao $\vee$ msg$_j$.direcao = 0)\;
          encSaltos := \textbf{falso}\;
          \eSe{casInt = \textbf{verdadeiro}}{
              encSaltos := msg$_j$.limSaltos $<$ tabelaInt$_i$[msg$_{j}$.interesse]\;
          }{
              encSaltos := msg$_j$.limSaltos $<$ limSaltosPadrao\;
          }
          \eSe{(encMsg $\wedge$ posNo $\wedge$ encSaltos) = \textbf{verdadeiro}}{
              \textbf{Espere } uniforme(0, 1)/dist\;
              \textbf{Enviar } msg$_j$ \textbf{para} \textbf{todos} vizinhos$_i$ \;
          }{
              \textbf{Descartar } msg$_j$\;
          }
        }{
          \textbf{Descartar } msg$_j$\;
        }
      }{
      	\textbf{Descartar} msg$_j$\;
      }
  	}
  \caption{Protocolo de comunicação da RAdNet-VE.}
  \label{alg:protocolo_radnet_ve}
\end{algorithm}

Ao receber uma mensagem ($msg_j$), o nó $i$ executa o Algoritmo \ref{alg:protocolo_radnet_ve}. O algoritmo primeiro checa se o identificador da mensagem recebida ($msg_j.id$) existe na tabela de prefixos de origens e identificadores de mensagens recebidas ($tabelaIds_i$). Se verdadeiro, o nó descarta $msg_j$. Se falso, o nó registra $msg_j.id$ em $tabelaIds_i$ e, em seguida, incrementa o valor do campo limite de saltos ($msg_j.limSaltos$). Se o valor do campo limite de saltos é igual a um, o algoritmo registra o prefixo e posição da origem da mensagem na tabela de posições relativas dos vizinhos ($tabelaPos_i$). Isto permite o nó atualizar as posições relativas dos vizinhos que estão a um salto de distância. Em seguida, o algoritmo checa se o valor do campo identificador de via ($msg_j.idVia$) existe na lista de identificadores de vias. Se não existir, o nó descarta $msg_j$. Caso contrário, o nó calcula o seu posicionamento em relação à origem da mensagem. O resultado deste cálculo deve ser -1 ou 1, tornando possível determinar se a fonte da mensagem está atrás ou à frente do nó. Baseado nas posições relativas em $tabelaPos_i$, o nó pode determinar se ele ou qualquer outro vizinho é o nó mais distante da origem da mensagem. O algoritmo extrai tal informação da $tabelaPos_i$ e a armazena em $encPrfx$. Nos dois próximos passos, o algoritmo realiza o processo de filtragem herdado do protocolo de comunicação da RAdNet. Primeiro, ele checa se a tabela de interesses do nó $i$ ($tabelaInt_i$) tem uma entrada igual ao interesse contido na mensagem recebida ($msg_j.interesse$). Após isto, o algoritmo checa se o prefixo da origem da mensagem ($msg_j.prfxOrg$) tem um ou mais pares de campos com valores iguais. Se o resultado da checagem relacionada aos interesses for igual a verdadeiro, o algoritmo checa se o nó $i$ é o destino de $msg_j$ ou se o prefixo de destino é nulo. Se verdadeiro, o algoritmo cria uma cópia de $msg_j$ e a encaminha para a aplicação subscritora do interesse contido na mensagem recebida.

Antes de encaminhar as mensagens, o algoritmo checa se o casamento de campos dos prefixos e interesses ocorreu. Se falso, ele descarta $msg_j$. Caso contrário, ele espera por um período de tempo antes de enviar $msg_j$ para todos os seus vizinhos. No entanto, três condições devem ser satisfeitas, são elas:

\begin{enumerate}
	\item O nó $i$ não é o destino de $msg_j$;
    \item O nó $i$ é o nó mais distante da origem da mensagem;
    \item $msg_j$ não excedeu o número máximo de saltos.
\end{enumerate}

Se estas condições não são satisfeitas, o nó descarta $msg_j$. Para a primeira condição, o algoritmo checa se o nó $i$ é o destino de $msg_j$. Quando esta condição não é satisfeita, uma aplicação associada a $msg_j.interesse$ recebeu uma cópia da mensagem recebida. Por isto, a mensagem deve ser descartada, pois ela alcançou o seu destino. Para a segunda condição, o algoritmo checa se o nó $i$ é o nó mais distante da origem da mensagem e se seu posicionamento em relação a esta o permite encaminhar a mensagem para seus vizinhos. Se esta condição não é satisfeita, é devido a existência de um nó mais distante da origem da mensagem. Para a terceira condição, o algoritmo checa se a mensagem alcançou o número máximo de saltos. Quando o interesse contido na mensagem recebida existe na tabela de interesses, o algoritmo checa se o valor do limite de saltos ($msg_j.limSaltos$) é menor que o número máximo de saltos registrado com o interesse. Caso contrário, o algoritmo checa se o valor do campo limite de saltos é menor que o número máximo de saltos padrão. Se uma destas condições é falsa, $msg_j$ é descartada.

Finalmente, o algoritmo determina o tempo de espera por meio do resultado da função de número aleatório da distribuição uniforme entre 0 e 1 dividido pela distância entre nó $i$ e a origem da mensagem. O tempo de espera é necessário, pois ele evita que muitos nós encaminhem mensagens em instantes muito próximos. À medida que a mensagem é encaminhada para longe de sua origem, o tempo de espera diminui.

\section{Rede \textit{Ad Hoc} Veicular Heterogênea Centrada em Interesses}

Devido à alta mobilidade dos veículos e à topologia dinâmica das redes \textit{ad hoc} veiculares, é difícil fornecer serviços de sistemas inteligentes de transporte somente por meio de uma rede, que tem como base uma única tecnologia de acesso à comunicação sem fio, especificamente rádios de comunicação dedicada de curto alcance ou \textit{Dedicated Short Range Communication} (DSRC). Atualmente, as tecnologias disponíveis de acesso à comunicação sem fio para ambientes veiculares são as que se baseiam em rádios de comunicação dedicada de curto alcance (IEEE 802.11 e IEEE 802.11p) e as que se baseiam em redes celulares (GSM, UMTS e LTE). No entanto, estas tecnologias têm suas próprias limitações quando usadas em ambientes veiculares. Em particular, tecnologias de acesso à comunicação sem fio baseadas em rádios de comunicação de curto alcance, foram inicialmente projetadas para fornecer comunicações sem fio, sem a necessidade de uma infraestrutura pervasiva em ambientes como os de rodovias, estradas e ruas. Por outro lado, embora redes celulares possam fornecer uma ampla cobertura geográfica, elas não podem fornecer de maneira eficiente trocas de informações de tempo real em áreas locais. Consequentemente, integrar redes baseadas em tecnologias de acesso à comunicação sem fio, tais como IEEE 802.11, IEEE 802.11p, GSM, UMTS e LTE, é de grande relevância para o desenvolvimento de aplicações para sistemas inteligentes de transporte. Por isto, \citet{Zheng:2015} argumenta que uma rede veicular heterogênea pode ser uma boa plataforma para atender os vários requisitos de comunicação dos serviços de sistemas inteligentes de transporte \cite{DOT:2016}.

Com base nessa demanda, resolveu-se estender o projeto da RAdNet-VE \cite{Goncalves:2016b}, a fim de criar uma rede \textit{ad hoc} veicular heterogênea centrada em interesses, pois o protocolo de comunicação da RAdNet-VE é capaz de satisfazer os requisitos de comunicação das categorias de aplicações para redes \textit{ad hoc} veiculares. Com isto, ao analisar o trabalho de \cite{Zheng:2015}, percebeu-se que o protocolo de comunicação da RAdNet-VE é capaz de satisfazer os requisitos de comunicação de parte dos serviços de sistemas inteligentes, devido aos bons resultados apresentados em \cite{Goncalves:2016b}. Tais resultados serão apresentados nesta tese no capítulo apropriado. Com base nestes resultados e nos requisitos de comunicação de aplicações de serviços de sistemas inteligentes, decidiu-se incluir a capacidade de os nós RAdNet-VE poderem se comunicar por meio de redes celulares, mais especificamente LTE, por se tratar de um padrão mais recente deste tipo de tecnologia de acesso à comunicação sem fio. Portanto, criou-se a \textit{\textbf{H}eterogeneous InteRest-Centric Mobile \textbf{Ad} Hoc \textbf{Net}work for \textbf{V}ehicular \textbf{E}nviornments} (HRAdNet-VE). 

Antes de apresentar como se deu a criação da HRadNet-VE, é necessário apresentar, ainda que brevemente, os requisitos de comunicação de aplicações de serviços de sistemas inteligentes de transporte. Portanto, tal tarefa cabe à seção seguinte.

\subsection{Requisitos de Serviços de Sistemas Inteligentes de Transporte}

O principal objetivo desta seção é resumir os requisitos dos serviços de sistemas inteligentes de transporte. Tal levantamento de requisitos é fruto do levantamento bibliográfico apresentado por \citet{Zheng:2015}. Os serviços de sistemas inteligentes de transporte podem ser categorizados em serviços relacionados à segurança e serviços não relacionados à segurança \cite{Zheng:2015}.

Os serviços relacionados à segurança visam a redução do risco de acidentes envolvendo veículos e a diminuição da perda vidas humanas. Nestes serviços, a oportunidade e a confiabilidade são considerados requisitos altamente exigentes. A frequência mínima de mensagens periódicas dos serviços de segurança varia de 1 Hz (uma mensagem por segundo) a 10 Hz (dez mensagens por segundo). Além disto, deve-se também ser levado em consideração o tempo de reação da maioria dos motoristas, que pode variar de 0.6 s a 1.4 s. Por isto, é razoável restringir a latência máxima em não mais que 100 ms. Por exemplo, a latência máxima do aviso de detecção pré-colisão é de 50 ms. Maiores detalhes acerca dos requisitos de comunicação de serviços de sistemas inteligentes de transporte não relacionados à segurança podem ser vistos em \cite{Zheng:2015}. Os requisitos de segurança e confiabilidade são muito rigorosos, devido às características dos serviços de segurança. Em serviços de segurança para sistemas inteligentes de transporte são considerados dois tipos de mensagens:

\begin{itemize}
	\item \textbf{Mensagens de ciência cooperativa:} são transmitidas periodicamente em uma área de interesse, principalmente para propósitos de avisos em estradas ou rodovias. Frequentemente, a troca de mensagens envolve dados como estado do veículo, posição, velocidade, entre outros;
    
    \item \textbf{Notificação ambiental descentralizada:} estas mensagens são frequentemente disparadas por eventos especiais. As mensagens deste tipo têm como objetivo notificar em uma área potencialmente perigosa.
\end{itemize}

Por fim, os serviços não relacionados à segurança são usados principalmente para gerenciamento de tráfego, controle de congestionamento, melhoria da fluidez do tráfego, entretenimento, entre outros. O principal objetivo dos serviços não relacionados à segurança é tornar possível uma experiência de direção mais eficiente e confortável. Estes serviços não têm requisitos rigorosos acerca da latência e confiabilidade. Comparados aos serviços relacionados à segurança, serviços não relacionados à segurança têm diferentes requisitos de qualidade de serviço. Para a maioria dos serviços não relacionados à segurança, a frequência máxima de mensagens periódicas é de 1 Hz (uma mensagem por segundo), enquanto a latência máxima é 500ms. Maiores detalhes acerca dos requisitos de comunicação de serviços de sistemas inteligentes de transporte não relacionados à segurança podem ser vistos em \cite{Zheng:2015}. 

No que diz respeito ao tratamento dos requisitos de comunicação das aplicações dos serviços de sistemas inteligentes de transporte, estes são tratados no projeto da RAdNet-VE, baseando-se nos requisitos de duas categorias de aplicações para redes \textit{ad hoc} veiculares, que são: serviços de informações gerais e serviços de informações de segurança. No entanto, estes requisitos foram definidos a partir de comunicações utilizando dispositivos de comunicação dedicada de curto alcance, como IEEE 802.11 e IEEE 802.11p. No trabalho de \cite{Willke:2009}, os autores não detalham as aplicações das duas categorias citadas acima. Portanto, o trabalho de \cite{Zheng:2015} fornece detalhes acerca dos requisitos de aplicações de serviços de sistemas inteligentes de transporte.

Com base nestes requisitos, foi possível, então, estender o projeto da RAdNet-VE e criar um porte da RAdNet-VE para redes \textit{ad hoc} veiculares heterogêneas. Portanto, a próxima seção detalhará como tal processo se deu.

\subsection{Projeto da Rede Veicular Heterogênea Centrada em Interesses e seu Protocolo de Comunicação}

Em uma rede veicular heterogênea, um mesmo nó deve ser capaz de lidar com diferentes tecnologias de acesso à comunicação sem fio, sem a necessidade de um protocolo de comunicação específico para cada uma das interfaces de comunicação sem fio instaladas no nó. Neste sentido, para que o protocolo de comunicação da RAdNet-VE seja utilizado em uma rede \textit{ad hoc} veicular heterogênea, é necessário inicialmente identificar as tecnologias de acesso à comunicação sem fio, para que o protocolo de comunicação da RAdNet-VE tenha que lidar somente com mensagens de rede, sem conhecer qualquer detalhe tecnológico acerca do envio e recebimento de mensagens. Para tanto, o cabeçalho de mensagens da RAdNet-VE deve sofrer uma extensão, de modo que um cabeçalho de rede específico para a HRAdNet-VE seja criado. Portanto, esta extensão consiste na adição de um campo cujo nome é tecnologia de acesso à comunicação the (8 bits). O campo tecnologia de acesso à comunicação deve ser utilizado para identificar a tecnologia de acesso à comunicação sem fio de uma interface de comunicação. 

Uma vez que as tecnologias de acesso à comunicação sem fio tenham sido identificadas, é possível isolar os detalhes de cada em componentes de software de responsabilidades bem definida. Tais responsabilidades estão relacionadas ao envio e recebimento de mensagens por meio das interfaces de acesso à comunicação sem fio instaladas nos nós. Para tanto, a camada de rede do nó deve ser capaz de instanciar tais componentes, conforme a disponibilidade das interfaces de acesso à comunicação sem fio. Tais componentes devem ser construídos a partir do uso de padrões de projeto de software tais como: \textit{Abstract Factory} \cite{Gamma:2000} e \textit{Forwarder and Receiver} \cite{Buschmann:1996}. O \textit{Abstract Factory} fornece uma interface para criação de famílias de objetos relacionados ou dependentes sem especificar suas classes concretas \cite{Gamma:2000}. O \textit{Forwarder and Receiver} fornece uma interface transparente de comunicação inter-processo para sistemas de software com um modelo de interação \textit{peer-to-peer}, introduzindo \textit{forwarders} e \textit{receivers} para desacoplar os \textit{peers} de detalhes específicos da comunicação entre os nós de rede \cite{Buschmann:1996}. A fim de encapsular todos estes componentes sob uma única interface de acesso, o padrão de projeto \textit{Wrapper Façade} deve ser utilizado, a fim de encapsular funções e dados de maneira mais concisa, robusta, portável, manutenível e coesa \cite{Schimidt:2001}.

\begin{algorithm}[!t]
  \SetKwData{Msg}{Msg}
  \SetAlgoLined
  	\footnotesize\Entrada{msg$_{j,tac}$}
  	\eSe{$msg_{j,tac}.id \in tabelaIds_{i,tac}[msg_{j,tac}.prfxOrg]$} {
	  \textbf{Descartar} msg$_{j,tac}$\;
  	}{
	  \textbf{Inserir} msg$_{j,tac}$.id \textbf{em} tabelaIds$_{i,tac}$[msg$_{j,tac}$.prfxOrg]\;
	  msg$_{j,tac}$.limSaltos := msg$_{j,tac}$.limSaltos + 1\;
      \Se{msg$_{j,tac}$.limSaltos = 1}{
      	\textbf{Inserir} msg$_{j,tac}$.posicao \textbf{em} tabelaPos$_{i,tac}$[msg$_{j, tac}$.prfxOrg]\;
      }
	  \eSe{msg$_{j,tac}$.idVia $\in$ lstIdVias$_{i,tac} \vee$ msg$_{j,tac}$.idVia = \textbf{nulo}}{
      	pos$_{i,tac}$ := calcPos(posicao$_{i,tac}$, msg$_{j,tac}$.posicao, msg$_{j,tac}$.idVia)\;
        prfxEnc := prfx$_{i,tac}$\;
        dist := calcDist(posicao$_{i,tac}$, msg$_{j,tac}$.posicao, msg$_{j,tac}$.idVia)\;
        \ParaCada{posicao $\in$ tabelaPos$_{i,tac}$ } {
          \Se{pos$_{i,tac}$ = msg$_{j,tac}$.direcao}{
              distViz := calcDist(position, msg$_{j,tac}$.position, msg$_{j,tac}$.roadId)\;
              \Se{distViz $>$ dist $\wedge$ distViz $\leq$ diametroRadio$_{tac}$/2}{
                  encPrfx := prefixo do vizinho\;
                  dist := distViz\;
              }
          }
        }
        casInt := msg$_{j}$.interesse $\in$ tabelaInt$_{i,tac}$\;
        casPrfx := $|$prefix$_{i,tac} \cap$ msg$_{j,tac}$.prfxOrg$| > 0$\;
        \Se{casInt = \textbf{verdadeiro} $\wedge$ (pos = msg$_{j,tac}$.direcao $\vee$ msg$_{j,tac}$.direcao = 0)}{
          \Se{msg$_{j}$.prfxDest = \textbf{nulo} $\vee$ msg$_{j}$.prfxDest = prefix$_{i,tac}$}{
            \textbf{Enviar } uma cópia de msg$_{j,tac}$ \textbf{para} aplicação\;
          }
        }
        \eSe{casPrfx = \textbf{verdadeiro} $\vee$ casInt = \textbf{verdadeiro}}{
          encMsg := msg$_{j}$.prfxDest = \textbf{nulo} $\vee$ msg$_{j}$.prfxDest $\neq$ prefixo$_{i,tac}$\;
          posNo := encPrfx = prefixo$_{i,tac}$ $\wedge$ (pos = msg$_{j,tac}$.direcao $\vee$ msg$_{j,tac}$.direcao = 0)\;
          encSaltos := \textbf{falso}\;
          \eSe{casInt = \textbf{verdadeiro}}{
              encSaltos := msg$_{j,tac}$.limSaltos $<$ tabelaInt$_{i,tac}$[msg$_{j}$.interesse]\;
          }{
              encSaltos := msg$_{j,tac}$.limSaltos $<$ limSaltosPadrao\;
          }
          \eSe{(encMsg $\wedge$ posNo $\wedge$ encSaltos) = \textbf{verdadeiro}}{
              \textbf{Espere } uniforme(0, 1)/dist\;
              \textbf{Enviar } msg$_{j,tac}$ \textbf{para} \textbf{todos} vizinhos$_{i,tac}$ \;
          }{
              \textbf{Descartar } msg$_{j,tac}$\;
          }
        }{
          \textbf{Descartar } msg$_{j,tac}$\;
        }
      }{
      	\textbf{Descartar} msg$_{j,tac}$\;
      }
  	}
  \caption{Protocolo de comunicação da HRAdNet-VE.}
  \label{alg:protocolo_hradnet_ve}
\end{algorithm}

Após o encapsulamento das interfaces de acesso à comunicação sem fio, foi necessário separar dados e estruturas de dados inerentes à operação do protocolo de comunicação. Neste sentido, para cada interface de acesso à comunicação sem fio devem existir as seguintes estruturas de dados: contador de mensagens enviadas, tabela de prefixos de origem e mensagens recebidas, tabela de interesses e números máximos de saltos, tabela de posições relativas de vizinhos dentro da área coberta pelo rádio de comunicação e lista de identificadores de vias. No que diz respeito ao prefixo do nó, este não varia em função do número de interfaces de acesso à comunicação sem fio instaladas no nó. Em outras palavras, o prefixo do nó é o mesmo independentemente do número de interfaces de acesso à comunicação sem fio instaladas no nó. Além disto, cada uma destas estruturas de dados citadas acima é inicializada da mesma maneira como a RAdNet-VE as inicializa. No entanto, tal inicialização deve ser realizada para cada interface de acesso à comunicação sem fio instalada no nó. Por fim, as aplicações, que irão executar sobre uma rede \textit{ad hoc} veicular heterogênea centrada em interesses, devem registrar os interesses e seus respectivos números máximos de saltos na estrutura de dados correspondente à tecnologia de acesso à comunicação sem fio da interface que deverá receber as mensagens.

Para o nó enviar uma mensagem, o protocolo de comunicação deve receber cinco itens, que são: prefixo de destino, interesse, direção, identificador de via e tecnologia de acesso. A definição de qual interface de comunicação deve ser depender das operações das aplicações de serviços de sistemas inteligentes de transporte. Antes de enviar a mensagem para seus vizinhos, o nó constrói a mensagem conforme os passos descritos na Seção \ref{subsec:radnet_ve}, além de configurar o campo tecnologia de acesso com o dado de entrada correspondente. 

Por fim, ao receber uma mensagem por meio de uma das interfaces de acesso à comunicação sem fio, o nó executa o Algoritmo \ref{alg:protocolo_radnet_ve}, mas utilizando as estruturas de dados correspondentes às tecnologias das interfaces de comunicação instaladas nos nós, como pode ser visto no Algoritmo \ref{alg:protocolo_hradnet_ve}. Desta forma, o protocolo de comunicação da RAdNet-VE foi estendido completamente, tornando-se um protocolo de comunicação para uma rede veicular heterogênea, sendo $i$ o nó que recebe uma mensagem de um nó $j$, que por sua vez pode ter enviado ou encaminhado a mensagem por meio de uma interface de uma tecnologia de acesso à comunicação sem fio, e $tac$ é o identificador da tecnologia de acesso à comunicação sem fio utilizada tanto pelo nó $j$ quanto pelo nó $i$.

Nesta seção, foram descritos os detalhes acerca do projeto da HRAdNet-VE e de seu protocolo de comunicação, aqui chamado de \textit{HRAdNet-VE Protocol} (HRVEP).

\section{Relação da RAdNets-VE e HRAdNet-VE com as Tecnologias de Acesso à Comunicação Sem Fio}

\begin{figure}[H]
	\centering
	\subfigure[]{
    	\includegraphics[width=4.35cm]{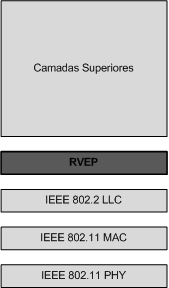}
        \label{fig:ieee80211}
    }
    \quad
    \subfigure[]{
    	\includegraphics[width=4.35cm]{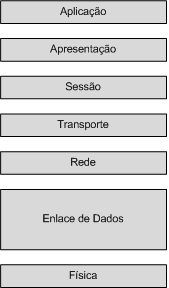}
    }
    \quad
    \subfigure[]{
    	\includegraphics[width=4.35cm]{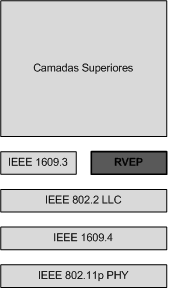}
        \label{fig:ieee80211p}
    }
    \caption{RVEP nas camadas de rede: (a) RVEP como um substituto do IP em nós equipados com rádios baseados no padrão IEEE 802.11; (b) modelo de referência OSI; (c) RVEP como um substituto do IP em nós equipados com rádios baseados no padrão IEEE 802.11p}
    \label{fig:relacao_rvep_dsrc}
\end{figure}

Em \cite{Goncalves:2016a,Goncalves:2016b}, o protocolo de comunicação da RAdNet-VE ou RVEP (\textit{RAdNet-VE Protocol)} foi projetado para executar sobre nós com uma única interface de acesso à comunicação sem fio, sendo esta do tipo IEEE 802.11n ou IEEE 802.11p. Além disto, a proposta de \citet{Goncalves:2016a} visou atender os requisitos de comunicação das categorias de aplicações para redes \textit{ad hoc} veiculares, criando o RVEP, de modo que este fosse um protocolo de comunicação mais genérico possível. Portanto, a relação do RVEP com as tecnologias de acesso à comunicação sem fio baseadas em comunicação dedicada de curto alcance é exibida na Figura \ref{fig:relacao_rvep_dsrc}. 

Como pode ser visto, o RVEP foi projetado para ser um protocolo de camada de rede, a fim de substituir o IP ou compensar alguma deficiência deste. Neste sentido, o RVEP substitui o IP tanto na Figura \ref{fig:ieee80211} quanto na Figura \ref{fig:ieee80211p}. Como pode ser visto na Figura \ref{fig:ieee80211p}, o RVEP divide a camada de rede do nó com um componente definido pelo padrão IEEE 1609.3. Segundo \cite{Sommer:2015}, o padrão IEEE 1609.3 foi desenvolvido para dar suporte acerca do fornecimento e uso de serviços em múltiplos canais, além de ser parte do padrão IEEE 1609. O padrão IEEE 1609 é chamado de \textit{Wireless Access Vehicular Environments} (WAVE). Além disto, o padrão também define uma pilha completa de protocolos para sistemas inteligentes de transporte sobre o IEEE 802.11p. Apesar disto, esta tese não aborda assuntos relacionados à disseminação de \textit{WAVE Service Advertisements} (WSA) e trocas de dados por meio de \textit{WAVE Short Messages} (WSMs), embora seja possível usar o RVEP como um único protocolo para a camada de rede sobre o IEEE 802.11p.

Com a criação da HRAdNet-VE e seu protocolo de comunicação, o HRVEP, este não somente herda completamente as estruturas de dados e mecanismos do RVEP, mas também o estende, na direção de fazer com que a HRAdNet-VE seja uma proposta de uma rede  veicular heterogênea centrada em interesses, tendo em vista que este assunto ainda é pouco explorado na comunidade de redes veiculares \cite{Zheng:2015}. Portanto, o HRVEP interage com as tecnologias de acesso à comunicação sem fio baseada em comunicação dedicada de curto alcance como o RVEP. No entanto, o nó não precisa possuir somente uma única interface de acesso à comunicação sem fio, ele pode ser equipado com múltiplas interfaces. Além de redes formadas por dispositivos de comunicação dedicada de curto alcance, o HRVEP é capaz de lidar com redes celulares, conforme detalhado na seção anterior. Portanto, o HRVEP atua como um substituto do IP na camada de rede dos nós equipados simultaneamente com interfaces de acesso à comunicação sem fio baseadas nos padrões IEEE 802.11, IEEE 802.11p e LTE.

No que diz respeito à comunicações sem fio, usando interfaces de acesso baseadas no padrão LTE,  a proposta do HRVEP não trata de comunicações veículo-a-veículo usando LTE \textit{Device-to-Device} (D2D) \cite{Zheng:2015}. Logo, os veículos não podem se comunicar diretamente uns com os outros por meio de redes celulares. Neste trabalho, os nós configurados como \textit{User Equipaments} (UEs) só podem se comunicar com nós configurados como \textit{Evolved Node B} (eNodeB).  Logo, os nós UE podem se movimentar e trocar dados com estações base, que são os nós eNodeB. No entanto, é possível que os nós UE se comuniquem uns com os outros de maneira indireta, desde que existam nós eNodeB, atuando como \textit{relayers} na rede.
    
\section{Considerações Finais}

Este capítulo apresentou duas contribuições desta tese, que são a RAdNet-VE e a HRAdNet. Ambos os trabalhos foram fundamentais para o desenvolvimento das propostas que serão apresentadas nos Capítulos \ref{cap:controle} e \ref{cap:orientacao}. 

No que diz respeito à proposta a ser apresentada pelo Capítulo \ref{cap:controle},  a RAdNet-VE forneceu todo o substrato de comunicação eficiente, permitindo a construção de uma estratégia de controle inteligente de tráfego para interseções isoladas, uma vez que tal estratégia demandava somente o uso de interfaces de acesso à comunicação sem fio baseadas no padrão IEEE 802.11n. A partir da criação da HRAdNet-VE, foi possível construir as estratégias para controlar interseções cujo acesso é compartilhado entre corredores e sincronizar as indicações de intervalos de verde das sinalizações semafóricas instaladas ao longo destes corredores.

Acerca da proposta a ser apresentada pelo Capítulo \ref{cap:orientacao}, a HRAdNet-VE permitiu que as sinalizações semafóricas pudessem se comunicar umas com as outras, a fim de compartilhar suas agendas de tempos. Partindo disto, foi possível construir uma estratégia para o problema de planejamento e orientação de rotas. Neste caso, a HRAdNet-VE foi fundamental, pois ela possibilitou que não somente veículos pudessem se comunicar com as sinalizações semafóricas, a fim de informar suas presenças nas vias e requisitar cálculos de rotas ótimas a elas, mas também permitiu que as mesmas pudessem se comunicar, a fim de atualizar o estado global do sistema de informações ao motorista, no que tange  alocação de espaços de nas vias, à medida que estes são alocados e desalocadas conforme os resultados dos cálculos de rotas ótimas. Por meio da HRAdNet-VE, as sinalizações semafóricas puderam notificar umas às outras sobre qualquer mudança e, cada sinalização semafórica, ao receber tal notificação, foi capaz de notificar os veículos que trafegam na via sob seu controle.  

Antes de apresentar os detalhes relativos às abordagens para os problemas de controle de tráfego e, planejamento e orientação de rotas, é necessário detalhar como o paradigma de sistemas multiagentes é utilizado nesta tese. Por isto, o próximo capítulo descreve os agentes e como eles interagem, a fim de formar um sistema multiagente.

  \chapter{Definição do Agentes dos Sistemas Multiagentes}\label{cap:agentes}

Este capítulo tem como objetivo introduzir e descrever de forma alto nível os agentes  utilizados tanto no sistema de controle de tráfego quanto no sistema de planejamento e orientação de rotas que são: Centro de Controle de Tráfego, Elemento Urbano, Veículo e Sinalização Semafórica. A descrições dos agentes busca tornar clara a participação e a contribuição dos mesmos, no âmbito do sistema de controle de tráfego e do sistema de planejamento e orientação de rotas. Portanto, este capítulo é uma preparação para os detalhamentos acerca destes sistemas, a partir dos dois capítulos subsequentes a este.

\section{Agente Centro de Controle de Tráfego} \label{sec:cct}

Nesta tese, o agente Centro de Controle de Tráfego foi projetado para monitorar o estado de controle das interseções isoladas e interseções que participam de sistemas coordenados de sinalizações semafóricas, conforme definido na \textit{National ITS Architecture 7.1} \cite{DOT:2016}. Para tanto, o agente Centro de Controle de Tráfego monitora as interações entre agentes Sinalização Semafórica, à medida que estes notificam uns aos outros sobre as mudanças de estado de seus mecanismos de controle de interseções e dos mecanismos de alocação de espaços nas vias de entrada das interseções, onde as sinalizações semafóricas que os embutem atuam. A partir deste monitoramento de interações, o agente Centro de Controle de Tráfego atualiza um sistema supervisório, que é responsável em monitorar o funcionamento do sistema de controle de tráfego. Para tanto, o agente Centro de Controle de Tráfego deve ser embutido em um nó de rede cuja função é atuar como um \textit{relayer} de mensagens enviadas por meio de interfaces de acesso à comunicação sem fio baseadas no padrão LTE \cite{Zheng:2015}. Nesta tese, considera-se que o sistema supervisório usa os dados obtidos pelo agente Centro de Controle de Tráfego para manter uma simulação do funcionamento do sistema real de controle de tráfego. 

O uso de um sistema supervisório tanto para o sistema de controle de tráfego quanto para o sistema de planejamento e orientação de rotas se faz necessário, pois os engenheiros de tráfego precisam interagir com o sistema real de controle de tráfego, seja recebendo informações acerca das condições de tráfego ou entrando com valores de parâmetros de configuração dos agentes Sinalização Semafórica (veja na Seção \ref{sec:sinalizacao}). Tais parâmetros são utilizados durante a inicialização dos agentes Sinalização Semafórica, quando as sinalizações semafóricas dos mesmos são ligadas. Além disto, levando em consideração a infraestrutura computacional disponível nos centros de controle de operações de tráfego, o acesso ao estado da simulação pode ser fornecido com alta disponibilidade. Ainda que não seja possível o fornecimento de uma infraestrutura computacional robusta, é possível usar a estratégia de uma simulação, pois ela somente levaria em consideração o controle de interseções isoladas e o controle de sistemas coordenados de sinalizações semafóricas. No que diz respeito aos veículos, a simulação apenas usa dados macroscópicos, em específico, a quantidade de veículos nas vias de entrada das interseções controladas pelo sistema de controle de tráfego. Nesta tese, não são abordados detalhes a respeito de uma implementação de um sistema supervisório e dos mecanismos para simulação do sistema de controle de tráfego mantida por ele. 

O estado da simulação mantida pelo sistema supervisório é fundamental para a recuperação tanto dos controles de interseções isoladas quanto dos controles de sistemas coordenados de sinalizações semafóricas. Neste caso, os estados, relativos ao controle de interseções isoladas ou de sistemas coordenados de sinalizações semafóricas, podem ser recuperados por meio de uma interação entre um agente Veículo, responsável pelo controle de uma interseção, e o agente Centro de Controle de Tráfego. Tal estado de controle consiste em parâmetros de configuração das sinalizações semafóricas, assim como, estruturas de dados de controle, como o multigrafo utilizado nas dinâmicas do SMER para controle de interseções isoladas, além das leituras referentes aos fluxos de tráfego das vias de entrada das interseções. Após a recuperação desse estado, os veículos podem ser utilizados para controlar autonomamente interseções isoladas ou aquelas que integram os sistemas coordenados de sinalizações semafóricas, criando sinalizações semafóricas virtuais \cite{Ferreira:2010}, que, por sua vez, operam de acordo com as dinâmicas do SMER para controle de interseções isoladas.  A condição para que isto aconteça é a detecção da ausência de funcionamento de sinalizações semafóricas. A ausência de funcionamento de sinalizações semafóricas pode ser causada por muitos fatores, dentre eles: falhas de equipamentos, falta de fornecimento de energia elétrica e acidentes. 

Uma vez que uma sinalização semafórica para de funcionar, isto é rapidamente detectado pelo agente Centro de Controle de Tráfego, pois a interação entre os agentes Sinalização Semafórica e o agente Centro de Controle de Tráfego é interrompida. Se isto acontece em uma interseção, onde uma das sinalizações semafóricas embute um agente Sinalização Semafórica responsável pelo controle de um sistema coordenado de sinalizações semafóricas, o controle não é afetado. Neste caso, o controle de sistemas coordenados de sinalizações semafóricas é assumido pelo agente Centro de Controle de Tráfego. Portanto, este agente se torna responsável por ativar e desativar sistemas coordenados de sinalizações semafóricas, bem como, ajustar os tamanhos dos intervalos de indicações de luzes verdes, a fim de sincronizar os inícios de tais indicações. Aquelas interseções cujas sinalizações semafóricas não funcionam, em específico, as que são líderes de seus sistemas coordenados de sinalizações semafóricas, passam a ser controladas pelos veículos que se aproximam delas. Quando as sinalizações semafóricas voltam a funcionar, o agente Centro de Controle de Tráfego devolve o controle de sistemas coordenados de sinalizações semafóricas para os agentes Sinalização Semafórica responsáveis por tal tarefa. Tais agentes são aqueles embutidos nas sinalizações semafóricas líderes de sistemas coordenados de sinalizações semafóricas. Além desta situação relacionada aos sistemas coordenados de sinalizações semafóricas, existe outra, que se relaciona com a ausência de funcionamento de sinalizações semafóricas que integram esses sistemas. Uma vez detectada a ausência de funcionamento destas sinalizações semafóricas, os agentes Veículo podem recuperar os estados de controle das interseções, interagindo com o agente Centro de Controle de Tráfego, e, em seguida, controlar as interseções autonomamente. No que tange a ausência de funcionamento das sinalizações semafóricas instaladas em interseções isoladas, os agentes Veículo também podem recuperar os dados relativos ao controle delas e, após isto, controlar autonomamente o acesso à estas regiões da rede viária. 

A estratégia de manter uma simulação do sistema real de tráfego permite a introdução de um mecanismo de recuperação de falhas. No que tange o controle de sistemas coordenados de sinalizações semafóricas, assim como, a coordenação do início dos intervalos destas, quando seus sistemas coordenados estão ativos, é fundamental para manutenção da fluidez do tráfego em uma rede viária. Embora as sinalizações semafóricas instaladas em interseções isoladas sejam de grande importância, pois estas, por meio dos agentes Sinalização Semafórica, são capazes de se adaptar às flutuações do tráfego das vias de entrada dessas interseções, o controle de sistemas coordenados de redes de sinalizações semafóricas e a coordenação do início das indicações de intervalo de verde das mesmas têm uma importância ainda maior. Com um mecanismo de recuperação de falhas em que agentes Veículos podem tomar o controle de interseções, caso as sinalizações destas apresentem ausência de funcionamento, pelotões de veículos têm a garantia de que não serão parados, enquanto atravessam os corredores em que tais interseções fazem parte. 

Além disso, outro ponto importante, no que tange a estratégia de manter uma simulação do sistema real de tráfego, é a manutenção das agendas de intervalos de indicações de luzes verdes cujas sinalizações semafóricas apresentam ausência de funcionamento. Uma vez que as sinalizações semafóricas apresentam ausência de funcionamento, seus agentes não podem gerar agendas de tempos a partir das configurações de controle vigentes, assim como, não podem compartilhar tais configurações com os demais agentes Sinalização Semafórica do sistema de controle de tráfego, de modo que estes últimos possam também gerar e manter as cópias dessas agendas. Uma vez que os agentes Veículo tomam o controle das interseções, criando sinalizações semafóricas virtuais, eles interagem com o agente Centro de Controle de Tráfego e os agentes Sinalização Semafórica. A partir disto, o agente Centro de Controle de Tráfego volta a monitorar o estado de controle das interseções isoladas e ou interseções participantes de sistemas coordenados de sinalizações semafóricas. Uma vez que as sinalizações semafóricas são criadas por meio da cooperação entre agentes Veículo, objetivando o controle de uma interseção, um agente Veículo se torna responsável pelo controle da interseção. Assim, este interage com o agente Centro de Controle de Tráfego, como se fosse um agente Sinalização Semafórica, e com os vários agentes Sinalização Semafórica do sistema de controle de tráfego. Por meio da interação entre o agente Veículo, responsável pelo controle de uma interação, com os demais agentes Sinalização Semafórica, o agente Centro de Controle de Tráfego monitora tais interações, e, a partir disto, atualiza o sistema supervisório, de modo que este mude o estado da simulação. Além disto, estas interações possibilitam que os agentes Sinalização Semafórica possam manter agendas de tempos relativas às sinalizações semafóricas virtuais. Além disto, os agentes Sinalização Semafórica também interagem com os agentes Veículo responsáveis pelas sinalizações semafóricas virtuais. Com isto, além de manterem as agendas de intervalos de indicações de luzes verdes relativas ao controle das interseções por meio de sinalizações semafóricas virtuais, os agentes Veículo também recebem configurações de controle de outras interseções, enviadas pelos agentes Sinalização Semafórica presentes nestas, e, com isto, gera e mantém cópias de agendas de tempos, enquanto são responsáveis pelo controle das interseções. Com isto, estes agentes Veículo podem também calcular rotas ótimas, alocar espaços de uso das vias de entradas das interseções e fornecer dados relativos às ondas verdes para os veículos que se aproximam das interseções por meio de suas vias de entrada. 

Com a ausência de veículos se aproximando de interseções, onde as sinalizações semafóricas apresentam ausência de funcionamento, o agente Centro de Controle de Tráfego para de monitorar os estados de controle relativos a estas interseções, pois não existes sinalizações semafóricas virtuais. Por isto, o agente Centro de Controle de Tráfego altera os estados de controle das interseções na simulação mantida pelo sistema supervisório, ajustando as representações dos mecanismos de controle dessas interseções, de modo que as representações das sinalizações semafóricas simulem o funcionamento destas, quando não existem fluxos de veículos nas vias de entrada se aproximando das interseções. Após isto, o agente Centro de Controle de Tráfego interage com os agentes Sinalização Semafórica, compartilhando as novas configurações de controle das interseções cujas sinalizações semafóricas apresentam ausência de funcionamento. Os agentes Sinalização Semafórica, então, geram novas agendas de tempos a partir das novas configurações de controle de interseções recebidas por eles. Este processo é valido somente para interseções isoladas e aquelas que participam de somente um sistema coordenado de sinalizações semafóricas, pois estas últimas só participam da coordenação de sinalizações semafóricas, quando o sistema coordenado em que elas estão inseridas está ativo. Caso contrário, a interseção é vista como uma interseção isolada e, por isto, é tratada como tal.

\section{Agente Elemento Urbano}

Nesta tese, o agente Elemento Urbano foi primeiramente projetado para representar entidades de mundo real, que estão presentes no ambiente urbano, tais como: instituições sociais e políticas; instalações educacionais e culturais; instalações comerciais e ou de serviços; instalações relativas ao estacionamento de veículos particulares; instalações de veículos comerciais; entre outros. Para tanto, este agente deve ser embutido em dispositivos capazes de se comunicar por meio de interfaces de acesso à comunicação sem fio, que permitam a comunicação indireta com sinalizações semafórica e veículos. Logo, tal interface deve ser baseada no padrão LTE \cite{Zheng:2015}.

Uma vez que os agentes Elemento Urbano podem representar entidades de mundo real, eles podem interagir periodicamente tanto com agentes Sinalização Semafórica quanto agentes Veículo, desde que este último esteja controlando uma interseção, compartilhando dados que os associam a interesses. Com estes dados, os agentes Sinalização Semafórica e Veículo, este último, de acordo com a condição descrita anteriormente, registram os interesses na camada de rede de seus ambientes e os associam às ações executadas por estes agentes. Desta forma, os ambientes (sinalizações semafóricas e veículos conectados) são configurados com os interesses relacionados aos agentes Elemento Urbano, que estão operando dentro de uma região, onde os sistemas de controle de tráfego e, planejamento e orientação de rotas atuam. Para tanto, o agente Elemento Urbano deve conter os seguintes parâmetros de configuração de performance: periodicidade de envio de mensagens interação e identificador da via onde se encontra.

Diferente dos ambientes relacionados aos agentes Sinalização Semafórica, os ambientes relativos aos agentes Veículo não mantêm, por muito tempo, os interesses registrados em suas camadas de rede, pois eles, à medida que seus agentes deixam de controlar interseções isoladas e cooperar com sistemas coordenados redes de sinalizações semafóricas, têm tais interesses removidos de suas camadas de redes. Outra finalidade relativa aos dados compartilhados pelos agentes Elemento Urbano é o uso destes para fins de atualização dos bancos de dados de mapas instalados nos dispositivos que embutem os agentes Veículo. Assim, nesta tese, parte-se do princípio que os motoristas solicitam rotas ótimas até os elementos urbanos, utilizando os interesses contidos nos dados compartilhados por seus agentes Elemento Urbano. Por uma questão de simplificação, nesta tese, os agentes Veículo utilizam como interesse o nome da via, onde o elemento urbano se encontra, para encontrar rotas ótimas. Logo, esta tese parte do princípio que os dispositivos que embutem os agentes Veículo possuem uma interface, em que os motoristas buscam os elementos urbanos e, em seguida, os selecionam, fazendo com que os agentes Veículos obtenham rotas ótimas a partir das interações com agentes Sinalização Semafórica, tendo como base disso a comunicação centrada em interesses provida por uma rede ad hoc veicular heterogênea centrada em interesses.

Embora esta tese não aborde questões relativas à segurança de redes \textit{ad hoc} veiculares centradas em interesses, é necessário ressaltar, que os interesses relativos aos agentes Elemento Urbano devem ser validados, a fim de evitar inconsistências nas associações entre os interesses e as ações executadas pelos agentes. No entanto, é importante indicar um possível caminho para tratar este problema. Neste sentido, esta tese aponta para a investigação do uso de técnicas baseadas em nomes auto-certificáveis \cite{Ghodsi:2011}, no que tange a validação de interesses. Nomes auto-certificáveis são nomes que permitem a verificação direta entre o nome e o objeto associado \cite{Ghodsi:2011}. Neste caso, o objeto associado pode ser o nome da ação a ser executada pelos agentes Veículo e Sinalização Semafórica. 

\section{Agente Veículo} \label{sec:veiculo}

Nesta tese, o agente Veículo foi projetado para interagir tanto com agentes Sinalização Semafórica quanto com outros agentes Veículo. Para tanto, tal agente precisar ser embutido em um veículo conectado. Um veículo conectado é aquele capaz de se comunicar tanto com outros veículos quanto com elementos relativos a infraestrutura viária, utilizando interfaces de acesso à comunicação sem fio baseadas em padrões tecnológicos, como IEEE 802.11 \cite{IEEE80211:2012}, IEEE 802.11p \cite{IEEE80211p:2010} e LTE \cite{Zheng:2015}. Além disto, tal tipo de veículo também oferece capacidade de realizar computações, permitindo que os mesmos sejam utilizados em aplicações de software para área de sistemas inteligentes de transporte. 

No que diz respeito à interação com agentes Sinalização Semafórica, no âmbito do sistema de controle de tráfego, os agentes Veículo interagem com estes agentes, a fim de informar a presença dos veículos nas vias de entrada de interseções controladas por sinalizações semafóricas. À medida que os agentes Veículo informam a presença de seus veículos conectados, interagindo com agentes Sinalização Semafórica, estes, por sua vez, passam a ter ciência dos veículos que estão se aproximando da interseção e, em seguida, confirmam isto, interagindo com os agentes Veículo. Para que cada agente possa realizar a ação descrita, eles contam com um parâmetro que contém a frequência de envios de mensagens desta interação para os agentes Sinalização Semafórica. A repetição de envios de mensagens da interação entre agentes, aqui em discussão, só é interrompida, após os agentes Veículo terem ciência de que as suas presenças foram reconhecidas pelos agentes Sinalização Semafórica. Vale ressaltar, que a interação em questão se dá sempre no escopo da via onde os veículos estejam trafegando e as sinalizações semafóricas estão instaladas. Portanto, tanto os agentes Veículo quanto os agentes Sinalização Semafórica têm ciência da via em que estão operando e, consequentemente, registram tal dado na camada de rede de seus ambientes. Por fim, os agentes Sinalização Semafórica precisam das interações com os agentes Veículos, para acumular conhecimento acerca das flutuações do fluxo de tráfego nas vias de entrada das interseções, onde suas sinalizações semafóricas estão instaladas. Os agentes Sinalização Semafórica utilizam este conhecimento para ajustar o tamanho dos intervalos de indicações das sinalizações semafóricas em que estão embutidos.

No que diz respeito a interação entre agentes Veículo e agentes Sinalização Semafórica, no âmbito do sistema de planejamento e orientação de rotas, os agentes Veículo interagem com os agentes Sinalização Semafórica, requisitando rotas ótimas, de acordo com os destinos informados pelos motoristas dos veículos conectados em que os agentes estão embutidos. Como os agentes Sinalização Semafórica têm a base de conhecimento necessária para compartilhar o conhecimento de rotas ótimas com os agentes Veículo, eles compartilham estas rotas com os agentes Veículo. No instante em que os agentes Sinalização Semafórica interagem com os agentes Veículo, em resposta a interação anterior, estes últimos passam a ter conhecimento de rotas ao longo de uma rede viária, que podem auxiliar o motorista, enquanto este dirige seu veículo conectado, objetivando uma viajem com o menor tempo possível. Para que todo esse processo de interação entre os agentes ocorra, é necessário que os agentes Veículo tenham como configuração de performance um parâmetro, no que diz respeito à frequência com que as mensagens de interação, requisitando rotas ótimas, são enviadas pelos agentes Veículo para os agentes Sinalizações Semafóricas. Vale ressaltar, que os envios das mensagens de interação, requisitando rotas ótimas, são cessados, à medida que os agentes Veículos passam a ter conhecimento de uma rota ótima. 

Ainda no âmbito do sistema de planejamento e orientação de rotas, à medida que os agentes Veículo orientam os motoristas dos veículos, estes precisam interagir com os agentes Sinalização Semafórica novamente. Desta vez, os agentes Veículo precisam ter ciência acerca do posicionamento, comprimento e velocidade dos seguimentos de ondas verdes nas vias em que as sinalizações semafóricas, contendo o tipo de agente correspondente as mesmas, estejam instaladas. Além disto, os agentes Veículo também precisam conhecer a duração do intervalo de verde disponibilizado pelo agente Sinalização Semafórica da via. O mecanismo das ondas verdes permite que os agentes Veículo orientem os motoristas dos veículos conectados, no que diz respeito ao controle de velocidade destes. Para tanto, os agentes Veículo usam o conhecimento acerca dos seguimentos de ondas verdes, a fim de se manterem atualizados sobre descolamento destes sobre a via em que estão posicionados. Logo, os agentes Veículo precisam ser configurados quanto à frequência de atualização de deslocamentos de seguimentos de onda verde. Por isto, os agentes Veículo possuem um parâmetro de configuração de performance para tal fim. É importante pontuar aqui também, que os agentes Veículo não automatizam a condução dos veículos. Portanto, eles apenas orientam os motoristas dos veículos a tomarem as seguintes decisões:

\begin{enumerate}
	\item Acelerar o veículo conectado, caso esteja posicionado atrás do ponto de término do seguimento de onda verde, a fim de alcançá-lo; 
    \item Manter a velocidade permitida da via, caso o veículo conectado esteja posicionado entre os pontos de início e fim do seguimento de onda verde;
    \item Reduzir a velocidade do veículo conectado, caso esteja posicionado à frente do ponto de início do seguimento de onda verde, a fim de esperar este último, até que ele alcance o veículo. Esta ação pode levar a retirada de veículos sobre o fim do seguimento de onda verde. 
\end{enumerate}

Embora os agentes Veículo possam orientar motoristas a seguirem as velocidades recomendadas por eles, os mesmos precisam ter conhecimento acerca de outros agentes Veículo, operando na mesma via em que trafegam. Este conhecimento é fundamental nas ações de orientação de velocidades, pois os agentes Veículos precisam conhecer o estado relativo do veículo que está imediatamente à frente do veículo conectado em que eles estão embutidos. Tal conhecimento é utilizado para auxiliar no cálculo da velocidade que um veículo conectado deve assumir, de acordo com as possibilidades de os motoristas alcançarem os seguimentos de ondas verdes, assim como, se manterem nestes ou esperá-los até que os veículos conectados sejam alcançados pelos mesmos. Portanto, os agentes Veículo interagem uns com outros, trocando dados sobre os estados relativos dos veículos em que estão embutidos. O estado relativo refere-se à velocidade e a posição geográfica do veículo conectado. Por isto, os agentes Veículo precisam ser configurados com um parâmetro de frequência para envio de mensagens de interação entre eles. Tal parâmetro deve compor juntamente com os outros parâmetros mencionados até aqui, as configurações dos agentes Veículo.

Os agentes Veículo podem assumir também a responsabilidade de controlar interseções, criando sinalizações semafóricas virtuais, quando eles detectam a ausência de um agente Sinalização Semafórica em uma das vias de entrada de uma interseção controlada por sinalizações semafóricas. Desde então, o agente Veículo interage com o agente Centro de Controle de Tráfego, a fim de obter estado de controle da interseção, tais como os descritos na Seção \ref{sec:sinalizacao}. Uma vez que estado de controle de uma interseção tenha sido obtido por um agente Veículo, ele sabe se ele ou outro agente Veículo em uma via conflitante será o responsável em manter uma sinalização semafórica virtual, pois ambos passam a simular o funcionamento das sinalizações semafóricas da interseção de onde eles se aproximam, utilizando as configurações de controle da interseção. Caso a interseção tenha uma sinalização semafórica cujo agente Sinalização Semafórica controlava um sistema coordenado de sinalizações semafóricas, o agente Veículo assume as configurações fornecidas pelo agente Centro de Controle de Tráfego, que neste momento está assumindo o controle de sistemas coordenados de sinalizações semafóricas. Dessa forma, um agente Veículo se torna responsável por uma interseção, quando a sinalização semafórica virtual está indicando luz vermelha e o seu veículo conectado é o mais próximo da faixa de retenção da via. A partir disto, ele adquire o comportamento de um agente Sinalização Semafórica, de acordo com as configurações de controle da interseção em questão, com exceção daquelas configurações de controle relativas ao controle de sistemas coordenados de redes de sinalizações semafóricas. Logo, durante intervalo de indicação de luz vermelha da sinalização semafórica virtual de sua via, ele é capaz de realizar as seguintes tarefas, de acordo com as configurações de controle da interseção em questão: manter agendas de tempos geradas com base em configurações de controle de interseções enviadas por sinalizações semafóricas; calcular rotas ótimas, à medida que os agentes Veículo às solicitam, alocar espaços de uso nas vias de entrada das interseções; manter e compartilhar dados acerca dos posicionamentos das ondas verdes das vias de entrada da interseção, onde ele é responsável; ajustar os intervalos de luzes verde, de acordo com o fluxo das vias de entrada; notificar os responsáveis por outras interseções, sejam estes agentes Sinalização Semafórica ou agentes Veículo, acerca das atualizações do estado de controle da interseção; e notificar os agentes Veículo, que estão embutidos nos veículos conectados que trafegam nas vias de entrada da interseção, acerca das atualizações do estado de controle da interseção. No que tange os demais agentes Veículo cujos veículos conectados trafegam pelas vias entrada da interseção, estes também simulam o funcionamento das sinalizações semafóricas da interseção para onde estão se dirigindo, de modo que os motoristas destes veículos tenham ciência das indicações das sinalizações semafóricas virtuais das vias de onde eles estão conduzindo seus veículos. Durante o intervalo de amarelo de uma via cuja sinalização semafórica virtual indicava luz verde, os agentes Veículos interagem entre si, a fim de elegerem um líder, que é aquele cujo veículo conectado mais próximo da faixa de retenção da via onde eles trafegam. Após o intervalo de amarelo, aquele agente cujo veículo é o mais próximo da faixa de retenção da via onde ele se encontra, passa a ser o novo responsável pelo controle da interseção. Vale ressaltar que tal controle varia de acordo com interseção. Se ela é uma interseção isolada, os agentes Veículo executarão o algoritmo SMER para controle de interseções isoladas. Caso contrário, se a interseção está participando de uma coordenação de redes de sinalização semafóricas em um sistema coordenado de sinalizações semafóricas ativo, os agentes Veículo seguirão as configurações de controle para tal situação. Por fim, se a interseção participa somente de um sistema coordenado de sinalização semafórica e este não se encontra ativo, os agentes Veículo executarão o algoritmo SMER para controle de interseções isoladas. 

\section{Agente Sinalização Semafórica} \label{sec:sinalizacao}

Nesta tese, o agente Sinalização Semafórica foi projetado com intuito de interagir tanto com os agentes Veículo quanto com outros agentes Sinalização Semafórica. Para isto, o agente deve ser embutido em um dispositivo que não somente permita controlar as indicações presentes nos grupos focais das sinalizações semafóricas, mas também de realizar computações e se comunicar tanto com veículos conectados quanto com outras sinalizações semafóricas, sendo estas participantes de uma mesma interseção ou não. Isto torna as sinalizações semafóricas inteligentes. Devido a esta característica, estes dispositivos devem ser equipados com interfaces de acesso à comunicação baseadas nos padrões IEEE 802.11 \cite{IEEE80211:2012}, IEEE 802.11p \cite{IEEE80211p:2010} e LTE \cite{Zheng:2015}.

No que diz respeito à interação dos agentes Sinalização Semafórica com os agentes Veículo, no âmbito do sistema de controle de tráfego, os agentes Sinalização Semafórica interagem com tais agentes, a fim de tornar mais precisa a medição do fluxo de tráfego das vias de entrada das interseções controladas por sinalizações semafóricas. Para tanto, os agentes Sinalização Semafórica reiniciam o processo de notificação de presença de veículos que se aproximam da interseção controlada por suas sinalizações semafóricas. Assim, os agentes Veículo dão início à interação descrita na seção anterior, sempre que recebem uma mensagem de interação dos agentes Sinalização Semafórica, para tal fim. Esta interação se dá de maneira periódica e, por isto, parâmetros, definindo esta periodicidade de obtenção de dados e a quantidade de obtenções de dados, são incluídos nas configurações dos agentes Sinalização Semafórica.

Ainda no âmbito do sistema de controle de tráfego, os agentes Sinalização Semafórica interagem uns com os outros para atingir três objetivos, que são: 

\begin{enumerate}
	\item Controlar interseções isoladas, levando em consideração as flutuações dos fluxos de tráfego das vias de entrada destes locais;
    \item Controlar interseções compartilhadas entre sistemas coordenados de sinalizações semafóricas, levando em consideração as flutuações de tráfego das vias que compõem os mesmos;
    \item Coordenar as redes de sinalizações semafóricas instaladas nos corredores, de modo que as sinalizações semafóricas tenham seus intervalos de indicações de luzes verdes reprogramados, a fim de permitir que pelotões de veículos atravessem os corredores, sem haver interrupções desnecessárias da locomoção do mesmos.
\end{enumerate}

No primeiro item da lista enumerada acima, os agentes Sinalização Semafórica interagem uns com os outros, primeiramente, trocando mensagens relacionadas à execução do algoritmo SMER para controle de interseções isoladas. Por meio destas mensagens, os agentes Sinalizações Semafóricas têm conhecimento do estado em que as interseções se encontram. Em outras palavras, este estado quer dizer quais vias têm permissão para escoar seus fluxos de tráfego e quais as vias não têm tal permissão. Uma segunda interação, relacionada ao primeiro item da lista enumerada acima, é quando os agentes Sinalização Semafórica interagem uns com os outros, compartilhando o conhecimento que cada um tem a respeito das condições de tráfego da via entrada controlada pela sinalização semafórica que o embute. Por meio desta segunda interação, todos os agentes Sinalização Semafórica têm conhecimento das demandas de cada via de entrada da interseção em que as sinalizações semafóricas que os embutem participam. Por fim, uma terceira interação, em que os agentes Sinalização Semafórica realizam uma última interação, que é a atualização das reversibilidades e, consequentemente, do número de arestas entre os vértices do grafo utilizado pelo algoritmo SMER para controle de interseções isoladas. 

Nesta tese, todos os agentes Sinalização Semafórica possuem um conjunto comum de parâmetros de configuração. Tal conjunto é formado pelos seguintes parâmetros: identificação da sinalização semafórica, identificação da interseção, tempo mínimo do intervalo de verde, tempo máximo do intervalo de verde, tempo do intervalo de amarelo, tempo de vermelho geral, via de entrada da interseção, grafo de controle utilizado durante a execução do algoritmo SMER para controle de interseções isoladas, comprimento da área para monitoramento fluxo de tráfego, número de sinalizações semafóricas da interseção, periodicidade de obtenção de quantidade de veículos de uma via e número de obtenções de quantidade de veículos de uma via.  

No segundo item da lista enumerada acima, os agentes Sinalização Semafórica interagem uns com os outros, trocando mensagens relacionadas à execução do algoritmo SMER, mas no contexto do controle do acesso às interseções compartilhadas entre sistemas coordenados de sinalizações semafóricas. Por meio das mensagens de interação, os agentes Sinalização Semafórica ativam e desativam os sistemas coordenados de sinalizações semafóricas. Uma vez que um sistema coordenado é ativado, todas as sinalizações semafóricas pertencentes a ele têm seus intervalos de indicações de luzes verdes modificados, de modo que cada uma possa operar de maneira síncrona em relação à outra. Para tanto, um agente Sinalização Semafórica é o líder do sistema coordenado de sinalizações semafóricas e, por isto, ele interage primeiramente com outros agentes com a mesma responsabilidade, a fim de executarem o algoritmo SMER para controle de sistemas coordenados de sinalizações semafóricas. Neste sentido, cada agente Sinalização Semafórica líder de um sistema coordenado de sinalizações semafóricas é representado como um vértice do grafo utilizado pelo SMER para controle de sistemas coordenados de sinalizações semafóricas, a fim de ativar e desativar a sincronização de redes de sinalizações semafóricas instaladas em corredores. Além disto, os agentes Sinalização Semafórica líderes de sistemas de sinalizações semafóricas interagem com os demais agentes embutidos nas sinalizações semafóricas do mesmo sistema coordenado. Cada agente líder interage com os demais agentes do mesmo sistema coordenado, atuando como um mestre e os demais agentes como escravos \cite{Buschmann:1996}. Isto, portanto, caracteriza uma segunda interação do agente Sinalização, dentro do contexto relacionado ao segundo item da lista enumerada acima. No que tange o controle de sistemas coordenados de sinalizações semafóricas, cada um vértice do grafo, que é utilizado durante a execução do algoritmo SMER para controle de sistemas coordenados de sinalização semafórica, representa uma agente Sinalização Semafórica. Por isto, estes agentes controladores precisam também ser configurados com os seguintes parâmetros de configuração: identificador do corredor contendo as sinalizações semafóricas, lista de participantes do sistema coordenado de sinalizações semafóricas, seguimentos de via componentes do corredor, grafo de controle utilizado pelo algoritmo SMER para controle de sistemas coordenados de sinalização semafórica, número mínimo de ciclos por operação de coordenação de sinalizações semafóricas e número máximo de ciclos operação de coordenação de sinalizações semafóricas, lista de identificações de corredores agrupados, periodicidade de compartilhamento de médias de quantidades de veículos, periodicidade de compartilhamento de médias de quantidades veículos de um grupo de sistemas coordenados de sinalizações semafóricas e periodicidade de atualização de demandas de corredores de sistemas coordenados de sinalizações semafóricas. 

Dessa forma, cada agente líder de um sistema coordenado de sinalizações semafóricas coordena todos os demais agentes participantes do mesmo sistema coordenado, com o intuito de sincronizar um a um, utilizando seu relógio como ponto de partida. Além disto, o agente Sinalização Semafórica líder de um sistema coordenado de sinalizações semafóricas também interage com o agente Centro de Controle de Tráfego. Isto, portanto, caracteriza o terceiro item da lista enumerada acima. Uma vez que um veículo conectado alcança uma onda verde de uma das sinalizações semafóricas, participantes do sistema coordenado cujas sinalizações semafóricas estevam sincronizadas, ele poderá alcançar o fim do corredor, sem a necessidade de parar atrás de uma sinalização semafórica com a luz vermelha acesa. À medida que um agente controlador de sistema coordenado de sinalizações semafóricas desativa um corredor em função da execução do algoritmo SMER para controle de sistemas coordenados de sinalizações semafóricas, ele realiza uma terceira interação, sendo esta com os demais agentes do mesmo sistema coordenado, liberando-os das operações sincronizadas. Neste caso, estes agentes passam a executar o SMER para controle de interseções isoladas. Além disso, os agentes Sinalização Semafórica controladores de sistemas coordenados de sinalizações semafóricas interagem com os demais agentes de mesma responsabilidade, compartilhando seus conhecimentos acerca das condições de tráfego de seus corredores. A partir desta quarta interação, todos os agentes líderes têm conhecimento das condições de tráfego de cada corredor existente em uma rede viária urbana. Por fim, os agentes Sinalização Semafórica líderes de sistemas coordenados de sinalizações semafóricas realizam uma quinta interação, que é a atualização das reversibilidades e, consequentemente, do número de arestas entre os vértices do grafo utilizado pelo algoritmo SMER para controle de sistemas coordenados de sinalizações semafóricas.  

Para que os agentes Sinalização Semafórica participantes de sistemas coordenados de sinalizações semafóricas interajam com os agentes Sinalização Semafórica controladores, compartilhando dados acerca das condições de tráfego das vias, onde suas sinalizações semafóricas estão instaladas, estes agentes precisam ser configurados com os seguintes parâmetros de configuração: identificadores dos corredores e identificadores dos controladores dos sistemas coordenados de sinalização semafóricas.

No âmbito do sistema de planejamento e orientação de rotas, os agentes Sinalização Semafórica interagem uns com os outros para atingir os seguintes objetivos, são eles:  

\begin{enumerate}
	\item Manter atualizadas as cópias de agendas de intervalos de indicações de luzes verdes geradas com base nos estados de controle enviados pelas sinalizações semafóricas instaladas em interseções isoladas;
    \item Manter atualizadas as cópias de agendas de intervalos de luzes verdes geradas com base nos estados de controle enviados pelas sinalizações semafóricas, que são instaladas em interseções participantes de sistemas coordenados de sinalizações semafóricas;
    \item Alocar e desalocar espaços de uso em vias de entradas de interseções controladas por sinalizações semafóricas.
\end{enumerate}

No primeiro item da lista enumerada acima, os agentes Sinalização Semafórica interagem uns com os outros, assim como, com agentes Veículo. Assim, à medida que os agentes Sinalização Semafórica reprogramam as sinalizações semafóricas, a fim de ajustar os intervalos de indicação de luzes verdes em função das flutuações dos fluxos de tráfego das vias de entrada das interseções, eles interagem com os agentes Sinalização Semafórica da mesma interseção e com os demais agentes Sinalização Semafórica do sistema de controle de tráfego, compartilhando o estado de controle da interseção. Como mencionado anteriormente, tal estado é compostos pelos parâmetros de configuração de performance de todos os agentes Sinalização Semafórica da interseção, assim como, o grafo de controle utilizado pelo algoritmo SMER para controle de interseções isoladas. Quando os agentes Sinalização Semafórica recebem tal estado de controle, eles o utilizam para gerar as entradas das agendas tempos relativas aos intervalos de indicação de verde de cada uma das sinalizações semafóricas instaladas na interseção. Após isto, cada um destes agentes Sinalização Semafórica interage com os agentes Veículo embutidos nos veículos que trafegam na via controlada pela sinalização semafórica do agente Sinalização Semafórica, notificando os agentes Veículo, de modo que estes possam interagir novamente com o agente Sinalização Semafórica, requisitando novos cálculos de rotas ótimas. 

No segundo item da lista enumerada acima, os agentes Sinalização Semafórica cujas sinalizações semafóricas participam de sistemas coordenados de sinalizações semafóricas interagem com os demais agentes Sinalização Semafórica pertencentes ao sistema de controle de tráfego e agentes Veículo. À medida que os agentes Sinalização Semafórica embutidos em sinalizações semafóricas integrantes de sistemas coordenados de sinalizações semafóricas participam da ativação destes últimos, a fim de ajustar os inícios dos intervalos de indicação de verde de tais sinalizações, de modo que as redes de sinalizações semafóricas destes corredores sejam sincronizadas, eles interagem com os agentes Sinalização Semafórica da mesma interseção. Esta interação faz com que o algoritmo SMER para controle de interseções isoladas controle a interseção, executando como se fosse o algoritmo SER. A partir disto, as agendas de tempos das sinalizações semafóricas de uma mesma interseção são modificadas. Neste caso, as entradas atuais das agendas de intervalos de indicações de luzes verdes das sinalizações semafóricas são removidas e, em seguida, novas entradas são geradas a partir do novo estado de controle da interseção. Após este procedimento, os agentes Sinalização Semafórica das sinalizações semafóricas instaladas na interseção interagem com os agentes Veículo dos veículos conectados que trafegam nas vias de entrada da interseção, de modo que estes possam interagir com os agentes Sinalização Semafórica, a fim de requisitar novos cálculos de rotas ótimas. O próximo passo é interagir com os demais agentes Sinalização Semafórica do sistema de controle de tráfego, compartilhando os novos estados de controle da interseção. O responsável por este passo é o agente Sinalização Semafórica embutido na sinalização semafórica participante do sistema coordenado de sinalizações semafóricas ativado. Ao receberem estes novos estados, os demais agentes Sinalização Semafórica do sistema de controle de tráfego reorganizam as entradas das cópias das agendas de intervalos de indicações de luzes verdes relativas às sinalizações semafóricas das interseções participantes de um sistema coordenado de sinalizações semafóricas ativo, conforme explicado neste parágrafo. Com isto, os demais agentes Sinalização Semafórica interagem com os agentes Veículo dos veículos conectados que trafegam nas vias de entradas das interseções, para que estes interajam com os agentes Sinalização Semafórica, requisitando novos cálculos de rotas ótimas. Por fim, à medida que os sistemas coordenados de sinalizações semafóricas são desativados, conforme as dinâmicas de execução do algoritmo SMER para controle de sistemas coordenados de sinalizações semafóricas, os agentes Sinalização Semafórica, participantes do sistema coordenado de sinalizações semafóricas, voltam a controlar as interseções, de acordo com as dinâmicas de execução do algoritmo SMER para controle de interseções isoladas. Após isto, todo o sistema de controle de tráfego sofre uma atualização, no que diz respeito às novas entradas das agendas de tempos relativas as sinalizações semafóricas que participaram das coordenações das redes de sinalização semafóricas promovidas pela ativação anterior de sistemas coordenados de sinalizações semafóricas. A partir disto, os agentes Sinalização Semafórica interagem com os agentes Veículo dos veículos nas vias de entrada das interseções, de modo que estes interajam com eles, requisitando novos cálculos de rotas ótimas.  

No terceiro item da lista enumerada acima, os agentes Sinalização Semafórica interagem com os demais agentes Sinalização Semafórica pertencentes ao sistema de controle de tráfego, manutenindo as estruturas de dados relativas às alocações de espaços de uso nas vias de entrada das interseções controladas pelas sinalizações semafóricas do sistema de controle de tráfego, após os agentes Veículo confirmarem o recebimento de um cálculo de rota ótima. À medida que os agentes Sinalização Semafórica atualizam localmente essas estruturas de dados, após receberem a confirmação do uso das rotas ótimas por parte dos agentes Veículo, eles interagem com os demais agentes Sinalização Semafórica, atualizando as estruturas de dados destes. Esta atualização consiste em remoções e inserções de entradas nas estruturas de dados relativas às alocações de espaços de uso as vias de entrada das interseções. Vale ressaltar, que para cada entrada de uma agenda de tempos de uma sinalização semafórica existe uma estrutura de dados própria para a alocação de espaços de uso na via de entrada durante o intervalo de indicação de verde, que é representado pela entrada da agenda de tempos. 

\section{Considerações Finais}

Este capítulo introduziu e descreveu os agentes utilizados tanto pelo sistema de controle de tráfego quanto pelo sistema de planejamento e orientação de rotas. A descrição dos agentes quanto dos agentes foi o mais alto nível possível, a fim de preparar para os detalhamentos acerca do sistema de controle de tráfego e do sistema de planejamento e orientação de rotas. 

A fim de dar prosseguimento ao detalhamento desta tese, o capítulo seguinte apresentará os detalhamentos acerca de um sistema de controle de tráfego, utilizando sinalizações semafóricas inteligentes e veículos conectados.

  \chapter{Controle Inteligente de Tráfego Utilizando Sinalizações Semafóricas e Veículos Conectados}\label{cap:controle}

Este capítulo tem como objetivo detalhar a implementação do sistema multiagente para controle de tráfego, que é proposto por esta tese. Este sistema multiagente faz uso de três dos quatro agentes apresentados no capítulo anterior, a saber: Centro de Controle Tráfego, Veículo e Sinalização Semafórica. Para tanto, este capítulo inicia, detalhando o problema de controle de tráfego, a fim de fornecer um esclarecimento maior acerca do assunto. Depois disto, discute-se conceitualmente sobre como tratar o problema de controle de tráfego, aplicando o algoritmo SMER e veículos conectados. Por fim, são apresentados os detalhamentos algorítmicos relativos às estratégias apresentadas para solução do problema de controle de tráfego.

\section{Descrevendo o Problema de Controle de Tráfego}

O problema de controle de tráfego por meio de sinalizações semafóricas está em reconfigurar o tamanho do intervalo de indicação de luz verde de cada sinalização semafórica, a fim de maximizar o fluxo de tráfego nas redes viárias controladas por tais sinalizações, levando em consideração regras de segurança e igualdade. Dessa forma, um conjunto de sinalizações semafóricas controla uma interseção, que é formado por um conjunto de vias. Tal conjunto de vias é dividido em dois subconjuntos, a saber: o subconjunto de vias onde os veículos entram na interseção e o subconjunto de vias onde os veículos deixam a interseção. As regras de segurança descrevem os princípios de funcionamento das sinalizações semafóricas. Maiores detalhes a respeito destes princípios são apresentados na Seção \ref{sec:controle_trafego}. 

Além destas regras, também existem as regras de igualdade cujo intuito é distribuir de maneira justa os tempos das fases das sinalizações semafóricas que controlam as vias de entrada de uma interseção. Portanto, para todas estas vias, deve ser fornecido um tempo mínimo de duração para fase cuja luz é verde. No que tange a intervalo de indicação cuja luz é amarela, esta tem um tempo fixo de duração. De acordo com os fluxos de tráfego que incidem em uma interseção, a fase de luz verde de cada uma das vias de entrada da interseção pode sofrer ajustes. Tais ajustes tem como objetivo aumentar ou diminuir o tamanho do intervalo de indicação de luz verde. A variação permitida para o tamanho do intervalo de indicação de luz verde é compreendida entre um tempo mínimo e um tempo máximo de duração. 

Com isto, fica a cargo de um controlador de sinalizações semafóricas distribuir de maneira ótima os tamanhos dos intervalos de indicação de luz verde para cada uma das vias de entrada de uma interseção, levando em consideração as flutuações do fluxo de tráfego em cada uma delas, sem que os tamanhos dos ciclos de fases das sinalizações ultrapassem o limite máximo de tempo recomendado pela literatura \cite{DENATRAN:2014}. Este problema de controle ótimo de tráfego é relativo às interseções isoladas, que são aquelas interseções, onde as sinalizações semafóricas operam a parte de sistema coordenado de sinalizações semafóricas. 
Um sistema coordenado de sinalizações semafóricas é um grupo de sinalizações semafóricas que têm seus inícios de intervalos de luz verde sincronizados, de modo que os veículos não precisem parar em cada uma interseção controlada por uma das sinalizações semafóricas deste sistema. Dessa forma, sistemas coordenados de sinalizações semafóricas são capazes de fornecer ondas verdes. Apesar disto, estes fluxos contínuos e ininterruptos de veículos podem ser interrompidos, caso os veículos não ganhem acesso exclusivo às interseções compartilhadas entre dois ou mais sistemas coordenados de sinalizações semafóricas, à medida que eles se aproximam delas com o intuito de atravessá-las. Neste caso, deve ficar a cargo de um controlador de interseções compartilhadas tal responsabilidade.

\section{Tratando o Problema de Controle de Tráfego}

Para tratar o problema de controle de tráfego, que consiste em controlar interseções isoladas e interseções compartilhadas entre dois ou mais sistemas coordenados e sinalizações semafóricas, esta tese faz uso do algoritmo SMER, tendo como base o trabalho realizado por \citet{Paiva:2012}. Abaixo, portanto, seguem as explicações de como a proposta de \citet{Paiva:2012} foi aplicada nesta tese. 

\subsection{Estratégia de Controle de Interseções Isoladas}

Para um melhor entendimento sobre como o SMER pode ser aplicado em um mecanismo multiagente de controle de interseções isoladas, é necessário pensar a área de uma interseção entre vias de uma rede viária como um recurso compartilhado entre os fluxos de tráfego oriundos das vias de entrada cujos grupos de movimento são conflitantes. Sendo assim, tais fluxos podem ser considerados processos cujas operações necessitam de acesso a um recurso compartilhado, que é a interseção. Tais processos, por sua vez, são agentes Sinalização Semafórica.

Dessa forma, quando uma via recebe luz verde da sinalização semafórica responsável pelo controle de seu fluxo, o agente Sinalização Semafórica toma para si o recurso compartilhado e este fica indisponível temporariamente para os agentes vizinhos até que a ação de controle termine. Passado o tempo de duração em que as luzes verdes das sinalizações semafóricas se mantêm acesas para as vias pertencentes ao mesmo grupo de movimento, o acesso por parte dos fluxos de tráfegos é interrompido pelo agente Sinalização Semafórica. Esta interrupção do acesso à interseção é sinalizada com a luz vermelha. Consequente, os fluxos de tráfego das vias cujos os grupos de movimento conflitam com aqueles das vias que anteriormente recebiam luz verde de suas sinalizações semafóricas passam a ter acesso à interseção, recebendo luz verde de suas sinalizações semafóricas. 

Para lidar com estas dinâmicas de controle, o SMER faz uso de um multigrafo direcionado, em que cada vértice deste, nesta tese, é um agente Sinalização Semafórica e as áreas onde os fluxos de tráfego se cruzam são representadas por meio de arestas, permitindo definir um relacionamento de exclusão mútua entre fluxos de tráfegos pertencentes a grupos de movimentos conflitantes. Para representar as permissões de acesso aos recursos compartilhados entre os agentes Sinalizações Semafóricas, o SMER utiliza as orientações das arestas do multigrafo para indicar quais agentes podem acessar os recursos compartilhados. Dessa forma, quando um agente Sinalização Semafórica tem o número suficiente de arestas direcionadas para ele em cada arco entre o mesmo e um vizinho, a sinalização semafórica controlada por ele indica o intervalo de luz verde para os veículos da via controlada por ela, acendendo a luz verde do grupo focal da sinalização semafórica. Consequentemente, os veículos da via passam a ter permissão para atravessar a interseção. Enquanto um agente Sinalização Semafórica não tem todas as suas arestas voltadas para ele, o mesmo não permite que os veículos da via controlada por sua sinalização semafórica atravessem a interseção. Passados os intervalos das indicações das luzes verdes de suas sinalizações semafóricas, os agentes embutidos interagem com os demais agentes vizinhos da mesma interseção, revertendo suas arestas por meio de um envio de mensagem de interação. Com isto, é possível definir como as sinalizações semafóricas poderão operar e controlar o acesso à interseção por parte dos fluxos de tráfego de vias de entrada pertencentes a grupos de movimentos conflitantes. 

No entanto, é preciso que o algoritmo SMER consiga lidar com as flutuações de fluxos de tráfego em cada uma das vias de entrada de a interseção. Por utilizar um multigrafo, que é um grafo que possui duas ou mais arestas entre pelo menos um par de vértices pertencentes ao conjunto de vértices, o algoritmo de escalonamento em questão faz uso de uma quantidade de arestas para indicar quais agentes Sinalização Semafórica devem ter suas sinalizações semafóricas indicando intervalos de luz verde. Para indicar a quantidade de arestas que devem incidir nos vértices, adota-se o conceito de reversibilidade para cada um dos agentes Sinalizações Semafórica. A reversibilidade é um número inteiro maior ou igual a um, que indica a quantidade de arestas que cada um dos agentes vizinhos de um agente Sinalização Semafórica precisa direcionar para o mesmo. Todos os agentes Sinalizações Semafóricas são iniciados com o mesmo valor de reversibilidade, que é um. O valor das reversibilidades de cada agente Sinalização Semafórica muda em função das flutuações do fluxo de tráfego da via controlada pela sinalização semafórica que embute o agente.

Por meio do mecanismo de reversibilidades, um agente Sinalização Semafórica permite que os veículos das vias controladas por sua sinalização semafórica atravessem a interseção, até que a quantidade de arestas direcionadas para ele seja menor que o valor de sua reversibilidade. Dessa forma, o algoritmo SMER consegue não pode somente controlar o acesso ao recurso compartilhado, que aqui é uma interseção isolada, mas também balancear o tempo em que os fluxos de tráfego podem fluir através da interseção em função das flutuações de fluxos de tráfego. O tempo que os fluxos de tráfego podem fluir através das interseções é sempre múltiplo do tempo mínimo do intervalo de verde e nunca deve ultrapassar o parâmetro tempo máximo do intervalo de verde.

Para descobrir a quantidade de veículos presentes nos fluxos de tráfego, os agentes Sinalização Semafórica contam com a cooperação dos agentes Veículos. Esta cooperação se dá, à medida que os agentes Veículo interagem com os agentes Sinalização Semafórica, informando a presença de seus veículos conectados nas vias controladas pelas sinalizações semafóricas por meio de mensagens de interação. Esta cooperação acontece em duas situações. Na primeira, o veículo do agente Veículo acabou de entrar em uma e, por isto, o agente é obrigado a interagir com o agente Sinalização Semafórica, a fim de notificá-lo de sua presença. Na última, o agente Sinalização Semafórica interage com os agentes Veículos, solicitando que os mesmos notifiquem novamente a presença de seus veículos. Isto acontece periodicamente, pois o agente Sinalização Semafórica precisa medir a quantidade de veículos que se aproximam da intersecção. Após um determinado número de vezes, o agente Sinalização Semafórica calcula uma média a partir das quantidades de veículos obtidas. Todos os agentes Sinalização Semafórica são configurados com mesmo valor de periodicidade de obtenção de dados e o mesmo valor de quantidade de obtenções de dados. Ao final de cada período de medição das quantidades de veículos, os agentes Sinalização Semafórica interagem com o agente Centro de Controle de Tráfego, compartilhando a quantidade de veículos medida durante o período de obtenção deste dado e o instante de tempo em que a medição foi realizada. Para evitar problemas de sincronização de relógios, esta tese considera o uso de um relógio global que, por sua vez, pode ser acessado por meio de dispositivos GPS. Em ambas as situações em que os agentes Veículo interagem com um agente Sinalização Semafórica, este último agente deve confirmar também sua presença via, de modo que os agentes Veículo possam saber se as sinalizações semafóricas da interseção apresentam ausência de funcionamento. Caso isto aconteça, os agentes Veículo devem adotar uma estratégia de controle de interseção com ausência de sinalizações semafóricas, que, por sua vez, é apresentada na Seção \ref{sec:ausencia}.

Uma vez que a quantidade de obtenções de dados é atingida, cada agente Sinalização Semafórica interage com os seus vizinhos de interseção, compartilhando as médias dos dados obtidos por eles. Após terem conhecimento de todas as médias de seus agentes vizinhos, o agente Sinalização Semafórica executa o cálculo de demandas proposto por \citet{Paiva:2012}. Uma vez conhecidas as demandas, cada agente Sinalização Semafórica pode mudar o valor da reversibilidade relativo a ele, bem como, o número de arestas entre ele e seus vizinhos de interseção. Porém, isto somente acontece, quando o agente Sinalização Semafórica tem todas as arestas de seus vizinhos apontadas para ele. Com isto, ele realiza o ajuste de reversibilidade proposto por \citet{Santos:2012}. Para tanto, o agente Sinalização Semafórica atualiza suas estruturas de dados locais e, em seguida, interage com os agentes vizinhos, notificando a mudança de reversibilidade por meio de mensagens de interação. Ao receberem estas mensagens, seus vizinhos atualizam suas estruturas de dados.   

Com tudo isso, esta tese trata o problema de controle e distribuição dos tamanhos dos intervalos de indicação de luzes verdes de sinalizações semafóricas utilizadas para controlar o acesso e limitar o tempo de uso de uma interseção isolada. No entanto, ainda é preciso tratar o problema de controle relativo às interseções compartilhadas entre dois ou mais sistemas coordenados de sinalizações semafóricas. 

\subsection{Estratégia de Controle de Sistemas Coordenados de Sinalizações Semafóricas}

Para tratar o problema de controle relativo às interseções compartilhadas entre dois ou mais sistemas coordenados de sinalizações semafóricas, é necessário que os corredores contendo redes de sinalizações semafóricas sejam identificados. A partir desta identificação, deve-se escolher a sinalização semafórica cujo agente também será responsável pela ativação e desativação do sistema coordenado de sinalizações semafóricas. Em outras palavras, o agente da sinalização semafórica atuará também como líder do sistema de coordenação de sinalizações semafóricas. Nesta tese, a sinalização semafórica escolhida para tais finalidades é sempre aquela que controla o primeiro seguimento de via do corredor contendo a rede de sinalizações semafóricas. Os líderes dos sistemas coordenados de sinalizações semafóricas devem conhecer todas as sinalizações semafóricas participantes do mesmo corredor, assim como, os identificadores das vias que elas controlam. Eles também devem conhecer outros líderes de sistemas coordenados de sinalizações, que, por sua vez, têm interseções compartilhadas entre eles. Com isto, é possível construir um multigrafo, de modo que este possa ser utilizado em um algoritmo SMER para controle de sistemas coordenados de sinalizações semafóricas. Neste contexto, os vértices do multigrafo representam os agentes líderes de sistemas coordenados de sinalizações semafóricas e cada conjunto de arestas representam o compartilhamento de uma interseção entre dois agentes líderes de sistemas coordenados de sinalizações semafóricas. Todos os agentes Sinalização Semafórica participantes de um sistema coordenado de sinalizações semafóricas devem conhecer os corredores em que participam, assim como, a identificação dos agentes Sinalização Semafórica líderes dos sistemas coordenado de sinalização semafóricas em que fazem parte. Por fim, cada agente líder deve saber quantos outros agentes líderes existem no sistema multiagente de controle de tráfego.

Assim como no algoritmo SMER para controle de interseções isoladas, o algoritmo SMER para controle de sistemas coordenados de sinalizações semafóricas precisa que o multigrafo esteja em um estado inicial acíclico. Além disto, cada agente Sinalização Semafórica líder de um sistema coordenado de sinalizações possui sua própria reversibilidade, que, por sua vez, é diferente da reversibilidade utilizada durante o controle de uma interseção isolada. A reversibilidade inicial para cada um dos agentes Sinalizações Semafóricas líderes de um sistema coordenados de sinalizações semafóricas é igual a um. Com base nisso, o controle de sistemas coordenados de sinalizações semafóricas é iniciado. 

Quando o agente líder de um sistema coordenado de sinalizações semafóricas possui um número suficiente de arestas direcionadas para ele, em cada um dos arcos entre o mesmo e outro agente líder de um sistema coordenado de sinalização semafóricas, ele ativa o seu sistema coordenado de sinalizações semafóricas. Para tanto, esta tese usa um método diferente daquele proposto por \citet{Paiva:2012}. Assim, o agente líder cria uma cópia do multigrafo do algoritmo SMER para controle de sistemas coordenados de sinalizações semafóricas e, em seguida, executa localmente o algoritmo SMER para controle de sistemas coordenados de sinalizações semafóricas. A execução local deste algoritmo simula as reversões das arestas direcionadas para o agente Sinalização Semafórica líder de um sistema coordenado de sinalizações semafóricas. Para cada reversão de arestas realizada, um número mínimo de ciclos é utilizado para gerar mensagens de sincronização das sinalizações semafóricas participantes de um sistema coordenado. Cada ciclo da sinalização semafórica do agente líder de um sistema coordenado de sinalizações semafóricas ativado equivale a geração de uma onda verde. A cada ciclo, o agente líder interage com os demais agentes participantes, configurando estes agentes, de modo que os inícios dos intervalos de indicação das luzes verdes das sinalizações semafóricas destes possam ser sincronizados. À medida que recebem a mensagem de interação pela primeira vez, os agentes participantes tomam ciência do número de ciclos configurado para sistema coordenado de sinalizações semafóricas e o instante em que a sinalização semafórica deve iniciar a sua participação no sistema coordenado de sinalizações semafóricas. A partir deste instante, os agentes das interseções participantes do sistema coordenado de sinalizações passam a controlá-las, como se eles estivessem executando o algoritmo SER.

Uma vez tomada a ciência de tais dados, o agente líder e os demais agentes do sistema coordenado de sinalizações, a cada ciclo de suas sinalizações semafóricas, decrementam em um o número de ciclos recebido. Quando este número chega a zero, agentes das interseções participantes voltam a controlá-las com o algoritmo SMER para controle de interseções isoladas, até que recebam uma nova mensagem de interação, requisitando a participação dos agentes no mesmo sistema coordenado de sinalizações semafóricas ou outro, caso a interseção seja compartilhada entre dois ou mais sistemas coordenados de sinalizações semafóricas. A desativação de um sistema coordenado de sinalizações semafóricas só ocorre efetivamente, quando o número de arestas de cada arco entre o agente líder do sistema coordenado e outro agente de mesma responsabilidade é menor que a sua reversibilidade. Quando isto acontece, a interseção do agente líder do sistema coordenado de interseções volta a ser controlada pelo algoritmo SMER para controle de interseções isoladas.

Por meio do mecanismo de reversibilidades, um agente Sinalização Semafórica líder de um sistema coordenado de sinalizações semafóricas permite que pelotões de veículos atrevem corredores formados por seguimentos de vias controlados por sinalizações semafóricas. Isto acontece, enquanto o número de arestas direcionadas para o agente, em cada um dos arcos entre ele e outro agente líder de um sistema coordenado de sinalizações, for maior ou igual a reversibilidade do agente. O tempo em que os pelotões de veículos podem fluir ao longo dos corredores varia em função das flutuações de tráfego nos seguimentos de vias componentes dos corredores. 

Para descobrir a quantidade de veículos dos seguimentos de vias de um corredor, o agente Sinalização Semafórica líder de um sistema de coordenado de sinalizações semafóricas conta com a interação entre os agentes Sinalização Semafórica participantes de seu sistema coordenado e ele. Esta interseção acontece, à medida que os agentes compartilham suas médias de quantidades de veículos com os agentes controladores dos sistemas coordenados de sinalizações semafóricas em que fazem parte. Uma vez que um número dessas interações é alcançado, o agente líder de um sistema coordenado de sinalizações semafóricas escolhe a maior média de quantidades de veículos como sua demanda e, em seguida, interage com os demais agentes controladores de sistemas coordenados de sinalizações semafóricas, compartilhando tal informação. Para aumentar o coeficiente de concorrência do algoritmo SMER para controle de sistemas coordenados de sinalizações semafóricas, \citet{Paiva:2012} criou um agrupamento de sistemas coordenados de sinalizações semafóricas, de modo que as reversibilidades de todos os agentes líderes de sistemas coordenados de sinalizações semafóricas de um mesmo grupo tivessem o mesmo valor. Para tanto, os agentes controladores de um mesmo grupo de sistemas coordenados de sinalizações semafóricas escolhem a maior média de quantidades de veículos do grupo e a utiliza para fins de cálculos e ajustes de reversibilidades, conforme descrito em \citet{Santos:2012}.

Antes de ajustarem suas reversibilidades, os agentes líderes precisam interagir com aqueles agentes líderes que concorrem pelas mesmas interseções, compartilhando a média de quantidade de veículos de seu grupo de sistemas coordenados de sinalizações semafóricas. Uma vez que isso tenha sido feito, os agentes controladores de sistemas coordenados de sinalizações semafóricas podem ajustar suas reversibilidades, quando estes ganharem o direito de ativas seus sistemas coordenados de sinalizações semafóricas. Dessa forma, o algoritmo SMER para controle de sistemas coordenados de sinalizações semafóricas balanceia os tempos em que os pelotões de veículos podem fluir ao longo de corredores formados por seguimentos de vias controladas por sinalizações semafóricas, levando em consideração as flutuações dos fluxos de tráfego desses corredores.

\subsection{Estratégia de Controle com Ausência de Sinalizações Semafóricas} \label{sec:ausencia}

Como mencionado anteriormente, os agentes Veículo podem detectar a ausência de funcionamento de sinalizações semafóricas tanto em interseções isoladas ou aquelas que participam de sistemas coordenados de sinalização semafórica. Uma vez detectada a ausência de funcionamento das sinalizações semafóricas de uma interseção, os agentes Veículos devem providenciar imediatamente o controle da mesma. 

Para detectar a ausência de funcionamento das sinalizações semafóricas de uma interseção, o banco de dados de mapas deve conter dados, que permitam os agentes Veículo verificarem se uma interseção é controlada ou não por um conjunto de sinalizações semafóricas \cite{Ferreira:2010}. Além disto, estes mesmos bancos de dados devem também fornecer dados acerca dos comprimentos das áreas para monitoramento de fluxo de tráfego. Por meio desta informação, os agentes Veículo se tornam cientes dos pontos dos seguimentos de via, onde eles podem iniciar interação com os agentes Sinalização Semafóricos embutidos nas sinalizações semafóricas instaladas sobre as vias em que trafegam. A partir do momento em que os veículos conectados dos agentes estejam em uma área de monitoramento de fluxo de tráfego, eles anunciam suas presenças na via de entrada da interseção para o agente Sinalização Semafórica, por meio de uma interação. Após um número de tentativas malsucedidas de anunciar a presença dos veículos, os agentes Veículo iniciam a estratégia de controle com a ausência de sinalizações semafóricas.

Para tanto, os agentes Veículos precisam inicialmente interagir com o agente Centro de Controle de Tráfego, a fim de recuperar o estado de controle das interseções, que é obtido a partir da simulação do sistema de controle de tráfego mantida pelo sistema supervisório. Juntamente com o estado de controle da interseção, o agente Veículo também recupera os parâmetros das sinalizações semafóricas da interseção. Tais parâmetros foram descritos na Seção \ref{sec:sinalizacao}. Esta tese considera que tal simulação está sincronizada com o sistema real de controle de tráfego. Dessa forma, os agentes Veículo recuperam os dados contendo o estado de controle das interseções e, em seguida, passam a executar localmente o algoritmo vigente de controle da interseção. O algoritmo vigente de controle de interseção pode ser uma versão local do algoritmo SMER para controle de interseções isoladas ou uma versão local do algoritmo SMER, executando como SER, quando a interseção está participando de um sistema coordenado de sinalizações ativo. Esta execução local cria sinalizações semafóricas virtuais, que, por sua vez, simulam o funcionamento das sinalizações semafóricas reais que apresentam ausência de funcionamento. Para que estas execuções locais se mantenham sincronizadas, elas fazem uso dos relógios dos dispositivos GPS, que, por sua vez, são embutidos nos veículos. Após isto, os agentes Veículos devem tomar ciência das indicações das sinalizações semafóricas virtuais das vias, onde os veículos que os embutem estão trafegando.

Se as sinalizações semafóricas virtuais indicarem luz amarela, os agentes Veículos, cujos veículos conectados estão na mesma via, precisam imediatamente iniciar uma eleição de líder entre eles. Para tanto, eles interagem entre si, compartilhando suas posições geográficas, enquanto o intervalo de indicação de luz amarela durar. À medida que os agentes Veículo recebem mensagens de interação, eles formam uma base de conhecimento acerca dos posicionamentos dos veículos em uma mesma via de entrada de uma interseção. Logo, à medida que esta base de conhecimento é formada, os agentes Veículo executam um algoritmo simples para identificar o líder. Este algoritmo identifica qual é o veículo mais próximo da faixa de retenção da via de entrada de uma interseção. Dessa forma, todos agentes Veículos sabem qual é o veículo conectado líder. 

Quando um veículo conectado é o líder, o agente embutido nele adquiri alguns comportamentos de um agente Sinalização Semafórica. O agente Veículo passa coletar dados acerca do fluxo de tráfego da via onde seu veículo conectado trafega e dos fluxos de tráfego das vias de grupos de movimentos conflitantes. Após o término período de medição das quantidades de veículos nas vias de entrada da interseção, o agente Veículo interage primeiramente com o agente Centro de Controle de Tráfego, compartilhando a quantidade veículos medida durante o período de medição, de modo que tal dado seja utilizado na atualização do estado de controle da interseção na simulação mantida pelo sistema supervisório. Por último, o agente Veículo atualiza sua base de conhecimento local e, em seguida, interagem com os agentes Veículos cujos veículos conectados trafegam pelas vias de entrada da interseção, compartilhando o mesmo dado. Desta forma, a versão local do algoritmo SMER para controle de interseções pode realizar os ajustes de reversibilidades e, com isto, reagir às flutuações de tráfego das vias de entrada da interseção. Se a interseção é participante de um sistema coordenado de sinalizações semafóricas, ele também interage com o (s) agente (s) Sinalização Semafórica controladores de sistema (s) coordenado (s) de sinalizações semafóricas, compartilhando a média das quantidades de veículos obtidas durante os períodos de medição. Desta forma, os líderes dos sistemas coordenados de sinalizações semafóricas atualizam os dados relativos aos seguimentos de via participantes de corredores. Todos esses comportamentos são abandonados pelo agente Veículo, após a sinalização semafórica virtual indicar luz verde.

\section{Sistema Multiagente de Controle de Tráfego}

Esta seção tem como objetivo apresentar os detalhamentos algorítmicos relativos a cada uma das estratégias apresentadas anteriormente. Primeiramente, são apresentados os detalhes acerca da inicialização do sistema multiagente de controle de tráfego. Após isto, são apresentados os detalhes do controle de tráfego em interseções isoladas. Em seguida, são apresentados os detalhes a respeito do controle de sistemas coordenados de sinalizações semafóricas. Por fim, são apresentados os detalhes de um controle de interseções baseado em veículos conectados.

\subsection{Inicialização do Sistema Multiagente de Controle de Tráfego}

Esta seção tem como objetivo descrever o processo de inicialização das instâncias dos agentes participantes do sistema multiagente de controle de tráfego. Tais agentes são os seguintes: Centro de Controle de Tráfego, Veículo e Sinalização Semafórica. Antes de iniciar a descrição dos processos de inicialização das instâncias destes agentes, é necessário definir uma URI para identificar o sistema multiagente de controle de tráfego. Sendo assim, a URI é definida da seguinte forma \textit{radnet://ttm/traffic\_control/signal\_control/}. A partir desta, as instâncias dos agentes podem utilizá-la para registrar interesses e vias na camada de rede de seus ambientes. Tais ambientes, no contexto da HRadNet-VE, são nós de rede. O termo ambiente está sendo usado aqui, por uma questão de coerência com o paradigma de sistemas multiagentes.

Para que um agente Centro de Controle de Tráfego possa entrar em operação, é necessário o registro de interesses na camada de rede do ambiente, a fim de preparar o ambiente para comunicações centradas em interesses e mapear as ações executadas pelo agente, à medida que o mesmo recebe uma mensagem de interação. Durante o registro de interesses na camada de rede do ambiente, devem ser informados os seguintes dados: interesse, número máximo de saltos (NMS) e a tecnologia de acesso à comunicação (TAC). Tal entrada de dados é comum para todos os agentes descritos nesta tese. A Tabela \ref{tab:interesses_cct} apresenta os interesses registrados por um agente Centro de Controle de Tráfego e as ações executadas pelo agente.

Para que um agente Veículo possa operar, é necessário atribuir um valor ao parâmetro de configuração de performance relativo à frequência de interações com um agente Sinalização Semafórica, no que tange a notificação da presença de um veículo em via controlada por uma sinalização semafórica. Segundo \citet{Zheng:2015}, o valor deste parâmetro deve ser igual a 1 Hz. Além desse parâmetro, outro que também é necessária a atribuição de um valor é aquele relativo a frequência de interações para realização de descobertas de veículos líderes em uma via de entrada de uma interseção. Segundo \citet{Zheng:2015}, o valor deste parâmetro deve ser igual a 10 Hz. No que diz respeito a via onde o veículo se encontra, esta é configurada dinamicamente pelo agente, à medida que o veículo trafega ao longo dos segmentos de via de uma rede viária. Para tanto, a cada segundo o agente utiliza as coordenadas do GPS embutido no veículo conectado e a banco de dados de mapas local. Por fim, também é necessário registrar os interesses na camada de rede do ambiente que suporta o agente, assim como, mapear as ações que deverão ser executadas, quando o agente receber uma mensagem de interação. A Tabela \ref{tab:interesses_veiculo} apresenta os dados relativos ao registro de interesses no ambiente de um agente Veículo e as descrições das ações associadas aos interesses. No que diz respeito às ações executadas pelo ambiente do agente Veículo, estas remetem aos procedimentos relativos à execução do protocolo de comunicação da HRAdNet-VE. Após o registro dos interesses, o agente Veículo escalona duas ações em seu mecanismo de planejamento de ações, são elas: monitoramento de faixas e notificação de presença na via onde ele está trafegando. 

Para que um agente Sinalização Semafórica possa operar, é necessário atribuir valores aos parâmetros de configuração de performance, de acordo com a atuação deste agente do sistema multiagente de controle de tráfego. Os valores para esses parâmetros devem ser informados pelo engenheiro de tráfego, quando o mesmo estiver configurando o sistema multiagente de controle de tráfego, utilizando um sistema supervisório. Sendo assim, o engenheiro de tráfego deve inicialmente atribuir os valores dos parâmetros de configuração de performance comuns a todos os agentes Sinalização Semafórica, que são: identificação da sinalização semafórica, identificação da interseção, periodicidade de obtenção de dados, quantidade de obtenções de dados, tempo mínimo do intervalo de verde, tempo máximo do intervalo de verde, tempo do intervalo de amarelo, tempo de vermelho geral, via de entrada da interseção, grafo de controle utilizado durante a execução do algoritmo SMER para controle de interseções isoladas, comprimento da área para monitoramento fluxo de tráfego, e número de sinalizações semafóricas da interseção. Após isto, a atenção do engenheiro de tráfego deve se voltar para a atribuição de valores dos parâmetros de configuração de performance dos agentes Sinalização Semafórica controladores de sistemas coordenados de sinalizações semafóricas. Neste sentido, ele deve atribuir valores aos seguintes parâmetros: identificador do corredor contendo as sinalizações semafóricas, grafo de controle utilizado pelo algoritmo SMER para controle de sistemas coordenados de sinalização semafórica, número mínimo de ciclos por operação de coordenação de sinalizações semafóricas e número máximo de ciclos operação de coordenação de sinalizações semafóricas, periodicidade de compartilhamento de médias de quantidades veículos de um grupo de sistemas coordenados de sinalizações semafóricas e periodicidade de atualização de demandas de corredores de sistemas coordenados de sinalizações semafórica. No que tange lista de participantes do sistema coordenado de sinalizações semafóricas, seguimentos de via componentes do corredor, e lista de identificações de corredores agrupados, estes são configurados automaticamente, à medida que os agentes envia mensagens de interação, contendo os interesses \textit{corridor\_participant\_traffic\_light} e \textit{group\_member}. Por fim, o engenheiro de tráfego finaliza a etapa de atribuição de valores de parâmetros de configuração de performance, atribuindo os valores dos parâmetros relativos aos agentes Sinalização Semafórica participantes de sistemas coordenados de sinalizações semafóricas. Sendo assim, o engenheiro de tráfego atribui valores ao atributo identificador do corredor. Os identificadores dos controladores dos sistemas coordenados de sinalização semafóricas são registrados automaticamente, à medida que os agentes controladores de sistemas coordenados de sinalizações semafóricas enviam mensagens de interação, contendo o interesse \textit{corridor\_controller\_traffic\_light}. Por fim, os agentes Sinalização Semafórica interagem com o agente Centro de Controle de Tráfego, enviando mensagens de interação com o interesse \textit{traffic\_light}, a fim de registrarem seus identificadores.

Além desses parâmetros de configuração de performance, os engenheiros de tráfego também precisam informar as vias, onde os agentes Sinalização Semafórica operarão. Tais dados devem ser extraídos de um banco de dados de mapas e, em seguida, registrados na camada de rede do ambiente. Após isto, um agente Sinalização Semafórica inicia o registro de interesses na camada de rede de seu ambiente. Existem interesses e ações que são comuns a todos os agentes Sinalização Semafórica. Estes, por sua vez, são apresentados na Tabela \ref{tab:interesses_sinalizacao}. Esses interesses garantem que os agentes Sinalização Semafórica possam executar o algoritmo SMER para controle de interseções isoladas. Além desses interesses, existem aqueles que também são adicionados pelos agentes Sinalização Semafórica que operam em sistemas coordenados de sinalizações semafóricas. Estes são apresentados nas Tabelas \ref{tab:interesses_controlador} e \ref{tab:interesses_participante}. Por fim, ainda existe um interesse que precisa ser registrado pelos agentes Sinalização Semafórica cujas sinalizações semafóricas não participam permanentemente (pertence a dois ou mais sistemas coordenados de sinalizações semafóricas) ou temporariamente (não é participante, mas integra uma interseção participante de um sistema coordenado de sinalizações semafóricas) de operações de coordenação de sinalizações semafóricas. Os interesses registrados por estes agentes são apresentados na Tabela \ref{tab:interesses_nao_participante}

Após a etapa de registro de interesses, um agente Sinalização Semafórica verifica o estado inicial do multigrafo utilizado no controle da interseção, a fim de verificar se ele poderá iniciar o intervalo de luz verde da sinalização semafórica que o embute. Se verdadeiro, a luz verde é acesa e, em seguida, o agente escalona uma ação para dar início ao intervalo de amarelo, utilizando o tempo mínimo de intervalo de luz verde para isto. Se falso, a luz vermelha é acesa e, em seguida, o agente espera por reversões de arestas por parte de seus vizinhos. Após isto, se o agente é um líder de um sistema coordenado de sinalizações semafóricas, ele deve verificar o estado inicial do multigrafo utilizado no controle de sistemas coordenados de sinalizações semafóricas, a fim de verificar se ele poderá ativar o seu sistema coordenado de sinalizações semafóricas. Se verdadeiro, o agente imediatamente executa o Algoritmo \ref{alg:inicia_coordenacao}, que é apresentado na Seção \ref{subsec:coordenacao}. Por fim, os agentes Sinalização Semafórica escalonam suas ações em seu mecanismo de planejamento de ações, a fim de obter as quantidades de veículos nas vias controladas por suas sinalizações semafóricas.

\subsection{Controlando Tráfego em Interseções Isoladas}

Esta seção tem como objetivo apresentar a abordagem algorítmicas para controle de tráfego em interseções isoladas, que, por sua vez, é proposta por esta tese. O detalhamento desta abordagem não serve somente para descrever os algoritmos utilizados no controle de interseções isoladas, mas também como ela tira proveito das características da HRAdNet-VE. 

\subsubsection{Monitoramento de Mudanças de Faixas ou Vias} \label{sec:monitoramento}

Após o início de sua operação, os agentes Veículo iniciam o monitoramento de faixas, tendo em vista que o agente deve ter ciência da via em que eles estão trafegando. Para tanto, eles escalonam periodicamente ações, de modo que eles possam detectar qualquer mudança de faixa ou via. Estas ações executam o Algoritmo \ref{alg:monitoramento_faixas}. O algoritmo inicia obtendo o identificador da via ($idVia$) onde o veículo conectado se encontra e, em seguida, verifica se houve uma mudança de faixa ou via. Se verdadeiro, o algoritmo remove o interesse \textit{vehicle\_out\_}$<$id. da via$>$, antes que aconteça a atualização da via atual e anterior. Após isto, ele envia uma mensagem   contendo o interesse \textit{vehicle\_out\_}$<$id. da via$>$, incrementa o número de tentativas de notificacao de mudança de faixa ou via ($numTentNotifMudanca_i$) e inicia a espera de uma confirmação por parte de um agente Sinalização Semafórica. O algoritmo envia $Msg_i$ para frente, utilizando uma interface de acesso à comunicação do tipo IEEE 802.11, a fim de notificar o agente Sinalização Semafórica da via anterior. Além disto, o algoritmo reconfigura a camada de rede do ambiente, removendo a faixa ou via anterior e adicionando a faixa ou via atual. Além disto, o algoritmo reconfigura o registro de interesses, de acordo com a atualização da via atual e anterior. Embora o veículo possa ter mudado, ele precisa continuar enviando mensagens de notificação acerca da mudança de faixa ou via, caso algum agente Sinalização Semafórica não tenha tomada ciência desta mudança. Enquanto o número de tentativas de notificação mudança ($numTentNotifMudanca_i$)for menor que o número máximo de tentativas de notificação de mudança ($paramNumMaxTentNotifMudanca_i$), o algoritmo envia mensagens de notificação e, em seguida, incrementa o número de tentativas de notificação de mudança.

\subsubsection{Notificação de Presença de Veículo na Via} \label{sec:notificacao}

À medida que um veículo conectado entra nos seguimentos de via de uma rede viária, o agente Veículo inicia uma tentativa de interação com algum agente Sinalização Semafórica, executando uma ação escalonada, conforme o parâmetro relativo a periodicidade de notificação de um veículo em uma via. A execução da ação é descrita pelo Algoritmo \ref{alg:notificacao_presenca}. O algoritmo inicia, verificando se a via possui sinalização semafórica. Se verdadeiro, ele verifica se o seu veículo conectado está dentro da área de monitoramento de tráfego relacionada a sinalização semafórica da instalada na via. Se verdadeiro, o ele verifica se algum agente Sinalização Semafórica confirmou sua presença na via ($presencaSinalizacao_i$). Se falso, ele verifica o se o número de tentativas de notificação ($numTentativasNotificacao_i$) é menor que o parâmetro de configuração de performance ($paramNumMaxTentativasNotificacao_i$). Se falso, ele inicia o controle de interseção utilizando veículos conectados. Se verdadeiro, ele cria o interesse \textit{vehicle\_on\_}$<$id. da via$>$ e, em seguida, cria uma mensagem de interação   com mesmo. Além do $interesse$, $Msg_i$ também é configurada com um destino igual a \textbf{nulo}, pois o agente Veículo não conhece a identificação do agente Sinalização Semafórica. Finalizando a configuração de $Msg_i$, o agente indica a via por onde a mensagem de interação deve trafegar, assim como, a direção de propagação da mensagem, que é para frente, e a tecnologia de acesso à comunicação da interface de comunicação sem fio. A mensagem também deve conter um parâmetro $posicao$, contendo a posição geográfica atual do veículo. Por fim, o agente envia a mensagem e, em seguida, incrementa $numTentativas_i$.

Ao receber uma mensagem mensagem com o interesse \textit{vehicle\_on\_}$<$id. da via$>$, o agente Sinalização Semafórica executa o Algoritmo \ref{alg:tratamento_vehicle_on}. O algoritmo inicia, verificando se a origem da mensagem está atrás da sinalização semafórica e se está dentro da área de monitoramento de tráfego da via. Se verdadeiro, ele verifica se o identificador da origem da mensagem ($Msg_j.origem$) não existe no conjunto de veículos da via ($veiculosVia_i$). Se verdadeiro, ele registra o identificador da origem da mensagem no conjunto de veículos da via. Em seguida, ele cria uma mensagem   contendo o interesse \textit{roadway\_presence\_confirmation}. Esta mensagem deve ser enviada diretamente para a origem da mensagem recebida ($Msg_j.origem$). Por isto, ela tem seu campo destino cofigurado com $Msg_j.origem$. A via indicada para envia de $Msg_i$ deve ser a mesma via da mensagem recebida ($Msg_j.via$) e, por isto, a mensagem deve ser propagada para trás, fazendo com o campo direção seja configurado com o valor -1. Ao final da configuração da mensagem, ela configurada com a tecnologia de acesso à comunicação da mensagem recebida ($Msg_j.tac$). Por fim, $Msg_i$ é enviada. 

Ao receber uma mensagem de interação, contendo o interesse \textit{roadway\_presence\_confirmation}, o agente Veículo trata tal mensagem, executando o Algoritmo \ref{alg:tratamento_roadway_presence_request}. O algoritmo inicia, verificando se o agente é o destino da mensagem. Se verdadeiro, ele verifica qual a interface de acesso à comunicação encaminhou a mensagem. Se for uma interface baseada no padrão IEEE 802.11, o algoritmo verifica se a mensagem possui um parâmetro \textit{ctrlIntersecao}. Se verdadeiro, o veículo inicia o processo de controle de intersecao com veículos e, em seguida, atribui o \textbf{falso} à variável \textit{presencaSinalizacao$_i$}. Caso contrário, o algoritmo somente atribui verdadeiro à variável \textit{presencaSinalizacao$_i$}. Caso a mensagem de interação chegue por uma interface baseada no padrão LTE, o algoritmo atribui o \textbf{falso} à variável \textit{presencaSinalizacao$_i$}. Por fim, o algoritmo atribui zero à variável responsável em acumular o número de tentativas de notificação de presença (\textit{numTentativaNotificacao$_i$}).  

Quando um agente Sinalização Semafórica recebe uma mensagem de interação, contendo o interesse \textit{vehicle\_out\_}$<$id. da via$>$, ele executa o Algoritmo \ref{alg:tratamento_vehicle_out}. O algoritmo inicia verificando se a origem da mensagem está atrás da sinalização semafória e se está dentro da área de monitoramento de tráfego da via. Se verdadeiro, ele verifica se o identificador da origem da mensagem ($Msg_j.origem$) existe no conjunto de veículos da via ($veiculosVia_i$). Se verdadeiro, ele remove o identificador da origem da mensagem no conjunto de veículos da via. Em seguida, ele cria uma mensagem   contendo o interesse \textit{roadway\_left\_confirmation}. Esta mensagem deve ser enviada diretamente para a origem da mensagem recebida ($Msg_j.origem$). Por isto, ela tem seu campo destino configurado com $Msg_j.origem$. A via indicada para envia de $Msg_i$ deve ser a mesma via da mensagem recebida ($Msg_j.via$) e, por isto, a mensagem deve ser propagada para trás, fazendo com que o campo direção seja configurado com o valor -1. Ao final da configuração da mensagem, ela configurada com a tecnologia de acesso à comunicação da mensagem recebida ($Msg_j.tac$). Por fim, $Msg_i$ é enviada. Ao receber esta mensagem, um agente Veículo atribui o valor \textbf{verdadeiro} para a variável 
\textit{confirmacaSaida}$_i$ e atribui o valor zero à variável \textit{numTentNotifMudanca$_i$} (veja Algoritmo \ref{alg:monitoramento_faixas}). Isto acontece indenpedente da tecnologia de acesso à comunicação utilizada para transmitir a mensagem.

\subsubsection{Obtenção de Quantidades de Veículos em Vias}

Para obter a quantidade de veículos de uma via, um agente Sinalização Semafórica precisa periodicamente acumular as quantidades de veículos que trafegaram na via durante intervalos de tempo para obtenção de tal dado. Após um determinado número de obtenções de quantidades de veículos, o agente precisa realizar uma agregação destes dados e, em seguida, compartilhar a sua demanda com outros agentes do sistema multiagente de controle de tráfego. Para tanto, de acordo com o valor do parâmetro de periodicidade de obtenção da quantidade de veículos de uma via, o agente escalona ações que executam o Algoritmo \ref{alg:coleta_compartilhamento_demanda}. Neste algoritmo, o agente Sinalização Semafórica compartilha a média das quantidades de veículos com seus vizinhos de interseção (Linha 10) e agente Centro de Controle de Tráfego (Linha 12).

Além disso, o agente Sinalização Semafórica periodicamente o Algoritmo \ref{alg:requisicao_notificacao_presenca_via}. A periodicidade de execução deste algoritmo é definida pelo parâmetro de periodicidade de obtenção de quantidades de veículos. O Algoritmo \ref{alg:requisicao_notificacao_presenca_via} tem como objetivo requisitar que os veículos dentro da área de monitoramento de tráfego reenviem suas notificações de presença. Este algoritmo envia para trás da sinalização semafórica uma mensagem com interesse \textit{roadway\_presence\_request} por meio de uma interface de comunicação baseada no padrão IEEE 802.11.

Ao receber uma mensagem de interação, contendo o interesse \textit{roadway\_presence\_request}, um agente Veículo trata esta mensagem, executando o Algoritmo \ref{alg:tratamento_roadway_presence_request}. O algoritmo inicia, verificando se a via da origem da mensagem é igual a via onde o veículo está trafegando. Se verdadeiro, ele verifica qual a tecnologia de acesso à comunicação foi utilizada pela origem, quando esta enviou $Msg_j$. De acordo com a tecnologia de acesso à comunicação, o agente Veículo configura uma mensagem de resposta e, em seguida, a envia para a origem da mensagem recebida.

Ao receber a mensagem de interação , contendo o interesse \textit{roadway\_vehicle\_amount}, um agente Sinalização Semafórica trata a mensagem, executando o Algoritmo \ref{alg:tratamento_roadway_vehicle_amount}. O algoritmo inicia, incrementando o número de médias de quantidades de veículos recebidas ($numMediasQtdeVeiculos_i$). Logo após, ele registra o valor do parâmetro \textit{mediaQtdeVeiculos} com o identificador do agente origem da mensagem ($Msg_j.origem$). Em seguida, ele verifica se o número de medias de quantidades de veículos recebidas é menor que o número de sinalizações semafóricas da interseção. Se verdadeiro, ele verifica se o agente recebeu as médias das quantidades de veículos de todos os seus vizinhos. Se verdadeiro, ele calcula a demanda de cada um dos agentes do multigrafo utilizado pelo algoritmo SMER para controle de interseções ($multigrafoSMERIntrsc_i$). Para tanto, ele utiliza o mesmo cálculo proposto por \citet{Paiva:2012}. Por fim, o algoritmo atribui zero ao número de médias de quantidades de veículos recebidas e calcula o mínimo múltiplo comum entre as demandas.

\subsubsection{Controle e Ajuste de Intervalos de Indicações de Sinalização}

À medida que o intervalo de indicação de luz verde termina, uma ação escalonada para indicação do intervalo de indicação de luz amarela entra em execução. Esta ação, portanto, executa o Algoritmo \ref{alg:acionamento_luz_amarela}. O algoritmo inicia, verificando se o agente está participando de um sistema coordenado de sinalizações semafóricas ativo. Caso não esteja, ele reverte as arestas do agente no multigrafo (\textit{multigrafoSMERIntrsc$_i$}) utilizado pelo algoritmo SMER de controle de interseções isoladas, utilizando seu identificador (\textit{idAgente$_i$}) e a sua reversibilidade \textit{revAgentesIntrsc$_i$[idAgente$_i$]}. Logo após, na Linha 27, o algoritmo verifica se a sinalização semafórica poderá trocar a indicação. Para tanto, é verificado se o agente não possui quantidades de arestas orientadas para ele maiores ou iguais a reversibilidade do mesmo. Se verdadeiro, o algoritmo acende a luz amarela e, em seguida, escalona o acendimento da luz vermelha no mecanismo de planejamento do agente, utilizando o parâmetro de intervalo de indicação de luz amarela. Caso contrário, o algoritmo escalona novamente o acendimento da luz amarela no mecanismo de planejamento do agente, utilizando o intervalo mínimo de indicação de luz verde. Por fim, ele envia uma mensagem de interação, contendo o interesse \textit{edge\_reversal}, para os demais agentes da interseção por meio de uma interface de acesso à comunicação baseada no padrão IEEE 802.11. As demais partes do Algoritmo \ref{alg:acionamento_luz_amarela} serão explicadas na Seção .

O escalonamento da ação relativa ao acendimento da luz vermelha tem como objetiva finalizar o processo de reversão de arestas durante a execução do algoritmo SMER para controle de interseções. Para tanto, a ação executa o Algoritmo \ref{alg:acionamento_luz_vermelha}. O algoritmo inicia, acendendo a luz vermelha da sinalização semafórica e, em seguida, ele finaliza, enviando uma mensagem de interação  , contendo o interesse \textit{edge\_reversal}, para os demais agentes da interseção por meio de uma interface de acesso à comunicação baseada no padrão IEEE 802.11.

Ao receber uma mensagem  , contendo o interesse \textit{edge\_reversal}, um agente Sinalização Semafórica excuta uma ação, a fim de tratar tal mensagem recebida. Esta ação executa o Algoritmo \ref{alg:tratamento_edge_reversal}. O algoritmo inicia, verificando se a origem da mensagem ($Msg_j.origem$)é um vizinho de interseção. Se verdadeiro, ele reverte as arestas referentes à origem da mensagem no multigrafo, utilizando a reversibilidade da origem da mensagem (\textit{revAgentesIntrsc$_i$[$Msg_j.origem$]}). Em seguida, se o agente tiver quantidades de arestas revertidas para ele maiores ou iguais a sua reversibilidade, o algoritmo escalona para execução imediata a ação para acendimento da luz verde.

A partir do escalonamento da ação de acendimento imediato da luz verde, o Algoritmo \ref{alg:acionamento_luz_verde} é executado. O algoritmo inicia, acendendo a luz verde da sinalização semafórica. Em seguida, ele escalona a ação para acendimento da luz amarela no mecanismo de planejamento de ações do agente, utilizando o parâmetro de intervalo mínimo de indicação de luz verde (\textit{paramIntMinIndLuzVerde$_i$}). Após isto, o algoritmo verifica se o agente está participando de um sistema coordenado de sinalizações semafóricas. Se falso, o algoritmo verifica se existe demanda calculada para a interseção (Linha 32). Se verdadeiro, ele calcula uma reversibilidade para o agente (\textit{revAgente}) e, em seguida, compara a reversibilidade calculada com a mantida pelo agente ($revAgentesIntrsc_i[idAgente_i]$). Se forem diferentes, o algoritmo ajusta o número de arestas em cada um dos arcos entre o agente e um vizinho. No próximo passo, o algoritmo envia uma mensagem de interação ($Msg_{i}^{Intrsc}$), contendo o interesse \textit{reversibility\_change}, para os agentes da interseção por meio de uma interface de acesso à comunicação baseado no padrão IEEE 802.11. Por fim, o algoritmo envia uma mensagem de interação ($Msg_{i}^{cor}$) para o agente Centro de Controle de Tráfego, contendo o mesmo interesse da mensagem anteriormente enviada. Esta última mensagem é enviada por meio da interface de acesso à comunicação baseada no padrão LTE. Neste algoritmo, é método para ajuste de reversibilidades proposto por \citet{Santos:2012} é aplicado.

Ao receber uma mensagem $MSg_j$, contendo o interesse \textit{reversibility\_change}, a partir de um vizinho de interseção, um agente Sinalização Semafórica trata a mensagem, executando o Algoritmo \ref{alg:tratamento_reversibility_change}. O algoritmo inicia, verificando se a origem da mensagem ($Msg_j.origem$) é vizinho na interseção. Se verdadeiro, o algoritmo calcula a reversibilidade para a origem da mensagem (\textit{revAgente}) e, em seguida, verifica se a reversibilidade calculada é diferente da reversibilidade conhecida pelo agente ($revAgentesIntrsc_i[Msg_j.origem]$). Se verdadeiro, o algoritmo atualiza a reversibilidade da origem da mensagem e, por fim, ele ajusta o número de arestas nos arcos entre a origem e seus vizinhos.

\subsection{Controlando Sistemas Coordenados de Sinalizações Semafóricas}\label{subsec:coordenacao}

Esta seção tem como objetivo apresentar a abordagem algorítmicas para controle de sistemas coordenados de sinalização semafórica, que, por sua vez, é proposta por esta tese. O detalhamento desta abordagem não serve somente para descrever os algoritmos utilizados no controle de sistemas coordenados de sinalizações semafóricas, mas também como ela tira proveito das características da HRAdNet-VE. 

\subsubsection{Ativação de um Sistema Coordenado de Sinalizações Semaforicas}

Ao iniciar sua operação, um agente Sinalização Semafórica líder de um sistema coordenado de sinalizações semafóricas deve verificar o estado inicial do multigrafo utilizado no controle de sistemas coordenados de sinalizações semafóricas. Esta verificação tem como objetivo saber se o agente poderá ativar o seu sistema coordenado de sinalizações semafóricas. Se verdadeiro, ele executa o Algoritmo \ref{alg:inicia_coordenacao}. O algoritmo inicia, atribuindo \textbf{verdadeiro} à variável para ativação e desativação de coordenação de sinalizações semafóricas (\textit{coordenacaoAtiva$_i$}). Em seguida, o algoritmo envia uma mensagem de interação (\textit{Msg$_i^{Intrsc}$}), contendo o interesse \textit{participation\_in\_traffic\_light\_coordination}, a fim de fazer com que as sinalizações semafóricas de uma interseção participem da coordenação de sinalizações semafóricas. Esta mensagem é enviada por meio de interface de acesso à comunicação sem fio baseada no padrão IEEE 802.11. Após isto, o algoritmo inicializa o número de ciclos do sistema coordenado de sinalizações semafóricas (\textit{numeroCiclos$_i$}). Em seguida, o algoritmo inicializa as variáveis de controle referentes aos \textit{offsets} do sistema coordenado de sinalizações semafóricas (\textit{somaOffSets}), instante do escalonamento para ação de participação da coordenação das sinalizações semafóricas (\textit{instanteEscAcao}) e o tempo acumulado durante a ativação do sistema coordenado de sinalizações semafóricas (\textit{instanteAtual}). Nos passos seguintes, o algoritmo atualiza a reversibilidade do corredor, caso exista uma demanda calculada. Além disto, ele também ajusta a quantidade de arestas em cada arco entre o corredor e um vizinho. Para atualizar a reversibilidade em outros agentes controladores de sistemas coordenados de sinalizações semafóricas, o algoritmo envia uma mensagem de interação, contendo o interesse \textit{corridor\_reversibility\_change}, utilizando uma interface de acesso à comunicação baseada no padrão LTE. Depois disto, o algoritmo simula a execução do algoritmo SMER para controle de tráfego. Durante esta simulação, o algoritmo reverte as arestas do agente líder do sistema coordenado de sinalizações semafóricas e, após isto, ele envia uma mensagem de interação (\textit{Msg$_i^{agente}$}), contendo o interesse \textit{traffic\_light\_coordination}, para cada agente cuja sinalização semafórica é parte do sistema coordenado de sinalizações semafóricas. Para cada mensagem enviada, o algoritmo calcula o \textit{offset} de cada segmento de via do corredor, de modo que ele seja utilizado para dar precisão ao instante em que as sinalizações de semafóricas iniciarão seus intervalos de indicação de luz verde, após a primeira onda verde ter sido gerada. No próximo passo, o algoritmo calcula a duração da ativação do sistema coordenado de sinalizações semafóricas. O valor desta variável é utilizado durante o envio de mensagem para cada agente, de modo que uma nova coordenação de sinalizações semafóricas possa ocorrer, caso ainda existam arestas direcionadas para o agente líder do sistema coordenado de sinalizações semafóricas. Por fim, o algoritmo reverte todas as arestas para o agente, no âmbito do controle da interseção, onde sua sinalização semafórica se encontra.

Ao receber uma mensagem de interação  , contendo o interesse \textit{participation\_in\_traffic\_light\_coordination}, um agente Sinalização Semafórica trata a mensagem, executando o Algoritmo \ref{alg:tratamento_participation_in_traffic_light_coordination}. O algoritmo inicia, verificando se o agente é vizinho da origem da mensagem. Se verdadeiro, o agente coopera com a coordenação de sinalizações semafóricas, atribuindo \textbf{verdadeiro} à variável \textit{coordenacaoAtiva$_i$} e, em seguida, reverte todas as arestas para o agente origem da mensagem. Após isto, a variável de controle de número de ciclos (numeroCiclos$_i$) é inicializada. Por uma questão de segurança, o algoritmo verifica qual a indicação atual da sinalização semafórica. Se a indicação da sinalização semafórica é verde, as seguintes ações são executadas: a luz amarela é acesa; ação de acendimento de luz amarela escalonada anteriormente é cancelada; e o acendimento da luz vermelha é escalonado, utilizando o parâmetro de intervalo de indicação de luz amarela. Se a indicação da sinalização semafórica é amarela, as seguintes ações são executadas: o acendimento da luz vermelha é cancelado; e um novo acendimento da luz vermelha é escalonado, utilizando o parâmetro de intervalo de indicação de luz amarela. Ao final, o algoritmo envia uma mensagem de interação, contendo o interesse \textit{confirmation\_in\_traffic\_light\_coordination}, para a origem da mensagem recebida.

Ao receber uma mensagem de interação  , contendo o interesse \textit{confirmation\_in\_traffic\_light\_coordination}, um agente Sinalização Semafórica trata a mensagem, executando o Algoritmo \ref{alg:tratamento_confirmation_in_traffic_light_coordination}. O algoritmo inicia, verificando se a origem da mensagem ($Msg_j.origem$) é um vizinho de interseção. Se verdadeiro, ele incrementa a variável número de vizinhos participantes da interseção (\textit{numeroVizinhosParticipantes$_i$}). Em seguida, ele cancela qualquer escalonamento de ação relativo ao acendimento de luzes da sinalização semafórica. Por fim, ele acende a luz verde da sinalização semafórica e, em seguida, escalona acendimento da luz amarela, tendo como base o parâmetro de intervalo de indicação de luz verde (\textit{paramIntervaloIndLuzVerde$_i$}). 

Ao receber uma mensagem de interação  , contendo o interesse \textit{traffic\_light\_coordination}, um agente Sinalização Semafórica trata a mensagem, executando o Algoritmo \ref{alg:traffic_light_coordination}. O algoritmo inicia, verificando se o valor do parâmetro \textit{corredor} (Msg$_j$.parametros[``corredor'']) pertence ao conjunto de identificadores de corredores (\textit{idCorredores$_i$}) e se o identificador da origem pertence ao conjunto de identificadores de controladores de sistemas coordenados de sinalizações semafóricas (\textit{idControladores$_i$}). Se verdadeiro, o algoritmo escalona uma ação no mecanismo de planejamento do agente, a fim de iniciar a participação da sinalização semafórica na coordenação de sinalizações semafóricas. Uma vez que o tempo de escalonamento desta ação é atingido, o Algoritmo \ref{alg:inicio_participacao_coordenacao} é executado. O algoritmo inicia, inicializando as váriáveis de controle de coordenação (\textit{coordenacaoAtiva$_i$} e \textit{numeroCiclos$_i$}). Após isto, ele inicializa com zero a variável número de vizinhos participantes da interseção (\textit{numeroVizinhosParticipantes$_i$}) e, em seguida, reverte todas as arestas para o agente. Por fim, uma mensagem de interação (\textit{Msg$_i$}), contendo o interesse \textit{participation\_in\_traffic\_light\_coordination}, é enviada para os agentes da interseção por meio de uma interface de acesso à comunicação baseada no padrão IEEE 802.11. Quando os agentes da interseção recebem $Msg_i$, eles executam o Algoritmo \ref{alg:tratamento_participation_in_traffic_light_coordination}.

\subsubsection{Controle da Ativação de Sistemas Coordenados de Sinalizações Semafóricas}

De acordo com o Algoritmo \ref{alg:acionamento_luz_amarela}, um agente Sinalização Semafórica líder de um sistema coordenado de sinalização semafórica, ao executar a ação de acionamento da luz amarela, reverte as arestas direcionadas para ele no âmbito do controle de sistemas coordenados de sinalizações semafóricas, caso o número de ciclos seja zero. Para tanto, uma mensagem de interação, contendo o interesse \textit{corridor\_edge\_reversal}, é enviada para todos os agentes controladores de sistemas coordenados de sinalizações semafóricas, utilizando uma interface de acesso à comunicação baseda no padrão LTE.

Ao receber uma mensagem de interação $Msg_j$, contendo o interesse \textit{corridor\_edge\_reversal}, um agente Sinalização Semafórica líder de um sistema coordenado de sinalizações semafóricas trata a mesma, executando o Algoritmo \ref{alg:tratamento_corridor_edge_reversal}. O algoritmo inicia, verificando se a mensagem de interação é proveniente de um corredor vizinho. Para isto, o algoritmo faz uso parâmetro \textit{corredor} contido na mensagem (\textit{Msg$_j$.parametros[``corredor'']}). Se verdadeiro, o algoritmo reverte as arestas do corredor cuja mensagem é proveniente. Em seguida, o algoritmo verifica se é possível escalonar o início de uma coordenação de sinalizações semafóricas. Para tanto, o agente líder de um sistema coordenado de sinalizações semafóricas precisar ter um número de arestas direcionadas nos arcos, que, por sua vez, seja maior ou igual a reversibilidade do corredor. Se verdadeiro, o algoritmo escalona para execução imediata o início da coordenação de sinalizações semafóricas. Isto faz com que um sistema coordenado de sinalizações semafóricas seja ativado. Portanto, o Algoritmo \ref{alg:inicia_coordenacao} é executado.

\subsubsection{Obtenção de Quantidades de Veículos em Corredores}

De acordo com o Algoritmo \ref{alg:coleta_compartilhamento_demanda}, se um agente Sinalização Semafórica é participante de um ou mais sistemas coordenados de sinalizações semafóricas, mas o mesmo não é um controlador, ele compartilha também a sua média de quantidades de veículos com o agente líder do sistema coordenado de sinalizações semafórica em que ele faz parte. Para tanto, o agente interage com cada líder de sistema coordenado de sinalizações semafóricas em que o mesmo é participante, enviando uma mensagem de interação cujo interesse contido nela é \textit{roadway\_segment\_vehicle\_amount}.

Ao receber uma mensagem de interação  , contendo o interesse \textit{roadway\_segment\_vehicle\_amount}, um agente Sinalização Semafórica líder de um sistema coordenado de sinalizações semafóricas trata esta mensagem, executando o Algoritmo \ref{alg:tratamento_roadway_segment_vehicle_amount}. Este algoritmo apenas registra as médias das quantidades de veículos obtidas por cada participante de um sistema coordenado de sinalizações semafóricas (\textit{mediasQtdeVeiculosCor$_i$}). A partir destas médias, um agente Sinalização Semafórica líder de um sistema coordenado de sinalizações semafóricas pode escolher a maior média entre elas e, em seguida, compartilhar com outros agentes controladores de sistemas coordenados de sinalizações semafóricas de um mesmo grupo. Isto é feito periodicamente, de acordo com o parâmetro de periodicidade de compartilhamento de medias de quantidades de veículos entre agentes de um mesmo grupo de sistemas coordenados de sinalizações semafóricas. O valor deste parâmetro deve ser superior à periodicidade de obtenção de médias de quantidades de veículos. Sendo assim, o Algoritmo \ref{alg:selecao_maior_media_sistema} é constantemente executado.

Ao receber uma mensagem de interação  , contendo o interesse \textit{group\_member\_vehicle\_amount}, um agente Sinalização Semafórica líder de um sistema coordenado de sinalizações semafóricas trata esta mensagem, executando o Algoritmo \ref{alg:tratamento_group_member_vehicle_amount}. Este algoritmo apenas registra as médias das quantidades de veículos obtidas em cada sistema coordenado de sinalizações semafóricas pertencente ao mesmo grupo (\textit{mediasQtdeVeiculosGrupo$_i$}). A partir destas médias, um agente Sinalização Semafórica líder de um sistema coordenado de sinalizações semafóricas pode escolher a maior média entre elas e, em seguida, assumir este valor como sua média de quantidade de veículos (Algoritmo \ref{alg:selecao_maior_media_grupo}). Isto faz com que todos os agentes controladores dos sistemas coordenados de sinalizações semafóricas de um mesmo grupo assumam o mesmo valor de média de quantidades de veículos. Após isto, o algoritmo envia uma mensagem de interação, contendo o interesse \textit{corridor\_vehicle\_amount}, a fim de compartilhar a média escolhida com os agentes controladores de sistemas coordenados de sinalizações semafóricas concorrentes. Esta mensagem é enviada, utilizando uma interface de acesso à comunicação baseada no padrão LTE. Vale ressaltar que, a execução do Algoritmo \ref{alg:selecao_maior_media_grupo} se dá de maneira periódica, tal como é para o Algoritmo \ref{alg:selecao_maior_media_sistema}. Porém, o valor do parâmetro de periodicidade de execução do Algoritmo \ref{alg:selecao_maior_media_grupo} deve ser maior que o do Algoritmo \ref{alg:selecao_maior_media_sistema}.

Ao receber uma mensagem de interação  , contendo o interesse \textit{corridor\_vehicle\_amount}, um agente Sinalização Semafórica líder de um sistema coordenado de sinalizações semafóricas trata a mesma, executando o Algoritmo \ref{alg:tratamento_corridor_vehicle_amount}. O algoritmo inicia, verificando se o valor do parâmetro \textit{corredor} corresponde a um identificador de corredor vizinho. Se verdadeiro, o algoritmo apenas associa o valor do parâmetro \textit{mediaQtdeVeiculos} ao identificador do corredor cujo agente é a origem da mensagem recebida. Com as médias das quantidades de veículos de todos os corredores, cada agente Sinalização Semafórica líder de um sistema coordenado de sinalizações semafóricas pode atualizar o conhecimento das demandas dos corredores dos sistemas coordenados de sinalizações semafóricas. Para tanto, o agente executa periodicamente o Algoritmo \ref{alg:atualizacao_demanda_corredores}.

O Algoritmo \ref{alg:atualizacao_demanda_corredores} inicia, verificando se o agente é um líder de sistema coordenado de sinalizações semafóricas. Se verdadeiro, ele calcula a soma de todas as médias de quantidades de veículos. Após isto, ele calcula a demanda de cada corredor do multigrafo utilizado pelo algoritmo SMER para controle de sistemas coordenados de sinalizações semafóricas. Em seguida, o mínimo múltiplo comum entre as demandas dos corredores é calculado e, por fim, a variável \textit{calculadaDemandaCor$_i$} recebe o valor \textbf{verdadeiro}. A partir disto, de acordo com o Algoritmo \ref{alg:inicia_coordenacao}, um agente Sinalização Semafórica líder de um sistema coordenado de sinalizações semafóricas pode reajustar a reversibilidade do corredor e, consequentemente, o número de arestas dos arcos entre o corredor e os vizinhos deste. Se isto acontecer, uma mensagem de interação, contendo o interesse \textit{corridor\_reversibility\_change}, é enviada.

Ao receber uma mensagem de interação  , contendo o interesse \textit{corridor\_reversibility\_change}, um agente Sinalização Semafórica líder de um sistema coordenado de sinalizações semafóricas trata esta mensagem, executando o Algoritmo \ref{alg:tratamento_corridor_reversibility_change}. O algoritmo inicia, inicializando a reversibilidade do corredor com o valor um. Em seguida, ele verifica se a reversibilidade do corredor é igual a zero. Se verdadeiro, a reversibilidade do corredor recebe o valor do mínimo múltiplo comum calculado a partir das demandas dos corredores. Caso contrário, uma nova reversibilidade é calculada. No próximo passo, se a reversibilidade calculada for diferente da reversibilidade atual do corredor, a reversibilidade do corredor é atualizada e, em seguida, os números de arestas nos arcos entre o corredor e seus vizinhos são atualizados.

\subsection{Notificando o Funcionamento das Sinalizações Semafóricas}

Nesta tese, periodicamente, os agentes Sinalização Semafórica precisam enviar mensagens de interação, contendo o interesse \textit{traffic\_light}, para o agente Centro de Controle de Tráfego, a fim de que este fique ciente do funcionamento das sinalizações semafóricas de uma interseção. À medida que o agente Centro de Controle de Tráfego recebe mensagens com este interesse, ele atualiza o tempo de vida da sinalização semafórica. Caso um agente Sinalização Semafórica pare de enviar tais mensagens, o tempo de vida deste chegará a zero. Desta forma, o agente Centro de Controle de Tráfego pode identificar a ausência de funcionamento das sinalizações semafóricas, em específico, aquelas responsáveis pelo controle de um sistema coordenado de sinalizações semafóricas. Sempre que isto acontece, o agente Centro de Controle de Tráfego assume o controle dos sistemas coordenados de sinalizações semafóricas cujo os agentes Sinalização Semafórica líder deste apresentam ausência de funcionamento. Quando um agente Sinalização Semafórica, responsável pelo controle de um sistema coordenado de sinalizações semafóricas, volta a operar, o agente Centro de Controle de Tráfego devolve o controle para o agente e os dados relacionados ao controle do sistema coordenado de sinalizações semafóricas.

\subsection{Controlando Interseções com Veículos Conectados}

De acordo com o Algoritmo \ref{alg:notificacao_presenca}, quando um agente Veículo detecta a ausência de funcionamento de uma sinalização semafórica na via onde ele trafega, ele inicia um controle de interseções utilizando veículos conectados. Para tanto, o agente Veículo precisa interagir com o agente Centro de Controle de Tráfego, a fim de obter os dados de controle das sinalizações semafóricas da interseção. Então, esta interação se dá, quando o agente Veículo envia uma mensagem de interação com o interesse \textit{intersection\_control\_data\_request}. Tal mensagem contém um parâmetro \textit{idIntersecao} e é enviada por meio de uma interface de acesso à comunicação baseada no padrão LTE. 

Ao receber uma mensagem de interação  , contendo o interesse \textit{intersection\_control\_data\_request}, o agente Centro de Controle de Tráfego trata a mesma, executando o Algoritmo \ref{alg:tratamento_intersection_control_data_request}. O algoritmo inicia, verificando a existência do identificador da interseção na base de dados se interseções (\textit{intersecoes$_i$}). Se verdadeiro, o algoritmo obtém os dados de controle da interseção. Estes dados contêm todos os parâmetros relativos às sinalizações semafóricas, o modo de controle da interseção, o estado atual das indicações das sinalizações semafóricas, que, por sua vez, é dado pelo estado do multigrafo utilizado pelo algoritmo SMER para controle de interseções isoladas. Vale ressaltar que, o estado do controle da interseção é fornecido por uma simulação do sistema de controle de tráfego real, que é mantida pelo sistema supervisório. Após os dados de controle da interseção terem sido obtidos, estes são enviados para o agente Veículo que os requisitou. Para tanto, o algoritmo cria uma mensagem de interação com o interesse \textit{intersection\_control\_data} e destino igual à origem da mensagem recebida ($Msg_j.origem$). Por fim, a mensagem é enviada por meio da interface de acesso à comunicação baseada no padrão LTE.

Ao receber uma mensagem de interação  , contendo o interesse \textit{intersection\_control\_data}, um agente Veículo a trata, executando o Algoritmo \ref{alg:tratamento_intersection_control_data}. O algoritmo inicia, obtendo a identificação interseção por meio da identificação da via onde o veículo está trafegando. Em seguida, ele compara esta identificação com o valor do parâmetro \textit{idIntersecao}, que é obtido a partir da mensagem recebida. Se os valores forem iguais, o algoritmo obtém os dados de controle de interseção (\textit{dadosControleIntrsc}) e, a partir deles, criar uma sinalização semafórica virtual (\textit{sinSemaforica$_i$}), que, por sua vez, possui as estruturas de dados necessárias tanto para executar localmente os algoritmos de controle de interseção quanto aquelas necessárias para interação com outros agentes. Uma vez criada a sinalização semafórica virtual, esta entra em operação. Se a indicação da sinalização semafórica for diferente de verde para via atual do veículo, por fim, ele inicia a identificação de líder. 

O processo de identificação de líder em uma via dura até a sinalização semafórica virtual indicar luz verde para via. Enquanto isto não acontece, um agente Veículo executa a cada segundo o Algoritmo \ref{alg:envio_posicao_veiculo}. O algoritmo inicia, verificando se a indicação da sinalização semafórica virtual é diferente de verde. Se verdadeiro, uma mensagem de interação com o interesse \textit{vehicle\_position} é enviada para todas as direções da via, utilizando uma interface de acesso à comunicação baseada no padrão IEEE 802.11.

Ao receber uma mensagem de interação  , contendo o interesse \textit{vehicle\_position}, o agente Veículo trata a mesma, executando o Algoritmo \ref{alg:tratamento_vehicle_position}. O algoritmo inicia, verificando se a indicação da sinalização semafórica virtual (sinSemaforica$_i$.indicacao) é diferente de verde. Se verdadeiro, o algoritmo prossegue, calculando a distância geométrica entre o veículo conectado do agente e a faixa de retenção da via. A partir disto, ele compara tal dado com as distâncias de todos os veículos conectados conhecidos. Se a distância de um veículo é menor, este passa a ser o líder. Ao final do algoritmo, ele compara se o identificador do líder é o mesmo identificador do agente Veículo. Se verdadeiro, o veículo conectado do agente é o líder e, por isto, precisa ser configurado como tal. Caso contrário, a configuração de líder é desfeita.

A configuração de um agente Veículo como um líder se dá por meio da associação de ações à interesses já registrados na camada de rede do ambiente, tais como \textit{vehicle\_on\_$<$id da via$>$} e \textit{vehicle\_out\_$<$id da via$>$}. Dessa forma, o líder pode registrar a presença de veículos conectados na via onde ele está trafegando. No entanto, o líder precisa ter ciência da quantidade de veículos de outras vias de entrada da interseção. Para tanto, o agente também registra os seguintes os interesses citados na interface de acesso à comunicação baseada no padrão LTE, adotando um número máximo de saltos igual a um. Ao fazer isto, o líder interage com o agente Centro de Controle de Tráfego, de modo que este último torne o agente Veículo responsável pela interseção, registre os mesmos interesses e, finalmente, associe ações aos mesmos, a fim de que as mensagens de interação enviadas por agentes Veículo, em outras vias de entrada da interseção, possam ser direcionadas para o líder. Além disto, o agente Centro de Controle tira proveito destes dados, a fim de atualizar o estado do controle da interseção na simulação mantida pelo sistema supervisório. 

Quando o agente Veículo, que é líder em uma via, recebe uma mensagem de interação, contendo o interesse \textit{intersection\_control\_data\_request}, isto significa que, as sinalizações semafóricas virtuais em uma via de movimento conflitante em relação ao da via do líder estão indicando amarelo. Logo, um novo líder foi identificado. Por isto, o agente receptor desta mensagem deve repassar todos os dados de controle da interseção para o novo líder, enviando uma mensagem de interação com o interesse \textit{intersection\_control\_data}. Durante a criação desta mensagem, o agente líder introduz dois parâmetros, que são: \textit{dadosControleIntrsc} e \textit{idAgenteLider}. Ao primeiro parâmetro, são atribuídos todos os dados relativos ao controle de da interseção. O último parâmetro mantém o identificador do novo líder, que, por sua vez, é a origem da mensagem de interação recebida pelo agente. Quando o agente Centro de Controle de Tráfego recebe a mensagem de interação, contendo o interesse \textit{intersection\_control\_data}, ele atualiza os dados de controle da interseção e, em seguida, encaminha a mensagem para o novo líder da interseção. Este, então, realiza todo o processo descrito anteriormente.

\section{Considerações Finais}

Este capítulo apresentou as estratégias, propostas por esta tese, para tratar o problema de controle de tráfego por meio de sinalizações, bem como, a implementação das mesmas sobre uma rede veicular heterogênea centrada em interesses. Tais estratégias lidam com o controle de interseções isoladas, coordenação de sinalizações semafóricas e o controle de interseções em situações em que as sinalizações semafóricas apresentam ausência de funcionamento. A partir da implementação destas estratégias sobre um a rede veicular heterogênea centrada em interesses, foi possível criar um sistema de controle de tráfego distribuído e descentralizado, utilizando o paradigma de sistemas multiagentes, em que sinalizações semafóricas inteligentes e veículos conectados cooperam uns com os outros, a fim de melhorar a fluidez do tráfego em uma rede viária. Esse sistema de controle de tráfego, portanto, fornece as bases necessárias para a construção de um sistema de planejamento e orientação de rotas. Os detalhes acerca deste sistema são apresentados no próximo capítulo.

  \chapter{Planejamento e Orientação Inteligentes de Rotas Baseado em Interesses de Motoristas}\label{cap:orientacao}

Este capítulo tem como objetivo apresentar a estratégia, proposta por esta tese, para lidar com o problema de planejamento e orientação de rotas, bem como, detalhar a implementação da mesma sobre a rede \textit{ad hoc} veicular heterogênea centrada em interesses. O sistema multiagente para planejamento e orientação de rotas faz uso dos quatro tipos agentes apresentados no Capítulo \ref{cap:agentes}, que são: Elemento Urbano, Centro de Controle Tráfego, Veículo e Sinalização Semafórica. Neste sentido, o sistema de planejamento e orientação de rotas tira proveito das implementações do sistema de controle de tráfego apresentado no capítulo anterior.

Este capítulo inicia, descrevendo brevemente o problema de planejamento e orientação de rotas. Em seguida, é discutido conceitualmente a proposta de uma estratégia de solução para o problema em questão. Por fim, são apresentados os detalhamentos algorítmicos relativos à implementação da estratégia sobre a rede \textit{ad hoc} veicular heterogênea centrada em interesses.

\section{Descrevendo o Problema de Planejamento e Orientação de Rotas}

O problema de planejamento e orientação de rotas está em dissipar rapidamente os fluxos de veículos e evitar congestionamentos, distribuindo os veículos de maneira uniforme ao longo de rotas alternativas. Tendo em vista que um controle de tráfego é capaz somente de lidar com flutuações de tráfego em interseções isoladas ou em corredores de sinalizações semafóricas, eles não são imunes ao desbalanceamento dos fluxos de tráfego, que é causado pelas escolhas equivocadas de rotas por parte dos motoristas ou por mecanismos de planejamento e orientação de rotas ineficientes, tais como \abbrev{TMC}{Traffic Message Channel} \cite{TMC:2016}, TomTom \cite{TomTom:2016}, Waze \cite{Waze:2016}, entre outros. Em outras palavras, Tais soluções reagem somente após os congestionamentos terem ocorrido e, com isso, acabam propagando os efeitos destes sobre uma parte substancial da rede viária. Isto se deve aos altos intervalos de atualização do estado do tráfego, que variam entre 2 a 30 minutos \cite{Wedde:2013}.

\section{Tratando o Problema de Planejamento e Orientação de Rotas}

Para tratar o problema relacionado à orientação de rotas, utilizando uma metodologia de sistemas multiagentes baseada em alocação de recursos é necessário o uso de uma abstração adequada para o tratamento deste problema. Neste sentido, é possível tratar o problema de orientação de rotas como um problema de Sistemas Flexíveis de Manufatura ou \textit{Flexible Manufacturing Systems} (FMS). 

Segundo \citet{Browne:1984}, Sistemas Flexíveis de Manufatura podem ser divididos em duas categorias: flexibilidade de máquina e flexibilidade de roteamento. A primeira categoria consiste na capacidade que o sistema tem de mudar, de modo que novos tipos de produtos sejam produzidos, bem como na capacidade de alterar a ordem das operações em uma parte deste sistema. A segunda categoria consiste na capacidade que sistema tem de usar várias máquinas para executar as mesmas operações em uma parte deste sistema, bem como a capacidade de absorver grandes mudanças em termos de volume, capacidade e capabilidade de processo. Dentro desta classificação, o problema de orientação de rotas pode ser tratado como um problema de Sistemas Flexíveis de Manufatura com flexibilidade de roteamento. 

Em Sistemas Flexíveis de Manufatura, existem diferentes processos com uma variedade de produtos que podem ou não seguir a mesma rota. Quando todos os grupos de produtos percorrem uma mesma rota para chegar às estações de trabalho, o Sistema Flexível de Manufatura é chamado de sistema de produção contínua ou \textit{Flow-Shop}. Diferente disto, quando cada tipo de produto tem sua própria rota, o Sistema Flexível de Manufatura é chamado de sistema de produção descontínua ou \textit{Job-Shop}. 

Dentro da perspectiva de um \textit{Job-Shop}, os destinos das viagens de cada motorista podem ser abstraídos na forma de diferentes produtos que precisam ser encaminhados para as diversas estações de trabalho do sistema de produção, utilizando rotas individuais, até que o processo de produção termine, ou seja, até que o motorista chegue ao destino de sua viagem sobre as vias de uma rede viária. Sendo assim, as viagens realizadas pelos motoristas podem ser abstraídas na forma de processos de produção, que são compostos por diversas estações de trabalho, de acordo com a rota a ser percorrida durante o processo de produção de um produto. Para que a viagem possa ser realizada, torna-se necessário o planejamento e ou o replanejamento de rotas. 

O planejamento e ou o replanejamento de rotas pode ser abstraído na forma de um escalonamento de operações de um produto em um \textit{Job-Shop}. Sendo assim, uma rota é abstraída na forma de um conjunto de operações a serem executadas em diversas estações de trabalhos do sistema. Com isto, cada operação representa um espaço de uso nas vias que conectam as interseções em uma rede viária. As diversas estações de trabalho do Sistema Flexível de Manufatura representam as interseções das vias de uma rede viária. Em uma rede viária, as interseções são conectadas umas às outras por meio vias que ao mesmo tempo são vias de saída de uma interseção e vias de entrada em uma outra interseção imediatamente a frente daquela cuja mesma via é uma via de saída. Transpondo isto para um modelo de \textit{Job-Shop}, as vias podem ser abstraídas na forma de filas, onde componentes de movimentação são organizados, de modo que estes se movam de estação de trabalho à estação de trabalho, de acordo com o planejamento de rotas de cada produto. 

Por fim, os componentes de movimentação abstraem os veículos que trafegam ao longo das vias de uma rede viária, sendo esta abstraída na forma de uma rede \textit{job-shop} composta por filas. Uma vez que o problema de orientação de rotas pode ser abstraído na forma de um problema de escalonamento de operações em um \textit{Job-Shop}, é possível definir as estratégias e mecanismos a serem adotadas na construção de uma possível solução para o problema de planejamento e orientação de rotas. Sendo assim, este trabalho propõe o uso de um algoritmo de escalonamento baseado em regras de despacho com as heurísticas de menor data de entrega (\textit{Earliest Due Date} - EDD) e menor tempo de processamento (\textit{Shortest Processing Time} - SPT). A construção deste algoritmo depende fortemente das agendas de intervalos de indicações de luzes verdes, que devem ser geradas com base nos estados de controle das interseções de uma rede viária controlada pelo sistema de controle de tráfego.

Conforme descrito até aqui, as sinalizações semafóricas são controladas por meio da execução de um algoritmo de escalonamento distribuído chamado SMER cujo funcionamento se baseia em um multigrafo. Para o correto funcionamento deste algoritmo, é necessário que cada sinalização semafórica mantenha uma cópia do multigrafo. Uma vez que tal grafo representa uma máquina de estados finitos, as sinalizações semafóricas tiram proveito desta abstração e, por isto, são capazes de utilizá-la para gerar localmente as agendas de intervalos de indicações de luzes verdes e das demais sinalizações semafóricas participantes da mesma interseção.  Ao receberem os dados de controle de uma interseção, os agentes responsáveis pelo controle de uma interseção, agentes Sinalização e agentes Veículo, devem ser capazes organizá-las, de modo que estas sejam utilizadas por um algoritmo local de geração de agendas de intervalos de indicações de luzes verdes. Desta forma, cada agente Sinalização Semafórica gera uma cópia das agendas de intervalos de indicações de luzes verdes, utilizando dados de controle de interseções, ainda que tais dados mudem em função das flutuações dos fluxos de tráfego. Neste caso, em específico, os novos dados de controle das interseções precisam ser compartilhados com as demais agentes responsáveis pelo controle de interseções, de modo que eles possam excluir as agendas e, em seguida, gerar novas agendas de intervalos de indicações de luzes verdes.

Com tudo isso, deve ser possível construir um algoritmo de escalonamento livre de \textit{deadlock} e \textit{starvation} baseado em regras de despacho com as heurísticas de menor data de entrega (\textit{Earliest Due Date} - EDD) e menor tempo de processamento (\textit{Shortest Processing Time} - SPT), uma vez que as agendas de intervalos de indicações de luzes verdes são geradas com base nas simulações de execuções do algoritmo de escalonamento SMER. Vale ressaltar que, uma das características do SMER é a garantia de acesso às interseções por parte dos fluxos de tráfego das vias de entrada de maneira mutuamente exclusiva, ou seja, dois fluxos de tráfego de vias de entrada de uma interseção, que sejam pertencentes a grupos de movimentos conflitantes, não podem ter acesso simultâneo à uma interseção. Esse algoritmo de escalonamento deve ser divido em quatro etapas: cálculo de uma rota ótima; desalocação de espaços nas vias e alocação de espaços nas vias em função na nova rota ótima; atualização das alocações dos espaços nas vias nos demais agentes responsáveis pelos controles de interseções; entrega da rota ótima calculada para o requerente do cálculo.

A primeira etapa deve executar, quando os agentes responsáveis pelos controles de interseções recebem uma requisição para o cálculo de rota ótima por meio dos agentes Veículo. Neste instante, os agentes responsáveis pelos controles de interseções devem executar um algoritmo capaz de calcular uma rota ótima para o veículo conectado requisitante até um determinado destino, levando em consideração as disponibilidades de acesso às interseções ao longo do tempo, limites de velocidade das vias de entrada destas interseções e os espaços físicos disponíveis nestas vias. Logo, o algoritmo deve iniciar com uma rota vazia, uma via de partida e um conjunto de vias de entrada das interseções vizinhas à interseção em que a sinalização semafórica está instalada. Em seguida, o algoritmo deve realizar a escolha de uma via que possa fornecer o menor tempo para que o veículo conectado alcance a próxima interseção. O algoritmo utiliza uma heurística de menor data de entrega (EDD), a fim de escolher as vias com menor tempo de espera por uma luz verde das sinalizações semafóricas que as controlam. Este tempo de espera equivale ao tempo de inicialização e é calculado com base nas entradas das agendas de intervalos de indicações de luzes verdes, que são mantidas pelos responsáveis pelos controles das interseções. Por fim, o algoritmo seleciona a via em que o veículo conectado gastará o menor tempo para alcançar a próxima interseção (SPT), levando em consideração o tempo acumulado das vias escolhidas anteriormente, o tempo de espera pela próxima luz verde, a disponibilidade de espaços de uso, o atraso causado pelos veículos conectados que possuem espaços alocados na via e o limite de velocidade da via. O menor tempo para um veículo conectado alcançar uma interseção equivale ao menor tempo de processamento de uma operação. Após o processo de escolha da melhor via, o algoritmo deve atualizar a rota parcial, adicionando a via escolhida e, em seguida, ele atualiza o custo parcial da rota e toma a via escolhida como ponto de partida. O algoritmo é encerrado, quando uma rota parcial ótima é calculada até o destino pretendido pelo motorista. 

Após isto, o algoritmo finaliza a primeira etapa e, em seguida, inicia a segunda etapa. Nesta etapa, se existirem alocações de espaços nas vias, que sejam relativas ao requerente do cálculo de rotas ótimas, elas são removidas. Após isto, novas alocações de espaços nas vias são realizadas, utilizando a rota ótima calculada. Após esta atualização local acerca das alocações de espaços nas vias, a segunda etapa é finalizada e, após isto, a terceira etapa é iniciada. Esta etapa consiste na atualização das alocações de espaços nas vias em outros agentes responsáveis pelos controles de interseções. Paralelo a isto, é realizado o envio da nova rota ótima ao requerente do cálculo de rota ótima.

Com base na modelagem do problema de planejamento e orientação de rotas na forma de um problema de escalonamento em Sistemas Flexíveis de Manufatura do tipo \textit{Job-Shop} e no algoritmo de escalonamento proposto neste parágrafo, apresentou-se uma solução para o problema de planejamento orientação de rotas descrito acima.

\section{Sistema Multiagente de Planejamento e Orientação de Rotas}

Esta seção tem como objetivo apresentar os detalhamentos algorítmicos relativos à estratégia apresentada anteriormente. Primeiramente, são apresentados os detalhes sobre a inicialização do sistema de planejamento e orientação de rotas. Após isto, são apresentados os detalhes relativos ao registro de interesses de usuários e interesses da rede. Em seguida, são apresentados os detalhes em torno do mecanismo de geração de agendas de intervalos de indicações de luzes verdes. Ao findar esta apresentação, inicia-se o detalhamento do mecanismo de cálculo de rotas ótimas e alocação de espaços nas vias. Logo após, é descrito como as agendas de intervalos de indicações de luzes verdes são compartilhadas entre veículos conectados, quando estes são os responsáveis pelo controle de uma interseção. Por fim, é detalhado como os veículos conectados tiram proveito dos intervalos de indicações de luzes verdes, à medida que entram nas vias.

\subsection{Inicialização do Sistema Multiagente}

Esta seção tem como objetivo descrever o processo de inicialização das instâncias dos agentes participantes do sistema multiagente de planejamento e orientação de rotas. Tais agentes são os seguintes: Elemento Urbano, Centro de Controle de Tráfego, Sinalização Semafórica e Veículo. Antes de iniciar a descrição dos processos de inicialização das instâncias destes agentes, é necessário definir uma URI para identificar o sistema multiagente de planejamento e orientação de rotas. Sendo assim, a URI é definida da seguinte forma: \textit{radnet://ttm/route\_guidance/trasit\_route}. Com base nesta, as instâncias dos agentes podem utilizá-la para registrar interesses e vias na camada de rede de seus ambientes. 

Para que um agente Elemento Urbano entre em operação, é necessário atribuir um valor ao parâmetro de configuração relativo à periodicidade de interações com os agentes do sistema multiagente de planejamento e orientação de rotas, objetivando o compartilhamento de dados do agente. Com estes dados, os agentes podem configurar a camada de rede de seus ambientes, registrando interesses e associando ações, se necessário. Desta forma, os ambientes dos agentes podem encaminhar mensagens de interações para os agentes, de modo que estes executem ações, desde que estas estejam associadas ao interesse registrado na camada de rede. Além disso, o registro de interesses, utilizando dos dados compartilhados pelos agentes Elemento Urbano, permite também que os ambientes dos agentes cooperem para o funcionamento da HRAdNet-VE, no que tange o encaminhamento de mensagens de rede. Nesta tese, o parâmetro em questão é configurado, de modo que os agentes Elemento Urbano enviem mensagens de interação a cada 60 segundos. Além disto, nesta tese, os agentes Elemento Urbano compartilham somente a via onde se encontram. 

Tendo em vista que o sistema multiagente para planejamento e orientação de rotas é construído sobre o sistema multiagente para controle de tráfego, as inicializações dos agentes Centro de Controle de Tráfego, Sinalização Semafóricas e Veículos são estendidas, de modo que estes agentes possam interagir no âmbito do sistema multiagente para planejamento e orientação de rotas. Neste sentido, as instâncias destes agentes registram seus interesses na camada de rede do ambiente. Os dados interesses registrados pelas instâncias dos agentes Centro de Controle de Tráfego, Sinalização Semafórica e Veículo e as ações executadas por elas ou pelo seus ambientes são apresentados respectivamente nas Tabelas \ref{tab:interesses_cct_2}, \ref{tab:interesses_sinalizacao_2} e \ref{tab:interesses_veiculo_2}.

\subsection{Registrando Interesses Relacionados aos Elementos Urbanos}

Após terem sido inicializados, os agentes Elemento Urbano interagem com os agentes Sinalização Semafórica, enviando, a cada minuto, mensagens de interação contendo o interesse \textit{urban\_element\_data}. Os detalhes acerca do envio desta mensagem são apresentados pelo Algoritmo \ref{alg:envio_urban_element_data}. Como pode ser visto, os agentes Elemento Urbano enviam uma lista de termos, aqui chamadados de interesses de usuário, e o identificador da via onde o elemento urbano relacionado a tais termos se encontra. Com base neste identificador de via, os demais tipos de agentes podem configurar a camada de rede de seus ambientes. O envio da mensagem de interação se dá por meio de uma interface de acesso à comunicação baseada no padrão LTE.

Ao receber uma mensagem de interação, contendo o interesse \textit{urban\_element\_data}, o agente Centro de Controle de Tráfego registra um interesse, que é a concatenação da \textit{string} \textit{route\_to\_} com o identificador da via contido no parâmetro \textit{idVia} da mensagem recebida. O agente registra o interesse em uma interface de acesso à comunicação baseada no padrão LTE com um número máximo de saltos igual a um. Além disto, a ação associada ao interesse registrado permite que o agente Centro de Controle de Tráfego encaminhe mensagens de interação com tal interesse para os agentes responsáveis pelos controles das interseções, sejam estes agentes Sinalização Semafórica e agentes Veículo. Além do interesse citado acima, o agente Centro de Controle de Tráfego também registra o interesse \textit{calculated\_route\_$<$id. da via$>$}, de modo que o agente possa encaminhar a mensagem para um agente Veículo requerente de um cálculo de rota ótima, se necessário. O agente registra o interesse em uma interface de acesso à comunicação baseada no padrão LTE com um número máximo de saltos igual a um. Por fim, o agente Centro de Controle de Tráfego encaminha a mensagem recebida para cada agente responsável em controlar uma interseção.

Quando um agente Sinalização Semafórica recebe a mensagem encaminhada pelo agente Centro de Controle de Tráfego, ele extrai o valor do parâmetro \textit{idVia} da mensagem de interação recebida e o concatena também com a \textit{string} \textit{route\_to\_}. O agente registra esta concatenação como um interesse em mais de uma interface de acesso à comunicação. Os dados relativos ao registro do interesse e ações executadas pelo agente são apresentados na Tabela \ref{tab:interesses_sinalizacao_2_1}. No que diz respeito a um agente Veículo responsável pelo controle de uma interseção, quando este recebe a mensagem encaminhada pelo agente Centro de Controle de Tráfego, utilizando os dados apresentados na Tabela \ref{tab:interesses_sinalizacao_2_1}. Após isto, ele associa as ações que permitem tratar mensagens contendo o interesse. Após os registros de interesses, tanto os agentes Sinalização Semafórica quanto os agentes Veículos encaminham a mensagem recebida por meio das interfaces de acesso à comunicação baseadas nos padrões IEEE 802.11 e IEEE 802.11p, de modo que a mensagem seja recebida pelos agentes Veículos que trafegam nas vias de entrada das interseções. Quando estes agentes Veículo recebem a mensagem de interação, eles extraem uma lista de interesses de usuários, utilizando parâmetro \textit{listaInteressesUsuario} e, em seguida, atualizam os dados relativos ao elemento urbano no sistema de planejamento e orientação de rotas. Além disto, eles também registram o interesse, conforme os dados apresentados na Tabela \ref{tab:interesses_veiculo_3}. Por fim, os agentes Veículo registram o interesse \textit{calculated\_route\_to\_$<$id. da via$>$}, utilizando os dados da Tabela \ref{tab:interesses_veiculo_2}. Vale ressaltar, quando um agente Veículo deixa de ser líder, ele desassocia a ação de cálculo de rotas ótimas dos interesses registrados e os reconfigura, de acordo com os dados da Tabela \ref{tab:interesses_veiculo_2}. Isto pode acontecer, de acordo com os Algoritmos \ref{alg:monitoramento_faixas} e \ref{alg:tratamento_vehicle_position}.

\subsection{Gerando Agendas de Intervalos de Indicações de Luzes Verdes}

A geração de uma agenda de intervalos de indicações de luzes verdes é de extrema importância para o sistema de planejamento e orientação de rotas, pois, por meio desta, é possível saber a disponibilidade das vias de entradas das interseções de uma rede viária, à medida que as sinalizações semafóricas que as controlam indicam luz verde durante intervalos de tempo determinados pela configuração do multigrafo, que é utilizado pelo algoritmo de controle de interseção. Para tanto, é preciso executar o Algoritmo \ref{alg:geracao_agenda_intersecao_isolada}. Este algoritmo deve receber os seguintes dados de entrada: tempo atual do GPS, dados de controle da interseção e um multiplicador.

Como pode ser visto nos Algoritmos \ref{alg:geracao_agenda_intersecao_isolada}, ele primeiramente inicializa tanto variáveis quanto estruturas de dados necessárias para execução do artigo. A variável \textit{tempo} é inicializada com o valor obtido pela multiplicação dos valores dos parâmetros de periodicidade de obtenção da quantidade de veículos (\textit{periodObtQtdeVeic}) e do número de obtenções de quantidades de veículos (\textit{numObtQtdeVeic}), e do multiplicador. O valor do \textit{tempo} é utilizado para definir uma janela de tempo, que é utilizada para limitar a geração de agendas de intervalos de indicações de luzes verdes. Em seguida, é inicializado o tempo inicial (\textit{tempoInicial}), que será utilizado na definição dos inícios de intervalos de indicações de luzes verdes. A próxima variável a ser inicializada é o acumulador de tempo (\textit{acumuladorTempo}), que, por sua vez, tem a finalidade de acumular as durações dos intervalos das indicações de luzes verdes. O valor desta variável é utilizado para controlar as iterações dos algoritmos, enquando o tempo acumulado seja menor que a janela de tempo calculada anteriormente. Agora, seguindo com a inicialização das estruturas de dados, a lista de tamanhos de intervalos de tempo ($tamanhosIntervalos_i$) é esvaziada. Esta lista é mantida pelos agentes responsáveis pelo controle da interseção. Cada entrada dela, possui um mapa, em que cada deste mapeia o identificador de uma via de entrada da interseção e o tamanho do intervalo de indicação de luz verde, que é disponibilizado para a via, de acordo com a configuração do multigrafo. Logo após, o mapa de alocações de espaços nas vias é esvaziado ($alocacoesEspacosVias_i$). Cada entrada deste mapa é um mapeamento entre o identificador de uma via e uma lista de mapas de alocações de espaços nestas vias. Cada entrada destes mapas de alocações de espaços nas vias mapeia um identificador de um agente veículo com a previsão do instante de tempo em que ele chegará na via e o tamanho do veículo. Em seguida, um mapa dos tempos iniciais para cada intervalo de indicação de luz verde (\textit{temposIniciais}) é inicializado. Cada entrada deste mapa corresponde ao mapeamento entre um vértice do multigrafo, que representa um agente Sinalização Semafórica, e o tempo de início do intervalo de indicação de luz verde da sinalização semafórica do agente. A próxima estrutura de dados a ser inicializada é um mapa dos tamanhos de intervalos de indicações de luzes verdes relativos a cada vértice em operação (\textit{tamIntVertices}), à medida que os algoritmos executam. Ao final de cada iteração dos algoritmos, este mapa é adicionado à lista de tamanhos de intervalos de indicações de luzes verdes ($tamanhosIntervalos_i$). Logo após essa última inicialização, dois conjuntos são inicializados: conjuntos dos vértices em operação (\textit{verticesOp}) e o conjunto dos vértices bloqueados (\textit{verticesBlc}.  Em seguida, as estruturas de dados relativas à agenda de intervalos das indicações de luzes verdes são inicializadas. Estas estruturas de dados são listas em que cada entrada corresponde a um mapa. Cada entrada deste mapa é o mapeamento entre uma via e um valor de tempo. De acordo com o objetivo da estrutura de dados, este valor de tempo determina o início de um intervalo de indicação de luz verde ou o fim deste. Por isto, os agentes responsáveis pelo controle da interseção devem manter duas listas: lista de inícios dos intervalos de indicações de luzes verdes ($iniciosIntervalos_i$) e lista de fins dos intervalos de indicações de luzes verdes ($finsIntervalos_i$). Para finalizar a inicialização das variáveis dos algoritmos, as seguintes estruturas de dados são recuperadas dos dados de controle da interseção: conjunto de vértices (\textit{vertices}), multigrafo utilizado pelo algoritmo de controle da interseção (\textit{multigrafo}), reversibilidades dos agentes Sinalização Semafórica (\textit{reversibilidades}), intervalos mínimos de indicações de luzes verdes (\textit{intMinIndVerdes}), intervalos mínimos de indicações de luzes amarelas (\textit{intMinIndAmarelas}) e o mapeamento entre os vértices do multigrafo e as vias controladas pelas sinalizações semafóricas (\textit{mapVerticesVias}).

Após a inicialização das variáveis e estruturas de dados, o algoritmo inicia um laço. Enquanto o tempo acumulado for menor que a janela de tempo, o algoritmo simula o funcionamento do controle da interseção. Para tanto, o algoritmo verifica se o conjunto de vértices em operação está vazio. Se verdadeiro, para cada vértice existente no conjunto de vértices, é verificado se ele possui arestas revertidas para ele. Se verdadeiro, o vértice é incluído no conjunto de vértices em operação, um mapeamento entre o vértice e o tempo de início de um intervalo de indicação de luz verde é adicionado ao mapa de tempos iniciais e, por fim, uma entrada no mapa dos tamanhos de intervalos de indicações de luzes verdes relativos a cada vértice em operação é adicionada. 

Após isso, para cada vértice em operação, é verificado se a interseção participa de uma coordenação de sinalizações semafóricas. Se falso, o algoritmo reverte as arestas do vértice para seus vizinhos. Caso contrário, o algoritmo reverte todas as arestas do vértice para seus vizinhos. Em seguida, é verificado se o vértice ainda possui arestas revertidas para ele. Se falso, o algoritmo realiza os seguintes passos: inclusão do vértice no conjunto de vértices bloqueados; inclusão de um início de intervalo de indicação de luz verde  na lista de inícios de intervalos de indicações de luzes verdes; inclusão de um fim de intervalo de luz verde na lista de fins de intervalos de indicações de luzes verdes; inclusão de uma lista de mapas de alocações de espaços na via; inclusão de um mapa de alocações à lista de mapas de alocações de espaços de uma via; e atualização de uma entrada no mapa dos tamanhos de intervalos de indicações de luzes verdes relativos a cada vértice em operação. 

Em seguida, para cada vértice existente no conjunto de vértices bloqueados, eles removem o mesmo do conjunto de vértices em operação. Após isto, o algoritmo verifica se o tamanho do conjunto de vértices em operação é igual a zero. Se verdadeiro, ele encontra o maior intervalo de indicação de luz verde de um vértice e o tempo de indicação de luz amarela do vértice. Em seguida, os seguintes passos são executados: atualização do tempo; atualização do tempo inicial; inserção do mapa de tamanhos de intervalos de indicações de luzes verdes relativos aos vértices na lista de tamanhos de intervalos de indicações de luzes verdes; incremento da fase da interseção.

Uma vez que o sistema de planejamento e orientação de rotas é projetado com base nos agentes utilizados no sistema de controle de tráfego, algumas modificações precisam ser feitas neste último sistema, de modo que a funcionalidade de geração de agendas de intervalos de indicações de luzes verdes seja incorporada aos agentes responsáveis pelo controle de interseções. Neste sentido, inicialmente todas os agentes Sinalização Semafórica devem executar o Algoritmo \ref{alg:geracao_agenda_intersecao_isolada}, quando os mesmos estiverem entrando em operação. Além disto, o mesmo algoritmo deve ser executado, quando um agente Sinalização Semafórica muda sua reversibilidade e os números de arestas nos arcos entre o agente e seus vizinhos. No caso dos agentes Sinalização Semafórica que iniciam a sua participação em uma coordenação de sinalizações semafóricas de um sistema coordenado de sinalizações semafóricas ativo, estes também devem executar o Algoritmo \ref{alg:geracao_agenda_intersecao_isolada}, pois a agenda de intervalos de indicações de luzes verde precisa refletir a execução do algoritmo SMER como se este fosse um algoritmo SER. No findar de tal participação, os agentes Sinalização Semafórica precisam executar o Algoritmo \ref{alg:geracao_agenda_intersecao_isolada} novamente, pois a interseção volta a operar com o algoritmo SMER para controle de interseções isoladas, se for o caso. No caso daqueles agentes Veículos, que por ausência do funcionamento das sinalizações semafóricas de uma interseção assumem o controle desta, executam o algoritmo, de acordo com os dados de controle de interseção obtidos do agente Centro de Controle de Tráfego. Por fim, sempre que o Algoritmo \ref{alg:geracao_agenda_intersecao_isolada} é executado, é necessário que todos os agentes responsáveis pelo controle de uma interseção, incluindo agentes Veículos, tomem ciência destas execuções. Para que um agente Veículo, responsável pelo controle de uma interseção, possa tomar ciência das agendas de intervalos de indicações de luzes verdes de outras interseções, eles registram o interesse \textit{new\_traffic\_light\_schedule}. O registro deste interesse utiliza os mesmos dados apresentados na Tabela \ref{tab:interesses_sinalizacao_2}.

Para tanto, os executores destes algoritmos precisam interagir com esses agentes, compartilhando os dados de controle de sua interseção. À medida que os agentes recebem estes dados, eles executam um algoritmo apropriado para a geração de agendas de intervalos de indicações de luzes verdes. Dessa forma, esses agentes passam a ter ciência dos intervalos de indicações de luzes verdes de todas as interseções de uma rede viária. A mensagem de interação enviada para os agentes é configurada da seguinte forma: o campo interesse é configurado com \textit{new\_traffic\_light\_schedule}; os campos destino e identificador da via são configurados com \textbf{nulo}; o campo direção é configurado com zero; o campo tecnologia de acesso à comunicação é configurado com o valor LTE; e é inserido um parâmetro \textit{dadosControleIntrsc}, que é configurado com os dados de controle da interseção. 

Ao receber uma mensagem de interação  , contendo o interesse \textit{new\_traffic\_light\_schedule}, os agentes responsáveis pelo controle de uma interseção tratam esta mensagem, executando o Algoritmo \ref{alg:geracao_agenda_dados_outras_intersecoes}. O algoritmo é uma variante do Algoritmo \ref{alg:geracao_agenda_intersecao_isolada}. A diferença entre os Algoritmos \ref{alg:geracao_agenda_intersecao_isolada} e \ref{alg:geracao_agenda_dados_outras_intersecoes} é que, no primeiro algoritmo, os tamanhos dos intervalos de indicações de luzes verdes são mantidos pelo agente participante da interseção cujos dados de controle serviram de entrada para o algoritmo. Os tamanhos dos intervalos de indicações de luzes verdes são utilizados para calcular tanto o avanço quanto o posicionamento dos seguimentos de ondas verdes das vias de entrada da interseção. No que diz respeito ao Algoritmo \ref{alg:geracao_agenda_dados_outras_intersecoes}, este apenas gera a agenda de intervalos de indicações de luzes verdes de outras interseções, a fim de que o agente tenha conhecimento dos intervalos de indicações de luzes verdes de outras interseções. Com base nestes intervalos, os agentes responsáveis pelo controle de uma interseção podem utilizá-los no cálculo de rotas ótimas, levando em consideração os espaços alocados nas vias em cada intervalo de indicação de luz verde. Por fim, o algoritmo envia uma mensagem de interação, contendo o mesmo interesse da mensagem recebida, para os veículos conectados da via, utilizando uma interface de acesso à comunicação baseada no padrão IEEE 802.11p. Além disto, a mensagem deve ser propagada no sentido oposto da via, ou seja, para trás dos veículos conectados. Todo agente Veículo, ao receber tal mensagem, requisita um novo cálculo de rotas, devido às modificações nas agendas de intervalos de luzes verde das interseções controladas pelo sistema de controle de tráfego.

\subsection{Calculando Rotas Ótimas}

Uma vez que os agentes responsáveis pelo controle de interseções tenham conhecimento de todas as agendas de intervalos de indicações de luzes verdes das interseções de uma rede viária, eles podem receber mensagens de interação oriundas de veículos conectados, desde que estejam na mesma via, cujo intuito é requisitar o cálculo de rota ótima para um determinado destino. Neste momento, fica mais clara a utilidade das listas de interesses de usuários, que são publicadas pelos agentes Elemento Urbano. Os termos contidos nestas listas auxiliam o motorista, quando este realiza a busca por um determinado local do mapa, que é a via onde se encontra o elemento urbano desejado. Esta busca pode ser motivada, por exemplo, pelo interesse do motorista em encontrar gasolina mais barata, promoções das mais diversas, eventos, entre outros. Uma vez que os destinos tenham sido escolhidos pelos motoristas, agentes Veículos iniciam o processo de cálculo de rotas ótimas, tentando interagir com os agentes responsáveis pelo controle das interseções. Para tanto, os esses agentes escalonam periodicamente ações que executam o Algoritmo \ref{alg:solicitacao_calculo_rotas}, de acordo com o valor do parâmetro de periodicidade de tentativas de requisição de cálculo de rotas ótimas. A execução deste algoritmo é interrompida, quando o veículo conectado alcança o destino desejado. 

Como pode ser visto no Algoritmo \ref{alg:solicitacao_calculo_rotas}, o algoritmo inicia, verificando se o agente já possui uma rota calculada. Se falso, o algoritmo envia uma mensagem de interação ($Msg_i$) na mesma via em que o veículo conectado se encontra. Tal mensagem é enviada para frente, de modo que o agente responsável pelo controle da interseção possa ser alcançado. Além disto, tal mensagem pode ser enviada por meio de interfaces de acesso à comunicação baseadas no padrão IEEE 802.11 ou no padrão IEEE 802.11p, de acordo com o valor da variável de controle da tecnologia de acesso à comunicação para cálculo de rotas ($tacCalcRotas_i$). Esta variável é mantida pelo agente Veículo. Após enviar a mensagem, o algoritmo incrementa o número de tentativas de requisição de cálculo de rotas ótimas. Tudo isso é feito, se o número de tentativas é menor que o valor do parâmetro de número máxima de tentativas de requisição de rotas ótimas ($paramNumTentativasCalcRotas_i$). Caso contrário, o algoritmo troca a tecnologia de acesso à comunicação, de modo que o agente comece a requisitar cálculos de rotas ótimas sempre pela tecnologia de menor alcance de comunicação. Caso o agente não receba uma rota calculada, o algoritmo troca a tecnologia de acesso à comunicação até que o padrão LTE seja utilizado. Por fim, se o agente possuir uma rota calculada, o algoritmo mantém o número de tentativas de requisições de cálculos de rotas ótimas com o valor zero e a tecnologia de acesso à comunicação no padrão IEEE 802.11.

Antes de entrar nos detalhes relativos ao tratamento de mensagens de interação cujo interesse é \textit{route\_to\_$<$id. da via$>$}. É preciso ressaltar que, ao receber uma mensagem com este interesse, o agente Centro de Controle de Tráfego a encaminha para um agente que esteja mais próximo do veículo conectado cujo agente é o requerente do cálculo de rota ótima. O agente cuja mensagem foi encaminhada deve ser responsável pelo controle de uma interseção.

Ao receber a mensagem de interação  , contendo o interesse \textit{route\_to\_$<$id. da via$>$}, os agentes responsáveis pelo controle de uma interseção tratam esta, executando o Algoritmo \ref{alg:tratamento_route_to}. O algoritmo verifica se a mensagem de interação recebida possui o parâmetro \textit{rota}. Se verdadeiro, o algoritmo verifica se os instantes de tempo relativos a cada componente da rota estão dentro de um limite de tempo tolerável, quando este são comparados com os tempos das alocações de espaços na via. Se verdadeiro, o algoritmo devolve a rota para o agente Veículo requerente. Caso contrário, ele calcula uma nova rota ótima para o requerente. Caso a mensagem não contenha uma rota, o algoritmo calcula uma nova rota ótima para o agente, que é a origem da mensagem de interação recebida. 

Para calcular uma rota ótima, o Algoritmo \ref{alg:calculo_rota_otima} é executado. O algoritmo inicia, extraindo os parâmetros \textit{destino} e \textit{posicaoGPS} da mensagem recebida. Em seguida, ele inicia uma lista cuja finalizada é armazenar os seguimentos de via componentes da rota ótima (\textit{rota}). Além desta estrutura de dados, também é inicializado um mapa para associar as vias e os instantes de tempo em que o veículo conectado atravessará as mesmas (\textit{instantes}). Após isto, o algoritmo calcula o custo inicial para o veículo conectado atravessar a via em que ele e o agente responsável pelo controle de interseção estão localizadas. Nos próximos passos, o algoritmo obtém as os identificadores das faixas da via (\textit{faixas}) e o conjunto de vias controladas por sinalizações semafóricas (\textit{viasControladas}). Além disto, ele inicializa um acumulador de tempo de viagem. Com base na inicialização das variáveis descritas até aqui, o algoritmo inicia o cálculo de rotas ótimas, tomando como início as faixas da via em que o veículo conectado está trafegando. Dessa forma, para cada faixa da via, o algoritmo calcula uma rota ótima até o destino. Este cálculo tem como base o algoritmo de caminho mais curto de Dijkstra, que é combinado com as premissas das heurísticas de despacho EDD (\textit{Earliest Due Date}) e SPT (\textit{Short Processing Time}). Para cada rota ótima calculada, o algoritmo a compara com outra rota anteriormente calculada pelo mesmo, caso esta última exista, a fim de encontrar a melhor rota ótima dentre as que foram calculadas. Com base na melhor rota ótima, o algoritmo aloca espaços nas vias nas agendas de intervalos de indicações de luzes verdes. Neste processo de alocação, se existir uma rota anteriormente alocada, ela é removida e, em seguida, dá lugar à alocação nova rota. 

\subsection{Alocando Espaços nas Vias}

Para notificar todos os responsáveis pelo controle de interseções, uma mensagem de interação, contendo o interesse \textit{roadway\_space\_allocation}, é enviada para tais agentes. Esta mensagem é configurada de modo que todos os agentes responsáveis pelos controles de interseções a recebam por meio de uma interface de acesso à comunicação baseada no padrão LTE. Além disto, esta mensagem é parametrizada com o identificador do agente veículo, rota e os instantes de tempos relativos às travessias de cada componente da rota. Ao receber tal mensagem, os agentes executam o procedimento de alocação temporário de rotas descrito anteriormente. 

Sempre que o Algoritmo \ref{alg:tratamento_route_to}, envia uma rota ótima para um agente Veículo, esta está contida em uma mensagem de interação cujo interesse é \textit{calculated\_route\_to$<$id. da via$>$}. O envio desta mensagem é detalhado no Algoritmo \ref{alg:envio_calculated_route_to}. Se a mensagem de interação   foi recebida por meio de uma interface de acesso à comunicação baseada no padrão IEEE 802.11 ou no IEEE 802.11p, ela deve ser enviada para sua origem, utilizando a mesma tecnologia de acesso à comunicação. Além disto, a mensagem deve ser enviada para trás, de modo que ela chegue a até a origem. Porém, se a mensagem de interação foi recebida por meio de interface de acesso à comunicação baseada no padrão LTE, esta deve receber mais um parâmetro, que é o identificador da origem. Este parâmetro é necessário, pois, quando o agente Centro de Controle de Tráfego receber a mensagem, ele poderá enviá-la diretamente para o agente Veículo requerente do cálculo de rota ótima.

Ao receber a mensagem de interação, contendo o interesse \textit{roadway\_space\_allocation}, o agente Veículo passa a ter conhecimento de uma rota e esta, por sua vez, deve ser mantida pelo mesmo até que uma nova requisição de cálculo de rota seja feito pelo agente. Além disto, a variável $possuiRotaCalculada_i$ tem seu valor alterado para \textbf{verdadeiro} (veja Algoritmo \ref{alg:solicitacao_calculo_rotas}). 

Por fim, como citado acima, os agentes Veículo requisitam novos cálculos de rotas ótimas, quando recebem mensagens de interação cujo interesse é \textit{new\_traffic\_light\_schedule}. Ao receber tal mensagem, o agente apenas atribui \textit{falso} à váriável de controle $possuiRotaCalculada_i$, tendo em vista que a rota alocada para o mesmo foi excluída em todas as agendas de intervalos de indicações de luzes verdes. Além disso, os agentes Veículos requisitam novos cálculos de rotas ótimas, quando seus veículos conectados atravessam as interseções. Neste caso, os agentes Veículo tomam como referência a via em que o seu veículo conectado se encontra.

\subsection{Agendas de Intervalos de Indicações de Luzes Verdes e o Controle de Interseções com Veículos Conectados}

No Capítulo \ref{cap:controle}, foi apresentada uma proposta de controle de interseções por meio de veículos conectados, quando as sinalizações destas apresentarem uma ausência de funcionamento. Nesta proposta, um veículo conectado é escolhido como o líder na via onde ele e outros veículos estão trafegando. Com base nisto, o agente deste veículo conectado assume a responsabilidade de controlar a interseção durante um intervalo de tempo. Após a sinalização semafórica virtual indicar luz verde, tal responsabilidade é transferido para outro agente Veículo, que, por sua vez, está em um veículo conectado cuja via está recebendo luz vermelha da sinalização virtual. Quando isto acontece, os agentes veículo conectado interagem, compartilhando os dados de controle da interseção, de modo que o próximo agente responsável pelo controle da interseção possa mantê-los, enquanto a sinalização semafórica virtual estiver indicando luz vermelha. Juntamente com os dados de controle da interseção, o agente veículo conectado também compartilha as agendas de intervalos de indicações de luzes verdes mantidas por ele. Desta forma, as agendas de intervalos de luzes verdes podem ser mantidas, sem necessariamente ter uma um agente Sinalização Semafórica responsável em mantê-las.

\subsection{Veículos Conectados Cientes de Intervalos de Indicações de Luzes Verdes}

Uma vez que os agentes responsáveis pelo controle de uma interseção são cientes dos estados de controle da interseção, eles podem tornar os agentes Veículos cientes dos estados dos intervalos de indicações de luzes verdes. Em outras palavras, os agentes Veículos podem saber quanto tempo falta para terminar ou iniciar um intervalo de indicação de luz verde na via em que ele está. Com esta informação, os agentes Veículo podem orientar os motoristas de seus veículos conectados, quanto a tomada de decisão, no que diz respeito a aumentar ou diminuir a velocidade destes, a fim de alcançar a janela de tempo definida por um intervalo de indicação de luz verde para uma via. Segundo \citet{Paiva:2012}, tal janela de tempo define uma onda verde.

Para representar esta onda verde, os agentes responsáveis pelo controle de uma interseção usam os seguintes dados: posições geográficas do início das ondas verdes (posição do início das áreas de monitoramento de tráfego) (\textit{posInicioOndasVerdes$_i$}), sinais de $x$ (sinaisX$_i$), sinais de $y$ (sinaisY$_i$) e ângulos das vias (angulosVias$_i$).  Com isto, os agentes escalonam ações periodicamente, de acordo com um parâmetro de periodicidade de atualização das ondas verdes cujo valor é 1 ms. Estas ações executam o Algoritmo \ref{alg:calculo_avancos_ondas_verdes}. De acordo com o algoritmo, as posições de início são atualizadas, bem como, a velocidade de deslocamento das ondas verdes (\textit{velocidadeOndasVerdes$_i$}), tempo para os próximos intervalos de indicações de luzes verdes (tempoParaProxIntVerde$_i$) e a duração da onda verde (\textit{tempoDurOndaVerde$_i$}).

Em embora os agentes responsáveis pelos controles de interseções escalonem periodicamente ações executoras do Algoritmo \ref{alg:calculo_avancos_ondas_verdes}, isto não é necessário para manter as atualizações descritas acima. Para tanto, sempre que a interseção muda de fase, ou seja, todas as sinalizações semafóricas mudam suas indicações em seus grupos focais, as ondas verdes são reposicionadas. À medida que as interseções mudam de fase, a variável de controle da fase em cada agente responsável pelos controles de interseções é incrementada ($fase_i$). Sendo mais específico, esta variável é incrementada, após a troca das indicações verdes e vermelhas nos grupos focais das sinalizações semafóricas da interseção, sejam estas reais ou virtuais. Esta variável de controle é de grande importância, pois ela é utilizada para acessar os tamanhos dos intervalos na lista de tamanhos de intervalos mantida pelo agente (\textit{tamanhosIntervalos$_i$}). Sempre que os tamanhos dos intervalos de indicações de luzes verdes de uma interseção são modificados em função das flutuações de tráfego, demanda a geração de uma nova agenda de intervalos de indicações de luzes verde na interseção, conforme mencionado antes. Quando isto acontece, a variável $fase_i$ é zerada em todos os agentes responsáveis pelo controle da interseção, a fim de que estes agentes acessem corretamente as estradas na nova agenda de intervalos de indicações de luzes verdes. Além disso, sempre que uma sinalização semafórica passa a indicar luz verde em seu grupo focal, o início da onda verde é posicionado no fim da via de entrada, especificamente, sobre a faixa de contenção da via. Por outro lado, sempre que uma sinalização semafórica passa a indicar luz vermelha em seu grupo focal, o início da onda verde é posicionado na posição geográfica do fim da área de monitoramento de tráfego da via. 

Para que um agente Veículo se torne ciente do intervalo de indicação de luz verde na via em que seu veículo conectado está trafegando, ele interagem com o agente responsável pelo controle da interseção, enviando uma mensagem de interação cujo interesse é \textit{green\_wave\_request}. Além deste interesse, a mensagem é configurada da seguinte forma: o campo destino é nulo; o campo via recebe o identificador da via onde o veículo conectado está trafegando; o campo direção recebe o valor um, pois a mensagem precisa ser enviada para frente do veículo, de modo que ela alcance o agente responsável pelo controle da interação; a tecnologia da interface de acesso à comunicação é IEEE 802.11. Essa mensagem de interação sempre é enviada, após o veículo conectado atravessar uma interseção e, em seguida, entrar em uma nova via.

Quando o agente reponsável pelo controle da interseção recebe uma mensagem de interação, contendo o interesse \textit{green\_wave\_request}, o agente responde com uma mensagem de interação, contendo o interesse \textit{green\_wave}. Esta mensagem tem como destino o agente Veículo, que é a origem da mensagem recebida. Além disto, a mensagem tem seu campo direção configurado com o valor -1, pois ela precisa ser encaminhada para trás dos veículos conectados. O campo via é configurado com o valor do campo via da mensagem recebida. A a tecnologia da interface de acesso à comunicação é IEEE 802.11. A mensagem é parametrizada com x parâmetros, são eles: posicao do início da onda verde ($posInicioOndaVerde$), velocidade da onda verde ($velocidadeOndaVerde$), tempo de duração da onda verde ($tempoDurOndaVerde$) e tempo para o próximo intervalo de verde ($tempoParaProxIntervaloVerde$). Após ter sido totalmente configurada, a mensagem de interação é enviada para a origem da mensagem recebida.

Por fim, ao receber uma mensagem de interação cujo o interesse é \textit{green\_wave}, o agente Veículo toma ciência do intervalo de indicação de luz verde da via. Após isto, o agente passa a executar a proposta de \citet{Faria:2013}. Para que a proposta de \citet{Faria:2013} funcione corretamente, é necessário que o agente Veículo tenha conhecimento dos veículos conectados próximos ao dele, de modo que o agente possa saber a velocidade do veículo conectado imediatamente à frente do seu. 

Para tanto, os agentes Veículo interagem uns com os outros periodicamente por meio do envio de mensagens de interação cujo o interesse é \textit{hello}, levando em consideração o valor de frequência de envio destas mensagens. O valor utilizado nesta tese é 10 Hz (10 mensagens por segundo), de acordo com \citet{Zheng:2015}. Estas mensagens são limitadas à vizinhança dos veículos conectados, conforme os dados utilizados no registro do interesse \textit{hello}. Além disto, estas mensagens são enviadas por meio de uma interface de acesso à comunicação baseada no padrão IEEE 802.11, contendo a velocidade e o identificador do agente Veículo como parâmetros. Ao receber a mensagem, os agentes Veículos atualizam suas bases de conhecimento acerca dos veículos conectados ao redor daquele que o embute. Utilizando esta base de conhecimento, os agentes extraem os veículos conectados imediatamente à frente dos seus e as velocidades dos mesmos.  

\section{Considerações Finais}

Este capítulo apresentou a estratégia, proposta por esta tese, para tratar o problema de planejamento e orientação de rotas, bem como, os detalhamentos acerca de sua implementação sobre uma rede \textit{ad hoc} veicular heterogênea centrada em interesses. Tal estratégia é capaz de criar agendas globais de intervalos  de indicações de intervalos de luzes verdes. Estes intervalos indicam a disponibilidade das vias de entrada das interseções de uma rede viária. Modelando as redes viárias como um Sistema Flexível de Manufatura do tipo \textit{Job-Shop} foi possível criar um algoritmo de roteamento de veículos conectados com base em regras de despacho SPT (\textit{Short Processing Time}) e EDD (\textit{Earliest Due Date}). Este algoritmo de escalonamento tira proveito das alocações de espaços nas vias, que são realizadas, à medida que as rotas ótimas para cada veículo conectado são inseridas no sistema de planejamento e orientação de rotas. Uma vez que este sistema usa como base o sistema de controle de tráfego, que escalona o acesso às interseções de uma rede viária, o algoritmo de roteamento de veículos conectados não precisa se preocupar com possíveis colisões de veículos. 

Por fim, prosseguindo com o detalhamento desta tese, o capítulo seguinte apresentará as avaliações experimentais das propostas apresentadas nesta tese, assim como, a análise em torno dos dados obtidos por meio dos experimentos.

  \chapter{Avaliação Experimental e Resultados}

Este capítulo tem como objetivo de apresentar as avaliações experimentais relativas às propostas apresentadas por esta tese.

\section{Metodologia Utilizada}

Para se obter os resultados apresentados neste capítulo, foi necessário o uso de ferramentas de software capazes de fornecer subsídios para o desenvolvimento de modelos de redes \textit{ad hoc} veiculares e cenários de uso relacionados às propostas apresentadas anteriormente nos Capítulos \ref{cap:radnet-ve}, \ref{cap:controle} e \ref{cap:orientacao}. Neste sentido, fez-se o uso de três \textit{frameworks}: Veins \cite{Veins:2013}, Inet \cite{Omnet:2013} e SimuLTE \cite{SimuLTE:2016}. Estes \textit{frameworks} permitiram a implementação tanto de modelos de redes \textit{ad hoc} veiculares quanto cenários de uso para avaliação das propostas desta tese. Para executar as simulações, este frameworks têm como base dois simuladores: Omnet++ \cite{Omnet:2013} e SUMO \cite{SUMO:2013}. O Omnet++ é um simulador de redes baseado em eventos discretos. O SUMO é um simulador que permite tanto a criação e importação de mapas rodoviários, assim como, a definição, configuração e simulação do tráfego de veículos em uma rede viária.

Com base nos \textit{frameworks} citados acima, foram desenvolvidos os protocolos de comunicação não somente da RAdNet-VE e da HRAdNet-VE, mas também da RAdNet \cite{Dutra:2012} e de dois modelos básicos de CCN, em que um deles é uma CCN baseada em roteamento reativo de dados (CCN$_R$) \cite{Amadeo:2013} o outro é uma CCN baseada em roteamento proativo de dados (CCN$_P$) \cite{Wang:2012b}. Vale ressaltar que estas CCNs, nesta tese, não fornecem serviços de ache de dados, pois os dados das aplicações de sistemas inteligentes propostas nesta tese são sensíveis a atrasos \cite{Yu:2013}. Esses protocolos de comunicação foram desenvolvidos com o intuito de comparar seus desempenhos com os dos protocolos de comunicação da RAdNet-VE e da HRadNet-VE. Vale ressaltar que, todos esses protocolos operam no nível da camada de rede. Além disso, considera-se que em todos os nós as aplicações podem acessar coordenadas geográficas obtidas por meio de dispositivos GPS e ter acesso a bancos de dados de mapas. Nos experimentos envolvendo a RAdNet, RAdNet-VE e ou HRAdNet-VE, os prefixos ativos foram compostos de oito campos com oito possibilidades. Por fim, para identificar os nós em experimentos envolvendo as CCNs, foi utilizado mecanismo de identificação de nós fornecido pelo Omnet++.

Para medir o desempenho dos protocolos de comunicação acima, foram utilizadas as seguintes métricas:

\begin{itemize}
	\item \textbf{Custo de Mensagens Trafegadas (CMT): } é o total de mensagens recebidas pelos nós (inclusos os nós destino e aqueles que as encaminharam);
    \item \textbf{Latência de Comunicação entre Nós (LCN): } é o tempo entre o envio de uma mensagem da camada de rede por um nó origem até a recepção da mensagem pela camada rede de um nó vizinho;  
    \item \textbf{Taxa de Entrega de Dados (TED): } total de mensagens de dados recebidas  dividido pelo total de mensagens dados enviadas;
    \item \textbf{Número de Saltos (NS): } consiste na número de vezes que as mensagens  foram encaminhadas pelos nós;
    \item \textbf{Alcance das Mensagens (AM): } consiste na distância que as mensagens percorreram, à medida que foram encaminhadas pelos nós;
    \item \textbf{Tempo de propagação de mensagens (TPM):} tempo gasto para uma mensagem alcançar uma determinada distância.
\end{itemize}

Além dos protocolos de comunicação, também foram desenvolvidos os modelos de agentes no nível da camada de aplicação, de modo que fosse possível simular o comportamento das aplicações de sistemas inteligentes de transporte propostas anteriormente nos Capítulos \ref{cap:controle} e \ref{cap:orientacao}, que são o sistema multiagente para controle de tráfego e o sistema multiagente para planejamento e orientação de rotas, respectivamente.

Para avaliar o desempenho do sistema multiagente para controle de tráfego, foi desenvolvido um modelo de sinalização semafórica pré-temporizada, de modo que este pudesse ter seu desempenho comparado com as estratégias adotadas pelo sistema multiagente para controle de tráfego. Para tanto, as seguintes métricas foram utilizadas:

\begin{itemize}
	\item \textbf{Taxa de vazão do sistema (TVS):} número de veículos que alcançaram seus destinos por hora;
	\item \textbf{Tempo médio de espera (TME):} a quantidade máxima de tempo que um veículo fica parado;
    \item \textbf{Tempo médio de viagem (TMV):} média dos tempos que os veículos gastam durante o percurso entre a origem de sua viagem e o destino da mesma;
    \item \textbf{Velocidade média (VM):} velocidade média dos veículos, à medida que viajam pela rede viária até alcançarem o destino de suas viagens.
\end{itemize}

No que diz respeito à avaliação do desempenho do sistema multiagente para planejamento e orientação de rotas, foram desenvolvidos algoritmos de escalonamento de veículos, a fim de simular as escolhas de caminhos realizadas pelos motoristas (Caminho Espacialmente mais Curto) e planejamento e orientação de rotas baseado nos sistemas atuais navegação (Caminho Temporalmente mais Curto). Para realizar as comparações de desempenho, as seguintes métricas foram utilizadas:

\begin{itemize}
	\item \textbf{Taxa de vazão do sistema (TVS):} número de veículos que alcançaram seus destinos por hora;
	\item \textbf{Tempo médio de espera (TME):} a quantidade de tempo que um veículo fica parado;
	\item \textbf{Tempo médio de viagem (TMV):} tempo que os veículos gastam durante o percurso entre a origem de sua viagem e o destino da mesma;
    \item \textbf{Velocidade média (VM):} velocidade média dos veículos, à medida que viajam pela rede viária até alcançarem o destino de suas viagens;
    
    \item \textbf{Consumo médio de combustível:} a quantidade de combustível gastou ao realizar uma viagem.
    \item \textbf{Emissões de CO (Monóxido de Carbono) (CO):} total de emissões de monóxido de carbono.
    \item \textbf{Emissões de CO$_2$ (Dióxido de Carbono) (CO$_2$):} total de emissões de dióxido de carbono.
    \item \textbf{Emissões de HC (Hidrocarbonetos) (HC):} total de emissões de hidrocarbonetos.
    \item \textbf{Emissões de NOx (Óxidos de Nitrogênio) (NOx):} total de emissões de óxidos de nitrogênio.
    \item \textbf{Emissões de PMx (Material Particulado)(PMx):} total de emissões de material particulado.
\end{itemize}

Uma vez que as simulações tiveram seus tempos de duração fixados em 3600 s, os valores relativos às métricas acima foram acumulados e calculados a cada 100 s. Desta forma, foi possível extrair os resultados, que serão apresentados nas Seções \ref{sec:radnetve}, \ref{sec:hradnetve}, \ref{sec:trafego} e \ref{sec:rotas}. Tais resultados são apresentados na forma de valores totais e médios. No que diz respeito aos valores totais, estes são os acumulados dos valores totais de cada intervalo de medição. Acerca dos valores médios, estes são as médias dos valores médios de cada intervalo de medição. Para os valores médios, foi adotado um intervalo de confiança de 95\%.  

\section{Ambiente Computacional para Realização dos Experimentos}

Embora o Omnet e seus \textit{frameworks} sejam amplamente utilizados para realizar simulações de redes veiculares, não é possível realizá-las em ambientes paralelos. Portanto, os seguintes equipamentos foram utilizados durante o período de realização de experimentos:

\begin{itemize}
	\item Desktop com Processador Intel$_\copyright$  Core i7-3770K CPU 3,5 GHz com quatro núcleos de processamento e 16 GB de memória RAM;
    \item Notebook Dell XPS com Processador Intel$_\copyright$  Core i7-2670QM CPU 2,2 GHz com quatro núcleos de processamento e 16 GB de memória RAM;
    \item Notebook Dell Inspiron com Processador Intel$_\copyright$  Core i7-6500U CPU 2,5GHz com dois núcleos de processamento e 16 GB de memória.
\end{itemize}

Nestes equipamentos, foi instalado o Linux Mint 17.1 Rebecca, pois não foi preciso instalar emuladores para execução do Omnet++. Além disto, é importante pontuar sobre a capacidade de armazenamento dos equipamentos. Nestes equipamentos, foram utilizados discos rígidos de 1TB. Embora o Omnet++ seja capaz de gerar arquivos, contendo os resultados das simulações, tais arquivos ocupam bastante espaço em disco,  podendo ultrapassar 30 GB de tamanho em experimentos com redes veiculares heterogêneas.

\section{Avaliando a Rede Ad Hoc Veicular Centrada em Interesses}\label{sec:radnetve}

Nesta seção, são apresentados os resultados obtidos nos primeiros experimentos com a RAdNet-VE e publicados em \citet{Goncalves:2016b}. Antes da apresentação dos resultados, é feita uma descrição dos cenários para os experimentos, utilizando as redes RAdNet-VE, RAdNet, CCN$_R$ e CCN$_P$. Em seguida, são apresentadas as configurações adotadas para os experimentos. Finalmente, são apresentados os resultados dos experimentos de acordo com os cenários e a análise comparativa destes, utilizando as métricas definidas anteriormente para comparar o desempenho dos protocolos de comunicação das redes citadas acima.

\subsection{Descrição dos Cenários para os Experimentos}

Para obtenção dos resultados da RAdNet-VE, foram criados dos cenários: cooperação entre veículos e sinalizações semafóricas e cooperação entre veículos.

Para o primeiro cenário, foram desenvolvidas suas aplicações, sendo elas um controlador de sinalizações semafóricas e um assistente de direção. O controlador de sinalizações semafóricas executou em cada uma das sinalizações semafóricas do mapa viário produzido para este cenário. Além disto, ela foi projetada para coletar dados de fluxos de tráfego nas vias de entrada de interseções e ajustar os intervalos de indicações de luzes verdes em função das flutuações de tráfego de tais vias. No que diz respeito ao assistente de direção, este é uma aplicação que foi projetada para executar nos veículos conectados e cooperar com instâncias do controlador de sinalizações semafóricas, enviando dados relacionados a entrada e saída do veículo de uma via de entrada de uma interseção. 

Além disso, para este cenário, foi criada uma grade manhattan 3 x 3 em que a distância entre quaisquer duas das dezesseis interseções foi de 300 m. Cada via que leva a uma interseção teve como velocidade máxima permitida 60 Km/h. Em cada interseção, foram instaladas sinalizações semafóricas equipadas com uma interface de acesso à comunicação sem fio. Desta forma, veículos conectados e sinalizações semafóricas puderam requisitar dados uns dos outros por meio das aplicações instaladas nos mesmos. 

Para obter dados relativos aos fluxos de tráfego das vias de entradas das interseções, os controladores das sinalizações semafóricas que as controlam requisitaram dados sobre a presença dos veículos conectados nessas vias. Ao receber as requisições, os assistentes de direção nos veículos conectados as respondem, enviando o identificador do veículo. Ao receber as respostas enviadas pelos assistentes de direção, os controladores de sinalizações registram o identificador do veículo em estruturas de dados apropriadas, que são conjuntos de identificadores de veículos conectados. Além disso, os controladores de sinalizações também requisitam dados relativos a partida de veículos conectados das interseções. Portanto, para conhecer o número de veículos que deixaram as interseções, os controladores requisitam dados aos veículos que deixaram imediatamente uma interseção. Ao receberem as requisições, os assistentes de direção nos veículos conectados as respondem, enviando o identificador do veículo. Ao receberem estas respostas, os controladores de sinalizações semafóricas registram os identificadores dos veículos em conjuntos de identificadores de veículos. De acordo com \citet{Goncalves:2016b}, a periodicidade de envio de requisições por dados de veículos entrando e deixando vias de entradas de interseções foi de 1s.

Para o segundo cenário, foram desenvolvidas duas aplicações, sendo elas um controle adaptativo e cooperativo de cruzeiro e um notificador de obstáculos. O controle adaptativo e cooperativo de cruzeiro executou em cada um dos veículos conectados, controlando as velocidades dos mesmos de acordo com os dados microscópicos dos veículos conectados vizinhos. O notificador de obstáculos executou em uma unidade de acostamento, notificando os veículos conectados sobre a presença de um obstáculo na estrada. 

Além disso, para esse cenário, foi criada um segmento de estrada e instalado um notificador de obstáculo em uma unidade de acostamento próxima ao final deste segmento. O objetivo disto é notificar os veículos conectados sobre a presença de um obstáculo ao final do segmento de estrada. Além disto, a unidade de acostamento foi posicionada, de modo que os motoristas dos veículos fossem notificados, quando seus veículos estivessem a 1 km de distância do obstáculo. Neste cenário, cada veículo conectado viajou a 80 km/h até alcançar a região do obstáculo. Ao atravessar tal região, os veículos conectados reduziram suas velocidades para 20 km/h.

Para receber dados sobre obstáculos na estrada, os controles adaptativos e cooperativos de cruzeiro nos veículos conectados enviaram requisições ao notificador de obstáculos. Ao receber estas requisições, o notificador de obstáculos as respondem, enviando os dados relativos ao obstáculo, tal como a posição geográfica e a velocidade máxima permitida na região do obstáculo. A receber tais dados, o controle adaptativo e cooperativo de cruzeiro iniciou o processo de descoberta de vizinhança.

Durante o processo de descoberta de vizinhança, os controles adaptativos e cooperativos de cruzeiro requisitaram dados relativos aos veículos dentro do alcance de comunicação da interface de acesso à comunicação sem fio. Ao receberem estas requisições, os controladores adaptativos e cooperativos de cruzeiro as respondem, enviando a posição geográfica e o identificador de seus veículos conectados. Ao receberem estas respostas, os controles adaptativos e cooperativos de cruzeiro as posições geográficas e identificadores dos veículos e os registram em estruturas de dados apropriadas. Uma vez iniciado o processo de descoberta de vizinhança, os controles adaptativos e cooperativos de cruzeiro requisitaram dados sobre os veículos conectados a cada 1 s.

Os controladores adaptativos e cooperativos de cruzeiro, ao identificarem os veículos conectados imediatamente à frente dos veículos que os executam, requisitam dados acerca destes veículos. Ao receber estas requisições, os controles adaptativos e cooperativos de cruzeiro as respondem, enviando as velocidades e posições geográficas de seus veículos conectados. Ao receberem estas respostas, os controles adaptativos e cooperativos de cruzeiro extraíram os dados e calcularam novas velocidades para seus veículos conectados, de acordo com o trabalho de \citet{Kato:2002}. Esse processo acontece a cada 0.1 s, ou seja, dez vezes por segundo.

\subsection{Configurações dos Experimentos}

Os tempos de simulação dos cenários foram fixados em 3600s. No primeiro cenário, os veículos conectados e sinalizações semafóricas foram equipados com interfaces de acesso à comunicação sem fio baseadas no padrão IEEE 802.11n. De acordo com \citet{Li:2007}, o padrão IEEE 802.11 tem sido utilizado em interseções para capturar dados de tráfego ou dados que podem auxiliar no roteamento e ou encaminhamento de mensagens ao longo de uma rede \textit{ad hoc} veicular. Os parâmetros da camada física das interfaces de acesso à comunicação sem fio são listados na Tabela \ref{tab:config_IEEE_80211n}. Para configurar a camada de enlace, o protocolo de controle acesso ao meio CSMA/CA (\textit{Carrier Sense Multiple Access with Collision Avoidance}) foi configurado de acordo com os parâmetros listados na Tabela \ref{tab:config_MAC_IEEE_80211n}.

No cenário dois, os veículos conectados e a unidade de acostamento foram equipados com interfaces de acesso à comunicação sem fio baseada no modelo IEEE 802.11p. Segundo \citet{Ploeg:2013}, o IEEE 802.11p é o padrão mais adequado para ambientes em que os veículos viajam em alta velocidade. Os parâmetros relativos à camada física das interfaces de acesso à comunicação sem fio são listados na Tabela \ref{tab:config_IEEE_80211p}.

Para identificar os dados nas CCN, foi adotada a seguinte estrutura de nomes: \textit{uri://type of provider/geolocation/application/data service name}. Esta estrutura de nomes baseia-se no trabalho de \citet{Wang:2012a}. O componente \textit{type of provider} define a entidade fornecedora do serviço, por exemplo, veículo conectado, sinalização semafórica, sinalizações verticais em rodovias, entre outros. O componente \textit{geolocation} usa o formado \textit{roadId/direction/section} definido por \citet{Wang:2012a}. As aplicações usaram o componente \textit{geolocation} para filtrar as mensagens de com a posição dos nós. O componente \textit{application} possui os dados fornecidos pelo nó. O componente \textit{data service name} é o nome de um determinado serviço. Dessa forma, as aplicações desenvolvidas para os cenários foram identificadas conforme os a listagem apresentada pela Tabela \ref{tab:aplicacoes_uri}. De acordo com o modelo de uma CCN, as aplicações desenvolvidas para os cenários são fornecedoras de dados. Estes dados são fornecidos por meio de um nome de serviço de dados criado com base na estrutura de nomes definida acima. Portanto, as aplicações fornecem dados por meio dos nomes de serviços de dados listados na Tabela \ref{tab:nomes_dados_CCNs}.

\begin{table}[!h]
    \centering
    \caption{Identificação das aplicações desenvoldidas para os cenários com base na estrutura de nomes defina para as CCNs.}
    \label{tab:aplicacoes_uri}
    \small \begin{tabular}{|p{6.5cm}|l|}
      \hline
      \textbf{Aplicação} & \textbf{Identificação da Aplicação} \\ \hline
      \centering Assistente de Direção (AD) & \textit{uri://vehicle/geolocation/da/} \\ \hline
      \centering Controlador de Sinalizações Semafóricas (CSS) & \multirow{2}{*}{\textit{uri://semaphore/geolocation/tlc/}} \\ \hline
      \centering Notificador de Obstáculos (NO) & \textit{uri://roadsidesig/geolocation/on/} \\ \hline
      \centering Controle Adaptativo e Cooperativo de Cruzeiro (CACC) & \multirow{2}{*}{\textit{uri://vehicle/geolocation/ccac/}} \\ \hline
    \end{tabular}
\end{table}

Tanto a RAdNet quanto a RAdNet-VE usam somente um tipo de mensagem de rede e ambas as redes foram projetadas com base no modelo de arquitetural Publisher/Subscriber. Portanto, para comparar o desempenho destas redes contra os das CCNs, foi necessário criar dois interesses para cada nome de serviço de dados listado na Tabela \ref{tab:nomes_dados_CCNs}. Com isto, nos experimentos com as redes RAdNet-VE e RAdNet, messagens com interesses terminados com a palavra \textit{req} atuaram como pacotes de Interesse e as mensagens com interesses terminados com a palavra \textit{data} atuaram como pacote de Dados. A Tabela \ref{tab:nomes_dados_interesses} lista os serviços de dados e os interesses relativos a tais serviços. Vale ressaltar que os interesses não foram definidos com base na estrutura de nomes para as CCNs. Neste caso, os interesses foram definidos apenas como termos. Com base nos interesses listados na Tabela \ref{tab:nomes_dados_interesses}, foram definidos seus números máximos de saltos, bem como, as direções de propagação das mensagens de rede que os contiveram. Estas definições foram exclusivas do protocolo de comunicação da RAdNet-VE e são listadas juntamente com seus interesses na Tabela \ref{tab:interesses_saltos_direcoes}. Por fim, no que diz respeito aos números máximos de saltos adotados nos experimentos com a RAdNet, CCN$_R$ e CCN$_P$, foram adotados um número máximo de cinco saltos para o primeiro cenário e um número máximo de 50 saltos para segundo cenário. Além disto, foi também definido um número máximo de saltos padrão para o protocolo de comunicação da RAdNet-VE, sendo este igual a 50 saltos.

Para finalizar as configurações das simulações, foram definidos os valores dos parâmetros das configurações de tráfego e de comportamento dos veículos. No primeiro cenário, a grade manhattan 3 x 3 recebeu em cada uma das suas oito entradas 1500 veículos/hora. No que diz respeito às configurações dos intervalos de tempo dos 32 semáforos presentes neste cenário, foram as seguintes: 40s para intervalos de verde, 5s para intervalos de amarelo e 1s para intervalos de vermelho total. No segundo cenário, o segmento de estrada recebeu 1500 veículos/hora. Além disto, a unidade de acostamento foi posicionada 500m antes do obstáculo. A área relativa ao obstáculo teve 10m de comprimento e velocidade máxima permitida de 20 Km/h. Para configurar o comportamento dos veículos, foi adotado \textit{Intelligent Driver Modelo} (IDM) \cite{Treiber:2000}. Os valores dos parâmetros em cada um dos cenários são listados na Tabela \ref{tab:configuracao_IDM_RAdNetVE}.

\begin{table}[!t]
    \centering
    \caption{Configurações do IDM para os dois cenários.}
    \label{tab:configuracao_IDM_RAdNetVE}
    \small \begin{tabular}{|l|c|c|}
      \hline
      \textbf{Parâmetro} & \textbf{Cenário 1} & \textbf{Cenário 2} \\ \hline
      Velocidade desejada (v$_0$) & 60 Km/h & 80 Km/h \\ \hline
      Tempo de reação do motorista (T) & 1.2s & 1.2s \\ \hline
      Espaçamento mínimo entre veículos (s$_0$) & 2.0m &2.0m \\ \hline
      Aceleração (a) & 1 m/s$^2$ & 1 m/s$^2$ \\ \hline
      Desaceleração (b) & 3 m/s$^2$ & 3 m/s$^2$ \\ \hline
    \end{tabular}
\end{table}

\subsection{Análise dos Resultados}

A seguir, a Tabela \ref{tab:cenario1} apresenta os resultados dos experimentos do cenário 1 e, em seguida, a Tabela \ref{tab:comp_cenario1} apresenta a análise dos mesmos.

\begin{table}[!h]
\centering
\caption{Resultados dos experimentos do cenário 1}
\label{tab:cenario1}
\small\begin{tabular}{|l|c|c|c|c|}
\hline
\multicolumn{1}{|c|}{\textbf{Métricas}} & \textbf{RAdNet-VE} & \textbf{RAdNet}    & \textbf{CCN$_R$}   & \textbf{CCN$_R$}    \\ \hline
\textbf{CMT (msgs.)}                    & 2,19 $\times 10^5$ & 1,12 $\times 10^6$ & 5,44 $\times 10^6$ & 1,49 $\times 10^7$  \\ \hline
\textbf{LCN (ms)}                       & 20,4 $\pm 0,05$    & 29,92 $\pm 0,05$   & 26,24 $\pm 0,06$   & 46,69 $\pm 0,08$    \\ \hline
\textbf{TED (\%)}                       & 83,25 $\pm 0,25$   & 79,19 $\pm 0,65$   & 84,9 $\pm 0,54$    & 73,09 $\pm 1,57$    \\ \hline
\textbf{NS (saltos)}                    & 4                  & 5                  & 4                  & 4                   \\ \hline
\textbf{AM (m)}                         & 305,92 $\pm 79,58$ & 576,39 $\pm 113,6$ & 293,99 $\pm 42,87$ & 429,39 $\pm 110,67$ \\ \hline
\textbf{TPM (ms)}                       & 229 $\pm 40,00$     & 231 $\pm 40,00$     & 237 $\pm 30,00$     & 252 $\pm 30,00$      \\ \hline
\end{tabular}
\end{table}

\begin{table}[!h]
\centering
\caption{Análise comparativa entre o desempenho da RAdNet-VE e os das demais redes no cenário 1.}
\label{tab:comp_cenario1}
\small\begin{tabular}{|l|c|c|c|}
\hline
\multicolumn{1}{|c|}{\textbf{Métricas}} & \textbf{RAdNet}     & \textbf{CCN$_R$}    & \textbf{CCN$_R$}    \\ \hline
\textbf{CMT (msgs.)}                    & \textless 80,44\%   & \textless 95,97\%   & \textless 98,53\%   \\ \hline
\textbf{LCN (ms)}                       & \textless 31,81\%   & \textless 22,25\%   & \textless 56,3\%    \\ \hline
\textbf{TED (\%)}                       & \textgreater 5,12\% & \textless 1,11\%    & \textgreater 13,9\% \\ \hline
\textbf{NS (saltos)}                    & \textless 20,00\%   & =                   & =                   \\ \hline
\textbf{AM (m)}                         & \textless 46,92\%   & \textgreater 4,05\% & \textless 28,75\%   \\ \hline
\textbf{TPM (ms)}                       & \textless 0,08\%    & \textless 3,37\%    & \textless 9,12\%    \\ \hline
\end{tabular}
\end{table}

\subsection{Análise dos Resultados do Cenário 2} 

A seguir, a Tabela \ref{tab:cenario1} apresenta os resultados dos experimentos do cenário 2 e, em seguida, a Tabela \ref{tab:comp_cenario1} apresenta a análise dos mesmos.

\begin{table}[!h]
\centering
\caption{Resultados dos experimentos do cenário 2}
\label{tab:cenario2}
\small\begin{tabular}{|l|c|c|c|c|}
\hline
\multicolumn{1}{|c|}{\textbf{Métricas}} & \textbf{RAdNet-VE} & \textbf{RAdNet}     & \textbf{CCN$_R$}    & \textbf{CCN$_R$}    \\ \hline
\textbf{CMT (msgs.)}                    & 7,86 $\times 10^6$ & 1,06 $\times 10^8$  & 2,01 $\times 10^7$  & 1,97 $\times 10^7$  \\ \hline
\textbf{LCN (ms)}                       & 2,87 $\pm 0,1$     & 36,52 $\pm 4,0$     & 20,01 $\pm 1,0$     & 23,04 $\pm 4,0$     \\ \hline
\textbf{TED (\%)}                       & 88,95 $\pm 0,4$    & 16,81 $\pm 0,4$     & 18,21 $\pm 1,0$     & 35,59 $\pm 0,15$    \\ \hline
\textbf{NS (saltos)}                    & 28                 & 30                  & 29                  & 29                  \\ \hline
\textbf{AM (m)}                         & 9957,62 $\pm 7,87$ & 9983,73 $\pm 28,14$ & 9932,73 $\pm 28,14$ & 9937,73 $\pm 37,78$ \\ \hline
\textbf{TPM (ms)}                       & 140 $\pm 30,00$    & 1290 $\pm 210,00$   & 640 $\pm 226,00$    & 730 $\pm 209,00$    \\ \hline
\end{tabular}
\end{table}

\begin{table}[!h]
\centering
\caption{Análise comparativa entre o desempenho da RAdNet-VE e os das demais redes no cenário 2.}
\label{tab:comp_cenario1}
\small\begin{tabular}{|l|c|c|c|}
\hline
\multicolumn{1}{|c|}{\textbf{Métricas}} & \textbf{RAdNet}    & \textbf{CCN$_R$}    & \textbf{CCN$_R$}    \\ \hline
\textbf{CMT (msgs.)}                    & \textless 92,58\%  & \textless 60,89\%   & \textless 60,1\%    \\ \hline
\textbf{LCN (ms)}                       & \textless 92,14\%  & \textless 85,65\%   & \textless 87,54\%   \\ \hline
\textbf{TED (\%)}                       & \textgreater 429\% & \textgreater 388\%  & \textgreater 149\%  \\ \hline
\textbf{NS (saltos)}                    & \textless 6,66\%   & \textless 3,44\%    & \textless 3,44\%    \\ \hline
\textbf{AM (m)}                         & \textless 0,02\%   & \textgreater 0,02\% & \textgreater 0,02\% \\ \hline
\textbf{TPM (ms)}                       & \textless 89,14\%  & \textless 78,12\%   & \textless 80,82\%   \\ \hline
\end{tabular}
\end{table}

\section{Avaliando a Rede  Veicular Heterogênea Centrada em Interesses}\label{sec:hradnetve}

Nesta seção, é apresentada a avaliação experimental da HRadNet-VE. Esta avaliação experimental consiste em dois cenários de avaliação, que foram criados com o intuito de comparar o desempenho da HRAdNet-VE contra o da RAdNet-VE, CCN$_R$ e CCN$_P$. Após a descrição dos cenários de avaliação, são apresentadas as configurações adotadas nos experimentos. Em seguida, são apresentadas as análises comparativas dos resultados obtidos por meio das simulações do primeiro cenário (Cenário 1). Por fim, são apresentadas as análises comparativas dos resultados obtidos por meio das simulações do segundo cenário (Cenário 2).

\subsection{Descrição dos Cenários para os Experimentos}

Para obtenção dos resultados da HRAdNet-VE, foram criados dois cenários: cooperação entre sistema multiagente de controle de tráfego e sistema multiagente de planejamento e orientação de rotas; e cooperação entre o sistema multiagente de controle de tráfego e sistema multiagente de planejamento e orientação de rotas com sinalizações semafóricas apresentando ausência de funcionamento.

Em ambos os cenários, são utilizados os agentes definidos nesta tese. A principal diferença entre os cenários se dá com base dos comportamentos dos agentes Veículo, Sinalização Semafórica e Centro de Controle de Tráfego, quando as sinalizações semafóricas instaladas sobre as interseções da rede viária apresentam ausência de funcionamento. Nesta situação, os agentes passam a exigir mais da rede heterogênea onde estão operando. Portanto, pretende-se avaliar não somente o desempenho da HRAdNet-VE contra os desempenhos da RAdNet, CCN$_R$ e CCN$_P$, mas também avaliar o desempenho da HRAdNet-VE nestes dois cenários. 

\begin{figure}
	\centering
    \subfigure[]{
    	\includegraphics[scale=0.6]{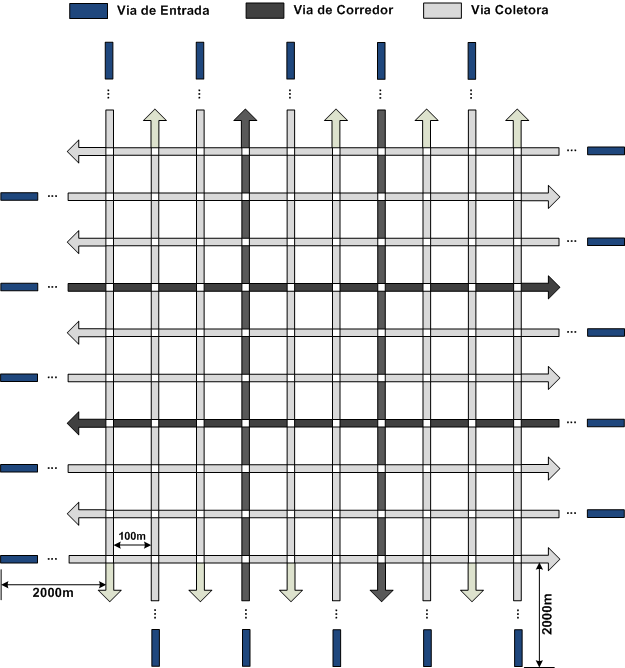}
        \label{fig:grade_1}
    }
    \quad
    \subfigure[]{
    	\includegraphics[scale=0.6]{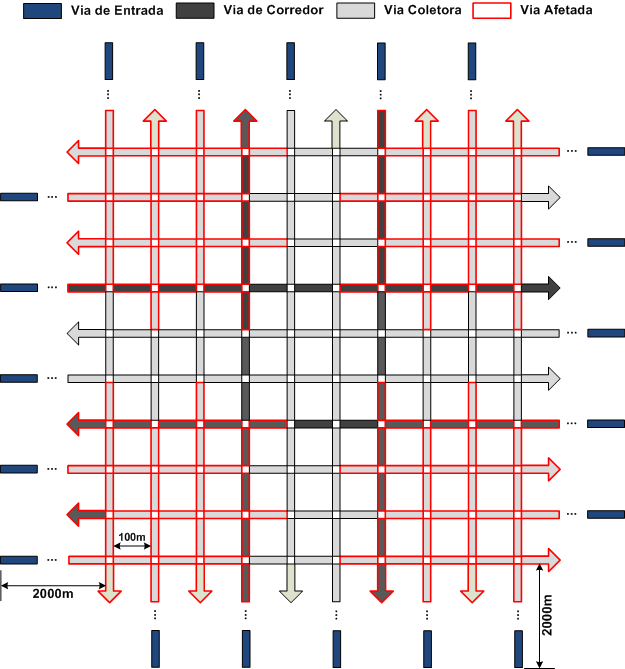}
        \label{fig:grade_2}
    }
    \caption{Grade manhattan 10 x 10: (a) grade utilizada no primeiro cenário; (b) grade utilizada no segundo cenário.}
\end{figure}

Para estes cenários, foi criada uma grade manhattan 10 x 10 em que a distância entre quaisquer das 100 interseções foi de 100m (veja Figura \ref{fig:grade_1}). Nesta grade, cada uma de suas 20 entradas tiveram 2000m de tamanho e velocidade máxima permitida igual a 80 Km/h. Além disto, a grade manhattan foi composta de quatro corredores (veja Figura \ref{fig:grade_1}). A velocidade máxima permitida em cada um dos corredores foi de 60 Km/h. As demais vias da grade tiveram suas velocidades máximas permitidas igual a 40 Km/h.

Por meio desta grade manhattan, foram estabelecidos os cenários citados acima. No primeiro cenário, todas as 200 sinalizações semafóricas não apresentaram ausência de funcionamento, ou seja, elas operaram do início ao fim das simulações. No segundo cenário, foram escolhidas quatro áreas da grade manhattan, a fim de que eles pudessem servir como regiões, onde as sinalizações semafóricas não funcionam, ou seja, apresentam ausência de funcionamento. Tais áreas podem ser vistas na Figura \ref{fig:grade_2}. Sendo assim, as sinalizações semafóricas das regiões escolhidas permaneceram inoperantes do início ao fim das simulações. Por esta razão, as vias de entrada das interseções pertencentes às áreas selecionadas são afetadas, pois nelas os agentes Veículo precisam se comportar com sinalizações semafóricas, a fim de controlar interseções isoladas ou cooperar com as coordenações de sinalizações semafóricas dos corredores.

\subsection{Configurações dos Experimentos}

Os tempos de simulação dos cenários descritos na seção anterior foram fixados em 3600s. No dos dois cenários, os veículos conectados e sinalizações semafóricas foram equipados com interfaces de acesso à comunicação sem fio baseadas no padrão IEEE 802.11n. Estas interfaces foram configuradas de acordo com os parâmetros listados nas Tabelas \ref{tab:config_IEEE_80211n} e \ref{tab:config_MAC_IEEE_80211n}. Além destas interfaces, esses mesmos nós também foram equipados com interfaces de acesso à comunicação sem fio baseadas no padrão IEEE 802.11p. As configurações adotadas para estas interfaces estão listadas nas Tabelas \ref{tab:config_IEEE_80211p} e \ref{tab:config_MAC_IEEE_80211p}. Para finalizar a configuração das interfaces de acesso à comunicação sem fio, veículos conectados, sinalizações semafóricas, elementos urbanos e centro de controle de tráfego foram equipados com interfaces baseadas no padrão LTE. De acordo com as definições dos sistemas multiagentes, veículos conectados, elementos urbanos e sinalizações semafóricas foram configurados como nós UE. As configurações das interfaces de acesso à comunicação sem fio destes nós são listadas na Tabela \ref{tab:LTEUEsettings}. No que diz respeito ao centro de controle de tráfego, este foi configurado com um nó eNodeB. As configurações da interface de acesso à comunicação sem fio deste nó é listada na Tabela \ref{tab:LTEeNodeBsettings}. Embora os cenários de avaliação remetam a um ambiente urbano, foi necessário usar o modelo de macrocélula RURAL\_MACROCELL, pois este modelo fornece um alcance de comunicação de 5000m, enquanto o modelo URBAN\_MACROCELL fornece apenas 2000m. Esses alcances de comunicação são impostos pelo framework SimuLTE \cite{SimuLTE:2016}. Além disso, o objetivo foi fornecer a maior área de cobertura possível.

Para comparar de maneira ainda mais justa o desempenho da HRadNet-VE contra o da RAdNet, CCN$_R$ e CCN$_P$, foi introduzida uma filtragem nos mecanismos de encaminhamento de mensagens dos protocolos de comunicação destas redes, utilizando o identificador da via. Esta filtragem é a mesma realizada pelo protocolo de comunicação da HRAdNet-VE. 

Para identificar as aplicações nos experimentos, foi utilizada uma estrutura de nomes similar a que foi definida no Capítulo \ref{cap:radnet-ve}. Uma vez que os experimentos envolvem redes diferentes, foi necessário introduzir apenas um parâmetro em tal estrutura, formando a seguinte URI: \textit{$<$rede$>$://user service bundle/user service/application name}, onde \textit{rede} é o nome da rede veicular heterogênea. Dessa forma, as identificações das aplicações de sistemas inteligentes de transporte definidas nos Capítulos \ref{cap:controle} e \ref{cap:orientacao} ficaram da seguinte forma: (i) sistema multiagente de controle de tráfego: \textit{$<$rede$>$://ttm/traffic\_control/signal\_control/}; (ii) sistema multiagente de planejamento e orientação de rotas:  \textit{$<$rede$>$://ttm/route\_guidance/trasit\_route}; (iii) controle adaptativo e cooperativo de cruzeiro: \textit{$<$rede$>$://lca/cruise\_control/ccac/}. Com base nestas identificações, foram definidas as configurações de interesses para os experimentos com HRAdNet-VE e RAdNet, levando em consideração as mesmas definições de identificação de interesses e tecnologias de acesso à comunicação sem fio utilizadas para enviar e receber mensagens de rede heterogênea. Nos experimentos com a HRAdNet-VE, também foram utilizados o mesmo número máximo de saltos definidos anteriormente nos Capítulos \ref{cap:controle} e \ref{cap:orientacao}. Os experimentos com as CCNs utilizaram as mesmas definições de identificação de interesses como nomes de dados, assim como, as mesmas definições de tecnologias de acesso à comunicação sem fio dos interesses. 

Nos experimentos com a RAdNet-VE, as aplicações foram desenvolvidas para operar de maneira reativa, de modo que elas pudessem requisitar dados e, em seguida, recebê-los. Nos experimentos com a HRAdNet-VE, foi necessário diferenciar as mensagens trocadas pelos agentes. Tal diferenciação consistiu em separar as mensagens trocadas entre os agentes em duas categorias, que são: controle e dados. Como pôde ser obervado nos Capítulos \ref{cap:controle} e \ref{cap:orientacao}, as mensagens relativas aos processos de controle em ambos os sistemas multiagentes são de caráter proativo. Dessa forma, a adoção dos mesmos procedimentos usados nos experimentos com a RAdNet-VE e RAdNet introduziria uma complexidade desnecessária aos experimentos com a HRAdNet-VE. Assim, tais procedimentos foram aplicados somente para mensagens contendo interesses relativos às trocas de dados entre os agentes, quando os desempenhos da HRAdNet-VE e RAdNet foram comparados com o desempenho da CCN$_R$. É importante ressaltar que, essa diferenciação entre mensagens de controle e de troca de dados foi apenas para facilitar a configuração dos experimentos desta seção. A diferenciação dos interesses pode ser vista nas Tabelas \ref{tab:interessesVeiculo}, \ref{tab:interessesSinalizacaoSemaforica}, \ref{tab:interessesSinalizacaoSemaforicaCoord}, \ref{tab:interessesSinalizacaoSemaforicaPart} e  \ref{tab:interessesCentroControleTrafego}.

Para evitar que os nós CCN$_R$ enviassem pacotes Interesse, requisitando mensagens contendo nomes de dados relativos às atividades de controle dos sistemas multiagentes, os nomes de dados foram registrados permanentemente nas \textit{Pendent Interest Bases} (PIBs) dos nós. Desta forma, as trocas de mensagens de controle entre os agentes foi realizada de acordo com as definições dos algoritmos apresentados nos Capítulos \ref{cap:controle} e \ref{cap:orientacao}. No que diz respeito às mensagens contendo interesses relativos às trocas de dados entre os agentes, estas geraram pacotes Interesse para requisitar dados. Consequentemente, os nós CCN$_R$, quando receberam tais pacotes, enviaram os dados para os agentes que os requereram. 

Nos experimentos com a RAdNet-VE, embora a CCN$_P$ tenha usado roteamento proativo de dados, os nós precisam enviar pacotes Interesse para que o envio proativo de dados pudesse acontecer \cite{Wang:2012b}. Nos experimentos com a CCN$_P$, este comportamento foi removido. Para que os nós pudessem trocar mensagens na CCN$_P$, os nomes dos dados foram registrados permanentemente nas PIBs dos mesmos. Com isto, as trocas de mensagens de controle e de dados entre os agentes foram realizadas de maneira proativa.

Com base nas modificações realizadas nas CCNs, foi possível criar dois grupos de comparação, são eles: reativo e proativo. No grupo reativo, os desempenhos da HRAdNet-VE e RAdNet foram comparados somente com o desempenho da CCN$_R$. Por outro lado, no grupo proativo, os desempenhos da HRAdNet-VE e RAdNet foram comparados somente com o desempenho da CCN$_P$. 

Para configurar o número máximo de saltos na RAdNet, CCN$_R$ e CCN$_P$, os seguintes valores saltos foram adotados: (i) oito saltos para comunicações com interfaces IEEE 802.11n; (ii) quatro saltos para comunicações com interfaces IEEE 802.11p; e (iii) um salto para comunicações com interfaces LTE. De acordo com o Algoritmo \ref{alg:protocolo_hradnet_ve}, o protocolo de comunicação da HRAdNet-VE prevê o uso de um número máximo de saltos padrão. Este parâmetro foi configurado para oito saltos.

Para os dois cenários, os tamanhos dos intervalos de indicações das sinalizações semafóricas foram configurados de acordo com os tipos de vias encontrados na grade 10 x 10. Dessa forma, agentes cujas sinalizações semafóricas estavam instaladas em vias de entradas de interseções participantes de corredores foram configurados com os seguintes tamanhos de intervalos mínimos de indicações de sinalização: (i) 33s para luzes verdes; (ii) 4s para luzes amarelas; e (iii) 1s para vermelho geral. Os demais agentes Sinalização Semafórica foram configurados com os seguintes tamanhos de intervalos mínimos de indicações de sinalização: (i) 22s para luzes verdes; (ii) 4s para luzes amarelas; e (iii) 1s para vermelho geral. No que diz respeito ao tamanho máximo dos intervalos de verde em cada agente Sinalização Semafórica, o valor deste parâmetro foi igual ao tempo mínimo de verdes vezes três.  

Além desses parâmetros de configurações, existem outros que precisam ser mencionadas aqui, que são aquelas relacionadas às performances dos agentes. Tais parâmetros foram citados no Capítulo \ref{cap:agentes}, á medida que os agentes Sinalização Semafórica e suas interações eram descritos. A Tabela \ref{tab:confPerfSinSemaf} lista esses parâmetros e seus respectivos valores. Tais valores foram obtidos por meio dos experimentos relativos ao sistema multiagente de controle de tráfego.

\begin{table}[H]
    \centering
    \caption{Configuração padrão dos agentes Sinalização Semafórica.}
    \label{tab:confPerfSinSemaf}
    \small \begin{tabular}{|l|c|}
      \hline
      \textbf{Parâmetro} & \textbf{Valor} \\
      \hline
      Periodicidade de obtenção de quantidade de veículos & 10s \\ \hline
      Número de obtenções de quantidade de veículos & 30\\ \hline
      Comprimento da área de monitoramento de fluxo de tráfego & 100m \\ \hline
    \end{tabular}
\end{table}

\begin{table}[H]
    \centering
    \caption{Configuração padrão dos agentes Sinalização Semafórica controladores de um sistema coordenado de sinalizações semafóricas.}
    \label{tab:confPerfCorredor}
    \small \begin{tabular}{|l|c|}
      \hline
      \textbf{Parâmetro} & \textbf{Valor} \\
      \hline
      Número mínimo de ciclos & 10 ciclos \\ \hline
      Número máximo de ciclos & 30 ciclos \\ \hline
      Periodicidade de compartilhamento de médias de quantidade de veículos & 300s \\ \hline
      Periodicidade de atualização de demandas de corredores & 300s \\ \hline
    \end{tabular}
\end{table}

\begin{figure}[H]
	\centering
    \includegraphics[scale=0.8]{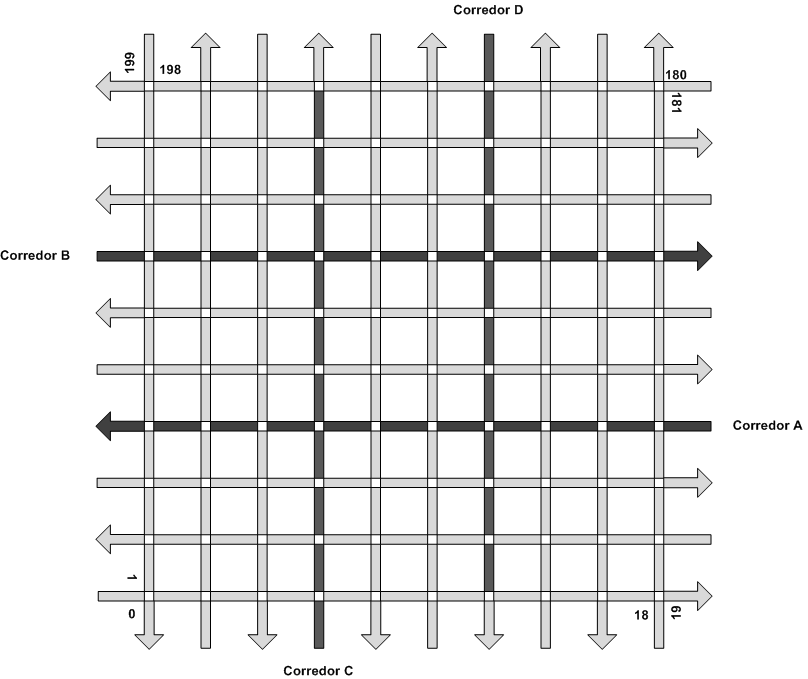}
    \caption{Identificações dos agentes Sinalização Semafórica e corredores de sinalizações semafóricas.}
    \label{fig:identificacaoSinSem}
\end{figure}

Como mencionado no Capítulo \ref{cap:agentes}, os agentes Sinalização Semafórica possuem um identificador único, que é fornecido pelo engenheiro de tráfego durante a configuração do sistema de controle de tráfego. Sendo assim, os agentes Sinalização Semafóricas foram numerados de 0 até 199, seguido o sentido das vias horizontais. Além disto, os corredores também foram identificados unicamente, como pode ser visto na Figura \ref{fig:identificacaoSinSem}. Todos os agentes Sinalização Semafórica controladores de um sistema coordenado de sinalizações semafóricas foram configurados com tais identificadores de corredor. Além destes agentes, aqueles cujas sinalizações semafóricas integraram corredores também tiveram conhecimento das identificações de corredores. Todos os agentes controladores de sistemas coordenados de sinalizações semafóricas tiveram conhecimento das identificações dos agentes embutidos nas sinalizações semafóricas integrantes de corredores e das vias em que estes agentes se encontravam. Após a identificação dos corredores, estes foram separados em grupos, que são: 

\begin{itemize}
	\item \textbf{Grupo Leste-Oeste-Leste}: formado pelos corredores A e B;
	\item \textbf{Grupo Norte-Sul-Norte:} formado pelos corredores C e D.
\end{itemize}

Além dessas desses parâmetros, os agentes Sinalização Semafórica controladores de um sistema coordenado de sinalizações semafóricas também tiveram outros que necessitaram de configuração. Tais parâmetros são listados na Tabela \ref{tab:confPerfCorredor}. Os valores desses parâmetros foram obtidos por meio dos experimentos relativos ao sistema multiagente de controle de tráfego.

Quando ao sistema multiagente de planejamento e orientação de rotas, o parâmetro penalidade por número de veículos foi configurado com o valor 3,5, que foi o mesmo valor de penalidade adotado por \citet{Faria:2013}. Além disto, o parâmetro de tolerância de rota foi configurado com o valor zero.

Assim como nos experimentos com a RAdNet-VE, o IDM também foi utilizado na definição do comportamento dos veículos durante a simulação. A Tabela \ref{tab:idm2} lista os valores para cada um dos parâmetros do desse. Esses valores podem ser encontrados em \textit{http://sumo.dlr.de/wiki/Vehicle\_Type\_Parameter\_Defaults}. Os destinos das viagens realizadas pelos veículos conectados foram escolhidos aleatoriamente e o algoritmo de roteamento de veículos adotados nos experimentos foi o algoritmo de roteamento proposto nesta tese. 

\begin{table}[t]
    \centering
    \caption{Configurações para o \textit{Intelligent Driver Model} (IDM)}
    \label{tab:idm2}
    \small \begin{tabular}{|l|l|}
      \hline
      \textbf{Parameter} & \textbf{Value}\\
      \hline
      Velocidade desejada (v$_0$) & 60 km/h \\ \hline
      Tempo de reação do motorista (T) & 1.2 s \\ \hline
      Espaçamento mínimo entre veículos (s$_0$)& 2.5 m \\ \hline
      Aceleração (a) & 2.9 m/s$^2$ \\ \hline
      Desaceleração (b) & 7.5 m/s$^2$ \\ \hline
    \end{tabular}
\end{table}

\begin{table}[H]
    \centering
    \caption{Configurações de fluxos de veículos para as entradas da grade manhattan 10 x 10.}
    \label{tab:vehicleflows}
    \small\begin{tabular}{|c|c|c|}
      \hline
      \textbf{Experimento} & \textbf{Coletoras} & \textbf{Corredores}\\
      \hline
      1 & 200 veículos/h & 300 veículos/h\\ \hline
      2 & 300 veículos/h& 450 veículos/h\\ \hline
      3 & 400 veículos/h& 600 veículos/h\\ \hline
    \end{tabular}
\end{table}

Por fim, foram criados três experimentos para cada um dos cenários. Cada um destes experimentos teve uma entrada de veículos específica, levando em consideração as primeiras vias imediatamente à frente das vias de entrada da grade 10 x 10. As configurações de fluxos de veículos para as entradas na grade manhattan 10 x 10 são listados na Tabela \ref{tab:vehicleflows}. 

\subsection{Análise dos Resultados do Cenário 1}

Esta seção tem como objetivo apresentar uma análise comparativa entre HRAdNet-VE, RAdNet, CCN$_R$ e CCN$_P$, tendo como base os resultados obtidos nos experimentos 1, 2 e 3 do cenário 1. 

\subsubsection{Experimento 1 com Grupo Reativo}

A seguir, as Tabelas \ref{tab:res_cen1_exp1_reativo_ieee80211n}, \ref{tab:res_cen1_exp1_reativo_ieee80211p} e \ref{tab:res_cen1_exp1_reativo_lte}   apresentam os resultados do grupo reativo e, em seguida, a Tabela \ref{tab:cen1_exp1_reativo} apresenta a análise comparativa destes.

\begin{table}[!h]
\centering
\caption{Resultados do grupo reativo no experimento 1 do cenário 1, utilizando interfaces de acesso à comunicação sem fio baseadas no padrão IEEE 802.11n}
\label{tab:res_cen1_exp1_reativo_ieee80211n}
\small\begin{tabular}{|l|c|c|c|}
\hline
\multicolumn{1}{|c|}{\textbf{Métricas}} & \multicolumn{1}{c|}{\textbf{HRAdNet-VE}} & \multicolumn{1}{c|}{\textbf{RAdNet}} & \multicolumn{1}{c|}{\textbf{CCN$_R$}} \\ \hline
\textbf{CMT (msgs.)}                            & 8,52 $\times 10^8$                       & 1,309 $\times 10^9$                  & 1,020 $\times 10^9$                   \\ \hline
\textbf{LCN (ms)}                            & 24,14 $\pm 0,32$                         & 44,44 $\pm 0,39$                     & 39,53 $\pm 0,58$                      \\ \hline
\textbf{TED (\%)}                            & 92,95 $\pm 0,11$                         & 91,71 $\pm 0,1$                      & 90,46 $\pm 0,1$                       \\ \hline
\textbf{NS (saltos)}                             & 7                                        & 7                                    & 7                                     \\ \hline
\textbf{AM (m)}                             & 1346 $\pm 17,63$                         & 1343 $\pm 18,61$                     & 1344 $\pm 18,29$                                 \\ \hline
\textbf{TPM (ms)}                            & 175  $\pm 18,61$                         & 338 $\pm 49,65$                      & 289 $\pm 31,03$                                  \\ \hline
\end{tabular}
\end{table}

\begin{table}[!h]
\centering
\caption{Resultados do grupo reativo no experimento 1 do cenário 1, utilizando interfaces de acesso à comunicação sem fio baseadas no padrão IEEE 802.11p}
\label{tab:res_cen1_exp1_reativo_ieee80211p}
\small\begin{tabular}{|l|c|c|c|}
\hline
\multicolumn{1}{|c|}{\textbf{Métricas}} & \textbf{HRAdNet-VE} & \textbf{RAdNet}     & \textbf{CCN$_R$}    \\ \hline
\textbf{CMT (msgs.)}                            & 8,3 $\times 10^8$   & 1,963 $\times 10^9$ & 1,513 $\times 10^9$ \\ \hline
\textbf{LCN (ms)}                            & 18,9 $\pm 0,48$               & 25,5 $\pm 0,68$               & 20,74 $\pm 0,55$              \\ \hline
\textbf{TED (\%)}                            & 62,44 $\pm 1,29$               & 47,76 $\pm 0,98$              & 30,72 $\pm 0,65$              \\ \hline
\textbf{NS (saltos)}                             & 4                   & 4                   & 4                   \\ \hline
\textbf{AM (m)}                             & 1977 $\pm 7,51$                & 1976 $\pm 7,83$               & 1977 $\pm 7,51$                \\ \hline
\textbf{TPM (ms)}                            & 122 $\pm 12,08$                & 447 $\pm 45,07$                & 299  $\pm 30,05$               \\ \hline
\end{tabular}
\end{table}

\begin{table}[!h]
\centering
\caption{Resultados do grupo reativo no experimento 1 do cenário 1, utilizando interfaces de acesso à comunicação sem fio baseadas no padrão LTE}
\label{tab:res_cen1_exp1_reativo_lte}
\small\begin{tabular}{|l|c|c|c|}
\hline
\multicolumn{1}{|c|}{\textbf{Métricas}} & \textbf{HRAdNet-VE} & \textbf{RAdNet}    & \textbf{CCN$_R$}   \\ \hline
\textbf{CMT (msgs.)}                            & 1,33 $\times 10^7$  & 2,82 $\times 10^7$ & 1,36 $\times 10^7$ \\ \hline
\textbf{LCN (ms)}                            & 43,3 $\pm 0,32$               & 44,36 $\pm 0,58$             & 43,9 $\pm 0,58$              \\ \hline
\textbf{TED (\%)}                            & 99,79 $\pm 0,13$              & 99,66 $\pm 0,003$             & 99,71  $\pm 0,003$            \\ \hline
\textbf{NS (saltos)}                             & 1                   & 1                  & 1                  \\ \hline
\textbf{AM (m)}                             & 2500                & 2500               & 2500               \\ \hline
\textbf{TPM (ms)}                            & 43,3 $\pm 0,32$                & 44,36 $\pm 0,58$             & 43,9 $\pm 0,58$               \\ \hline
\end{tabular}
\end{table}

\begin{table}[!h]
\centering
\caption{Análise comparativa entre o desempenho da HRAdNet-VE e os das demais redes do grupo reativo, utilizando os resultados do experimento 1 do cenário 1.}
\label{tab:cen1_exp1_reativo}
\small\begin{tabular}{l|l|l|l|l|l|l|}
\cline{2-7}
                                        & \multicolumn{2}{c|}{\textbf{IEEE 802.11n}}      & \multicolumn{2}{c|}{\textbf{IEEE 802.11p}}      & \multicolumn{2}{c|}{\textbf{LTE}}               \\ \hline
\multicolumn{1}{|l|}{\textbf{Métricas}} & \textbf{RAdNet}        & \textbf{CCN$_R$}       & \textbf{RAdNet}        & \textbf{CCN$_R$}       & \textbf{RAdNet}        & \textbf{CCN$_R$}       \\ \hline
\multicolumn{1}{|l|}{\textbf{CMT}}      & \textless 34,9\%       & \textless 16,47\%      & \textless 57,7\%       & \textless 45,14\%      & \textless 52,83\%      & \textless 2,2\%        \\ \hline
\multicolumn{1}{|l|}{\textbf{LCN}}      & \textless 45,67\%      & \textless 38,9\%       & \textless 25,8\%       & \textless 8,87\%       & \textless 2,38\%       & \textless 0,013\%       \\ \hline
\multicolumn{1}{|l|}{\textbf{TED}}      & \textgreater 1,35\%       & \textgreater 2,75\%       & \textgreater 30,73\%      & \textgreater 103,25\%      & \textgreater 1,07\%       & \textgreater 0,008\%       \\ \hline
\multicolumn{1}{|l|}{\textbf{NS}}       & \multicolumn{1}{c|}{=} & \multicolumn{1}{c|}{=} & \multicolumn{1}{c|}{=} & \multicolumn{1}{c|}{=} & \multicolumn{1}{c|}{=} & \multicolumn{1}{c|}{=} \\ \hline
\multicolumn{1}{|l|}{\textbf{AM}}       & \textgreater 0,02\%    & \textgreater 0,01\%    & \textgreater 0,005\%   & \multicolumn{1}{c|}{=} & \multicolumn{1}{c|}{=} & \multicolumn{1}{c|}{=} \\ \hline
\multicolumn{1}{|l|}{\textbf{TPM}}      & \textless 48,22\%      & \textless 39,44\%      & \textless 72,7\%       & \textless 59,19\%      & \textless 1,38\%       & \textless 0,002\%      \\ \hline
\end{tabular}
\end{table}

\newpage

\subsubsection{Experimento 1 com Grupo Proativo}

A seguir, as Tabelas \ref{tab:res_cen1_exp1_proativo_ieee80211n}, \ref{tab:res_cen1_exp1_proativo_ieee80211p} e \ref{tab:res_cen1_exp1_proativo_lte} apresentam os resultados do grupo proativo e, em seguida, a Tabela \ref{tab:cen1_exp1_proativo} apresenta a análise comparativa destes.

\begin{table}[!h]
\centering
\caption{Resultados do grupo proativo no experimento 1 do cenário 1, utilizando interfaces de acesso à comunicação sem fio baseadas no padrão IEEE 802.11n}
\label{tab:res_cen1_exp1_proativo_ieee80211n}
\small\begin{tabular}{|l|c|c|c|}
\hline
\multicolumn{1}{|c|}{\textbf{Métricas}} & \textbf{HRAdNet-VE} & \textbf{RAdNet}    & \textbf{CCN$_P$}   \\ \hline
\textbf{CMT (msgs.)}                            & 7,14 $\times 10^8$  & 7,79 $\times 10^8$ & 7,34 $\times 10^8$ \\ \hline
\textbf{LCN (ms)}                            & 9,03 $\pm 0,06$     & 38,73 $\pm 0,39$   & 15,5 $\pm 0,13$    \\ \hline
\textbf{TED (\%)}                            & 98,84 $\pm 0,11$    & 98,78 $\pm 0,11$   & 97,79 $\pm 0,11$   \\ \hline
\textbf{NS (saltos)}                             & 7                   & 7                  & 7                  \\ \hline
\textbf{AM (m)}                             & 1348 $\pm 16,98$    & 1343 $\pm 18,61$   & 1347 $\pm 17,31$   \\ \hline
\textbf{TPM (ms)}                            & 78,00 $\pm 3,91$    & 301,00 $\pm 41,48$ & 113,4 $\pm 12,08$  \\ \hline
\end{tabular}
\end{table}

\begin{table}[!h]
\centering
\caption{Resultados do grupo proativo no experimento 1 do cenário 1, utilizando interfaces de acesso à comunicação sem fio baseadas no padrão IEEE 802.11p}
\label{tab:res_cen1_exp1_proativo_ieee80211p}
\small\begin{tabular}{|l|c|c|c|}
\hline
\multicolumn{1}{|c|}{\textbf{Métricas}} & \textbf{HRAdNet-VE} & \textbf{RAdNet}    & \textbf{CCN$_P$}   \\ \hline
\textbf{CMT (msgs.)}                            & 3,51 $\times 10^8$  & 7,54 $\times 10^8$ & 6,39 $\times 10^8$ \\ \hline
\textbf{LCN (ms)}                            & 13,93 $\pm 0,32$    & 22,8 $\pm 0,62$    & 15,12 $\pm 0,39$   \\ \hline
\textbf{TED (\%)}                            & 86,38 $\pm 1,72$    & 65,91 $\pm 1,07$   & 81,94 $\pm 1,63$   \\ \hline
\textbf{NS (saltos)}                             & 4                   & 4                  & 4                  \\ \hline
\textbf{AM (m)}                             & 1978 $\pm 7,18$     & 1976 $\pm 7,83$    & 1978 $\pm 7,18$    \\ \hline
\textbf{TPM (ms)}                            & 92,94 $\pm 9,14$    & 178,27 $\pm 17,96$ & 96,48 $\pm 9,79$   \\ \hline
\end{tabular}
\end{table}

\begin{table}[!h]
\centering
\caption{Resultados do grupo proativo no experimento 1 do cenário 1, utilizando interfaces de acesso à comunicação sem fio baseadas no padrão LTE}
\label{tab:res_cen1_exp1_proativo_lte}
\small\begin{tabular}{|l|c|c|c|}
\hline
\multicolumn{1}{|c|}{\textbf{Métricas}} & \textbf{HRAdNet-VE} & \textbf{RAdNet}    & \textbf{CCN$_P$}   \\ \hline
\textbf{CMT (msgs.)}                            & 1,21 $\times 10^7$  & 1,69 $\times 10^7$ & 1,32 $\times 10^7$ \\ \hline
\textbf{LCN (ms)}                            & 43,1 $\pm 0,58$     & 44,8 $\pm 0,58$    & 44,00 $\pm 0,32$   \\ \hline
\textbf{TED (\%)}                            & 99,99 $\pm 0,006$   & 99,91 $\pm 0,003$  & 99,84 $\pm 0,03$   \\ \hline
\textbf{NS (saltos)}                             & 1                   & 1                  & 1                  \\ \hline
\textbf{AM (m)}                             & 2500                & 2500               & 2500               \\ \hline
\textbf{TPM (ms)}                            & 43,1 $\pm 0,58$     & 44,8 $\pm 0,58$    & 44,00 $\pm 0,32$   \\ \hline
\end{tabular}
\end{table}

\begin{table}[!h]
\centering
\caption{Análise comparativa entre o desempenho da HRAdNet-VE e os das demais redes do grupo proativo, utilizando os resultados do experimento 1 do cenário 1.}
\label{tab:cen1_exp1_proativo}
\small\begin{tabular}{l|l|l|l|l|l|l|}
\cline{2-7}
                                        & \multicolumn{2}{c|}{\textbf{IEEE 802.11n}}      & \multicolumn{2}{c|}{\textbf{IEEE 802.11p}}      & \multicolumn{2}{c|}{\textbf{LTE}}               \\ \hline
\multicolumn{1}{|l|}{\textbf{Métricas}} & \textbf{RAdNet}        & \textbf{CCN$_P$}       & \textbf{RAdNet}        & \textbf{CCN$_P$}       & \textbf{RAdNet}        & \textbf{CCN$_P$}       \\ \hline
\multicolumn{1}{|l|}{\textbf{CMT}}      & \textless 8,34\%       & \textless 2,72\%       & \textless 53,44\%      & \textless 45,07\%      & \textless 28,4\%       & \textless 8,33\%       \\ \hline
\multicolumn{1}{|l|}{\textbf{LCN}}      & \textless 76,68\%      & \textless 41,74\%      & \textless 38,9\%       & \textless 7,87\%       & \textless 3,79\%       & \textless 2,04\%       \\ \hline
\multicolumn{1}{|l|}{\textbf{TED}}      & \textgreater 0,005\%   & \textgreater 0,015\%    & \textgreater 31,05\%   & \textgreater 5,41\%    & \textgreater 0,008\%   & \textgreater 0,01\%   \\ \hline
\multicolumn{1}{|l|}{\textbf{NS}}       & \multicolumn{1}{c|}{=} & \multicolumn{1}{c|}{=} & \multicolumn{1}{c|}{=} & \multicolumn{1}{c|}{=} & \multicolumn{1}{c|}{=} & \multicolumn{1}{c|}{=} \\ \hline
\multicolumn{1}{|l|}{\textbf{AM}}       & \textgreater 0,03\%    & \textgreater 0,007\%   & \textgreater 0,01\%   &  \multicolumn{1}{c|}{=} & \multicolumn{1}{c|}{=} & \multicolumn{1}{c|}{=} \\ \hline
\multicolumn{1}{|l|}{\textbf{TPM}}      & \textless 74,08\%      & \textless 30,97\%      & \textless 47,89\%      & \textless 3,66\%       & \textless 1,82\%       & \textless 0,071\%      \\ \hline
\end{tabular}
\end{table}

\newpage

\subsubsection{Experimento 2 com Grupo Reativo}

A seguir, as Tabelas \ref{tab:res_cen1_exp2_reativo_ieee80211n}, \ref{tab:res_cen1_exp2_reativo_ieee80211p} e \ref{tab:res_cen1_exp2_reativo_lte} apresentam os resultados do grupo reativo e, em seguida, a Tabela  \ref{tab:cen1_exp2_reativo} apresenta a análise comparativa destes.

\begin{table}[!h]
\centering
\caption{Resultados do grupo reativo no experimento 2 do cenário 1, utilizando interfaces de acesso à comunicação sem fio baseadas no padrão IEEE 802.11n}
\label{tab:res_cen1_exp2_reativo_ieee80211n}
\small\begin{tabular}{|l|c|c|c|}
\hline
\multicolumn{1}{|c|}{\textbf{Métricas}} & \textbf{HRAdNet-VE} & \textbf{RAdNet}    & \textbf{CCN$_R$}   \\ \hline
\textbf{CMT (msgs.)}                            & 1,142 $\times 10^9$ & 3,09 $\times 10^9$ & 3,02 $\times 10^9$ \\ \hline
\textbf{LCN (ms)}                            & 25,31 $\pm 0,44$    & 41,12 $\pm 0,65$   & 37,39 $\pm 0,71$   \\ \hline
\textbf{TED (\%)}                            & 91,61 $\pm 0,14$    & 87,5 $\pm 0,13$    & 82,73 $\pm 0,12$   \\ \hline
\textbf{NS (saltos)}                             & 7                   & 7                  & 7                  \\ \hline
\textbf{AM (m)}                             & 1346 $\pm 17,63$    & 1344 $\pm 18,29$   & 1344 $\pm 18,29$   \\ \hline
\textbf{TPM (ms)}                            & 190 $\pm 22,21$     & 329 $\pm 33,97$    & 301 $\pm 35,27$    \\ \hline
\end{tabular}
\end{table}

\begin{table}[!h]
\centering
\caption{Resultados do grupo reativo no experimento 2 do cenário 1, utilizando interfaces de acesso à comunicação sem fio baseadas no padrão IEEE 802.11p}
\label{tab:res_cen1_exp2_reativo_ieee80211p}
\small\begin{tabular}{|l|c|c|c|}
\hline
\multicolumn{1}{|c|}{\textbf{Métricas}} & \textbf{HRAdNet-VE} & \textbf{RAdNet}    & \textbf{CCN$_R$}    \\ \hline
\textbf{CMT (msgs.)}                            & 1,738 $\times 10^9$ & 5,93 $\times 10^9$ & 2,797 $\times 10^9$ \\ \hline
\textbf{LCN (ms)}                            & 23,02 $\pm 0,58$    & 35,88 $\pm 0,58$   & 37,39 $\pm 0,48$    \\ \hline
\textbf{TED (\%)}                            & 49,31 $\pm 1,04$    & 42,12 $\pm 0,89$   & 28,84 $\pm 0,61$    \\ \hline
\textbf{NS (saltos)}                             & 4                   & 4                  & 4                   \\ \hline
\textbf{AM (m)}                             & 1976 $\pm 7,83$     & 1974 $\pm 8,49$    & 1976 $\pm 7,83$     \\ \hline
\textbf{TPM (ms)}                            & 152 $\pm 8,81$      & 766 $\pm 68,27$    & 690 $\pm 22,53$     \\ \hline
\end{tabular}
\end{table}

\begin{table}[!h]
\centering
\caption{Resultados do grupo reativo no experimento 2 do cenário 1, utilizando interfaces de acesso à comunicação sem fio baseadas no padrão LTE}
\label{tab:res_cen1_exp2_reativo_lte}
\small\begin{tabular}{|l|c|c|c|}
\hline
\multicolumn{1}{|c|}{\textbf{Métricas}} & \textbf{HRAdNet-VE} & \textbf{RAdNet}    & \textbf{CCN$_R$}   \\ \hline
\textbf{CMT (msgs.)}                            & 2,59 $\times 10^7$  & 6,04 $\times 10^7$ & 2,87 $\times 10^7$ \\ \hline
\textbf{LCN (ms)}                            & 49,1 $\pm 0,68$     & 52,58 $\pm 0,65$   & 50,24 $\pm 0,68$   \\ \hline
\textbf{TED (\%)}                            & 99,24 $\pm 0,25$    & 99,16 $\pm 0,002$  & 98,83 $\pm 0,002$  \\ \hline
\textbf{NS (saltos)}                             & 1                   & 1                  & 1                  \\ \hline
\textbf{AM (m)}                             & 2500                & 2500               & 2500               \\ \hline
\textbf{TPM (ms)}                            & 49,1 $\pm 0,68$     & 52,58 $\pm 0,65$   & 50,24 $\pm 0,68$   \\ \hline
\end{tabular}
\end{table}

\begin{table}[!h]
\centering
\caption{Análise comparativa entre o desempenho da HRAdNet-VE e os das demais redes do grupo reativo, utilizando os resultados do experimento 2 do cenário 1.}
\label{tab:cen1_exp2_reativo}
\small\begin{tabular}{l|l|l|l|l|l|l|}
\cline{2-7}
                                        & \multicolumn{2}{c|}{\textbf{IEEE 802.11n}}      & \multicolumn{2}{c|}{\textbf{IEEE 802.11p}}      & \multicolumn{2}{c|}{\textbf{LTE}}               \\ \hline
\multicolumn{1}{|l|}{\textbf{Métricas}} & \textbf{RAdNet}        & \textbf{CCN$_R$}       & \textbf{RAdNet}        & \textbf{CCN$_R$}       & \textbf{RAdNet}        & \textbf{CCN$_R$}       \\ \hline
\multicolumn{1}{|l|}{\textbf{CMT}}      & \textless 63,04\%      & \textless 62,18        & \textless 70,69\%      & \textless 37,86\%      & \textless 57,11\%      & \textless 9,75\%       \\ \hline
\multicolumn{1}{|l|}{\textbf{LCN}}      & \textless 38,44\%      & \textless 32,3\%       & \textless 35,84\%      & \textless 12,5\%       & \textless 6,66\%       & \textless 2,26\%       \\ \hline
\multicolumn{1}{|l|}{\textbf{TED}}      & \textgreater 4,69\%    & \textgreater 10,73\%    & \textgreater 17,70\%   & \textgreater 70,15\%   & \textgreater 0,008\%   & \textgreater 0,04\%    \\ \hline
\multicolumn{1}{|l|}{\textbf{NS}}       & \multicolumn{1}{c|}{=} & \multicolumn{1}{c|}{=} & \multicolumn{1}{c|}{=} & \multicolumn{1}{c|}{=} & \multicolumn{1}{c|}{=} & \multicolumn{1}{c|}{=} \\ \hline
\multicolumn{1}{|l|}{\textbf{AM}}       & \textgreater 0,01\%    & \textgreater 0,01\%    & \textgreater 0,01\%    & \multicolumn{1}{c|}{=} & \multicolumn{1}{c|}{=} & \multicolumn{1}{c|}{=} \\ \hline
\multicolumn{1}{|l|}{\textbf{TPM}}      & \textless 42,24\%      & \textless 36,87\%      & \textless 80,15\%      & \textless 77,97\%      & \textless 6,61\%       & \textless 2,26\%       \\ \hline
\end{tabular}
\end{table}

\newpage

\subsubsection{Experimento 2 com Grupo Proativo}

A seguir, as Tabelas \ref{tab:res_cen1_exp2_proativo_ieee80211n}, \ref{tab:res_cen1_exp2_proativo_ieee80211p} e \ref{tab:res_cen1_exp2_proativo_lte} apresentam os resultados do grupo proativo e, em seguida, a Tabela \ref{tab:cen1_exp2_proativo} apresenta a análise comparativa destes.

\begin{table}[!h]
\centering
\caption{Resultados do grupo proativo no experimento 2 do cenário 1, utilizando interfaces de acesso à comunicação sem fio baseadas no padrão IEEE 802.11n}
\label{tab:res_cen1_exp2_proativo_ieee80211n}
\small\begin{tabular}{|l|c|c|c|}
\hline
\multicolumn{1}{|c|}{\textbf{Métricas}} & \textbf{HRAdNet-VE} & \textbf{RAdNet}    & \textbf{CCN$_P$}    \\ \hline
\textbf{CMT (msgs.)}                            & 1,076 $\times 10^9$ & 1,21 $\times 10^9$ & 1,115 $\times 10^9$ \\ \hline
\textbf{LCN (ms)}                            & 9,09 $\pm 0,06$     & 11,12 $\pm 0,006$  & 10,7 $\pm 0,16$     \\ \hline
\textbf{TED (\%)}                            & 98,83 $\pm 0,15$    & 98,72 $\pm 0,15$   & 96,56 $\pm 0,14$    \\ \hline
\textbf{NS (saltos)}                             & 7                   & 7                  & 7                   \\ \hline
\textbf{AM (m)}                             & 1349 $\pm 16,65$    & 1348 $\pm 16,98$   & 1348 $\pm 16,98$    \\ \hline
\textbf{TPM (ms)}                            & 75 $\pm 3,26$       & 95 $\pm 13,39$     & 86 $\pm 9,79$       \\ \hline
\end{tabular}
\end{table}

\begin{table}[!h]
\centering
\caption{Resultados do grupo proativo no experimento 2  do cenário 1, utilizando interfaces de acesso à comunicação sem fio baseadas no padrão IEEE 802.11p}
\label{tab:res_cen1_exp2_proativo_ieee80211p}
\small\begin{tabular}{|l|c|c|c|}
\hline
\multicolumn{1}{|c|}{\textbf{Métricas}} & \textbf{HRAdNet-VE} & \textbf{RAdNet}     & \textbf{CCN$_P$}    \\ \hline
\textbf{CMT (msgs.)}                            & 7,49 $\times 10^8$  & 1,858 $\times 10^9$ & 1,013 $\times 10^9$ \\ \hline
\textbf{LCN (ms)}                            & 12,23 $\pm 0,32$    & 16,68 $\pm 0,42$    & 14,29 $\pm 0,37$    \\ \hline
\textbf{TED (\%)}                            & 84,41 $\pm 1,79$    & 66,65 $\pm 1,41$    & 67,15 $\pm 1,41$    \\ \hline
\textbf{NS (saltos)}                             & 4                   & 4                   & 4                   \\ \hline
\textbf{AM (m)}                             & 1978 $\pm 7,18$     & 1977 $\pm 7,51$     & 1978 $\pm 7,18$     \\ \hline
\textbf{TPM (ms)}                            & 89 $\pm 8,81$       & 178 $\pm 17,96$     & 105 $\pm 10,45$     \\ \hline
\end{tabular}
\end{table}

\begin{table}[!h]
\centering
\caption{Resultados do grupo proativo no experimento 2 do cenário 1, utilizando interfaces de acesso à comunicação sem fio baseadas no padrão LTE}
\label{tab:res_cen1_exp2_proativo_lte}
\small\begin{tabular}{|l|c|c|c|}
\hline
\multicolumn{1}{|c|}{\textbf{Métricas}} & \textbf{HRAdNet-VE} & \textbf{RAdNet}   & \textbf{CCN$_P$}  \\ \hline
\textbf{CMT (msgs.)}                            & 1,57 $\times 10^7$  & 1,9 $\times 10^7$ & 1,7 $\times 10^7$ \\ \hline
\textbf{LCN (ms)}                            & 43,2 $\pm 0,58$     & 45 $\pm 0,68$     & 44,6 $\pm 0,78$   \\ \hline
\textbf{TED (\%)}                            & 99,99 $\pm 0,03$    & 99,97 $\pm 0,009$ & 99,93 $\pm 0,02$  \\ \hline
\textbf{NS (saltos)}                             & 1                   & 1                 & 1                 \\ \hline
\textbf{AM (m)}                             & 2500                & 2500              & 2500              \\ \hline
\textbf{TPM (ms)}                            & 43,2 $\pm 0,58$     & 45 $\pm 0,68$     & 44,6 $\pm 0,78$   \\ \hline
\end{tabular}
\end{table}

\begin{table}[!h]
\centering
\caption{Análise comparativa entre o desempenho da HRAdNet-VE e os das demais redes do grupo proativo, utilizando os resultados do experimento 2  do cenário 1.}
\label{tab:cen1_exp2_proativo}
\small\begin{tabular}{l|l|l|l|l|l|l|}
\cline{2-7}
                                        & \multicolumn{2}{c|}{\textbf{IEEE 802.11n}}      & \multicolumn{2}{c|}{\textbf{IEEE 802.11p}}      & \multicolumn{2}{c|}{\textbf{LTE}}               \\ \hline
\multicolumn{1}{|l|}{\textbf{Métricas}} & \textbf{RAdNet}        & \textbf{CCN$_P$}       & \textbf{RAdNet}        & \textbf{CCN$_P$}       & \textbf{RAdNet}        & \textbf{CCN$_P$}       \\ \hline
\multicolumn{1}{|l|}{\textbf{CMT}}      & \textless 11,07\%      & \textless 3,49\%       & \textless 59,14\%      & \textless 25,07\%      & \textless 17,36\%      & \textless 11,29\%      \\ \hline
\multicolumn{1}{|l|}{\textbf{LCN}}      & \textless 18,25\%      & \textless 15,04\%      & \textless 26,67\%      & \textless 14,41\%      & \textless 4,22\%       & \textless 3,36\%       \\ \hline
\multicolumn{1}{|l|}{\textbf{TED}}      & \textgreater 0,01\%    & \textgreater 2,35\%    & \textgreater 26,64\%   & \textgreater 25,70\%   & \textgreater 0,002\%   & \textgreater 0,006\%   \\ \hline
\multicolumn{1}{|l|}{\textbf{NS}}       & \multicolumn{1}{c|}{=} & \multicolumn{1}{c|}{=} & \multicolumn{1}{c|}{=} & \multicolumn{1}{c|}{=} & \multicolumn{1}{c|}{=} & \multicolumn{1}{c|}{=} \\ \hline
\multicolumn{1}{|l|}{\textbf{AM}}       & \textgreater 0,007\%    & \textgreater 0,007\%    & \textgreater 0,005\%    & \multicolumn{1}{c|}{=} & \multicolumn{1}{c|}{=} & \multicolumn{1}{c|}{=} \\ \hline
\multicolumn{1}{|l|}{\textbf{TPM}}      & \textless 21,05\%      & \textless 9,47\%       & \textless 50\%         & 15,23\%                & \textless 4,22\%       & \textless 3,36         \\ \hline
\end{tabular}
\end{table}

\subsubsection{Experimento 3 com Grupo Reativo}

A seguir, as Tabelas \ref{tab:res_cen1_exp3_reativo_ieee80211n}, \ref{tab:res_cen1_exp3_reativo_ieee80211p} e \ref{tab:res_cen1_exp3_reativo_lte} apresentam os resultados do grupo reativo e, em seguida, a Tabela  \ref{tab:cen1_exp3_reativo} apresenta a análise comparativa destes.

\begin{table}[!h]
\centering
\caption{Resultados do grupo reativo no experimento 3  do cenário 1 utilizando interfaces de acesso à comunicação sem fio baseadas no padrão IEEE 802.11n}
\label{tab:res_cen1_exp3_reativo_ieee80211n}
\small\begin{tabular}{|l|c|c|c|}
\hline
\multicolumn{1}{|c|}{\textbf{Métricas}} & \textbf{HRAdNet-VE} & \textbf{RAdNet}     & \textbf{CCN$_R$}   \\ \hline
\textbf{CMT (msgs.)}                            & 1,981 $\times 10^9$ & 4,209 $\times 10^9$ & 3,17 $\times 10^9$ \\ \hline
\textbf{LCN (ms)}                            & 27,99 $\pm 0,49$    & 47,7 $\pm 0,89$     & 41,12 $\pm 0,71$   \\ \hline
\textbf{TED (\%)}                            & 91,95 $\pm 0,14$    & 90,77 $\pm 0,13$    & 88,83 $\pm 0,13$   \\ \hline
\textbf{NS (saltos)}                             & 7                   & 7                   & 7                  \\ \hline
\textbf{AM (m)}                             & 1566 $\pm 11,1$     & 1563 $\pm 12,08$    & 1564 $\pm 11,75$   \\ \hline
\textbf{TPM (ms)}                            & 188 $\pm 21,88$     & 465 $\pm 56,83$     & 410 $\pm 48,01$    \\ \hline
\end{tabular}
\end{table}

\begin{table}[!h]
\centering
\caption{Resultados do grupo reativo no experimento 3  do cenário 1, utilizando interfaces de acesso à comunicação sem fio baseadas no padrão IEEE 802.11p}
\label{tab:res_cen1_exp3_reativo_ieee80211p}
\small\begin{tabular}{|l|c|c|c|}
\hline
\multicolumn{1}{|c|}{\textbf{Métricas}} & \textbf{HRAdNet-VE} & \textbf{RAdNet}     & \textbf{CCN$_R$}    \\ \hline
\textbf{CMT (msgs.)}                            & 4,115 $\times 10^9$ & 6,417 $\times 10^9$ & 4,496 $\times 10^9$ \\ \hline
\textbf{LCN (ms)}                            & 21,12 $\pm 0,52$    & 43,61 $\pm 0,68$    & 39,61 $\pm 0,48$    \\ \hline
\textbf{TED (\%)}                            & 49,35 $\pm 1,04$    & 31,7 $\pm 0,74$     & 20,41 $\pm 0,43$    \\ \hline
\textbf{NS (saltos)}                             & 4                   & 4                   & 4                   \\ \hline
\textbf{AM (m)}                             & 1987 $\pm 4,24$     & 1983 $\pm 5,55$     & 1984 $\pm 5,22$     \\ \hline
\textbf{TPM (ms)}                            & 160 $\pm 11,1$      & 947 $\pm 8,16$      & 781 $\pm 6,85$      \\ \hline
\end{tabular}
\end{table}

\begin{table}[!h]
\centering
\caption{Resultados do grupo reativo no experimento 3  do cenário 1, utilizando interfaces de acesso à comunicação sem fio baseadas no padrão LTE}
\label{tab:res_cen1_exp3_reativo_lte}
\small\begin{tabular}{|l|c|c|c|}
\hline
\multicolumn{1}{|c|}{\textbf{Métricas}} & \textbf{HRAdNet-VE} & \textbf{RAdNet}    & \textbf{CCN$_R$}   \\ \hline
\textbf{CMT (msgs.)}                            & 3,89 $\times 10^7$  & 8,19 $\times 10^7$ & 4,28 $\times 10^7$ \\ \hline
\textbf{LCN (ms)}                            & 51,4 $\pm 0,78$     & 54,6 $\pm 0,65$    & 52,28 $\pm 0,65$   \\ \hline
\textbf{TED (\%)}                            & 97,95 $\pm 0,32$    & 96,91 $\pm 0,32$   & 97,86 $\pm 0,32$   \\ \hline
\textbf{NS (saltos)}                             & 1                   & 1                  & 1                  \\ \hline
\textbf{AM (m)}                             & 2500                & 2500               & 2500               \\ \hline
\textbf{TPM (ms)}                            & 51,4 $\pm 0,78$     & 54,6 $\pm 0,65$    & 52,28 $\pm 0,65$   \\ \hline
\end{tabular}
\end{table}

\begin{table}[!h]
\centering
\caption{Análise comparativa entre desempenho da HRAdNet-VE e os das demais redes do grupo reativo, utilizando os resultados do experimento 3 do cenário 1.}
\label{tab:cen1_exp3_reativo}
\small\begin{tabular}{l|l|l|l|l|l|l|}
\cline{2-7}
                                        & \multicolumn{2}{c|}{\textbf{IEEE 802.11n}}      & \multicolumn{2}{c|}{\textbf{IEEE 802.11p}}                        & \multicolumn{2}{c|}{\textbf{LTE}}               \\ \hline
\multicolumn{1}{|l|}{\textbf{Métricas}} & \textbf{RAdNet}        & \textbf{CCN$_R$}       & \textbf{RAdNet}        & \textbf{CCN$_R$}                         & \textbf{RAdNet}        & \textbf{CCN$_R$}       \\ \hline
\multicolumn{1}{|l|}{\textbf{CMT}}      & \textless 52,93\%      & \textless37,50\%       & \textless 35,87\%      & \textless 8,47\%                         & \textless 52,5\%       & \textless 9,11\%       \\ \hline
\multicolumn{1}{|l|}{\textbf{LCN}}      & \textless 41,03\%      & \textless 31,93\%      & \textless 51,57\%      & \textless 46,68\%                        & \textless 5,86\%       & \textless 1,68\%       \\ \hline
\multicolumn{1}{|l|}{\textbf{TED}}      & \textgreater 1,29\%    & \textgreater 3,51\%    & \textgreater 55,67\%     & \textgreater 141,79\%                     & \textgreater 1,07\%    & \textgreater 0,009\%   \\ \hline
\multicolumn{1}{|l|}{\textbf{NS}}       & \multicolumn{1}{c|}{=} & \multicolumn{1}{c|}{=} & \multicolumn{1}{c|}{=} & \multicolumn{1}{c|}{=}                   & \multicolumn{1}{c|}{=} & \multicolumn{1}{c|}{=} \\ \hline
\multicolumn{1}{|l|}{\textbf{AM}}       & \textgreater 0,01\%   & \textgreater 0,01\%   & \textgreater 0,02\%    & \multicolumn{1}{c|}{\textgreater 0,01\%} & \multicolumn{1}{c|}{=} & \multicolumn{1}{c|}{=} \\ \hline
\multicolumn{1}{|l|}{\textbf{TPM}}      & \textless 61,23\%      & \textless 54,14\%      & \textless 83,10\%      & \textless 79,51\%                        & \textless 0,003\%      & \textless 0,03\%       \\ \hline
\end{tabular}
\end{table}

\subsubsection{Experimento 3 com Grupo Proativo}

A seguir, as Tabelas \ref{tab:res_cen1_exp3_proativo_ieee80211n}, \ref{tab:res_cen1_exp3_proativo_ieee80211p} e \ref{tab:res_cen1_exp3_proativo_lte} apresentam os resultados do grupo proativo e, em seguida, a Tabela \ref{tab:cen1_exp3_proativo} apresenta a análise comparativa destes.

\begin{table}[!h]
\centering
\caption{Resultados do grupo proativo no experimento 3  do cenário 1, utilizando interfaces de acesso à comunicação sem fio baseadas no padrão IEEE 802.11n}
\label{tab:res_cen1_exp3_proativo_ieee80211n}
\small\begin{tabular}{|l|c|c|c|}
\hline
\multicolumn{1}{|c|}{\textbf{Métricas}} & \textbf{HRAdNet-VE} & \textbf{RAdNet}     & \textbf{CCN$_P$}    \\ \hline
\textbf{CMT (msgs.)}                            & 1,806 $\times 10^9$ & 2,064 $\times 10^9$ & 2,046 $\times 10^9$ \\ \hline
\textbf{LCN (ms)}                            & 9,1 $\pm 0,13$      & 11,17 $\pm 0,19$    & 11,03 $\pm 0,45$    \\ \hline
\textbf{TED (\%)}                            & 98,36 $\pm 0,15$    & 97,82 $\pm 0,15$    & 96,38 $\pm 0,14$    \\ \hline
\textbf{NS (saltos)}                             & 7                   & 7                   & 7                   \\ \hline
\textbf{AM (m)}                             & 1569 $\pm 10,12$    & 1568 $\pm 10,45$    & 1568 $\pm 10,45$    \\ \hline
\textbf{TPM (ms)}                            & 94 $\pm 10,77$      & 159 $\pm 18,61$     & 97 $\pm 11,1$       \\ \hline
\end{tabular}
\end{table}

\begin{table}[!h]
\centering
\caption{Resultados do grupo proativo no experimento 3  do cenário 1, utilizando interfaces de acesso à comunicação sem fio baseadas no padrão IEEE 802.11p}
\label{tab:res_cen1_exp3_proativo_ieee80211p}
\small\begin{tabular}{|l|c|c|c|}
\hline
\multicolumn{1}{|c|}{\textbf{Métricas}} & \textbf{HRAdNet-VE} & \textbf{RAdNet}     & \textbf{CCN$_P$}    \\ \hline
\textbf{CMT (msgs.)}                            & 1,657 $\times 10^9$ & 3,506 $\times 10^9$ & 2,106 $\times 10^9$ \\ \hline
\textbf{LCN (ms)}                            & 14,16 $\pm 0,35$    & 23,18 $\pm 0,58$    & 19,85 $\pm 0,48$    \\ \hline
\textbf{TED (\%)}                            & 83,39 $\pm 1,77$    & 66,88 $\pm 1,42$    & 72,63 $\pm 1,54$    \\ \hline
\textbf{NS (saltos)}                             & 4                   & 4                   & 4                   \\ \hline
\textbf{AM (m)}                             & 1988 $\pm 3,91$     & 1986 $\pm 4,57$     & 1987 $\pm 14,24$    \\ \hline
\textbf{TPM (ms)}                            & 74,11 $\pm 6,53$    & 188 $\pm 16,55$     & 95,67 $\pm 8,49$    \\ \hline
\end{tabular}
\end{table}

\begin{table}[!h]
\centering
\caption{Resultados do grupo proativo no experimento 3  do cenário 1, utilizando interfaces de acesso à comunicação sem fio baseadas no padrão LTE}
\label{tab:res_cen1_exp3_proativo_lte}
\small\begin{tabular}{|l|c|c|c|}
\hline
\multicolumn{1}{|c|}{\textbf{Métricas}} & \textbf{HRAdNet-VE} & \textbf{RAdNet}    & \textbf{CCN$_P$}   \\ \hline
\textbf{CMT (msgs.)}                            & 3,63 $\times 10^7$  & 4,83 $\times 10^7$ & 4,02 $\times 10^7$ \\ \hline
\textbf{LCN (ms)}                            & 43,6 $\pm 0,58$     & 47,8 $\pm 0,78$    & 44,3 $\pm 0,65$    \\ \hline
\textbf{TED (\%)}                            & 99,96 $\pm 0,03$    & 99,87 $\pm 0,03$   & 99,89 $\pm 0,03$   \\ \hline
\textbf{NS (saltos)}                             & 1                   & 1                  & 1                  \\ \hline
\textbf{AM (m)}                             & 2500                & 2500               & 2500               \\ \hline
\textbf{TPM (ms)}                            & 43,6 $\pm 0,58$     & 47,8 $\pm 0,78$    & 44,3 $\pm 0,65$    \\ \hline
\end{tabular}
\end{table}

\begin{table}[!h]
\centering
\caption{Análise comparativa entre desempenho da HRAdNet-VE e os das demais redes do grupo proativo, utilizando os resultados do experimento 3 do cenário 1.}
\label{tab:cen1_exp3_proativo}
\small\begin{tabular}{l|l|l|l|l|l|l|}
\cline{2-7}
                                        & \multicolumn{2}{c|}{\textbf{IEEE 802.11n}}      & \multicolumn{2}{c|}{\textbf{IEEE 802.11p}}                      & \multicolumn{2}{c|}{\textbf{LTE}}               \\ \hline
\multicolumn{1}{|l|}{\textbf{Métricas}} & \textbf{RAdNet}        & \textbf{CCN$_P$}       & \textbf{RAdNet}        & \textbf{CCN$_P$}                       & \textbf{RAdNet}        & \textbf{CCN$_P$}       \\ \hline
\multicolumn{1}{|l|}{\textbf{CMT}}      & \textless 12,5\%       & \textless 11,73\%      & \textless 52,73\%      & \textless 21,32\%                      & \textless 24,84\%      & \textless 9,7\%        \\ \hline
\multicolumn{1}{|l|}{\textbf{LCN}}      & \textless 18,53\%      & \textless 17,49\%      & \textless 38,91\%      & \textless 28,66\%                      & \textless 8,78\%       & \textless 1,58\%       \\ \hline
\multicolumn{1}{|l|}{\textbf{TED}}      & \textgreater 0,05\%    & \textgreater 2,05\%    & \textgreater 24,68\%   & \textgreater 14,81\%                    & \textgreater 0,009\%   & \textgreater 0,007\%   \\ \hline
\multicolumn{1}{|l|}{\textbf{NS}}       & \multicolumn{1}{c|}{=} & \multicolumn{1}{c|}{=} & \multicolumn{1}{c|}{=} & \multicolumn{1}{c|}{=}                 & \multicolumn{1}{c|}{=} & \multicolumn{1}{c|}{=} \\ \hline
\multicolumn{1}{|l|}{\textbf{AM}}       & \textless 0,006\%      & \textless 0,006\%      & \textless 0,001\%      & \multicolumn{1}{c|}{\textless 0,005\%} & \multicolumn{1}{c|}{=} & \multicolumn{1}{c|}{=} \\ \hline
\multicolumn{1}{|l|}{\textbf{TPM}}      & \textless 40,88\%      & \textless 3,09\%       & \textless 60,57\%      & \textless 22,53                        & \textless 24,93        & \textless 10\%         \\ \hline
\end{tabular}
\end{table}

\subsection{Análise dos Resultados do Cenário 2}

Esta seção tem como objetivo apresentar uma análise comparativa entre HRAdNet-VE, RAdNet, CCN$_R$ e CCN$_P$, tendo como base os resultados obtidos nos experimentos 1, 2 e 3 do cenário 2.

\subsubsection{Experimento 1 com Grupo Reativo}

A seguir, as Tabelas \ref{tab:res_cen2_exp1_reativo_ieee80211n}, \ref{tab:res_cen2_exp1_reativo_ieee80211p} e \ref{tab:res_cen2_exp1_reativo_lte} apresentam os resultados do grupo reativo e, em seguida, a Tabela \ref{tab:cen2_exp1_reativo} apresenta a análise comparativa destes.

\begin{table}[!h]
\centering
\caption{Resultados do grupo reativo no experimento 1  do cenário 2, utilizando interfaces de acesso à comunicação sem fio baseadas no padrão IEEE 802.11n}
\label{tab:res_cen2_exp1_reativo_ieee80211n}
\small\begin{tabular}{|l|c|c|c|}
\hline
\multicolumn{1}{|c|}{\textbf{Métricas}} & \textbf{HRAdNet-VE} & \textbf{RAdNet}     & \textbf{CCN$_R$}    \\ \hline
\textbf{CMT}                            & 9,5 $\times 10^8$   & 1,772 $\times 10^9$ & 1,245 $\times 10^9$ \\ \hline
\textbf{LCN}                            & 28,53 $\pm 0,32$    & 97,04 $\pm 0,39$    & 65,88 $\pm 0,58$    \\ \hline
\textbf{TED}                            & 91,95 $\pm 0,1$     & 85,6 $\pm 0,1$      & 88,19 $\pm 0,1$     \\ \hline
\textbf{NS}                             & 7                   & 7                   & 7                   \\ \hline
\textbf{AM}                             & 1346 $\pm 17,63$    & 1335 $\pm 21,23$    & 1340 $\pm 19,59$    \\ \hline
\textbf{TPM}                            & 197 $\pm 18,61$     & 800 $\pm 85,91$     & 487 $\pm 52,26$     \\ \hline
\end{tabular}
\end{table}

\begin{table}[!h]
\centering
\caption{Resultados do grupo reativo no experimento 1  do cenário 2, utilizando interfaces de acesso à comunicação sem fio baseadas no padrão IEEE 802.11p}
\label{tab:res_cen2_exp1_reativo_ieee80211p}
\small\begin{tabular}{|l|c|c|c|}
\hline
\multicolumn{1}{|c|}{\textbf{Métricas}} & \textbf{HRAdNet-VE} & \textbf{RAdNet}     & \textbf{CCN$_R$}    \\ \hline
\textbf{CMT (msgs.)}                    & 8,35 $\times 10^8$  & 1,963 $\times 10^9$ & 1,513 $\times 10^9$ \\ \hline
\textbf{LCN (ms)}                       & 18,93 $\pm 0,52$    & 25,5 $\pm 0,68$     & 21,05 $\pm 0,62$    \\ \hline
\textbf{TED (\%)}                       & 61,63 $\pm 1,3$     & 46,85 $\pm 0,93$    & 30,22 $\pm 0,6$     \\ \hline
\textbf{NS (saltos)}                    & 4                   & 4                   & 4                   \\ \hline
\textbf{AM (m)}                         & 1977 $\pm 7,51$     & 1976 $\pm 7,83$     & 1977 $\pm 7,51$     \\ \hline
\textbf{TPM (ms)}                       & 122 $\pm 12,41$     & 447 $\pm 45,07$     & 300 $\pm 31,68$     \\ \hline
\end{tabular}
\end{table}

\begin{table}[!h]
\centering
\caption{Resultados do grupo reativo no experimento 1  do cenário 2, utilizando interfaces de acesso à comunicação sem fio baseadas no padrão LTE}
\label{tab:res_cen2_exp1_reativo_lte}
\small\begin{tabular}{|l|c|c|c|}
\hline
\multicolumn{1}{|c|}{\textbf{Métricas}} & \textbf{HRAdNet-VE} & \textbf{RAdNet}   & \textbf{CCN$_R$}   \\ \hline
\textbf{CMT (msgs.)}                    & 1,53 $\times 10^7$  & 3,6 $\times 10^7$ & 1,74 $\times 10^7$ \\ \hline
\textbf{LCN (ms)}                       & 43,56 $\pm 0,32$    & 54,03 $\pm 0,58$  & 51,18 $\pm 0,58$   \\ \hline
\textbf{TED (\%)}                       & 98,81 $\pm 0,13$    & 98,68 $\pm 0,003$ & 98,73 $\pm 0,003$  \\ \hline
\textbf{NS (saltos)}                    & 1                   & 1                 & 1                  \\ \hline
\textbf{AM (m)}                         & 2500                & 2500              & 2500               \\ \hline
\textbf{TPM (ms)}                       & 43,56 $\pm 0,32$    & 54,03 $\pm 0,58$  & 51,18 $\pm 0,58$   \\ \hline
\end{tabular}
\end{table}

\begin{table}[!h]
\centering
\caption{Análise comparativa entre o desempenho da HRAdNet-VE e os das demais redes do grupo reativo, utilizando os resultados o experimento 1.}
\label{tab:cen2_exp1_reativo}
\small\begin{tabular}{l|l|l|l|l|l|l|}
\cline{2-7}
                                        & \multicolumn{2}{c|}{\textbf{IEEE 802.11n}}      & \multicolumn{2}{c|}{\textbf{IEEE 802.11p}}                         & \multicolumn{2}{c|}{\textbf{LTE}}               \\ \hline
\multicolumn{1}{|l|}{\textbf{Métricas}} & \textbf{RAdNet}        & \textbf{CCN$_R$}       & \textbf{RAdNet}        & \textbf{CCN$_R$}                          & \textbf{RAdNet}        & \textbf{CCN$_R$}       \\ \hline
\multicolumn{1}{|l|}{\textbf{CMT}}      & \textless 46,38\%      & \textless 23,69\%      & \textless 57,89\%      & \textless 44,81\%                         & \textless 57,5\%       & \textless 12,06\%      \\ \hline
\multicolumn{1}{|l|}{\textbf{LCN}}      & \textless 70,59\%      & \textless 56,69\%      & \textless 25,76\%      & \textless 10,07\%                         & \textless 19,37\%      & \textless 14,88\%      \\ \hline
\multicolumn{1}{|l|}{\textbf{TED}}      & \textgreater 7,41\%     & \textgreater 4,26\%    & \textgreater 31,54\%   & \textgreater 103,93\%                      & \textgreater 0,01\%    & \textgreater 0,008\%   \\ \hline
\multicolumn{1}{|l|}{\textbf{NS}}       & \multicolumn{1}{c|}{=} & \multicolumn{1}{c|}{=} & \multicolumn{1}{c|}{=} & \multicolumn{1}{c|}{=}                    & \multicolumn{1}{c|}{=} & \multicolumn{1}{c|}{=} \\ \hline
\multicolumn{1}{|l|}{\textbf{AM}}       & \textgreater 0,08\%    & \textgreater 0,04\%    & \textgreater 0,005\%   & \multicolumn{1}{c|}{=} & \multicolumn{1}{c|}{=} & \multicolumn{1}{c|}{=} \\ \hline
\multicolumn{1}{|l|}{\textbf{TPM}}      & \textless 75,37\%      & \textless 59,54\%      & \textless 72,7\%       & \textless 59,33\%                         & \textless 9,12\%       & \textless 4,06\%       \\ \hline
\end{tabular}
\end{table}

\newpage

\subsubsection{Experimento 1 com Grupo Proativo}

A seguir, as Tabelas \ref{tab:res_cen2_exp1_proativo_ieee80211n}, \ref{tab:res_cen2_exp1_proativo_ieee80211p} e \ref{tab:res_cen2_exp1_proativo_lte} apresentam os resultados do grupo proativo e, em seguida, a Tabela 
\ref{tab:cen2_exp1_proativo} apresenta a análise comparativa destes.

\begin{table}[!h]
\centering
\caption{Resultados do grupo proativo no experimento 1  do cenário 2, utilizando interfaces de acesso à comunicação sem fio baseadas no padrão IEEE 802.11n}
\label{tab:res_cen2_exp1_proativo_ieee80211n}
\small\begin{tabular}{|l|c|c|c|}
\hline
\multicolumn{1}{|c|}{\textbf{Métricas}} & \textbf{HRAdNet-VE} & \textbf{RAdNet}    & \textbf{CCN$_P$}   \\ \hline
\textbf{CMT (msgs.)}                    & 8,23 $\times 10^8$  & 9,14 $\times 10^8$ & 7,34 $\times 10^8$ \\ \hline
\textbf{LCN (ms)}                       & 9,4 $\pm 0,06$      & 55,9 $\pm 0,97$    & 23,57 $\pm 0,19$   \\ \hline
\textbf{TED (\%)}                       & 97,55 $\pm 0,11$    & 90,73 $\pm 0,1$    & 90,27 $\pm 0,1$    \\ \hline
\textbf{NS (saltos)}                    & 7                   & 7                  & 7                  \\ \hline
\textbf{AM (m)}                         & 1349 $\pm 16,51$    & 1341 $\pm 19,27$   & 1346 $\pm 17,63$   \\ \hline
\textbf{TPM (ms)}                       & 98,62 $\pm 10,45$   & 403 $\pm 43,11$    & 176,4 $\pm 18,94$  \\ \hline
\end{tabular}
\end{table}

\begin{table}[!h]
\centering
\caption{Resultados do grupo proativo no experimento 1  do cenário 2, utilizando interfaces de acesso à comunicação sem fio baseadas no padrão IEEE 802.11p}
\label{tab:res_cen2_exp1_proativo_ieee80211p}
\small\begin{tabular}{|l|c|c|c|}
\hline
\multicolumn{1}{|c|}{\textbf{Métricas}} & \textbf{HRAdNet-VE} & \textbf{RAdNet}    & \textbf{CCN$_P$}   \\ \hline
\textbf{CMT (msgs.)}                    & 3,52 $\times 10^8$  & 7,58 $\times 10^8$ & 6,39 $\times 10^8$ \\ \hline
\textbf{LCN (ms)}                       & 14,33 $\pm 0,39$    & 22,9 $\pm 0,62$    & 16,49 $\pm 0,42$   \\ \hline
\textbf{TED (\%)}                       & 85,51 $\pm 1,7$     & 63,24 $\pm 1,23$   & 80,94 $\pm 1,6$    \\ \hline
\textbf{NS (saltos)}                    & 4                   & 4                  & 4                  \\ \hline
\textbf{AM (m)}                         & 1978 $\pm 7,18$     & 1976 $\pm 7,83$    & 1977 $\pm 7,51$    \\ \hline
\textbf{TPM (ms)}                       & 93,23 $\pm 9,47$    & 180 $\pm 18,61$    & 97,33 $\pm 9,79$   \\ \hline
\end{tabular}
\end{table}

\begin{table}[!h]
\centering
\caption{Resultados do grupo proativo no experimento 1  do cenário 2, utilizando interfaces de acesso à comunicação sem fio baseadas no padrão LTE}
\label{tab:res_cen2_exp1_proativo_lte}
\small\begin{tabular}{|l|c|c|c|}
\hline
\multicolumn{1}{|c|}{\textbf{Métricas}} & \textbf{HRAdNet-VE} & \textbf{RAdNet}   & \textbf{CCN$_P$}  \\ \hline
\textbf{CMT (msgs.)}                    & 1,5 $\times 10^7$   & 2,1 $\times 10^7$ & 1,6 $\times 10^7$ \\ \hline
\textbf{LCN (ms)}                       & 43,44 $\pm 0,58$    & 44,96 $\pm 0,58$  & 43,48 $\pm 0,65$  \\ \hline
\textbf{TED (\%)}                       & 99 $\pm 0,3$        & 98,92 $\pm 0,3$   & 98,85 $\pm 0,3$   \\ \hline
\textbf{NS (saltos)}                    & 1                   & 1                 & 1                 \\ \hline
\textbf{AM (m)}                         & 2500                & 2500              & 2500              \\ \hline
\textbf{TPM (ms)}                       & 43,44 $\pm 0,58$    & 44,96 $\pm 0,58$  & 43,48 $\pm 0,65$  \\ \hline
\end{tabular}
\end{table}

\begin{table}[!h]
\centering
\caption{Análise comparativa entre o desempenho da HRAdNet-VE e os das demais redes do grupo proativo, utilizando os resultados do experimento 1.}
\label{tab:cen2_exp1_proativo}
\small\begin{tabular}{l|l|l|l|l|l|l|}
\cline{2-7}
                                        & \multicolumn{2}{c|}{\textbf{IEEE 802.11n}}      & \multicolumn{2}{c|}{\textbf{IEEE 802.11p}}                         & \multicolumn{2}{c|}{\textbf{LTE}}               \\ \hline
\multicolumn{1}{|l|}{\textbf{Métricas}} & \textbf{RAdNet}        & \textbf{CCN$_P$}       & \textbf{RAdNet}        & \textbf{CCN$_P$}                          & \textbf{RAdNet}        & \textbf{CCN$_P$}       \\ \hline
\multicolumn{1}{|l|}{\textbf{CMT}}      & \textless 9,95\%      & \textgreater 12,12\%      & \textless 53,56\%      & \textless 44,91\%                         & \textless 57,5\%       & \textless 12,06\%      \\ \hline
\multicolumn{1}{|l|}{\textbf{LCN}}      & \textless 83,18\%      & \textless 60,11\%      & \textless 37,42\%      & \textless 13,09\%                         & \textless 3,38\%       & \textless 0,009\%      \\ \hline
\multicolumn{1}{|l|}{\textbf{TED}}      & \textgreater 7,51\%    & \textgreater 8,06\%    & \textgreater 35,21\%   & \textgreater 5,64\%                       & \textgreater 0,008\%   & \textgreater 0,01\%    \\ \hline
\multicolumn{1}{|l|}{\textbf{NS}}       & \multicolumn{1}{c|}{=} & \multicolumn{1}{c|}{=} & \multicolumn{1}{c|}{=} & \multicolumn{1}{c|}{=}                    & \multicolumn{1}{c|}{=} & \multicolumn{1}{c|}{=} \\ \hline
\multicolumn{1}{|l|}{\textbf{AM}}       & \textgreater 0,05\%    & \textgreater 0,02\%    & \textgreater 0,01\%    & \multicolumn{1}{c|}{\textgreater 0,005\%} & \multicolumn{1}{c|}{=} & \multicolumn{1}{c|}{=} \\ \hline
\multicolumn{1}{|l|}{\textbf{TPM}}      & \textless 75,52\%      & \textless 44,09\%      & \textless 48,2\%       & \textless 4,21\%                          & \textless 3,38\%       & \textless 1,06\%       \\ \hline
\end{tabular}
\end{table}

\newpage

\subsubsection{Experimento 2 com Grupo Reativo}

A seguir, as Tabelas \ref{tab:res_cen2_exp2_reativo_ieee80211n}, \ref{tab:res_cen2_exp2_reativo_ieee80211p} e \ref{tab:res_cen2_exp2_reativo_lte} apresentam os resultados do grupo reativo e, em seguida, a Tabela  \ref{tab:cen2_exp2_reativo} apresenta a análise comparativa destes.

\begin{table}[!h]
\centering
\caption{Resultados do grupo reativo no experimento 2  do cenário 2, utilizando interfaces de acesso à comunicação sem fio baseadas no padrão IEEE 802.11n}
\label{tab:res_cen2_exp2_reativo_ieee80211n}
\small\begin{tabular}{|l|c|c|c|}
\hline
\multicolumn{1}{|c|}{\textbf{Métricas}} & \textbf{HRAdNet-VE} & \textbf{RAdNet}    & \textbf{CCN$_R$}   \\ \hline
\textbf{CMT (msgs.)}                    & 1,286 $\times 10^9$ & 6,31 $\times 10^9$ & 6,12 $\times 10^9$ \\ \hline
\textbf{LCN (ms)}                       & 29,95 $\pm 0,52$    & 112 $\pm 1,95$     & 88,21 $\pm 0,71$   \\ \hline
\textbf{TED (\%)}                       & 90,75 $\pm 0,13$    & 80,7 $\pm 0,12$    & 80,89 $\pm 0,12$   \\ \hline
\textbf{NS (saltos)}                    & 7                   & 7                  & 7                  \\ \hline
\textbf{AM (m)}                         & 1345 $\pm 17,96$    & 1333 $\pm 21,88$   & 1336 $\pm 20,9$    \\ \hline
\textbf{TPM (ms)}                       & 239 $\pm 27,76$     & 898 $\pm 105,18$   & 682 $\pm 80,68$    \\ \hline
\end{tabular}
\end{table}

\begin{table}[!h]
\centering
\caption{Resultados do grupo reativo no experimento 2  do cenário 2, utilizando interfaces de acesso à comunicação sem fio baseadas no padrão IEEE 802.11p}
\label{tab:res_cen2_exp2_reativo_ieee80211p}
\small\begin{tabular}{|l|c|c|c|}
\hline
\multicolumn{1}{|c|}{\textbf{Métricas}} & \textbf{HRAdNet-VE} & \textbf{RAdNet}    & \textbf{CCN$_R$}    \\ \hline
\textbf{CMT (msgs.)}                    & 1,738 $\times 10^9$ & 5,93 $\times 10^9$ & 2,797 $\times 10^9$ \\ \hline
\textbf{LCN (ms)}                       & 23,02 $\pm 0,58$    & 35,88 $\pm 0,58$   & 26,31 $\pm 0,48$    \\ \hline
\textbf{TED (\%)}                       & 49,31 $\pm 1,04$    & 42,12 $\pm 0,89$   & 28,84 $\pm 0,61$    \\ \hline
\textbf{NS (saltos)}                    & 4                   & 4                  & 4                   \\ \hline
\textbf{AM (m)}                         & 1976 $\pm 7,83$     & 1974 $\pm 8,49$    & 1976 $\pm 7,83$     \\ \hline
\textbf{TPM (ms)}                       & 152 $\pm 8,81$      & 766 $\pm 68,27$    & 690 $\pm 22,53$     \\ \hline
\end{tabular}
\end{table}

\begin{table}[!h]
\centering
\caption{Resultados do grupo reativo no experimento 2  do cenário 2, utilizando interfaces de acesso à comunicação sem fio baseadas no padrão LTE}
\label{tab:res_cen2_exp2_reativo_lte}
\small\begin{tabular}{|l|c|c|c|}
\hline
\multicolumn{1}{|c|}{\textbf{Métricas}} & \textbf{HRAdNet-VE} & \textbf{RAdNet}    & \textbf{CCN$_R$}  \\ \hline
\textbf{CMT (msgs.)}                    & 3,22 $\times 10^7$  & 8,02 $\times 10^7$ & 3,4 $\times 10^7$ \\ \hline
\textbf{LCN (ms)}                       & 50,4 $\pm 0,68$     & 61,34 $\pm 0,65$   & 52,6 $\pm 0,65$   \\ \hline
\textbf{TED (\%)}                       & 98,14 $\pm 0,03$    & 97,97 $\pm 0,03$   & 98,01 $\pm 0,03$  \\ \hline
\textbf{NS (saltos)}                    & 1                   & 1                  & 1                 \\ \hline
\textbf{AM (m)}                         & 2500                & 2500               & 2500              \\ \hline
\textbf{TPM (ms)}                       & 50,4 $\pm 0,68$     & 61,34 $\pm 0,65$   & 52,6 $\pm 0,65$   \\ \hline
\end{tabular}
\end{table}

\begin{table}[!h]
\centering
\caption{Análise comparativa entre o desempenho da HRAdNet-VE e os desempenhos das demais redes do grupo reativo, utilizando os dados obtidos no experimento 2 do cenário 2.}
\label{tab:cen2_exp2_reativo}
\small\begin{tabular}{l|l|l|l|l|l|l|}
\cline{2-7}
                                        & \multicolumn{2}{c|}{\textbf{IEEE 802.11n}}      & \multicolumn{2}{c|}{\textbf{IEEE 802.11p}}                        & \multicolumn{2}{c|}{\textbf{LTE}}               \\ \hline
\multicolumn{1}{|l|}{\textbf{Métricas}} & \textbf{RAdNet}        & \textbf{CCN$_R$}       & \textbf{RAdNet}        & \textbf{CCN$_R$}                         & \textbf{RAdNet}        & \textbf{CCN$_R$}       \\ \hline
\multicolumn{1}{|l|}{\textbf{CMT}}      & \textless 79,61\%      & \textless 78,98\%      & \textless 70,79\%      & \textless 37,86\%                        & \textless 59,85\%      & \textless 5,29\%       \\ \hline
\multicolumn{1}{|l|}{\textbf{LCN}}      & \textless 73,25\%      & \textless 66,04\%      & \textless 35,84\%      & \textless 12,5                           & \textless 17,83\%      & \textless 4,18\%       \\ \hline
\multicolumn{1}{|l|}{\textbf{TED}}      & \textgreater 12,45\%  & \textgreater 12,21\%   & \textgreater 17,07\%   & \textgreater 70,97\%                     & \textgreater 0,01\%    & \textgreater 0,01\%    \\ \hline
\multicolumn{1}{|l|}{\textbf{NS}}       & \multicolumn{1}{c|}{=} & \multicolumn{1}{c|}{=} & \multicolumn{1}{c|}{=} & \multicolumn{1}{c|}{=}                   & \multicolumn{1}{c|}{=} & \multicolumn{1}{c|}{=} \\ \hline
\multicolumn{1}{|l|}{\textbf{AM}}       & \textgreater 0,09\%    & \textgreater 0,06\%    & \textgreater 0,01\%    & \multicolumn{1}{c|}{=} & \multicolumn{1}{c|}{=} & \multicolumn{1}{c|}{=} \\ \hline
\multicolumn{1}{|l|}{\textbf{TPM}}      & \textless 73,38\%      & \textless 64,95\%      & \textless 80,15\%      & \textless 77,97\%                        & \textless 17,83\%      & \textless 4,18\%       \\ \hline
\end{tabular}
\end{table}

\newpage

\subsubsection{Experimento 2 com Grupo Proativo}

A seguir, as Tabelas \ref{tab:res_cen2_exp2_proativo_ieee80211n}, \ref{tab:res_cen2_exp2_proativo_ieee80211p} e \ref{tab:res_cen2_exp2_proativo_lte} 
apresentam os resultados do grupo proativo e, em seguida, a Tabela \ref{tab:cen2_exp2_proativo} apresenta a análise comparativa destes.

\begin{table}[!h]
\centering
\caption{Resultados do grupo proativo no experimento 2  do cenário 2, utilizando interfaces de acesso à comunicação sem fio baseadas no padrão IEEE 802.11n}
\label{tab:res_cen2_exp2_proativo_ieee80211n}
\small\begin{tabular}{|l|c|c|c|}
\hline
\multicolumn{1}{|c|}{\textbf{Métricas}} & \textbf{HRAdNet-VE} & \textbf{RAdNet}     & \textbf{CCN$_P$}    \\ \hline
\textbf{CMT (msgs.)}                    & 1,121 $\times 10^9$ & 1,493 $\times 10^9$ & 1,235 $\times 10^9$ \\ \hline
\textbf{LCN (ms)}                       & 9,6 $\pm 0,16$      & 60,24 $\pm 1,04$    & 25,00 $\pm 0,42$    \\ \hline
\textbf{TED (\%)}                       & 97,94 $\pm 0,15$    & 92,61 $\pm 0,14$    & 96,56 $\pm 0,14$    \\ \hline
\textbf{NS (saltos)}                    & 7                   & 7                   & 7                   \\ \hline
\textbf{AM (m)}                         & 1349 $\pm 16,65$    & 1341 $\pm 19,27$    & 1346 $\pm 17,63$    \\ \hline
\textbf{TPM (ms)}                       & 96,2 $\pm 4,63$     & 450 $\pm 52,59$     & 196 $\pm 1,95$      \\ \hline
\end{tabular}
\end{table}

\begin{table}[!h]
\centering
\caption{Resultados do grupo proativo no experimento 2  do cenário 2, utilizando interfaces de acesso à comunicação sem fio baseadas no padrão IEEE 802.11p}
\label{tab:res_cen2_exp2_proativo_ieee80211p}
\small\begin{tabular}{|l|c|c|c|}
\hline
\multicolumn{1}{|c|}{\textbf{Métricas}} & \textbf{HRAdNet-VE} & \textbf{RAdNet}     & \textbf{CCN$_P$}    \\ \hline
\textbf{CMT (msgs.)}                    & 7,49 $\times 10^8$  & 1,858 $\times 10^9$ & 1,013 $\times 10^9$ \\ \hline
\textbf{LCN (ms)}                       & 12,4 $\pm 0,32$     & 16,69 $\pm 0,42$    & 15,18 $\pm 0,39$    \\ \hline
\textbf{TED (\%)}                       & 84,31 $\pm 1,79$    & 66,48 $\pm 1,41$    & 96,56 $\pm 1,41$    \\ \hline
\textbf{NS (saltos)}                    & 4                   & 4                   & 4                   \\ \hline
\textbf{AM (m)}                         & 1978 $\pm 7,18$     & 1977 $\pm 7,51$     & 1978 $\pm 7,18$     \\ \hline
\textbf{TPM (ms)}                       & 89 $\pm 8,85$       & 178 $\pm 18,29$     & 105 $\pm 10,45$     \\ \hline
\end{tabular}
\end{table}

\begin{table}[!h]
\centering
\caption{Resultados do grupo proativo no experimento 2  do cenário 2, utilizando interfaces de acesso à comunicação sem fio baseadas no padrão LTE}
\label{tab:res_cen2_exp2_proativo_lte}
\small\begin{tabular}{|l|c|c|c|}
\hline
\multicolumn{1}{|c|}{\textbf{Métricas}} & \textbf{HRAdNet-VE} & \textbf{RAdNet}    & \textbf{CCN$_P$}   \\ \hline
\textbf{CMT (msgs.)}                    & 1,94 $\times 10^7$  & 2,25 $\times 10^7$ & 2,07 $\times 10^7$ \\ \hline
\textbf{LCN (ms)}                       & 46,2 $\pm 0,65$     & 60 $\pm 0,68$      & 51 $\pm 0,71$      \\ \hline
\textbf{TED (\%)}                       & 99,00 $\pm 0,03$    & 98,94 $\pm 0,03$   & 98,98 $\pm 1,41$   \\ \hline
\textbf{NS (saltos)}                    & 1                   & 1                  & 1                  \\ \hline
\textbf{AM (m)}                         & 2500                & 2500               & 2500               \\ \hline
\textbf{TPM (ms)}                       & 46,2 $\pm 0,65$     & 60 $\pm 0,68$      & 51 $\pm 0,71$      \\ \hline
\end{tabular}
\end{table}

\begin{table}[!h]
\centering
\caption{Análise comparativa entre o desempenho da HRAdNet-VE e os desempenhos das demais redes do grupo proativo, utilizando os dados obtidos no experimento 2 do cenário 2.}
\label{tab:cen2_exp2_proativo}
\small\begin{tabular}{l|l|l|l|l|l|l|}
\cline{2-7}
                                        & \multicolumn{2}{c|}{\textbf{IEEE 802.11n}}      & \multicolumn{2}{c|}{\textbf{IEEE 802.11p}}      & \multicolumn{2}{c|}{\textbf{LTE}}               \\ \hline
\multicolumn{1}{|l|}{\textbf{Métricas}} & \textbf{RAdNet}        & \textbf{CCN$_P$}       & \textbf{RAdNet}        & \textbf{CCN$_P$}       & \textbf{RAdNet}        & \textbf{CCN$_P$}       \\ \hline
\multicolumn{1}{|l|}{\textbf{CMT}}      & \textless 24,9\%       & \textless 9,23\%       & \textless 59,68\%      & \textless 26,06\%      & \textless 13,77\%      & \textless 6,28\%       \\ \hline
\multicolumn{1}{|l|}{\textbf{LCN}}      & \textless 84,06\%      & \textless 61,6\%       & \textless 25,7\%       & \textless 25,68\%      & \textless 23\%         & \textless 9,41\%       \\ \hline
\multicolumn{1}{|l|}{\textbf{TED}}      & \textgreater 5,53\%    & \textgreater 1,22\%     & \textgreater 26,82\%   & \textgreater 20,43\%   & \textgreater 0,006\%   & \textgreater 0,002\%   \\ \hline
\multicolumn{1}{|l|}{\textbf{NS}}       & \multicolumn{1}{c|}{=} & \multicolumn{1}{c|}{=} & \multicolumn{1}{c|}{=} & \multicolumn{1}{c|}{=} & \multicolumn{1}{c|}{=} & \multicolumn{1}{c|}{=} \\ \hline
\multicolumn{1}{|l|}{\textbf{AM}}       & \textgreater 0,005\%    & \textgreater 0,02\%    & \textgreater 0,005\%    & \multicolumn{1}{c|}{=} & \multicolumn{1}{c|}{=} & \multicolumn{1}{c|}{=} \\ \hline
\multicolumn{1}{|l|}{\textbf{TPM}}      & \textless 78,62\%      & \textless 50,91\%      & \textless 50\%         & \textless 15,23\%      & \textless 23\%         & \textless 9,41\%       \\ \hline
\end{tabular}
\end{table}

\newpage

\subsubsection{Experimento 3 com Grupo Reativo}

A seguir, as Tabelas \ref{tab:res_cen2_exp3_reativo_ieee80211n}, \ref{tab:res_cen2_exp3_reativo_ieee80211p} e \ref{tab:res_cen2_exp3_reativo_lte} 
apresentam os resultados do grupo reativo e, em seguida, a Tabela \ref{tab:cen2_exp3_reativo} apresenta a análise comparativa destes.

\begin{table}[!h]
\centering
\caption{Resultados do grupo reativo no experimento 3  do cenário 2, utilizando interfaces de acesso à comunicação sem fio baseadas no padrão IEEE 802.11n}
\label{tab:res_cen2_exp3_reativo_ieee80211n}
\small\begin{tabular}{|l|c|c|c|}
\hline
\multicolumn{1}{|c|}{\textbf{Métricas}} & \textbf{HRAdNet-VE} & \textbf{RAdNet}     & \textbf{CCN$_R$}   \\ \hline
\textbf{CMT (msgs.)}                    & 2,288 $\times 10^9$ & 8,439 $\times 10^9$ & 5,83 $\times 10^9$ \\ \hline
\textbf{LCN (ms)}                       & 33,08 $\pm 0,58$    & 128 $\pm 2,28$      & 64,43 $\pm 1,11$   \\ \hline
\textbf{TED (\%)}                       & 91,12 $\pm 0,14$    & 86,64 $\pm 0,13$    & 84,96 $\pm 0,13$   \\ \hline
\textbf{NS (saltos)}                    & 7                   & 7                   & 7                  \\ \hline
\textbf{AM (m)}                         & 1575 $\pm 8,16$     & 1560 $\pm 13,06$    & 1570 $\pm 9,79$    \\ \hline
\textbf{TPM (ms)}                       & 193 $\pm 22,53$     & 1115 $\pm 130,33$   & 592 $\pm 69,25$    \\ \hline
\end{tabular}
\end{table}

\begin{table}[!h]
\centering
\caption{Resultados do grupo reativo no experimento 3  do cenário 2, utilizando interfaces de acesso à comunicação sem fio baseadas no padrão IEEE 802.11p}
\label{tab:res_cen2_exp3_reativo_ieee80211p}
\small\begin{tabular}{|l|c|c|c|}
\hline
\multicolumn{1}{|c|}{\textbf{Métricas}} & \textbf{HRAdNet-VE} & \textbf{RAdNet}     & \textbf{CCN$_R$}    \\ \hline
\textbf{CMT (msgs.)}                    & 4,115 $\times 10^9$ & 6,417 $\times 10^9$ & 4,496 $\times 10^9$ \\ \hline
\textbf{LCN (ms)}                       & 21,12 $\pm 0,52$    & 43,61 $\pm 0,58$    & 39,61 $\pm 0,48$    \\ \hline
\textbf{TED (\%)}                       & 49,35 $\pm 0,96$    & 31,5 $\pm 0,71$     & 20,41 $\pm 0,43$    \\ \hline
\textbf{NS (saltos)}                    & 4                   & 4                   & 4                   \\ \hline
\textbf{AM (m)}                         & 1987 $\pm 4,24$     & 1983 $\pm 5,55$     & 1984 $\pm 5,22$     \\ \hline
\textbf{TPM (ms)}                       & 160 $\pm 11,1$      & 974 $\pm 8,16$      & 781 $\pm 6,85$      \\ \hline
\end{tabular}
\end{table}

\begin{table}[!h]
\centering
\caption{Resultados do grupo reativo no experimento 3  do cenário 2, utilizando interfaces de acesso à comunicação sem fio baseadas no padrão LTE}
\label{tab:res_cen2_exp3_reativo_lte}
\small\begin{tabular}{|l|c|c|c|}
\hline
\multicolumn{1}{|c|}{\textbf{Métricas}} & \textbf{HRAdNet-VE} & \textbf{RAdNet}    & \textbf{CCN$_R$}   \\ \hline
\textbf{CMT (msgs.)}                    & 4,69 $\times 10^7$  & 1,09 $\times 10^8$ & 5,91 $\times 10^7$ \\ \hline
\textbf{LCN (ms)}                       & 64,71 $\pm 0,97$    & 79,38 $\pm 0,97$   & 69,53 $\pm 0,97$   \\ \hline
\textbf{TED (\%)}                       & 97,93 $\pm 0,32$    & 97,86 $\pm 0,32$   & 97,23 $\pm 0,32$   \\ \hline
\textbf{NS (saltos)}                    & 1                   & 1                  & 1                  \\ \hline
\textbf{AM (m)}                         & 2500                & 2500               & 2500               \\ \hline
\textbf{TPM (ms)}                       & 64,71 $\pm 0,97$    & 79,38 $\pm 0,97$   & 69,53 $\pm 0,97$   \\ \hline
\end{tabular}
\end{table}

\begin{table}[!h]
\centering
\caption{Análise comparativa entre o desempenho da HRAdNet-VE e os desempenhos das demais redes do grupo reativo, utilizando os dados obtidos no experimento 3 do cenário 2.}
\label{tab:cen2_exp3_reativo}
\small\begin{tabular}{l|l|l|l|l|l|l|}
\cline{2-7}
                                        & \multicolumn{2}{c|}{\textbf{IEEE 802.11n}}      & \multicolumn{2}{c|}{\textbf{IEEE 802.11p}}                        & \multicolumn{2}{c|}{\textbf{LTE}}               \\ \hline
\multicolumn{1}{|l|}{\textbf{Métricas}} & \textbf{RAdNet}        & \textbf{CCN$_R$}       & \textbf{RAdNet}        & \textbf{CCN$_R$}                         & \textbf{RAdNet}        & \textbf{CCN$_R$}       \\ \hline
\multicolumn{1}{|l|}{\textbf{CMT}}      & \textless 72,88\%      & \textless 60,75\%      & \textless 35,87\%      & \textless 8,47\%                         & \textless 52,5\%       & \textless 9,32\%       \\ \hline
\multicolumn{1}{|l|}{\textbf{LCN}}      & \textless 74,15\%      & \textless 48,65\%      & \textless 51,57\%      & \textless 46,68\%                        & \textless 18,48\%      & \textless 6,93\%       \\ \hline
\multicolumn{1}{|l|}{\textbf{TED}}      & \textgreater 5,17\%    & \textgreater 7,25\%    & \textgreater 56,66\%   & \textgreater 141,17\%                     & \textgreater 0,007\%   & \textgreater 0,07\%    \\ \hline
\multicolumn{1}{|l|}{\textbf{NS}}       & \multicolumn{1}{c|}{=} & \multicolumn{1}{c|}{=} & \multicolumn{1}{c|}{=} & \multicolumn{1}{c|}{=}                   & \multicolumn{1}{c|}{=} & \multicolumn{1}{c|}{=} \\ \hline
\multicolumn{1}{|l|}{\textbf{AM}}       & \textgreater 0,09\%    & \textgreater 0,03\%    & \textgreater 0,02\%    & \multicolumn{1}{c|}{\textgreater 0,01\%} & \multicolumn{1}{c|}{=} & \multicolumn{1}{c|}{=} \\ \hline
\multicolumn{1}{|l|}{\textbf{TPM}}      & \textless 82,69\%      & \textless 67,39\%      & \textless 83,1\%       & \textless 79,51\%                        & \textless 18,48\%      & \textless 6,93\%       \\ \hline
\end{tabular}
\end{table}

\newpage

\subsubsection{Experimento 3 com Grupo Proativo}

A seguir, as Tabelas \ref{tab:res_cen2_exp3_proativo_ieee80211n}, \ref{tab:res_cen2_exp3_proativo_ieee80211p} e \ref{tab:res_cen2_exp3_proativo_lte} 
apresentam os resultados do grupo reativo e, em seguida, a Tabela  \ref{tab:cen2_exp3_proativo} apresenta a análise comparativa destes.

\begin{table}[!h]
\centering
\caption{Resultados do grupo proativo no experimento 3  do cenário 2, utilizando interfaces de acesso à comunicação sem fio baseadas no padrão IEEE 802.11n}
\label{tab:res_cen2_exp3_proativo_ieee80211n}
\small\begin{tabular}{|l|c|c|c|}
\hline
\multicolumn{1}{|c|}{\textbf{Métricas}} & \textbf{HRAdNet-VE} & \textbf{RAdNet}    & \textbf{CCN$_P$}   \\ \hline
\textbf{CMT (msgs.)}                    & 4,69 $\times 10^9$  & 1,09 $\times 10^9$ & 5,91 $\times 10^9$ \\ \hline
\textbf{LCN (ms)}                       & 9,21 $\pm 0,13$     & 95,82 $\pm 1,66$   & 40,47 $pm 0,71$   \\ \hline
\textbf{TED (\%)}                       & 97,02 $\pm 0,14$    & 77,65 $\pm 0,11$   & 79,29 $\pm 0,11$   \\ \hline
\textbf{NS (saltos)}                    & 7                   & 7                  & 7                  \\ \hline
\textbf{AM (m)}                         & 1579 $\pm 6,85$     & 1565 $\pm 11,43$   & 1574 $\pm 8,49$    \\ \hline
\textbf{TPM (ms)}                       & 116 $\pm 13,39$     & 825 $\pm 96,69$    & 449 $\pm 52,59$    \\ \hline
\end{tabular}
\end{table}

\begin{table}[!h]
\centering
\caption{Resultados do grupo proativo no experimento 3  do cenário 2, utilizando interfaces de acesso à comunicação sem fio baseadas no padrão IEEE 802.11p}
\label{tab:res_cen2_exp3_proativo_ieee80211p}
\small\begin{tabular}{|l|c|c|c|}
\hline
\multicolumn{1}{|c|}{\textbf{Métricas}} & \textbf{HRAdNet-VE} & \textbf{RAdNet}     & \textbf{CCN$_P$}    \\ \hline
\textbf{CMT (msgs.)}                    & 1,657 $\times 10^9$ & 3,506 $\times 10^9$ & 2,106 $\times 10^9$ \\ \hline
\textbf{LCN (ms)}                       & 14,36 $\pm 0,37$    & 23,68 $\pm 0,58$    & 18,85 $\pm 0,48$    \\ \hline
\textbf{TED (\%)}                       & 83,39 $\pm 1,75$    & 66,71 $\pm 1,41$    & 72,63 $\pm 1,5$     \\ \hline
\textbf{NS (saltos)}                    & 4                   & 4                   & 4                   \\ \hline
\textbf{AM (m)}                         & 1988 $\pm 3,91$     & 1986 $\pm 4,57$     & 1987 $\pm 4,24$     \\ \hline
\textbf{TPM (ms)}                       & 74,11 $\pm 6,53$    & 188 $\pm 16,65$     & 95,97 $\pm 8,49$    \\ \hline
\end{tabular}
\end{table}

\begin{table}[!h]
\centering
\caption{Resultados do grupo proativo no experimento 3  do cenário 2, utilizando interfaces de acesso à comunicação sem fio baseadas no padrão LTE}
\label{tab:res_cen2_exp3_proativo_lte}
\small\begin{tabular}{|l|c|c|c|}
\hline
\multicolumn{1}{|c|}{\textbf{Métricas}} & \textbf{HRAdNet-VE} & \textbf{RAdNet}    & \textbf{CCN$_P$}   \\ \hline
\textbf{CMT (msgs.)}                    & 3,63 $\times 10^7$  & 6,31 $\times 10^7$ & 4,72 $\times 10^7$ \\ \hline
\textbf{LCN (ms)}                       & 44,32 $\pm 0,58$    & 66,86 $\pm 0,78$   & 49,71 $\pm 0,658$  \\ \hline
\textbf{TED (\%)}                       & 98,95 $\pm 0,03$    & 97,88 $\pm 0,03$   & 97,93 $\pm 0,03$   \\ \hline
\textbf{NS (saltos)}                    & 1                   & 1                  & 1                  \\ \hline
\textbf{AM (m)}                         & 2500                & 2500               & 2500               \\ \hline
\textbf{TPM (ms)}                       & 44,32 $\pm 0,58$    & 66,86 $\pm 0,78$   & 49,71 $\pm 0,658$  \\ \hline
\end{tabular}
\end{table}

\begin{table}[!h]
\centering
\caption{Análise comparativa entre o desempenho da HRAdNet-VE e os desempenhos das demais redes do grupo proativo, utilizando os dados obtidos no experimento 3 do cenário 2.}
\label{tab:cen2_exp3_proativo}
\small\begin{tabular}{l|l|l|l|l|l|l|}
\cline{2-7}
                                        & \multicolumn{2}{c|}{\textbf{IEEE 802.11n}}      & \multicolumn{2}{c|}{\textbf{IEEE 802.11p}}                         & \multicolumn{2}{c|}{\textbf{LTE}}               \\ \hline
\multicolumn{1}{|l|}{\textbf{Métricas}} & \textbf{RAdNet}        & \textbf{CCN$_P$}       & \textbf{RAdNet}        & \textbf{CCN$_P$}                          & \textbf{RAdNet}        & \textbf{CCN$_P$}       \\ \hline
\multicolumn{1}{|l|}{\textbf{CMT}}      & \textless 27,52\%      & \textless 25,96\%      & \textless 52,73\%      & \textless 21,32\%                         & \textless 27,57\%      & \textless 3,17\%       \\ \hline
\multicolumn{1}{|l|}{\textbf{LCN}}      & \textless 90,38\%      & \textless 77,24\%      & \textless 39,35\%      & \textless 27,65\%                         & \textless 33,71\%      & \textless 10,84\%      \\ \hline
\multicolumn{1}{|l|}{\textbf{TED}}      & \textgreater 24,94\%   & \textgreater 22,36\%   & \textgreater 25\%      & \textgreater 14,81\%                       & \textgreater 1,09\%    & \textgreater 1,04\%    \\ \hline
\multicolumn{1}{|l|}{\textbf{NS}}       & \multicolumn{1}{c|}{=} & \multicolumn{1}{c|}{=} & \multicolumn{1}{c|}{=} & \multicolumn{1}{c|}{=}                    & \multicolumn{1}{c|}{=} & \multicolumn{1}{c|}{=} \\ \hline
\multicolumn{1}{|l|}{\textbf{AM}}       & \textgreater 0,08\%    & \textgreater 0,03\%    & \textgreater 0,01\%    & \multicolumn{1}{c|}{\textgreater 0,005\%} & \multicolumn{1}{c|}{=} & \multicolumn{1}{c|}{=} \\ \hline
\multicolumn{1}{|l|}{\textbf{TPM}}      & \textless 85,93\%      & \textless 0,25,83\%    & \textless 60,57\%      & \textless 22,45\%                         & \textless 33,71\%      & \textless 10,84\%      \\ \hline
\end{tabular}
\end{table}

\section{Avaliando o Sistema Multiagente de Controle de Tráfego}\label{sec:trafego} 

Nesta seção, é apresentada a avaliação experimental acerca do sistema multiagente de controle de tráfego. Para tanto, primeiramente, é realizada uma descrição dos cenários adotados utilizados para os experimentos. Em seguida, são apresentadas as configurações adotadas nos experimentos. Por fim, são apresentados os resultados e as análises comparativas sobre estes. 

\subsection{Descrição dos Cenários para os Experimentos}

Para a obtenção dos resultados referentes ao sistema multiagente de controle de tráfego, foram criados dois cenários: operação do sistema multiagente de controle de tráfego com sinalizações semafóricas em total funcionamento; e operação do sistema de controle de tráfego com sinalizações semafóricas apresentando ausência de funcionamento. Para cada um destes cenários, foram criados dois subcenários, a saber: controle de tráfego em interseções isoladas (SMER$_I$); e controle de tráfego envolvendo interseções isoladas mais controle de sistemas coordenados de sinalizações semafóricas (SMER$_{I+C}$). 

Em ambos os cenários, foram utilizados os seguintes agentes definidos nesta tese: Veículo, Sinalização Semafórica e Centro de Controle de Tráfego. Com estes cenários, pretendeu-se avaliar não somente o desempenho das estratégias de controle de tráfego definidas anteriormente, mas também comparar os resultados obtidos por meio dos experimentos realizados com elas. Neste sentido, no primeiro cenário tanto SMER$_I$ quanto SMER$_{I+C}$ tiveram seus desempenhos comparados contra um sistema de controle de tráfego baseado em sinalizações semafóricas. Em seguida, os resultados obtidos no primeiro cenário, foram comparados com os resultados do segundo cenário, a fim de comparar o desempenho do sistema multiagente de controle de tráfego em dois cenários distintos, ou seja, comparar os resultados obtidos com sinalizações semafóricas em total funcionamento contra os obtidos com sinalizações semafóricas apresentando ausências de funcionamento.

Para os dois cenários, foram utilizadas a grades manhattan 10 x 10, que foram utilizadas nos experimentos com a HRAdNet. Estas grades podem ser vistas nas Figuras \ref{fig:grade_1} e \ref{fig:grade_2}. 

\subsection{Configurações dos Experimentos}

Os tempos de simulação dos cenários descritos na seção anterior foram fixados em 3600s. No dois cenários, os veículos conectados e sinalizações semafóricas foram equipados com interfaces de acesso à comunicação sem fio baseadas no padrão IEEE 802.11n. Estas interfaces foram configuradas, seguindo os valores dos parâmetros listados nas Tabelas \ref{tab:config_IEEE_80211n} e \ref{tab:config_MAC_IEEE_80211n}. Além disto, veículos conectados, sinalizações semafóricas e o centro de controle de tráfego foram equipadas com interfaces de acesso à comunicação sem fio baseadas no padrão LTE. De acordo com as definições anteriores, veículos conectados e sinalizações semafóricas são nós UE. Por isto, suas interfaces de acesso a comunicação sem fio baseadas no padrão LTE foram configuradas, de acordo com os valores dos parâmetros listados na Tabela \ref{tab:LTEUEsettings}. No que diz respeito ao centro de controle de tráfego, este atuou como um nó eNodeB. Portanto, sua interface de acesso a comunicação sem fio baseada no padrão LTE foi configurada, de acordo com os valores dos parâmetros listados na Tabela \ref{tab:LTEeNodeBsettings}. No que tange as configurações da camada de rede, todos os nós foram configuradas com o protocolo de comunicação da HRAdNet-VE. Por esta razão, as configurações de interesses utilizadas nesses cenários foram  aquelas definidas no Capítulo \ref{cap:controle}.

Para os dois subcenários, foram adotadas as mesmas configurações dos agentes Sinalização Semafórica descritas na seção anterior, ou seja, eles foram configurados com os mesmos tamanhos de intervalos de indicações de luzes das sinalizações semafóricas e os valores dos parâmetros listados na Tabela \ref{tab:confPerfSinSemaf}. Em específico, no subcenário 2, os agentes Sinalização Semafórica também foram identificados unicamente, conforme explicado na seção anterior, assim como, os corredores e grupos de corredores. Conforme a Figura \ref{fig:identificacaoSinSem}, os agentes sinalizações semafóricas foram enumerados de 0 a 199 e os corredores foram identificados de A a D. Todos os agentes Sinalização Semafórica controladores de um sistema coordenado de sinalizações semafóricas foram configurados com tais identificadores de corredor. Além destes agentes, aqueles cujas sinalizações semafóricas integraram corredores também tiveram conhecimento das identificações de corredores. Todos os agentes controladores de sistemas coordenados de sinalizações semafóricas tiveram conhecimento das identificações dos agentes embutidos nas sinalizações semafóricas integrantes de corredores e as vias onde estes agentes se encontravam. A identificação dos grupos de corredores seguiu a mesma lógica descrita na seção anterior. No que diz respeito aos parâmetros exclusivos dos agentes Sinalização Semafórica controladores de um sistema coordenado de sinalizações semafóricas, os números mínimo e máximo e ciclos foram 10 e 30, respectivamente. 

No que diz respeito à configuração do comportamento dos veículos, em ambos os cenários o IDM também foi utilizado e teve seus parâmetros configurados com os mesmos valores utilizados nos experimentos descritos na seção anterior. Tais valores foram listados na Tabela \ref{tab:configuracao_IDM_RAdNetVE}. Além disto, os destinos das viagens realizadas pelo veículos foram escolhidos aleatoriamente e o algoritmo de roteamento de veículos adotado foi o algoritmo de caminho mínimo espacialmente mais curto.

Para cada um dos cenários e seus subcenários, foram criados três experimentos. Em cada um destes experimentos, as grades manhattam receberam diferentes fluxos de veículos em suas vias de entrada. Estes fluxos de veículos foram definidos, de acordo com as primeiras vias imediatamente à frente das vias de entrada existentes nas grades. As definições de fluxos de veículos nas vias de entrada das grades manhattan 10 x 10 seguiram os valores listados pela Tabela \ref{tab:vehicleflows}.

Para encontrar o melhor valor para o parâmetro número de obtenções de quantidade de veículos, no cenário 1, para cada experimento, variou-se o valor desse parâmetro, utilizando três valores: 18, 24 e 30. Dessa forma, com o valor do parâmetro periodicidade de obtenção de quantidade de veículos sendo 10s, os agentes Sinalização Semafórica atualizaram as demandas de suas vias de entrada a cada 3, 4 e 5 minutos. Esses três valores também influenciaram na configuração de parâmetros exclusivos dos agentes Sinalização Semafórica controladores de um sistema coordenado de sinalizações semafóricas. Dessa forma, de acordo com os valores citados anteriormente, os parâmetros periodicidade de compartilhamento de médias de quantidades de veículos e periodicidade de atualização de demandas de corredores foram configurados com os seguintes valores: 180, 240 e 300. 

Com base nos resultados obtidos com as variações de valores dos parâmetros citados acima, foi possível escolher o melhor valor para a realização dos experimentos do cenário 2. 

\subsection{Análise dos Resultados do Cenário 1}

Esta seção tem como objetivo apresentar os resultados obtidos por meio dos experimentos 1, 2 e 3 do cenário 1, assim como, as análises comparativas dos mesmos.

\subsubsection{Experimento 1}

A seguir, a Tabela \ref{tab:smer_cen1_exp1} apresenta os resultados dos experimento 1 do cenário 1 e, em seguida, a Tabela \ref{tab:comp_smer_cen1_exp1} apresenta a análise dos mesmos.

\begin{table}[!h]
\centering
\caption{Desempenhos do sistema de controle de tráfego baseado em sinalizações semafóricas pré-temporizadas, SMER$_I$ e SMER$_{I+C}$ no experimento 1 do cenário 1.}
\label{tab:smer_cen1_exp1}
\small\begin{tabular}{cc|p{1.0cm}|p{1.0cm}|p{1.0cm}|p{1.0cm}|p{1.0cm}|p{1.0cm}|}
\cline{3-8}
\multicolumn{1}{l}{}                     & \multicolumn{1}{l|}{}    & \multicolumn{3}{c|}{\textbf{SMER$_I$}}        & \multicolumn{3}{c|}{\textbf{SMER$_{I+C}$}}    \\ \hline
\multicolumn{1}{|c|}{\textbf{Métricas}}  & \textbf{Pré-Temp.} & \multicolumn{1}{c|}{\textbf{18}} & \multicolumn{1}{c|}{\textbf{24}} & \multicolumn{1}{c|}{\textbf{30}} & \multicolumn{1}{c|}{\textbf{18}} & \multicolumn{1}{c|}{\textbf{24}} & \multicolumn{1}{c|}{\textbf{30}} \\ \hline
\multicolumn{1}{|c|}{\textbf{TVS (veic./h)}}   & 3969               & 4163        & 4175        & 4200        & 4243        & 4262        & 4280          \\ \hline
\multicolumn{1}{|c|}{\textbf{TME (s)}}   & 14,67 $\pm 0,55$                   & 14,55 $\pm 0,51$      & 13,58 $\pm 0,51$      & 12,61 $\pm 0,49$      & 14,01 $\pm 0,48$      & 13,1 $\pm 0,48$       & 12,21 $\pm 0,47$           \\ \hline
\multicolumn{1}{|c|}{\textbf{TMV (s)}}   & 262,91 $\pm 9,89$                  & 266,92 $\pm 7,74$     & 265,83 $\pm 7,51$     & 261,8 $\pm 7,42$      & 259,77 $\pm 6,8$     & 257,23 $\pm 6,79$      & 254,00 $\pm 6,62$        \\ \hline
\multicolumn{1}{|c|}{\textbf{VM (km/h)}} & 21,01 $\pm 0,32$                   & 20,92 $\pm 0,32$      & 20,65 $\pm 0,32$      & 21,18 $\pm 0,32$      & 21,21 $\pm 0,48$      & 22,06 $\pm 0,48$      & 22,97 $\pm 0,48$        \\ \hline
\end{tabular}
\end{table}

\begin{table}[!h]
\centering
\caption{Comparação dos desempenhos de SMER$_I$ e SMER$_{I+C}$ contra o desempenho do sistema de controle de tráfego baseado em sinalizações semafóricas pré-temporizadas, utilizando os resultados obtidos no experimento 1 do cenário 1.}
\label{tab:comp_smer_cen1_exp1}
\small\begin{tabular}{c|c|c|c|c|c|c|}
\cline{2-7}
\multicolumn{1}{l|}{}                   & \multicolumn{3}{c|}{\textbf{SMER$_I$}}                          & \multicolumn{3}{c|}{\textbf{SMER$_{I+C}$}}                      \\ \hline
\multicolumn{1}{|c|}{\textbf{Métricas}} & \textbf{18}       & \textbf{24}       & \textbf{30}       & \textbf{18}       & \textbf{24}       & \textbf{30}       \\ \hline
\multicolumn{1}{|c|}{\textbf{TVS}}      & \textgreater 4,88\%  & \textgreater 5,19\% & \textgreater 5,82\% & \textgreater 6,9\%  & \textgreater 7,38\% & \textgreater 7,83\% \\ \hline
\multicolumn{1}{|c|}{\textbf{TME}}      & \textless 0,08\%    & \textless 7,43\%    & \textless 14,04\%   & \textless 4,49\%    & \textless 10,7\%    & \textless 16,76\%    \\ \hline
\multicolumn{1}{|c|}{\textbf{TMV}}      & \textgreater 1,52\% & \textgreater 1,11\% & \textless 0,04\%    & \textless 1,19\% & \textless 2,26\%   & \textless 3,38\%    \\ \hline
\multicolumn{1}{|c|}{\textbf{VM}}       & \textless 0,04\%    & \textless 1,71\%    & \textgreater 0,08\% & \textgreater 0,09\% & \textgreater 4,99\% & \textgreater 9,32\% \\ \hline
\end{tabular}
\end{table}

\newpage

\subsubsection{Experimento 2}

A seguir, a Tabela \ref{tab:smer_cen1_exp2} apresenta os resultados dos experimento 2 do cenário 1 e, em seguida, a Tabela \ref{tab:comp_smer_cen1_exp2} apresenta a análise dos mesmos.

\begin{table}[!h]
\centering
\caption{Resultados obtidos com o sistema de controle de tráfego baseado em sinalizações semafóricas pré-temporizadas, SMER$_I$ e SMER$_{I+C}$ no experimento 2 do cenário 1.}
\label{tab:smer_cen1_exp2}
\small\begin{tabular}{cc|p{1cm}|p{1cm}|p{1cm}|p{1cm}|p{1cm}|p{1cm}|}
\cline{3-8}
\multicolumn{1}{l}{}                         & \multicolumn{1}{l|}{}    & \multicolumn{3}{c|}{\textbf{SMER$_I$}}        & \multicolumn{3}{c|}{\textbf{SMER$_{I+C}$}}    \\ \hline
\multicolumn{1}{|c|}{\textbf{Métricas}}      & \textbf{Pré-Temp.} & \multicolumn{1}{c|}{\textbf{18}} & \multicolumn{1}{c|}{\textbf{24}} & \multicolumn{1}{c|}{\textbf{30}} & \multicolumn{1}{c|}{\textbf{18}} & \multicolumn{1}{c|}{\textbf{24}} & \multicolumn{1}{c|}{\textbf{30}} \\ \hline
\multicolumn{1}{|c|}{\textbf{TVS (veic./h)}} & 5378                     & 5531          & 5545          & 5735          & 5958          & 5973          & 6024          \\ \hline
\multicolumn{1}{|c|}{\textbf{TME (s)}}       & 29,12 $\pm 2,11$                   & 28,09 $\pm 1,92$        & 26,33 $\pm 1,88$        & 24,59 $\pm 1,85$        & 26,02 $\pm 1,86$        & 25,24 $\pm 1,83$        & 23,85 $\pm 1,79$        \\ \hline
\multicolumn{1}{|c|}{\textbf{TMV (s)}}       & 392,04 $\pm 36,32$                  & 361,61 $\pm 25,15$        & 354,3 $\pm 23,51$         & 347,00 $\pm 22,43$          & 370,8 $\pm 26,47$       & 350,2 $\pm 24,26$       & 340,00 $\pm 21,93$          \\ \hline
\multicolumn{1}{|c|}{\textbf{VM (Km/h)}}     & 14,24 $\pm 1,21$                    & 14,17 $\pm 1,08$        & 14,82 $\pm 1,07$        & 15,53 $\pm 1,05$        & 15,45 $\pm 1,1$         & 15,82 $\pm 1,08$        & 16,01 $\pm 1,07$        \\ \hline
\end{tabular}
\end{table}

\begin{table}[!h]
\centering
\caption{Comparação dos desempenhos de SMER$_I$ e SMER$_{I+C}$ contra o desempenho do sistema de controle de tráfego baseado em sinalizações semafóricas pré-temporizadas, utilizando os resultados obtidos no experimento 2 do cenário 1.}
\label{tab:comp_smer_cen1_exp2}
\small\begin{tabular}{c|c|c|c|c|c|c|}
\cline{2-7}
\multicolumn{1}{l|}{}                   & \multicolumn{3}{c|}{\textbf{SMER$_I$}}                          & \multicolumn{3}{c|}{\textbf{SMER$_{I+C}$}}                         \\ \hline
\multicolumn{1}{|c|}{\textbf{Métricas}} & \textbf{18}       & \textbf{24}       & \textbf{30}       & \textbf{18}        & \textbf{24}        & \textbf{30}        \\ \hline
\multicolumn{1}{|c|}{\textbf{TVS}}      & \textgreater 2,84\% & \textgreater 3,1\%  & \textgreater 6,63\% & \textgreater 10,04\% & \textgreater 11,06\%  & \textgreater 12,01\% \\ \hline
\multicolumn{1}{|c|}{\textbf{TME}}      & \textless 3,35\%   & \textless 9,58\%   & \textless 15,55\%   & \textless 10,64\%     & \textless 13,32\%    & \textless 18,09\%    \\ \hline
\multicolumn{1}{|c|}{\textbf{TMV}}      & \textless 7,76\%    & \textless 9,03\%    & \textless 11,48\%   & \textless 5,4\%     & \textless 10,66\%    & \textless 13,26\%    \\ \hline
\multicolumn{1}{|c|}{\textbf{VM}}       & \textgreater 0,04\% & \textgreater 4,07\% & \textgreater 9,05\% & \textgreater 8,49\%  & \textgreater 11,09\% & \textless 12,42\%    \\ \hline
\end{tabular}
\end{table}

\newpage

\subsubsection{Experimento 3}

A seguir, a Tabela \ref{tab:smer_cen1_exp3} apresenta os resultados dos experimento 3 do cenário 1 e, em seguida, a Tabela \ref{tab:comp_smer_cen1_exp3} apresenta a análise dos mesmos.
 
\begin{table}[!h]
\centering
\caption{Resultados obtidos com o sistema de controle de tráfego baseado em sinalizações semafóricas pré-temporizadas, SMER$_I$ e SMER$_{I+C}$ no experimento 3 do cenário 1.}
\label{tab:smer_cen1_exp3}
\small\begin{tabular}{cc|p{1cm}|p{1cm}|p{1cm}|p{1cm}|p{1cm}|p{1cm}|}
\cline{3-8}
\multicolumn{1}{l}{}                         & \multicolumn{1}{l|}{}    & \multicolumn{3}{c|}{\textbf{SMER$_I$}}        & \multicolumn{3}{c|}{\textbf{SMER$_{I+C}$}}    \\ \hline
\multicolumn{1}{|c|}{\textbf{Métricas}}      & \textbf{Pré-Temp.} & \multicolumn{1}{c|}{\textbf{18}} & \multicolumn{1}{c|}{\textbf{24}} & \multicolumn{1}{c|}{\textbf{30}} & \multicolumn{1}{c|}{\textbf{18}} & \multicolumn{1}{c|}{\textbf{24}} & \multicolumn{1}{c|}{\textbf{30}} \\ \hline
\multicolumn{1}{|c|}{\textbf{TVS (veic./h)}} & 4404                     & 4852          & 5087          & 5557          & 5252          & 5482          & 5805          \\ \hline
\multicolumn{1}{|c|}{\textbf{TME (s)}}       & 39,28 $\pm 4,65$                   & 29,62 $\pm 1,07$        & 26,84 $\pm 1,05$         & 25,17 $\pm 1,03$        & 24,35 $\pm 1,13$         & 21,62 $\pm 1,11$       & 20,08 $\pm 1,07$        \\ \hline
\multicolumn{1}{|c|}{\textbf{TMV (s)}}       & 670,50 $\pm 73,82$                  & 533,14 $\pm 44,75$       & 522,96 $\pm 42,46$        & 509,20 $\pm 40,5$       & 533,86 $\pm 44,09$        & 518,62 $\pm 41,81$        & 500,11 $\pm 39,85$       \\ \hline
\multicolumn{1}{|c|}{\textbf{VM (Km/h)}}     & 9,06 $\pm 1,3$                     & 12,29 $\pm 1,3$        & 12,41 $\pm 1,3$        & 12,91 $\pm 1,3$        & 13,05 $\pm 1,63$        & 13,17 $\pm 1,63$        & 13,43 $\pm 1,63$        \\ \hline
\end{tabular}
\end{table}

\begin{table}[!h]
\centering
\small\caption{Comparação dos desempenhos de SMER$_I$ e SMER$_{I+C}$ contra o desempenho do sistema de controle de tráfego baseado em sinalizações semafóricas pré-temporizadas, utilizando os resultados obtidos no experimento 3 do cenário 1.}
\label{tab:comp_smer_cen1_exp3}
\begin{tabular}{c|c|c|c|c|c|c|}
\cline{2-7}
\multicolumn{1}{l|}{}                   & \multicolumn{3}{c|}{\textbf{SMER$_I$}}                             & \multicolumn{3}{c|}{\textbf{SMER$_{I+C}$}}                         \\ \hline
\multicolumn{1}{|c|}{\textbf{Métricas}} & \textbf{18}        & \textbf{24}        & \textbf{30}        & \textbf{18}        & \textbf{24}        & \textbf{30}        \\ \hline
\multicolumn{1}{|c|}{\textbf{TVS}}      & \textgreater 10,17\% & \textgreater 15,5\%  & \textgreater 26,18\% & \textgreater 19,25\% & \textgreater 24,47\% & \textgreater 31,81\% \\ \hline
\multicolumn{1}{|c|}{\textbf{TME}}      & \textless 24,59\%    & \textless 31,67\%    & \textless 35,92\%    & \textless 38,00\%     & \textless 44,95\%    & \textless 48,87\%    \\ \hline
\multicolumn{1}{|c|}{\textbf{TMV}}      & \textless 20,48\%    & \textless 22,00\%    & \textless 24,05\%    & \textless 20,37\%    & \textless 22,65\%    & \textless 25,41\%    \\ \hline
\multicolumn{1}{|c|}{\textbf{VM}}       & \textgreater 35,65\% & \textgreater 36,97\% & \textgreater 42,49\% & \textgreater 44,03\% & \textgreater 45,36\% & \textgreater 48,23\% \\ \hline
\end{tabular}
\end{table}

\subsection{Análise dos Resultados do Cenário 2}

Analisando as Tabelas \ref{tab:comp_smer_cen1_exp1}, \ref{tab:comp_smer_cen1_exp2} e \ref{tab:comp_smer_cen1_exp2}, foi possível identificar que os melhores resultados de SMER$_I$ e SMER$_{I+C}$ foram obtidos, utilizando números de obtenções de quantidade de veículos iguais a 30. Este resultados, por sua vez, serão comparados com os resultados dos experimentos 1, 2 e 3 do cenário 2. 

Para evitar qualquer dúvida na leitura dos resultados, os mesmos foram separados, de acordo com os cenários em que foram obtidos. Para tanto, os nomes SMER$_I$ e SMER$_{I+C}$ receberam sobrescritos, que são os números de cada um dos cenários definidos para a avaliação do sistema multiagente de controle de tráfego. 

\subsubsection{Experimento 1}

A seguir, a Tabela \ref{tab:smer_cen2_exp1} apresenta os resultados dos experimento 1 do cenário 2 e, em seguida, a Tabela \ref{tab:comp_smer_cen2_exp1} apresenta a análise dos mesmos.
 
\begin{table}[!h]
\centering
\caption{Resultados obtidos com SMER$^2_I$ e SMER$^2_{I+C}$ no experimento 1 do cenário 2.}
\label{tab:smer_cen2_exp1}
\small\begin{tabular}{c|c|c|c|c|}
\cline{2-5}
\multicolumn{1}{l|}{}                        & \multicolumn{2}{c|}{\textbf{Cenário 1}}           & \multicolumn{2}{c|}{\textbf{Cenário 2}}             \\ \hline
\multicolumn{1}{|c|}{\textbf{Métricas}}      & \textbf{SMER$^{1}_I$} & \textbf{SMER$^{1}_{I+C}$} & \textbf{SMER$^{2}_{I}$} & \textbf{SMER$^{2}_{I+C}$} \\ \hline
\multicolumn{1}{|c|}{\textbf{TVS (veic./h)}} & 4200                  & 4280                      & 3853                    & 4116                      \\ \hline
\multicolumn{1}{|c|}{\textbf{TME (s)}}       & 12,61 $\pm 0,49$      & 12,21 $\pm 0,47$          & 14,89 $\pm 0,65$        & 13,13 $\pm 0,5$           \\ \hline
\multicolumn{1}{|c|}{\textbf{TMV (s)}}       & 261,96 $\pm 7,42$     & 254,00 $\pm 6,62$         & 288,69 $\pm 8,51$       & 284,01 $\pm 6,88$         \\ \hline
\multicolumn{1}{|c|}{\textbf{VM (Km/h)}}     & 21,18 $\pm 0,38$      & 22,97 $\pm 0,48$          & 19,93 $\pm 0,48$        & 20,73 $\pm 0,68$                    \\ \hline
\end{tabular}
\end{table}

\begin{table}[!h]
\centering
\caption{Comparação dos desempenhos de SMER$^2_I$ e SMER$^2_{I+C}$ contra os desempenhos de SMER$^1_I$ e SMER$^1_{I+C}$, utilizando os resultados obtidos no experimento 1 do cenário 1.}
\label{tab:comp_smer_cen2_exp1}
\small\begin{tabular}{|c|c|c|}
\hline
\textbf{Métricas} & \textbf{SMER$^{2}_{I}$} & \textbf{SMER$^{2}_{I+C}$} \\ \hline
\textbf{TVS}      & \textless 8,26\%        & \textless 3,83\%          \\ \hline
\textbf{TME}      & \textgreater 18,08\%    & \textgreater 7,53\%       \\ \hline
\textbf{TMV}      & \textgreater 10,20\%    & \textgreater 11,81\%       \\ \hline
\textbf{VM}       & \textless 5,90\%        & \textless 9,75\%          \\ \hline
\end{tabular}
\end{table}

\subsubsection{Experimento 2}

A seguir, a Tabela \ref{tab:smer_cen2_exp2} apresenta os resultados dos experimento 2 do cenário 2 e, em seguida, a Tabela \ref{tab:comp_smer_cen2_exp2} apresenta a análise dos mesmos.

\begin{table}[!h]
\centering
\caption{Resultados obtidos com SMER$^2_I$ e SMER$^2_{I+C}$ no experimento 2 do cenário 2.}
\label{tab:smer_cen2_exp2}
\small\begin{tabular}{c|c|c|c|c|}
\cline{2-5}
\multicolumn{1}{l|}{}                        & \multicolumn{2}{c|}{\textbf{Cenário 1}}           & \multicolumn{2}{c|}{\textbf{Cenário 2}}             \\ \hline
\multicolumn{1}{|c|}{\textbf{Métricas}}      & \textbf{SMER$^{1}_I$} & \textbf{SMER$^{1}_{I+C}$} & \textbf{SMER$^{2}_{I}$} & \textbf{SMER$^{2}_{I+C}$} \\ \hline
\multicolumn{1}{|c|}{\textbf{TVS (veic./h)}} & 5735                  & 6024                      & 5382                    & 5662                      \\ \hline
\multicolumn{1}{|c|}{\textbf{TME (s)}}       & 24,59 $\pm 1,85$      & 23,85 $\pm 1,79$          & 29 $\pm 1,99$           & 25,05 $\pm 1,96$          \\ \hline
\multicolumn{1}{|c|}{\textbf{TMV (s)}}       & 347,00 $\pm 22,33$    & 340,00 $\pm 21,93$        & 374,86 $\pm 24,01$      & 353,5 $\pm 23,19$         \\ \hline
\multicolumn{1}{|c|}{\textbf{VM (Km/h)}}     & 15,53 $\pm 1,05$      & 16,01 $\pm 1,07$          & 14,91 $\pm 1,3$         & 15,18 $\pm 1,27$          \\ \hline
\end{tabular}
\end{table}

\begin{table}[!h]
\centering
\caption{Comparação dos desempenhos de SMER$^2_I$ e SMER$^2_{I+C}$ contra os desempenhos de SMER$^1_I$ e SMER$^1_{I+C}$, utilizando os resultados obtidos no experimento 2 do cenário 1.}
\label{tab:comp_smer_cen2_exp2}
\small\begin{tabular}{|c|c|c|}
\hline
\textbf{Métricas} & \textbf{SMER$^{2}_{I}$} & \textbf{SMER$^{2}_{I+C}$} \\ \hline
\textbf{TVS}      & \textless 6,15\%        & \textless 6,39\%          \\ \hline
\textbf{TME}      & \textgreater 17.93\%    & \textgreater 5,03\%      \\ \hline
\textbf{TMV}      & \textgreater 8,02\%     & \textgreater 3,97\%      \\ \hline
\textbf{VM}       & \textless 3,99\%        & \textless 5,18\%          \\ \hline
\end{tabular}
\end{table}

\newpage

\subsubsection{Experimento 3}

A seguir, a Tabela \ref{tab:smer_cen2_exp3} apresenta os resultados dos experimento 3 do cenário 2 e, em seguida, a Tabela \ref{tab:comp_smer_cen2_exp3} apresenta a análise dos mesmos.

\begin{table}[!h]
\centering
\caption{Resultados obtidos com SMER$^2_I$ e SMER$^2_{I+C}$ no experimento 3 do cenário 2.}
\label{tab:smer_cen2_exp3}
\small\begin{tabular}{c|c|c|c|c|}
\cline{2-5}
\multicolumn{1}{l|}{}                        & \multicolumn{2}{c|}{\textbf{Cenário 1}}           & \multicolumn{2}{c|}{\textbf{Cenário 2}}             \\ \hline
\multicolumn{1}{|c|}{\textbf{Métricas}}      & \textbf{SMER$^{1}_I$} & \textbf{SMER$^{1}_{I+C}$} & \textbf{SMER$^{2}_{I}$} & \textbf{SMER$^{2}_{I+C}$} \\ \hline
\multicolumn{1}{|c|}{\textbf{TVS (veic./h)}} & 5557                  & 5805                      & 5065                        & 5375                      \\ \hline
\multicolumn{1}{|c|}{\textbf{TME (s)}}       & 25,17 $\pm 1,03$      & 20,8 $\pm 1,07$           & 30,7 $\pm 1,43$             & 27,63 $\pm 1,7$           \\ \hline
\multicolumn{1}{|c|}{\textbf{TMV (s)}}       & 509,86 $\pm 40,5$     & 500,17 $\pm 39,85$        & 561,79 $\pm 42,46$          & 536,4 $\pm 39,85$                         \\ \hline
\multicolumn{1}{|c|}{\textbf{VM (Km/h)}}     & 12,91 $\pm 1,3$       & 12,95 $\pm 1,63$          & 10,52 $\pm 1,3$             & 12,5 $\pm 1,63$                         \\ \hline
\end{tabular}
\end{table}

\begin{table}[!h]
\centering
\caption{Comparação dos desempenhos de SMER$^2_I$ e SMER$^2_{I+C}$ contra os desempenhos de SMER$^1_I$ e SMER$^1_{I+C}$, utilizando os resultados obtidos no experimento 3 do cenário 1.}
\label{tab:comp_smer_cen2_exp3}
\small\begin{tabular}{|c|c|c|}
\hline
\textbf{Métricas} & \textbf{SMER$^{2}_{I}$} & \textbf{SMER$^{2}_{I+C}$} \\ \hline
\textbf{TVS}      & \textless 8,85\%        & \textless 8,0\%           \\ \hline
\textbf{TME}      & \textgreater 21,97\%    & \textgreater 32,83\%       \\ \hline
\textbf{TMV}      & \textgreater 10,18\%    & \textgreater 7,24\%       \\ \hline
\textbf{VM}       & \textless 18,51\%       & \textless 3,47\%          \\ \hline
\end{tabular}
\end{table}

\section{Avaliando o Sistema Multiagente de Planejamento e Orientação de Rotas}\label{sec:rotas}

Nesta seção, é apresentada a avaliação experimental acerca do sistema multiagente de planejamento e orientação de rotas.

\subsection{Descrição dos Cenários para os Experimentos}

Para a obtenção dos resultados referentes ao sistema multiagente de planejamento e orientação de rotas, foram criados dois cenários: operação do sistema multiagente de planejamento e rotas sobre o sistema multiagente de controle de tráfego com sinalizações semafóricas em total funcionamento; e operação do sistema multiagente de planejamento e rotas sobre o sistema multiagente de controle de tráfego com sinalizações semafóricas apresentando ausência de funcionamento. 

Em ambos os cenários, foram utilizados todos os agentes definidos nesta tese. Com esses cenários, pretendeu-se avaliar o desempenho do algoritmo de roteamento orientado a ondas verdes (ROOV), comparando-os com os desempenhos de dois algoritmos de roteamento. O primeiro foi um algoritmo de roteamento de caminho espacialmente mais curto (RCEMC). O segundo foi um algoritmo de roteamento de caminho temporalmente mais curto (RCTMC). Essa comparação de desempenho foi realizada com os resultados do cenário 1.  

Para os dois cenários, foram utilizadas a grades manhattan 10 x 10, que foram utilizadas nos experimentos com a HRAdNet. Estas grades podem ser vistas nas Figuras \ref{fig:grade_1} e \ref{fig:grade_2}. Além disto, Para cada um destes cenários, foram criados dois subcenários, a saber: controle de tráfego em interseções isoladas (SMER$_I$); e controle de tráfego envolvendo interseções isoladas mais controle de sistemas coordenados de sinalizações semafóricas (SMER$_{I+C}$). 

\subsection{Configurações dos Experimentos}

As configurações adotadas nos experimentos relativos ao sistema multiagente de planejamento e orientação de rotas foram as mesmas adotadas nos experimentos com a HRAdNet-VE. Além disto, o parâmetro penalidade por número de veículos foi configurado com o valor 3,5, que foi o mesmo valor de penalidade adotado por \citet{Faria:2013}. Além disto, o parâmetro de tolerância de rota foi configurado com o valor zero.

\subsection{Análise dos Resultados do Cenário 1}

Esta seção tem como objetivo apresentar os resultados obtidos por meio dos experimentos 1, 2 e 3 do cenário, assim como, as análises comparativas dos mesmos.

\subsubsection{Experimento 1}

A seguir, a Tabela \ref{tab:roov_cen_1_exp_1} apresenta os resultados dos experimento 1 do cenário 1 e, em seguida, a Tabela \ref{tab:comp_roov_cen_1_exp_1} apresenta a análise dos mesmos.

\begin{table}[!h]
\centering
\caption{Resultados obtidos com o ROOV, RCEMC e RCTMC no experimento 1 do cenário 1.}
\label{tab:roov_cen_1_exp_1}
\small\begin{tabular}{c|p{1.3cm}|p{1.5cm}|p{1.5cm}|p{1.3cm}|p{1.5cm}|p{1.5cm}|}
\cline{2-7}
\multicolumn{1}{l|}{}                        & \multicolumn{3}{c|}{\textbf{SMER$_I$}}                                             & \multicolumn{3}{c|}{\textbf{SMER$_{I+C}$}}                                                                     \\ \hline
\multicolumn{1}{|c|}{\textbf{Métricas}}      & \textbf{ROOV}               & \textbf{RCEMC} & \multicolumn{1}{l|}{\textbf{RCTMC}} & \multicolumn{1}{l|}{\textbf{ROOV}} & \multicolumn{1}{l|}{\textbf{RCEMC}} & \multicolumn{1}{l|}{\textbf{RCTMC}} \\ \hline
\multicolumn{1}{|l|}{\textbf{TVS (veic./h)}} & 4291                        & 4200           & 4227                                & 4400                               & 4280                                & 4400                                \\ \hline
\multicolumn{1}{|c|}{\textbf{TME (s)}}       & 5,6 $\pm 0,57$             & 12,61 $\pm 0,49$         & 14,89 $\pm 0,46$          & 5 $\pm 0,39$                        & 12,21 $\pm 0,47$                              & 14,25 $\pm 0,44$                              \\ \hline
\multicolumn{1}{|c|}{\textbf{TMV (s)}}       & 257 $\pm 7,51$             & 261,8 $\pm 7,51$        & 221,09 $\pm 7,83$          & 224 $\pm 6,53$                      & 254 $\pm 7,18$                                & 212 $\pm 8,16$                                \\ \hline
\multicolumn{1}{|c|}{\textbf{VM (Km/h)}}     & 22,29 $\pm 0,42$           & 21,18 $\pm 0,39$         & 24,22 $\pm 0,65$          & 26,65 $\pm 1,45$                    & 22,97 $\pm 0,39$                              & 25,93 $\pm 0,65$                              \\ \hline
\multicolumn{1}{|c|}{\textbf{CMC (ml)}}      & \multicolumn{1}{l|}{416,49} & 417,21         & 339,00                              & 432,28                             & 481,3                               & 351,74                              \\ \hline
\multicolumn{1}{|c|}{\textbf{CO(g)}}         & 12,75                       & 13,84          & 11,06                               & 13,57                              & 15,86                               & 11,48                               \\ \hline
\multicolumn{1}{|c|}{\textbf{CO$_2$(g)}}     & 1044,67                     & 1094,46        & 850,23                              & 1086,25                            & 1262,83                             & 869,61                              \\ \hline
\multicolumn{1}{|c|}{\textbf{HC(g)}}         & 0,367                       & 0,3918         & 0,2994                              & 0,372                              & 0,481                               & 0,309                               \\ \hline
\multicolumn{1}{|c|}{\textbf{NOx(g)}}        & 2,08                        & 2,201          & 1,729                               & 2,21                               & 2,44                                & 1,46                                \\ \hline
\multicolumn{1}{|c|}{\textbf{PMx(g)}}        & 0,1277                      & 0,1351         & 0,106                               & 0,139                              & 0,143                               & 0,109                               \\ \hline
\end{tabular}
\end{table}

\begin{table}[!h]
\centering
\caption{Comparação do desempenho do ROOV contra os desempenhos do RCEMC e RCTMC, utilizando os dados obtidos no experimento 1 do cenário 1.}
\label{tab:comp_roov_cen_1_exp_1}
\small\begin{tabular}{c|c|c|c|c|}
\cline{2-5}
\multicolumn{1}{l|}{}                        & \multicolumn{2}{c|}{\textbf{SMER$_I$}}                    & \multicolumn{2}{c|}{\textbf{SMER$_{I+C}$}}                                \\ \hline
\multicolumn{1}{|c|}{\textbf{Métricas}}      & \textbf{RCEMC}      & \multicolumn{1}{l|}{\textbf{RCTMC}} & \multicolumn{1}{l|}{\textbf{RCEMC}} & \multicolumn{1}{l|}{\textbf{RCTMC}} \\ \hline
\multicolumn{1}{|l|}{\textbf{TVS (veic./h)}} & \textgreater 2,16\% & \textgreater 1,51\%                 & \textgreater 2,80\%                 & =                                   \\ \hline
\multicolumn{1}{|c|}{\textbf{TME (s)}}       & \textless 55,59\%   & \textless 62,39\%                   & \textless 59,04\%                   & \textless 64,91\%                   \\ \hline
\multicolumn{1}{|c|}{\textbf{TMV (s)}}       & \textless 1,83\%    & \textgreater 16,24\%                & \textless 11,81\%                   & \textgreater 5,66\%                 \\ \hline
\multicolumn{1}{|c|}{\textbf{VM (Km/h)}}     & \textgreater 5,24\% & \textless 7,96\%                    & \textgreater 16,02\%                & \textgreater 2,77\%                 \\ \hline
\multicolumn{1}{|c|}{\textbf{CMC (ml)}}      & \textless 0,017\%   & \textgreater 22,85\%                & \textless 10,18\%                   & \textgreater 22,89\%                \\ \hline
\multicolumn{1}{|c|}{\textbf{CO(g)}}         & \textless 7,87\%    & \textgreater 15,28\%                & \textless 14,43\%                   & \textgreater 18,20\%                \\ \hline
\multicolumn{1}{|c|}{\textbf{CO$_2$(g)}}     & \textless 4,54\%    & \textgreater 22,86\%                & \textless 13,98\%                   & \textgreater 24,91\%                \\ \hline
\multicolumn{1}{|c|}{\textbf{HC(g)}}         & \textless 6,13\%    & \textgreater 22,74\%                & \textless 22,66\%                   & \textgreater 20,38\%                \\ \hline
\multicolumn{1}{|c|}{\textbf{NOx(g)}}        & \textless 5,49\%    & \textgreater 20,30                  & \textless 9,42\%                    & \textgreater 51,36\%                \\ \hline
\multicolumn{1}{|c|}{\textbf{PMx(g)}}        & \textless 5,47\%    & \textgreater 20,47\%                & \textless 2,79\%                    & \textgreater 27,52\%                \\ \hline
\end{tabular}
\end{table}

\newpage

\subsubsection{Experimento 2}

A seguir, a Tabela \ref{tab:roov_cen_1_exp_2} apresenta os resultados dos experimento 2 do cenário 1 e, em seguida, a Tabela \ref{tab:comp_roov_cen_1_exp_2} apresenta a análise dos mesmos.

\begin{table}[!h]
\centering
\caption{Resultados obtidos com o ROOV, RCEMC e RCTMC no experimento 2 do cenário 1.}
\label{tab:roov_cen_1_exp_2}
\small\begin{tabular}{c|p{1.3cm}|p{1.5cm}|p{1.5cm}|p{1.3cm}|p{1.5cm}|p{1.5cm}|}
\cline{2-7}
\multicolumn{1}{l|}{}                        & \multicolumn{3}{c|}{\textbf{SMER$_I$}}                                             & \multicolumn{3}{c|}{\textbf{SMER$_{I+C}$}}                                                                     \\ \hline
\multicolumn{1}{|c|}{\textbf{Métricas}}      & \textbf{ROOV}               & \textbf{RCEMC} & \multicolumn{1}{l|}{\textbf{RCTMC}} & \multicolumn{1}{l|}{\textbf{ROOV}} & \multicolumn{1}{l|}{\textbf{RCEMC}} & \multicolumn{1}{l|}{\textbf{RCTMC}} \\ \hline
\multicolumn{1}{|l|}{\textbf{TVS (veic./h)}} & 6254                        & 5735           & 5871                                & 6600                               & 6024                                & 6155                                \\ \hline
\multicolumn{1}{|c|}{\textbf{TME (s)}}       & 5,68 $\pm 0,47$                       & 24,59 $\pm 1,85$         & 14,96 $\pm 0,59$                              & 5 $\pm 0,39$                                & 23,85 $\pm 1,79$                              & 14,78 $\pm 0,59$                               \\ \hline
\multicolumn{1}{|c|}{\textbf{TMV (s)}}       & 299,06 $\pm 14,37$     & 347,09 $\pm 22,33$       & 322,71 $\pm 21,88$ & 288,11 $\pm 13,71$ & 340,86 $\pm 22,86$ & 313,88 $\pm 23,19$                             \\ \hline
\multicolumn{1}{|c|}{\textbf{VM (Km/h)}}     & 20,06 $\pm 0,65$       & 14,53 $\pm 1,05$         & 16,40 $\pm 0,84$   & 21,15 $\pm 0,65$   & 16,01 $\pm 1,05$  & 17,21 $\pm 0,88$ \\ \hline
\multicolumn{1}{|c|}{\textbf{CMC (ml)}}      & \multicolumn{1}{l|}{575,19} & 646,67         & 590,69                              & 624,76                             & 669,21                              & 637,35                              \\ \hline
\multicolumn{1}{|c|}{\textbf{CO(g)}}         & 16,87                       & 22,53          & 18,23                               & 18,52                              & 23,44                               & 18,7                                \\ \hline
\multicolumn{1}{|c|}{\textbf{CO$_2$(g)}}     & 1362,25                     & 1613,81        & 1481,45                             & 1479,54                            & 1669,96                             & 1507,97                             \\ \hline
\multicolumn{1}{|c|}{\textbf{HC(g)}}         & 0,527                       & 0,616          & 0,553                               & 0,570                              & 0,640                               & 0,561                               \\ \hline
\multicolumn{1}{|c|}{\textbf{NOx(g)}}        & 2,81                        & 3,01           & 2,88                                & 3,06                               & 3,11                                & 2,94                                \\ \hline
\multicolumn{1}{|c|}{\textbf{PMx(g)}}        & 0,164                       & 0,154          & 0,164                               & 0,179                              & 0,160                               & 0,169                               \\ \hline
\end{tabular}
\end{table}

\begin{table}[!h]
\centering
\caption{Comparação do desempenho do ROOV contra os desempenhos do RCEMC e RCTMC, utilizando os dados obtidos no experimento 2 do cenário 1.}
\label{tab:comp_roov_cen_1_exp_2}
\small\begin{tabular}{c|c|c|c|c|}
\cline{2-5}
\multicolumn{1}{l|}{}                        & \multicolumn{2}{c|}{\textbf{SMER$_I$}}                     & \multicolumn{2}{c|}{\textbf{SMER$_{I+C}$}}                                \\ \hline
\multicolumn{1}{|c|}{\textbf{Métricas}}      & \textbf{RCEMC}       & \multicolumn{1}{l|}{\textbf{RCTMC}} & \multicolumn{1}{l|}{\textbf{RCEMC}} & \multicolumn{1}{l|}{\textbf{RCTMC}} \\ \hline
\multicolumn{1}{|l|}{\textbf{TVS (veic./h)}} & \textgreater 9,04\%  & 6,52\%                              & \textgreater 9,56\%                 & \textgreater 7,22\%                 \\ \hline
\multicolumn{1}{|c|}{\textbf{TME (s)}}       & \textless 76,90\%    & \textless 62,03\%                   & \textless 79,03\%                   & \textless 66,17\%                   \\ \hline
\multicolumn{1}{|c|}{\textbf{TMV (s)}}       & \textless 13,83\%    & \textless 7,32\%                    & \textless 15,47\%                   & \textless 8,21\%                    \\ \hline
\multicolumn{1}{|c|}{\textbf{VM (Km/h)}}     & \textgreater 29,16\% & \textgreater 22,31\%                & \textgreater 32,1\%                & \textgreater 22,89\%                \\ \hline
\multicolumn{1}{|c|}{\textbf{CMC (ml)}}      & \textless 11,05\%    & \textless 2,62\%                    & \textless 6,64\%                    & \textless 1,97\%                    \\ \hline
\multicolumn{1}{|c|}{\textbf{CO(g)}}         & \textless 25,12\%    & \textless 7,46\%                    & \textless 20,98\%                   & \textless 0,09\%                    \\ \hline
\multicolumn{1}{|c|}{\textbf{CO$_2$(g)}}     & \textless 15,58\%    & \textless 8,04\%                    & \textless 11,40\%                   & \textless 1,88\%                    \\ \hline
\multicolumn{1}{|c|}{\textbf{HC(g)}}         & \textless 14,44\%    & \textless 4,70\%                    & \textless 10,93\%                   & \textgreater 1,60\%                 \\ \hline
\multicolumn{1}{|c|}{\textbf{NOx(g)}}        & \textless 6,64\%     & \textless 2,43\%                    & \textless 1,60\%                    & \textgreater 4,08\%                 \\ \hline
\multicolumn{1}{|c|}{\textbf{PMx(g)}}        & \textgreater 6,49\%  & =                                   & \textgreater 11,87\%                & \textgreater 5,91\%                 \\ \hline
\end{tabular}
\end{table}

\newpage

\subsubsection{Experimento 3}

A seguir, a Tabela \ref{tab:roov_cen_1_exp_3} apresenta os resultados dos experimento 3 do cenário 1 e, em seguida, a Tabela \ref{tab:comp_roov_cen_1_exp_3} apresenta a análise dos mesmos.

\begin{table}[!h]
\centering
\caption{Resultados obtidos com o ROOV, RCEMC e RCTMC no experimento 3 do cenário 1.}
\label{tab:roov_cen_1_exp_3}
\small\begin{tabular}{c|p{1.3cm}|p{1.5cm}|p{1.5cm}|p{1.3cm}|p{1.5cm}|p{1.5cm}|}
\cline{2-7}
\multicolumn{1}{l|}{}                        & \multicolumn{3}{c|}{\textbf{SMER$_I$}}                                             & \multicolumn{3}{c|}{\textbf{SMER$_{I+C}$}}                                                                     \\ \hline
\multicolumn{1}{|c|}{\textbf{Métricas}}      & \textbf{ROOV}               & \textbf{RCEMC} & \multicolumn{1}{l|}{\textbf{RCTMC}} & \multicolumn{1}{l|}{\textbf{ROOV}} & \multicolumn{1}{l|}{\textbf{RCEMC}} & \multicolumn{1}{l|}{\textbf{RCTMC}} \\ \hline
\multicolumn{1}{|l|}{\textbf{TVS (veic./h)}} & 6970                        & 5557           & 4347                                & 7057                               & 5806                                & 4506                                \\ \hline
\multicolumn{1}{|c|}{\textbf{TME (s)}}       & 9,22 $\pm 0,87$                       & 25,17 $\pm 1,03$         & 38,78 $\pm 5,63$                              & 9,00 $\pm 0,89$                              & 20,08 $\pm 1,07$                              & 37,87 $\pm 5,5$                               \\ \hline
\multicolumn{1}{|c|}{\textbf{TMV (s)}}       & 368,1 $\pm 17,96$                      & 509,20 $\pm 40,5$        & 749,00 $\pm 106,81$                             & 365,92 $\pm 18,29$                            & 500,11 $\pm 40,17$                             & 733,64 $\pm 104,53$                             \\ \hline
\multicolumn{1}{|c|}{\textbf{VM (Km/h)}}     & 16,95 $\pm 1,3$                      & 12,91 $\pm 1,3$          & 13,43 $\pm 1,63$                               & 16,98 $\pm 1,3$                             & 12,95 $\pm 1,3$                               & 13,98 $\pm 1,63$                               \\ \hline
\multicolumn{1}{|c|}{\textbf{CMC (ml)}}      & \multicolumn{1}{l|}{917,55} & 599,35         & 513,06                              & 917,83                             & 622,00                              & 537,99                              \\ \hline
\multicolumn{1}{|c|}{\textbf{CO(g)}}         & 27,02                       & 16,66          & 14,99                               & 27,25                              & 25,18                               & 15,42                               \\ \hline
\multicolumn{1}{|c|}{\textbf{CO$_2$(g)}}     & 2138,40                     & 1503,41        & 1286,58                             & 2139,02                            & 1560,20                             & 1409,96                             \\ \hline
\multicolumn{1}{|c|}{\textbf{HC(g)}}         & 0,777                       & 0,631          & 0,544                               & 0,779                              & 0,657                               & 0,564                               \\ \hline
\multicolumn{1}{|c|}{\textbf{NOx(g)}}        & 4,83                        & 2,66           & 2,3                                 & 4,27                               & 2,77                                & 1,22                                \\ \hline
\multicolumn{1}{|c|}{\textbf{PMx(g)}}        & 0,257                       & 0,130          & 0,114                               & 0,259                              & 0,136                               & 0,119                               \\ \hline
\end{tabular}
\end{table}

\begin{table}[!h]
\centering
\caption{Comparação do desempenho do ROOV contra os desempenhos do RCEMC e RCTMC, utilizando os dados obtidos no experimento 3 do cenário 1.}
\label{tab:comp_roov_cen_1_exp_3}
\small\begin{tabular}{c|c|c|c|c|}
\cline{2-5}
\multicolumn{1}{l|}{}                        & \multicolumn{2}{c|}{\textbf{SMER$_I$}}                     & \multicolumn{2}{c|}{\textbf{SMER$_{I+C}$}}                                \\ \hline
\multicolumn{1}{|c|}{\textbf{Métricas}}      & \textbf{RCEMC}       & \multicolumn{1}{l|}{\textbf{RCTMC}} & \multicolumn{1}{l|}{\textbf{RCEMC}} & \multicolumn{1}{l|}{\textbf{RCTMC}} \\ \hline
\multicolumn{1}{|l|}{\textbf{TVS (veic./h)}} & \textgreater 25,42\% & \textgreater 60,34\%                & \textgreater 21,54\%                & \textgreater 56,61\%                \\ \hline
\multicolumn{1}{|c|}{\textbf{TME (s)}}       & \textless 63,36\%    & \textless 23,77\%                   & \textless 55,17\%                   & \textless 76,23\%                   \\ \hline
\multicolumn{1}{|c|}{\textbf{TMV (s)}}       & \textless 27,77\%    & \textless 50,85\%                   & \textless 26,83\%                   & \textless 50,12\%                   \\ \hline
\multicolumn{1}{|c|}{\textbf{VM (Km/h)}}     & \textgreater 31,29\% & \textgreater 26,20\%                & \textgreater 31,11\%                & \textgreater 21,45\%                \\ \hline
\multicolumn{1}{|c|}{\textbf{CMC (ml)}}      & \textgreater 53,09\% & \textgreater 78,83\%                & \textgreater 47,56\%                & \textgreater 70,60\%                \\ \hline
\multicolumn{1}{|c|}{\textbf{CO(g)}}         & \textgreater 62,18\% & \textgreater 80,25\%                & \textgreater 8,22\%                 & \textgreater 76,71\%                \\ \hline
\multicolumn{1}{|c|}{\textbf{CO$_2$(g)}}     & \textgreater 42,23\% & \textgreater 66,20\%                & \textgreater 37,09\%                & \textgreater 51,70\%                \\ \hline
\multicolumn{1}{|c|}{\textbf{HC(g)}}         & \textgreater 23,13\% & \textgreater 42,83\%                & \textgreater 18,56\%                & \textgreater 38,12\%                \\ \hline
\multicolumn{1}{|c|}{\textbf{NOx(g)}}        & \textgreater 81,57\% & \textgreater 110\%                  & \textgreater 54,15\%                & \textgreater 250\%                  \\ \hline
\multicolumn{1}{|c|}{\textbf{PMx(g)}}        & \textgreater 97,69\% & \textgreater125,43\%                & \textgreater 90,44\%                & \textgreater 117,64\%               \\ \hline
\end{tabular}
\end{table}

\newpage

\subsection{Análise dos Resultados do Cenário 2}

Esta seção tem como objetivo apresentar os resultados obtidos com o ROOV no cenário 2. Além disto, ela também visa apresentar uma comparação dos desempenhos do ROOV no cenário 1 (ROOV$^1$) contra os desempenhos do ROOV no cenário 2 (ROOV$^2$). As apresentações dos resultados e análises serão feitas de acordo com os experimentos 1, 2 e 3. 

\subsubsection{Experimento 1}

A seguir, a Tabela \ref{tab:roov_cen_2_exp_1} apresenta os resultados dos experimento 1 do cenário 2 e, em seguida, a Tabela \ref{tab:comp_roov_cen_2_exp_1} apresenta a análise dos mesmos.

\begin{table}[!h] 
\centering
\caption{Resultados obtidos com ROOV nos experimentos 1 dos cenários 1 e 2.}
\label{tab:roov_cen_2_exp_1}
\small\begin{tabular}{c|c|c|c|c|}
\cline{2-5}
\multicolumn{1}{l|}{}                        & \multicolumn{2}{c|}{\textbf{SMER$_I$}}                                                 & \multicolumn{2}{c|}{\textbf{SMER$_{I+C}$}}                                                                  \\ \hline
\multicolumn{1}{|c|}{\textbf{Métricas}}      & \textbf{ROOV$^1$} & \multicolumn{1}{l|}{\textbf{ROOV$^2$}} & \multicolumn{1}{l|}{\textbf{ROOV$^1$}} & \multicolumn{1}{l|}{\textbf{ROOV$^2$}} \\ \hline
\multicolumn{1}{|l|}{\textbf{TVS (veic./h)}} & 4291                            & 4157                                                 & 4400                                                 & 4400                                                 \\ \hline
\multicolumn{1}{|c|}{\textbf{TME (s)}}       & 5,6 $\pm 0,57$                            & 5,71 $\pm 0,85$                                                & 5,00 $\pm 0,39$                                                & 5,26 $\pm 0,87$                                                \\ \hline
\multicolumn{1}{|c|}{\textbf{TMV (s)}}       & 257,00 $\pm 7,51$                         & 263,98 $\pm 7,51$                                              & 224,00 $\pm 6,53$                                              & 233,03 $\pm 7,18$                                              \\ \hline
\multicolumn{1}{|c|}{\textbf{VM (Km/h)}}     & 22,29 $\pm 0,39$                          & 21,40 $\pm 0,48$                                               & 26,65 $\pm 0,42$                                               & 24,9 $\pm 0,48$                                                \\ \hline
\multicolumn{1}{|c|}{\textbf{CMC (ml)}}      & 416,49                          & 405,42                                               & 432,28                                               & 416,61                                               \\ \hline
\multicolumn{1}{|c|}{\textbf{CO(g)}}         & 12,75                           & 13,11                                                & 13,57                                                & 14,08                                                \\ \hline
\multicolumn{1}{|c|}{\textbf{CO$_2$(g)}}     & 1044,67                         & 1040,35                                              & 1086,25                                              & 1062,04                                              \\ \hline
\multicolumn{1}{|c|}{\textbf{HC(g)}}         & 0,367                           & 0,362                                                & 0,372                                                & 0,360                                                \\ \hline
\multicolumn{1}{|c|}{\textbf{NOx(g)}}        & 2,08                            & 2,13                                                 & 2,21                                                 & 2,25                                                 \\ \hline
\multicolumn{1}{|c|}{\textbf{PMx(g)}}        & 0,127                           & 0,134                                                & 0,139                                                & 0,143                                               \\ \hline
\end{tabular}
\end{table}

\begin{table}[!h]
\centering
\caption{Comparação dos desempenhos do ROOV$^2$ contra os do ROOV$^1$, utilizando os resultados obtidos no experimento 1 do cenário 2.}
\label{tab:comp_roov_cen_2_exp_1}
\small\begin{tabular}{|c|c|c|}
\hline
\textbf{Métricas} & \multicolumn{1}{l|}{\textbf{ROOV$^2_{SMER_I}$}} & \multicolumn{1}{l|}{\textbf{ROOV$^2_{SMER_{I+C}}$}} \\ \hline
\textbf{TVS}      & \textless 3,12\%                                & =                                                   \\ \hline
\textbf{TME}      & \textgreater 1,96\%                             & \textgreater 5,20\%                                 \\ \hline
\textbf{TMV}      & \textgreater 2,71\%                             & \textgreater 4,03\%                                 \\ \hline
\textbf{VM}       & \textless 3,99\%                                & \textless 6,56\%                                    \\ \hline
\textbf{CMC}      & \textless 2,65\%                                & \textless 3,62\%                                    \\ \hline
\textbf{CO}       & \textgreater 2,82\%                             & \textgreater 3,75\%                                 \\ \hline
\textbf{CO$_2$}   & \textless 0,04\%                                & \textless 2,22\%                                    \\ \hline
\textbf{HC}       & \textless 1,36\%                                & \textless 3,22\%                                    \\ \hline
\textbf{NOx}      & \textgreater 2,40\%                             & \textgreater 1,80\%                                 \\ \hline
\textbf{PMx}      & \textgreater 5,51\%                             & \textgreater 0,01\%                                 \\ \hline
\end{tabular}
\end{table}

\newpage

\subsubsection{Experimento 2}

A seguir, a Tabela \ref{tab:roov_cen_2_exp_2} apresenta os resultados dos experimento 2 do cenário 2 e, em seguida, a Tabela \ref{tab:comp_roov_cen_2_exp_2} apresenta a análise dos mesmos.

\begin{table}[!h]
\centering
\caption{Resultados obtidos com ROOV nos experimentos 2 dos cenários 1 e 2.}
\label{tab:roov_cen_2_exp_2}
\small\begin{tabular}{c|c|c|c|c|}
\cline{2-5}
\multicolumn{1}{l|}{}                        & \multicolumn{2}{c|}{\textbf{SMER$_I$}}                     & \multicolumn{2}{c|}{\textbf{SMER$_{I+C}$}}                                      \\ \hline
\multicolumn{1}{|c|}{\textbf{Métricas}}      & \textbf{ROOV$^1$} & \multicolumn{1}{l|}{\textbf{ROOV$^2$}} & \multicolumn{1}{l|}{\textbf{ROOV$^1$}} & \multicolumn{1}{l|}{\textbf{ROOV$^2$}} \\ \hline
\multicolumn{1}{|l|}{\textbf{TVS (veic./h)}} & 6254              & 6254                                   & 6600                                   & 6557                                   \\ \hline
\multicolumn{1}{|c|}{\textbf{TME (s)}}       & 5,68 $\pm 0,57$             & 6,18 $\pm 0,47$                                  & 5  $\pm 0,39$                                    & 6,10 $\pm 0,46$                                  \\ \hline
\multicolumn{1}{|c|}{\textbf{TMV (s)}}       & 299,06 $\pm 14,37$           & 300,52 $\pm 9,47$                                & 288,11 $\pm 13,71$                                & 292,30 $\pm 9,14$                                \\ \hline
\multicolumn{1}{|c|}{\textbf{VM (Km/h)}}     & 20,06 $\pm 0,65$            & 19,93 $\pm 0,65$                                 & 21,15 $\pm 0,65$                                 & 20,92 $\pm 0,65$                                 \\ \hline
\multicolumn{1}{|c|}{\textbf{CMC (ml)}}      & 575,19            & 660,43                                 & 624,76                                 & 612,00                                    \\ \hline
\multicolumn{1}{|c|}{\textbf{CO(g)}}         & 18,87             & 18,70                                  & 18,52                                  & 19,18                                \\ \hline
\multicolumn{1}{|c|}{\textbf{CO$_2$(g)}}     & 1362,25           & 1446,64                                & 1479,54                                & 1472,53                                \\ \hline
\multicolumn{1}{|c|}{\textbf{HC(g)}}         & 0,527             & 0,535                                  & 0,570                                  & 0,542                                  \\ \hline
\multicolumn{1}{|c|}{\textbf{NOx(g)}}        & 2,81              & 3,10                                   & 3,06                                   & 3,16                                   \\ \hline
\multicolumn{1}{|c|}{\textbf{PMx(g)}}        & 0,164             & 0,189                                  & 0,179                                  & 0,193                                  \\ \hline
\end{tabular}
\end{table}

\begin{table}[!h]
\centering
\caption{Comparação dos desempenhos do ROOV$^2$ contra os do ROOV$^1$, utilizando os resultados obtidos no experimento 2 do cenário 2.}
\label{tab:comp_roov_cen_2_exp_2}
\small\begin{tabular}{|c|c|c|}
\hline
\textbf{Métricas} & \multicolumn{1}{l|}{\textbf{ROOV$^2_{SMER_I}$}} & \multicolumn{1}{l|}{\textbf{ROOV$^2_{SMER_{I+C}}$}} \\ \hline
\textbf{TVS}      & =                                               & \textless 0,06\%                                    \\ \hline
\textbf{TME}      & \textgreater 8,80\%                             & \textgreater 22,00\%                                \\ \hline
\textbf{TMV}      & \textgreater 0,04\%                             & \textgreater 1,45\%                                 \\ \hline
\textbf{VM}       & \textless 0,06\%                                & \textgreater 1,08\%                                 \\ \hline
\textbf{CMC}      & \textgreater 14,81\%                            & \textless 3,64\%                                    \\ \hline
\textbf{CO}       & \textless 0,09\%                                & \textless 3,64\%                                    \\ \hline
\textbf{CO$_2$}   & \textgreater 6,19\%                             & \textless 0,04\%                                    \\ \hline
\textbf{HC}       & \textgreater 1,51\%                             & \textless 4,91\%                                    \\ \hline
\textbf{NOx}      & \textgreater 10,32\%                            & \textgreater 3,26\%                                 \\ \hline
\textbf{PMx}      & \textgreater 15,24\%                            & \textgreater 7,82\%                                 \\ \hline
\end{tabular}
\end{table}

\newpage

\subsubsection{Experimento 3}

A seguir, a Tabela \ref{tab:roov_cen_2_exp_3} apresenta os resultados dos experimento 3 do cenário 2 e, em seguida, a Tabela \ref{tab:comp_roov_cen_2_exp_3} apresenta a análise dos mesmos.

\begin{table}[!h]
\centering
\caption{Resultados obtidos com ROOV nos experimentos 3 dos cenários 1 e 2.}
\label{tab:roov_cen_2_exp_3}
\small\begin{tabular}{c|c|c|c|c|}
\cline{2-5}
\multicolumn{1}{l|}{}                        & \multicolumn{2}{c|}{\textbf{SMER$_I$}}                     & \multicolumn{2}{c|}{\textbf{SMER$_{I+C}$}}                                      \\ \hline
\multicolumn{1}{|c|}{\textbf{Métricas}}      & \textbf{ROOV$^1$} & \multicolumn{1}{l|}{\textbf{ROOV$^2$}} & \multicolumn{1}{l|}{\textbf{ROOV$^1$}} & \multicolumn{1}{l|}{\textbf{ROOV$^2$}} \\ \hline
\multicolumn{1}{|l|}{\textbf{TVS (veic./h)}} & 6970              & 6603                                   & 7057                                   & 6899                                   \\ \hline
\multicolumn{1}{|c|}{\textbf{TME (s)}}       & 9,22 $\pm 0,87$             & 10,71 $\pm 0,86$                                 & 9 $\pm 0,89$                                     & 10,33 $\pm 0,89$                                 \\ \hline
\multicolumn{1}{|c|}{\textbf{TMV (s)}}       & 368,1 $\pm 18,29$            & 379,25 $\pm 18,61$                                & 365,92 $\pm 17,96$                                & 376,16 $\pm 18,29$                                \\ \hline
\multicolumn{1}{|c|}{\textbf{VM (Km/h)}}     & 16,95 $\pm 1,3$            & 14,44 $\pm 1,3$                                  & 16,98 $\pm 1,3$                                 & 14,49 $\pm 1,3$                                 \\ \hline
\multicolumn{1}{|c|}{\textbf{CMC (ml)}}      & 917,55            & 948,95                                 & 917,83                                 & 914,381                                \\ \hline
\multicolumn{1}{|c|}{\textbf{CO(g)}}         & 27,02             & 32,99                                  & 27,25                                  & 46,89                                  \\ \hline
\multicolumn{1}{|c|}{\textbf{CO$_2$(g)}}     & 2138,4            & 2333,79                                & 2139,02                                & 2421,95                                \\ \hline
\multicolumn{1}{|c|}{\textbf{HC(g)}}         & 0,777             & 0,859                                  & 0,779                                  & 0,895                                  \\ \hline
\multicolumn{1}{|c|}{\textbf{NOx(g)}}        & 4,83              & 4,82                                   & 4,27                                   & 5,02                                   \\ \hline
\multicolumn{1}{|c|}{\textbf{PMx(g)}}        & 0,257             & 0,307                                 & 0,259                                  & 0,322                                  \\ \hline
\end{tabular}
\end{table}

\begin{table}[!h]
\centering
\caption{Comparação dos desempenhos do ROOV$^2$ contra os do ROOV$^1$, utilizando os resultados obtidos no experimento 3 do cenário 2.}
\label{tab:comp_roov_cen_2_exp_3}
\small\begin{tabular}{|c|c|c|}
\hline
\textbf{Métricas} & \multicolumn{1}{l|}{\textbf{ROOV$^2_{SMER_I}$}} & \multicolumn{1}{l|}{\textbf{ROOV$^2_{SMER_{I+C}}$}} \\ \hline
\textbf{TVS}      & \textless 5,26\%                                & \textless 2,22\%                                    \\ \hline
\textbf{TME}      & \textgreater 16,16\%                            & \textgreater 14,77\%                                \\ \hline
\textbf{TMV}      & \textgreater 3,02\%                             & \textgreater 2,79\%                                 \\ \hline
\textbf{VM}       & \textless 14,80\%                               & \textless 14,66\%                                   \\ \hline
\textbf{CMC}      & \textgreater 3,42\%                             & \textless 0,03\%                                    \\ \hline
\textbf{CO}       & \textgreater 22,09\%                            & \textgreater 72,07\%                                \\ \hline
\textbf{CO$_2$}   & \textgreater 9,13\%                             & \textgreater 13,22\%                                \\ \hline
\textbf{HC}       & \textgreater 10,55\%                            & \textgreater 14,89\%                                \\ \hline
\textbf{NOx}      & \textless 0,02\%                                & \textgreater 17,56\%                                \\ \hline
\textbf{PMx}      & \textgreater 19,45\%                            & \textgreater 24,32\%                                \\ \hline
\end{tabular}
\end{table}

\section{Discussão}

Nesta seção, são apresentadas as discussões acerca da RAdNet-VE e HRAdNet-VE. Além destas discussões, também são apresentadas as discussões em torno do sistema multiagente de controle de tráfego e do sistema multiagente de planejamento e orientação de rotas. Essas discussões são apresentadas nas próximas seções.

\subsection{RAdNet-VE e HRAdNet-VE}

No que diz respeito à RAdNet-VE e à HRAdNet-VE, percebe-se que o baixo custo de mensagem trafegadas foi consequência do mecanismo de encaminhamento de mensagens. Tal mecanismo fez uso dos campos posição relativa da origem da mensagem de rede, direção de propagação de mensagens e identificador de via para filtrar mensagens. Este baixo custo de mensagens trafegadas também resultou também da estratégia adotada no mecanismo de registro de interesses. Tal mecanismo registrou os interesses definidos pelas aplicações e limitou o número de saltos que as mensagens puderam alcançar, à medida que eram encaminhadas pelos nós. Com isto, os nós transmitiram mensagens dentro de um escopo bem definido, resultando na redução do custo de mensagens trafegadas. Devido ao baixo custo de mensagens trafegadas, as mensagens não congestionaram os canais de comunicação dos nós. Consequentemente, isto levou a uma baixa latência de comunicação entre os nós da rede. Com o baixo custo de mensagens trafegadas e uma baixa latência de comunicação entre os nós, foi possível alcançar altas taxas de entrega de dados. Essas mesmas características puderam também ser observadas nos resultados da HRAdNet-VE. 

Embora o protocolo de comunicação da HRAdNet-VE tenha fornecido altas taxas de entrega de dados por meio de interfaces de acesso à comunicação sem fio baseadas nos padrões IEEE 802.11n e LTE, ele não forneceu altas taxas de entrega de dados com interfaces baseadas no padrão IEEE 802.11p, quando as aplicações operaram em modo reativo. As baixas taxas de entrega de dados é uma consequência do alto número de mensagens produzidas pelo modo de comunicação do tipo requisição-resposta e as altas densidades de nós dentro dos alcances de comunicação das interfaces baseadas no padrão IEEE 802.11p. Embora os nós tenham trafegado em diferentes vias, suas interfaces de acesso à comunicação sem fio compartilharam o mesmo meio. Por isto, o desempenho do protocolo de comunicação foi fortemente afetado. Por esta razão, neste trabalho, argumenta-se que interfaces de acesso a comunicação sem fio baseadas no padrão IEEE 802.11p não devem ser usadas para fornecer comunicações baseadas em comunicações do tipo requisição-resposta em ambientes com altas densidades de nós. 

Além disso, também é importante fazer uma discussão dos protocolos de comunicação da RAdNet-VE e HRAdNet-VE em torno dos requisitos de comunicação das categorias de aplicações para VANETs \cite{Willke:2009}. Embora \citet{Zheng:2015} tenha definidos os requisitos de comunicação de aplicações de serviços de sistemas inteligentes de transporte, observa-se que estes estão inclusos nas definições de requisitos de comunicação de aplicações para VANETs, como pôde ser observado em \citet{Willke:2009}. 

Quando aplicações precisam de comunicações com baixa latência, o protocolo de comunicação deve permitir que os nós se comunicação com o menor atraso possível \cite{Willke:2009}. De acordo com os resultados apresentados nas seções anteriores, os protocolos de comunicação da RAdNet-VE e da HRAdNet-VE forneceram comunicações com baixo atraso. Tanto na RAdNet-VE quanto na HRAdNet-VE não sofreram com o dinamismo topológico inerente às redes veiculares, pois o mecanismo de comunicação centrada em interesse não leva em conta a topologia da rede. Além disto, os protocolos de comunicação da RAdNet-VE e da HRAdNet-VE não usam mensagens de controle para manter ou atualizar dados de roteamento de mensagens. Por este motivo, um número menor de mensagens trafegadas foi produzido durante as comunicações entre os nós.

Segundo \citet{Willke:2009}, aplicações precisam de um protocolo que entregue mensagens a um grupo de nós. Tanto o protocolo da RAdNet-VE quanto o da HRAdNet-VE, satisfazem parte deste requisito por meio do mecanismo de comunicação centrada em interesses herdado do protocolo de comunicação da RAdNet \cite{Dutra:2012}. Porém, o protocolo deve assegurar uma alta probabilidade de entrega de mensagens \cite{Willke:2009}. Para tanto, os protocolos de comunicação da RAdNet-VE e da HRAdNet-VE usam dois campos presentes em seus cabeçalhos de mensagens para encaminhar mensagens, usando os nós mais distantes da origem das mensagens e direções para propagações de mensagens, que são definidas pelas aplicações. Os benefícios disto foram a redução do custo de mensagens trafegadas e redução do tempo de propagação de mensagens em longas distâncias. Outro campo importante no processo de encaminhamento de mensagens é o identificador de via. Por meio deste campo, foi possível limitar o escopo de comunicação nas vias. 

Segundo \citet{Willke:2009}, aplicações para controlar movimentos individuais (por exemplo, controle adaptativo e cooperativo de cruzeiro) operam em um escopo bem definido, que pode ser uma vizinhaça de veículos ou uma pequena região dentro de uma rede. Portanto, o protocolo de comunicação deve assegurar a entrega seletiva de mensagens, que pode ser baseada em trajetória, proximidade de veículo ou identificação do veículo \cite{Willke:2009}. Além disto, aplicações de serviços de sistemas inteligentes de transporte podem necessitar de escopos bem definidos de comunicação, pois os nós de infraestrutura precisam se comunicar uns com os outros para capturar, processar e distribuir informações de segurança e dados relativos às condições de tráfego. Para tanto, esses nós precisam estabelecer comunicações de um salto. Com base no registro do número máximo de saltos juntamente com os interesses, foi possível estabelecer um escopo de comunicação limitado na RAdNet-VE e na HRAdNet-VE. Os benefícios destas estratégias puderam ser observados nos experimentos do segundo cenário de avaliação da HRAdNet-VE, pois foi possível observar que o custo de mensagens aumentou muito pouco, quando comparado com os experimentos do primeiro cenário. 

No que tange o baixo custo de mensagens para comunicações de estrutura de grupos, aplicações para controlar movimentos de grupo precisam de um protocolo que as permitam manter e atualizar estruturas persistentes de grupos \cite{Willke:2009}. Além disto, aplicações de serviços de sistemas inteligentes de transporte para controle de tráfego em áreas compartilhadas entre duas ou mais vias baseiam-se em comunicações de estrutura de grupos para capturar dados de condições de tráfego. Neste caso, cada via forma uma estrutura persistente de grupo. Tanto a RAdNet-VE quanto a HRAdNet-VE são redes centradas em interesses e, por isto, elas não precisam de mecanismos centralizados para gerenciar grupos. Por meio do mecanismo de comunicação centrada em interesses, essas redes implementam comunicações de estrutura de grupos de maneira distribuído e com um baixo custo de mensagens.

\subsection{Sistema Multiagente de Controle de Tráfego}

O sistema multiagente de controle de tráfego teve como ponto de partida o trabalho desenvolvido por \citet{Paiva:2012}, que forneceu um estudo inicial acerca do uso do algoritmo de escalonamento distribuído SMER para controlar sinalizações semafóricas em interseções isoladas ou sistemas coordenados de sinalizações semafóricas. Esse trabalho, por sua vez, foi estendido por esta tese, introduzindo elementos relativos à comunicação veicular, a fim de substituir a leitura das flutuações de tráfego, que antes eram feitas por meio de sensores de pressão instalados nas entradas das vias controladas pelas sinalizações semafóricas, pela troca de mensagens de dados entre veículos conectados e sinalizações semafóricas inteligentes. 

Com base nesta extensão, é possível completar uma lacuna encontrada no trabalho de \citet{Paiva:2012}, que é a contagem de veículos realizada somente, quando os veículos entram em uma via cujo fluxo é controlado por uma sinalização semafórica inteligente. Isto poderia comprometer o ajuste dos intervalos de indicações de luzes verdes nas sinalizações semafóricas, uma vez que alguns veículos poderiam ficar retidos na via, quando os mesmos não conseguissem atravessar a interseção durante o intervalo de indicação de luz verde dedicado para a via em que estão localizados. Nesta tese, essa deficiência foi corrigida por meio de trocas de mensagens dados, que permitiram as sinalizações semafóricas constantemente requisitarem a presença de veículos nas vias de entradas de interseções, onde elas foram instaladas. 

Devido ao tempo dedicado ao projeto, desenvolvimento e teste das redes veiculares centradas em interesses, não foi possível realizar experimentos com o sistema multiagente de controle de tráfego, utilizando tipos diferentes de veículos, tais como caminhões e ônibus. De acordo com o estado atual deste trabalho, as leituras dos fluxos de tráfego levam somente em consideração a entrada de automóveis nas vias. Assim, em experimentos com tipos diferentes de veículos, as leituras dos fluxos de tráfego das vias de entrada das interseções serão equivocadas, no que tange a quantidade equivalente de veículos presentes nestas vias. Segundo \cite{DENATRAN:2014}, caminhões de dois eixos, caminhões de três eixos e ônibus têm fatores de equivalência de tráfego iguais a dois, três e dois, respectivamente. Os fatores de equivalência de tráfego correspondem ao número de veículos que uma determinada classe de veículo corresponde. 

Outra lacuna relativa ao trabalho de \citet{Paiva:2012}, que foi preenchida por por esta tese, é a definição de primitivas acerca do controle de interseções isoladas e do controle de sistemas coordenados de sinalizações semafóricas. Nesta tese, tais primitivas foram definidas na forma de interesses, que foram utilizados nas interações e trocas dados entre os agentes Sinalização Semafóricas. Juntamente com as primitivas de controle, também foram definidos os algoritmos para tratamento das mesmas. 

De acordo com os resultados obtidos, por meio dos experimentos com a HRAdNet-VE, foi possível constatar, durante a avaliação experimental do sistema multiagente de controle de tráfego, que as estratégias relativas ao controle de interseções isoladas e ao controle de sistemas coordenados de sinalizações semafóricas funcionaram corretamente em um ambiente de rede veicular heterogênea. Até a conclusão dos experimentos, existiu uma preocupação quanto às operações do sistema multiagente de controle de tráfego sobre a HRAdNet, pois os agentes Sinalização Semafórica precisariam se adaptar rapidamente às flutuações de fluxos de tráfego. Para tanto, a periodicidade de obtenções de médias de quantidades de veículos nas vias de entrada das interseções deveria pequena. Isto poderia comprometer as operações do sistema multiagente de controle de tráfego. Embora tenha acontecido uma troca intensiva de mensagens de interação entre os agentes, o sistema multiagente de controle de tráfego operou sem apresentar qualquer problema sobre a HRAdNet-VE. Isto, por sua vez, levanta a possibilidade de desenvolvimento de protótipos futuramente, tendo em vista que a indústria local de sistemas inteligentes de transporte é verdadeiramente tímida, no que diz respeito à inovação tecnológica.

Quanto a operação do sistema multiagente de controle de tráfego, como pôde ser observado anteriormente, esta pode se dá tanto em condições favoráveis ou desfavoráveis, quando se trata do funcionamento das sinalizações semafóricas. Esta tese não somente estendeu e aperfeiçoou o trabalho de \citet{Paiva:2012}, mas também propôs estratégias para o controle de interseções isoladas e o controle de sistemas coordenados de sinalizações semafóricas, utilizando a presença de veículos conectados nas vias cujas sinalizações semafóricas apresentaram ausência de funcionamento. Tal estratégia dependeu de um centro de controle de tráfego hipotético, que é capaz de manter um sistema supervisor, que, por sua vez, é alimentado com dados capturados pelo agente Centro de Controle de Tráfego. Embora um centro de controle de tráfego possa oferecer alta disponibilidade para recuperar dados de controle de interseções isoladas e interseções participantes de sistemas coordenados de sinalizações semafóricas, a ausência deste compromete a estratégia, pois veículos conectados e sinalizações semafóricas dependem do encaminhamento de mensagens da rede celular para que a estratégia funcionem corretamente. Uma maneira de resolver este problema é a utilização de comunicação D2D em redes celulares LTE. No entanto, nesta tese, não foi possível tratar disto e, por isto, uma investigação nesta direção deve ser feita no futuro.

Analisando os resultados obtidos por meio dos experimentos dos cenários definidos para avaliação, percebeu-se que o sistema multiagente de controle de tráfego cumpre o seu objetivo, pois ele foi capaz de melhorar a fluidez do tráfego, maximizando o número de veículos que chegam aos destinos de viagem e a velocidade média dos veículos. Além disto, o sistema multiagente de controle de tráfego também foi capaz de minimizar o tempo médio de espera e tempo médio de viagem. Mesmo em um cenário, utilizando interseções com sinalizações semafóricas com ausência de funcionamento, o sistema multiagente de controle de tráfego mostrou-se viável, uma vez que, ao se comparar os resultados obtidos em condições como esta, ele apresentou resultados numéricos próximos aos que foram obtidos nos experimentos do cenário em que todas as sinalizações semafóricas fucionam. Com isto, pode-se afirmar que as estratégias de controle, utilizando o arcabouço teórico do algoritmo distribuído SMER, são promissoras. No entanto, é importante que estas estratégias também sejam avaliadas em cenários cujos os mapas viários sejam diferentes de uma grade manhattan, a fim de aprimorá-las cada vez mais, de modo que elas possam estar preparadas para condições reais de tráfego.

Embora o sistema multiagente de controle de tráfego tenha se saído melhor, quando comparado com um sistema de controle de tráfego baseado em sinalizações semafóricas pré-temporizadas, os resultados poderiam ter sido melhores, caso o sistema fosse imune ao desbalanceamento severo das densidades de veículos em algumas vias. Isto confirmou uma hipótese levantada na introdução deste trabalho. O desempenho de um sistema avançado de gerenciamento de tráfego cai, quando um desbalanceamento severo nas densidades das vias acontece. No mundo real, este desbalanceamento pode ser causado por escolhas equivocadas de rotas por parte dos motoristas ou o uso de sistemas de navegação baseados em mapas estáticos, que, por não terem ciência de rotas alternativas, acabam escolhendo rotas que levam ao aumento da concentração de veículos em determinadas vias, enquanto outras ficam subutilizadas ou até mesmo ociosas.

\subsection{Sistema Multiagente de Planejamento e Orientação de Rotas}

O sistema multiagente de planejamento e orientação de rotas teve como base o trabalho desenvolvido por \citet{Faria:2013}, que forneceu um estudo inicial sobre uma abordagem para planejamento e orientação de rotas. De acordo com a literatura de sistemas multiagentes, tal abordagem pode ser entendida como um sistema multiagente baseado em alocação de recursos. Portanto, a abordagem se baseia no compartilhamento dos planos de temporização das sinalizações semafóricas pertencentes a um sistema de controle de tráfego. No entanto, \citet{Faria:2013} não descreve como tal compartilhamento acontece, pois o mesmo parte do pressuposto de que exista uma agenda global de intervalos de luz verde, que é conhecida por todas as sinalizações semafóricas de um sistema avançado de controle de tráfego. 

Partindo de um ambiente suportado por uma rede veicular heterogênea, esta tese estendeu o trabalho de \citet{Faria:2013}, adicionando um mecanismo de compartilhamento de dados de controle das interseções, que permitiu a cada agente Sinalização Semafórica gerar localmente entradas em sua visão da agenda global de intervalos de indicações de luzes verdes. Os dados de controle das interseções foram compostos dos valores de parâmetros das configurações de desempenho das sinalizações semafóricas e os multigrafos de controle utilizados pelo algoritmo de escalonamento distribuído SMER. Dessa forma, cada agente pôde ter sua visão particular sobre a agenda global produzida pelo sistema avançado de gerenciamento de tráfego. Nesta tese, o sistema gerenciamento de controle de tráfego é o sistema multiagente de controle de tráfego. O maior benefício alcançado por esta técnica é a dispensa do uso de uma infraestrutura computacional de alto custo. No entanto, mais estudos precisam ser realizados no futuro, pois os tamanhos das cópias das agendas crescem, de acordo com o tamanho da rede viária controlada por sinalizações semafóricas. Neste caso, precisam ser realizadas avaliações quanto ao armazenamento das agendas em sinalizações semafóricas inteligentes.

Com base nas disponibilidades das entradas das agendas de intervalos de luz verde, foi possível construir um algoritmo de cálculo de rotas ótimas, que é executado pelos agentes Sinalização Semafórica ou pelos agentes Veículo, quando estes estão controlando o tráfego em interseções. Este ponto põe uma luz sobre o trabalho de \citet{Faria:2013}, havia uma indefinição acerca de quem era o responsável em calcular as rotas. Além disto, nesta tese, com as disponibilidades citadas acima, o mapa viário pôde ser visto como um grande sistema flexível de manufatura do tipo \textit{job-shop}. Desta forma, o algoritmo de cálculo de rotas teve como base heurísticas para geração de regras de despacho. Este algoritmo tirou proveito dos espaços alocados nas vias para os veículos, à medida que estes requisitam cálculos de rotas aos responsáveis pelos controles das interseções. Nesta tese, os espaços são alocados conforme os tamanhos dos veículos e o tamanho das vias. Como só foram considerados automóveis nos experimentos, faz-se necessário investigações, utilizando diferentes classes de veículos.

Nesta tese, sempre que um agente Veículo requisitou um replanejamento de rota, quando o veículo que o embute deixava uma interseção, o sistema de planeamento e orientação de rotas sofreu alterações, no que diz respeito a manutenção de rotas nas entradas das agendas de interlavos de luzes verdes. Com base nos experimentos com a HRAdNet-VE, é possível afirmar que as trocas de mensagens não afetaram o desempenho da rede veicular heterogênea. No entanto, alguns experimentos precisam ser realizados no futuro. Tais experimentos consistem em avaliar o tempo de resposta para os pedidos de cálculo de rotas ótimas, pois os agentes responsáveis pelo controle das interseções precisam efetuar este cálculo localmente. Além disto, também é importante avaliar o tempo gasto para atualizar o estado do sistema de planejamento e orientação de rotas, à medida que novas alocações de rotas são feitas e desfeitas.

Outro ponto importante a ser discutido é o impacto das mudanças das agendas de intervalos de luzes verdes, à medida que os agentes responsáveis pelo controle das interseções mudam as configurações de controle das interseções em função das flutuações dos fluxos do tráfego de uma via. Tais mudanças não afetaram o desempenho da HRAdNet-VE, apesar do alto custo de comunicação gerado por elas.

Nesta tese, foi introduzido o conceito de elemento urbano. Na prática, elementos urbanos podem enriquecer o processo de seleção de rotas para os veículos, pois eles foram capazes de usar a HRAdNet-VE para publicar dados acerca de suas localizações. Com base nestes dados, as camadas de rede dos veículos conectados, sinalizações semafóricas inteligentes e centro de controle de tráfego foram configuradas. Nesta tese, não foi considerado o custo de recuperação de interesses de baixa popularidade na rede centrada em interesses. Como mencionado anteriormente, os interesses com maior popularidade são mantidos em memória. Os interesses com baixa popularidade devem ficar armazenados em disco. Por isto, estudos futuros devem investigar o impacto do custo de recuperação de interesses de baixa popularidade durante a recepção e o encaminhamento de mensagens de rede.

\begin{figure}[H]
	\centering
    \includegraphics[scale=0.75]{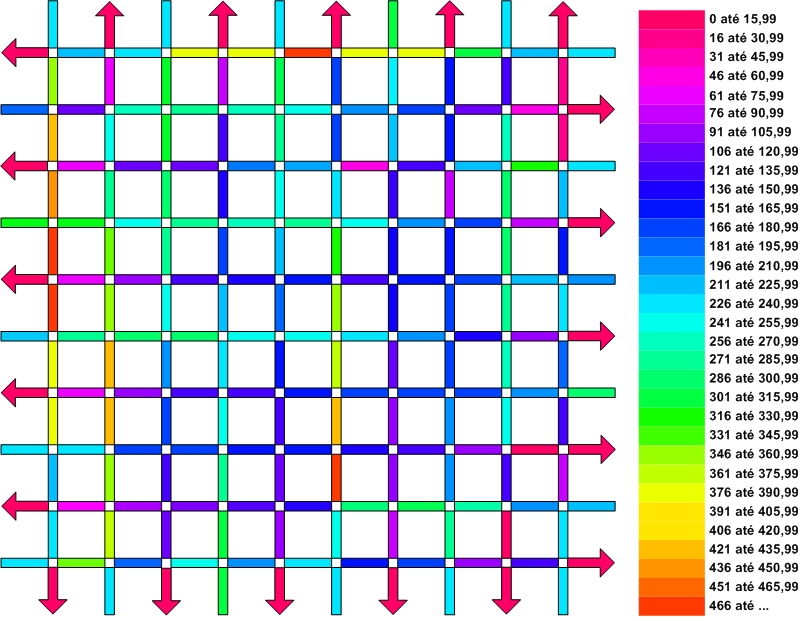}
    \caption{Distribuição do volume de tráfego realizada pelo algoritmo de roteamento baseado no caminho espacialmente mais curto durante experimento 1.}
    \label{fig:dist_traf_caminho_exp_1}
\end{figure}

\begin{figure}[H]
	\centering
    \includegraphics[scale=0.75]{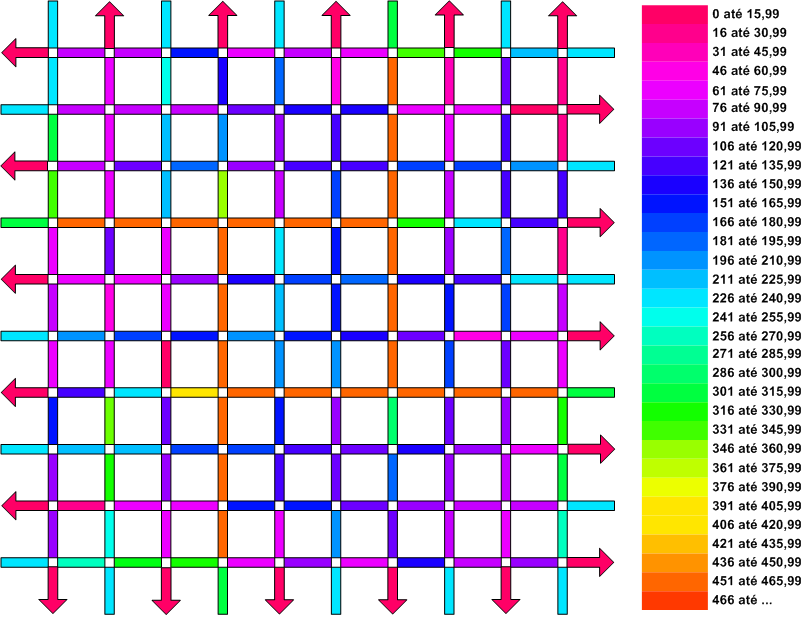}
    \caption{Distribuição do volume de tráfego realizada pelo algoritmo de roteamento baseado no caminho temporalmente mais curto durante experimento 1.}
    \label{fig:dist_traf_tempo_exp_1}
\end{figure}

\begin{figure}[H]
	\centering
    \includegraphics[scale=0.75]{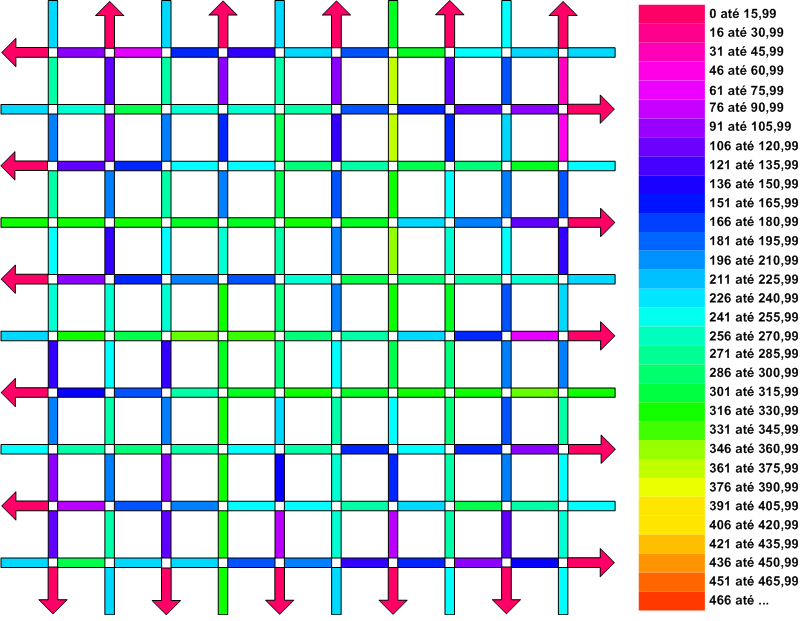}
    \caption{Distribuição do volume de tráfego realizada pelo algoritmo de roteamento orientado à ondas verdes mais curto durante experimento 1.}
    \label{fig:dist_traf_onda_exp_1}
\end{figure}

\begin{figure}[H]
	\centering
    \includegraphics[scale=0.75]{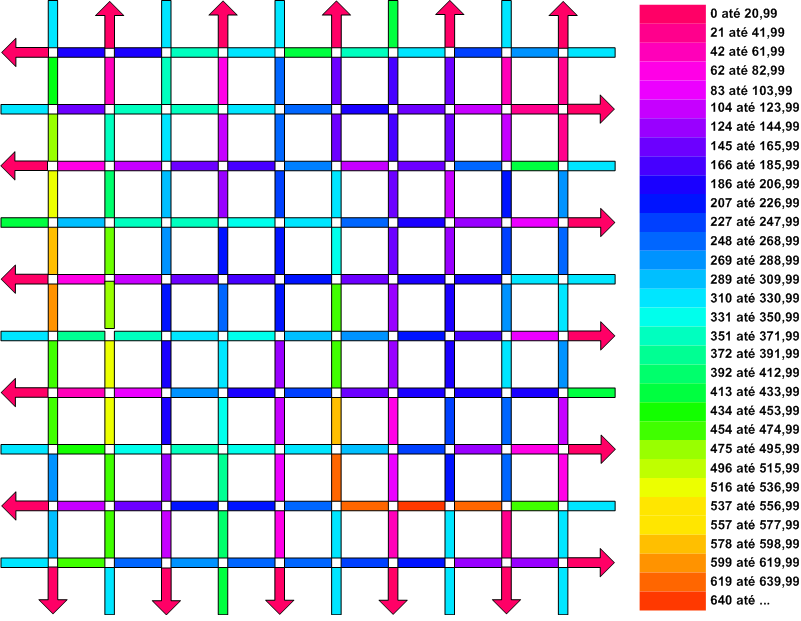}
    \caption{Distribuição do volume de tráfego realizada pelo algoritmo de roteamento baseado no caminho espacialmente mais curto durante experimento 2.}
    \label{fig:dist_traf_caminho_exp_2}
\end{figure}

\begin{figure}[H]
	\centering
    \includegraphics[scale=0.75]{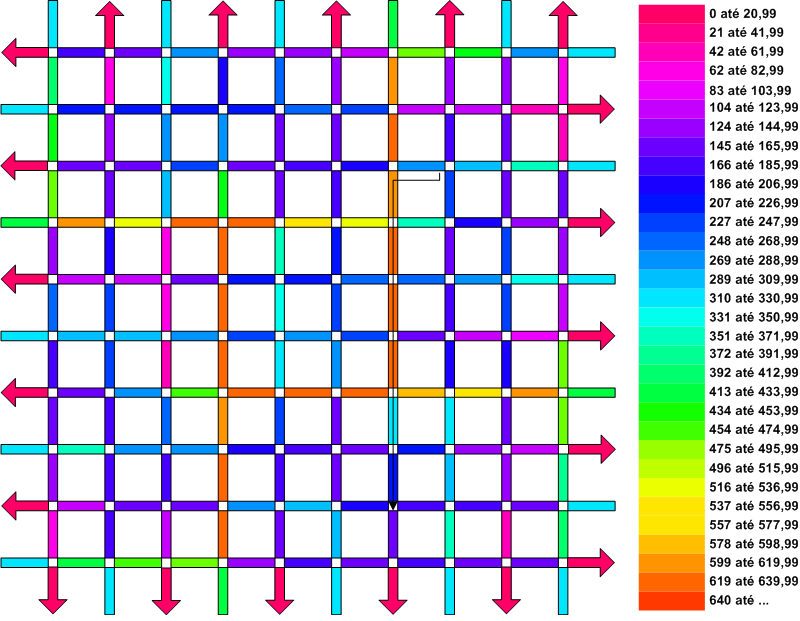}
    \caption{Distribuição do volume de tráfego realizada pelo algoritmo de roteamento baseado no caminho temporalmente mais curto durante experimento 2.}
    \label{fig:dist_traf_tempo_exp_2}
\end{figure}

\begin{figure}[H]
	\centering
    \includegraphics[scale=0.75]{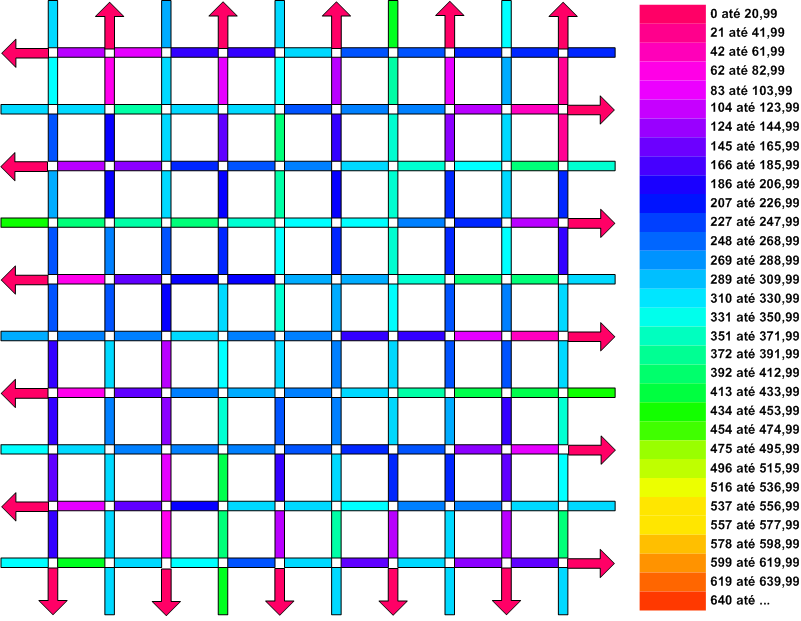}
    \caption{Distribuição do volume de tráfego realizada pelo algoritmo de roteamento orientado à ondas verdes mais curto durante experimento 2.}
    \label{fig:dist_traf_onda_exp_2}
\end{figure}

\begin{figure}[H]
	\centering
    \includegraphics[scale=0.75]{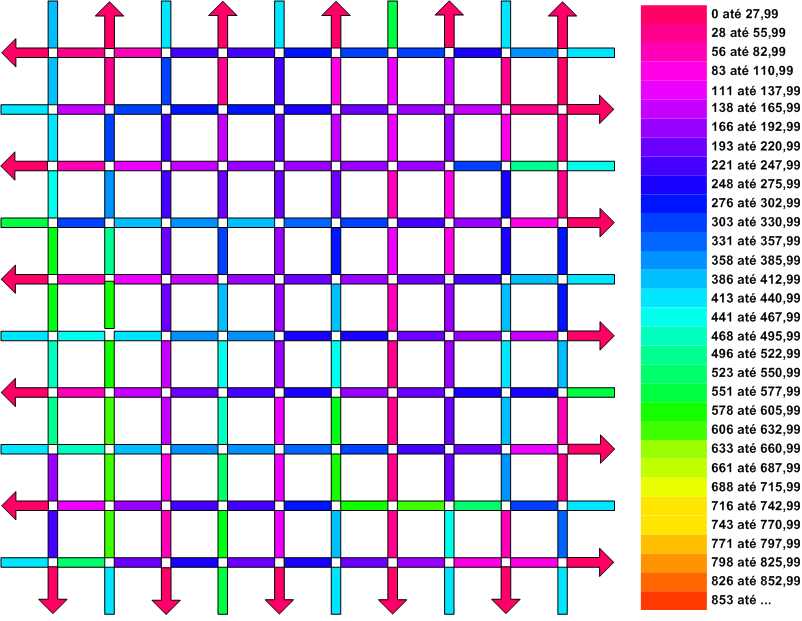}
    \caption{Distribuição do volume de tráfego realizada pelo algoritmo de roteamento baseado no caminho espacialmente mais curto durante experimento 3.}
    \label{fig:dist_traf_caminho_exp_3}
\end{figure}

\begin{figure}[H]
	\centering
    \includegraphics[scale=0.75]{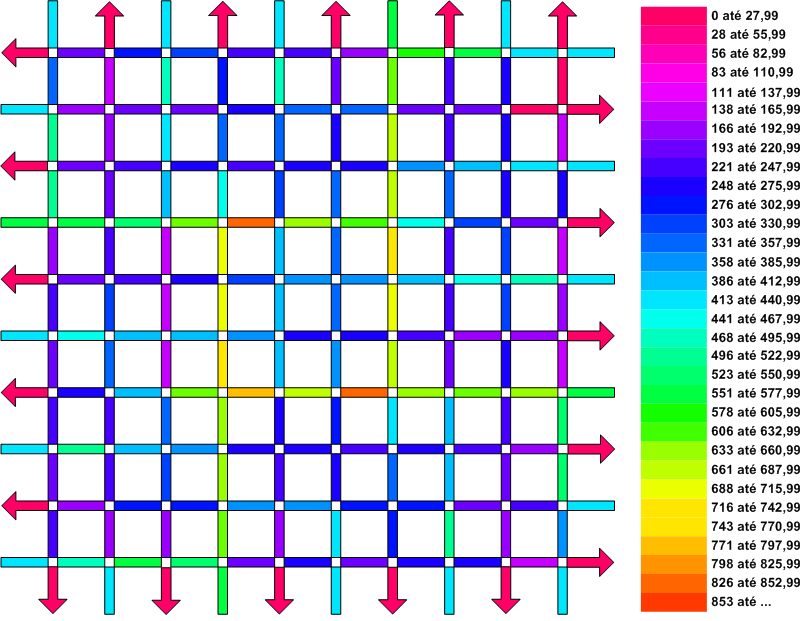}
    \caption{Distribuição do volume de tráfego realizada pelo algoritmo de roteamento baseado no caminho temporalmente mais curto durante experimento 3.}
    \label{fig:dist_traf_tempo_exp_3}
\end{figure}

\begin{figure}[H]
	\centering
    \includegraphics[scale=0.75]{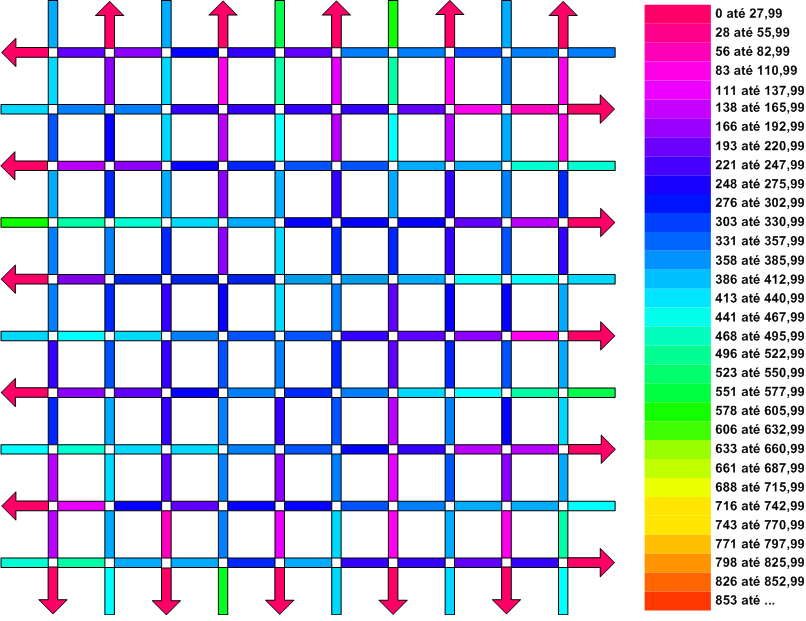}
    \caption{Distribuição do volume de tráfego realizada pelo algoritmo de roteamento orientado à ondas verdes mais curto durante experimento 3.}
    \label{fig:dist_traf_onda_exp_3}
\end{figure}

Analisando os resultados dos experimentos com o sistema de planejamento e orientação de rotas, é possível afirmar que o mesmo atingiu seus objetivos. Neste sentido, ele foi capaz de maximizar o número de veículos que chegam ao destino planejado, assim como, a velocidade média dos veículos. Além destas otimizações, o sistema também foi capaz de minimizar o tempo médio de espera, tempo médio de viagem, consumo de combustível e emissões (CO, CO$_2$, HC, NOx e PMx). Estes resultados só foram alcançados, pois o algoritmo de planejamento e roteamento orientado a ondas verde foi capaz de distribuir os volumes de tráfego ao longo de uma rede viária. Tal distribuição de volume de tráfego pôde ser observada nos três experimentos definidos no cenário 1. As figuras seguintes apresentarão as distribuições de volume de tráfego geradas pelo roteamento baseado em caminho espacialmente mais curto, roteamento baseado em caminho temporalmente mais curto e roteamento orientado a ondas verdes. As distribuições de volume de tráfego geradas durante o experimento 1 são apresentadas nesta ordem pelas Figuras \ref{fig:dist_traf_caminho_exp_1}, \ref{fig:dist_traf_tempo_exp_1} e \ref{fig:dist_traf_onda_exp_1}. As distribuições de volume de tráfego geradas durante o experimento 2 são apresentadas nesta ordem pelas Figuras \ref{fig:dist_traf_caminho_exp_2}, \ref{fig:dist_traf_tempo_exp_2} e \ref{fig:dist_traf_onda_exp_2}. As distribuições de volume de tráfego geradas durante o experimento 3 são apresentadas nesta ordem pelas Figuras \ref{fig:dist_traf_caminho_exp_3}, \ref{fig:dist_traf_tempo_exp_3} e \ref{fig:dist_traf_onda_exp_3}.

Como pode ser observado nas figuras apresentadas anteriormente, o algoritmo de roteamento orientado a ondas verdes foi capaz de distribuir de maneira uniforme o volume de tráfego sobre o mapa viário utilizado nos experimentos, enquanto os demais algoritmos de roteamento acumularam veículos em determinadas áreas do mapa. Como pode ser observado nas Figuras \ref{fig:dist_traf_onda_exp_1}, \ref{fig:dist_traf_onda_exp_2} e \ref{fig:dist_traf_onda_exp_3}, as cores com maior predominância nos mapas são aquelas mais próximas do centro das escalas de cores. É importante perceber que o algoritmo de roteamento orientado a ondas verdes é capaz de montar rotas que minimizam o tempo de espera do veículos, quando eles chegam em uma via, onde a sinalização semafórica indica luz vermelha. No entanto, os resultados numéricos apresentados anteriormente também foram obtidos com uso do mecanismo de controle de velocidade orientado à ondas verdes proposto no trabalho de \citet{Faria:2013}. 

Nesta tese, o mecanismo de controle de velocidade orientado a ondas verdes foi viabilizado em conjunto com um controle adaptativo e cooperativo de cruzeiro. Embora \citet{Faria:2013} tenha proposto o mecanismo em questão, ele não foi testado em ambiente de rede veicular. Portanto, esta tese mostrou que a proposta de \citet{Faria:2013} funcionou adequadamente, quando esta foi executado sobre uma rede veicular centrada em interesses. 

\section{Considerações Finais}

Este capítulo apresentou os procedimentos adotados durante o processo de avaliação experimental de cada uma das propostas apresentadas nesta tese. Na seção de metodologia, foram apresentadas as ferramentas utilizadas para no desenvolvimento e avaliação das propostas. Em seguida, foi descrito o ambiente computacional utilizado durante a realização dos experimentos. Após esta descrição, as seções subsequentes apresentaram as avaliações de cada uma das propostas apresentas por esta tese. Em cada uma das apresentações, foram descritos os cenários e configurações dos experimentos. Após isto, foram apresentados os resultados obtidos por meio dos experimentos. Com base nestes resultados, as avaliações terminaram com uma análise dos mesmos.

  \chapter{Conclusões e Trabalhos Futuros}

Este capítulo apresenta as conclusões e trabalhos futuros relacionados à pesquisa desenvolvida nesta tese.

\section{Conclusões}

Esta tese apresentou um controle distribuído de tráfego baseado em veículos conectados. Devido aos requisitos de comunicação deste tipo de aplicação para sistemas inteligentes de transporte, tornou-se imperativo o desenvolvimento de uma rede \textit{ad hoc} veicular cujo protocolo de comunicação fosse capaz de fornecer baixo custo de mensagens trafegadas, comunicações de baixa latência, altas taxas de entrega, escopo de comunicação bem definido e escalabilidade.

Nesse sentido, este estudo propôs inicialmente uma rede \textit{ad hoc} veicular centrada em interesses chamada RAdNet-VE, que é uma extensão da rede \textit{ad hoc} móvel centrada em interesses para ambientes veiculares. Devido à necessidade de realizar comunicações com diversos tipos de tecnologias de acesso à comunicação sem fio, o projeto da RAdNet-VE foi estendido, objetivando a criação de uma rede veicular heterogênea centrada em interesses, que foi chamada de HRAdNet-VE. Tanto a RAdNet-VE quanto a HRadNet-VE foram capazes de satisfazer os requisitos citados acima. 

Após a conclusão do projeto da HRAdNet-VE, deu-se início ao projeto de um sistema multiagente de controle de tráfego. As estratégias de controle distribuído propostas neste sistema, tiveram como base os algoritmos de escalonamento distribuído propostos por \citet{Paiva:2012}. Os resultados obtidos por meio dos experimentos mostraram que tal sistema é capaz de melhorar a fluidez do tráfego em uma área controlada por sinalizações semafóricas inteligentes. No entanto, tal sistema de controle de tráfego, assim como os existentes na literatura, não foi imune ao desbalanceamento na distribuição dos fluxos de tráfego. 

Para tratar este problema, este estudo propôs um sistema multiagente de planejamento e orientação de rotas. Tal sistema foi construído sobre o sistema multiagente de controle de tráfego, a fim de tirar proveito das disponibilidades de intervalos de indicações de luzes verdes das sinalizações semafóricas e, com isto, criar rotas ótimas para o roteamento de veículos. Para lidar com o problema de roteamento de veículos, esta tese se apropriou das teorias e heurísticas de despacho de sistemas flexíveis de manufatura do tipo job-shop, assim como, das propostas apresentadas por \citet{Faria:2013}. Com isto, foi possível construir um algoritmo de roteamento de veículos orientado a ondas verdes, que tira proveito das disponibilidades de intervalos de luz verde e espaços nas vias de uma rede viária.  Por meio do sistema multiagente de planejamento e orientação de rotas, e um mecanismo de controle de velocidade orientado à ondas verdes, foi possível melhorar um pouco mais a fluidez do tráfego de veículos.

Por fim, conclui-se que, de acordo com os resultados apresentados no capítulo anterior a este, esta pesquisa alcançou os objetivos traçados inicialmente na introdução deste trabalho. Além disto, até o término deste estudo, não foram encontradas abordagens similares às propostas aqui. Por isto, conclui-se que o controle distribuído de tráfego baseado em veículos conectados, utilizando comunicações centradas em interesses em uma rede heterogênea veicular, é inédito. Embora existam trabalhos relacionados à redes centradas em informação/conteúdo, estratégias de controle distribuído de tráfego e estratégias distribuídas de planejamento e orientação de rotas, ainda não foram direcionados esforços na direção proposta neste estudo.

\section{Publicações e Submissões}

Durante a condução deste trabalho de doutorado, obteve-se uma publicação, que é a seguinte: \textit{Interest-Centric Vehicular Ad Hoc Network}, aceito para publicação na \textit{12th IEEE International Conference on Wireless and Mobile Computing, Networking and Communications}, realizada em outubro de 2016. Este artigo apresenta a proposta da RAdNet-VE. Além disto, também foi submetido um artigo entitulado \textit{Heterogeneous Interest-Centric Vehicular Network} para o periódico \textit{IEEE Transactions on Intelligent Transportation Systems.} Segundo a última avaliação do Qualis, a conferência e o periódico foram avaliados como A2 e A1, respectivamente. 

\section{Trabalhos Futuros}

As propostas de trabalhos futuros relacionados a este estudo foram agrupadas, de acordo com as propostas apresentadas nos Capítulos \ref{cap:radnet-ve}, \ref{cap:controle} e \ref{cap:orientacao}. Com base nisto, seguem os trabalhos futuros:

\begin{itemize}
	\item \textbf{Redes Veiculares Centradas em Interesses:}
    
    \begin{itemize}
	    \item Pesquisar e desenvolver novos cenários relacionados à área de sistemas inteligentes de transporte, de modo que, com base nestes, sejam criados novos experimentos para realização de testes com a RAdNet-VE e a HRAdNet-VE;
        \item Projeto e desenvolvimento de uma arquitetura de software cujo objetivo é servir como um \textit{framework} para desenvolvimento de aplicações de serviços de sistemas inteligentes de transporte;
        \item Projeto e desenvolvimento de um \textit{framework} para desenvolvimento de cenários de avaliação de sistemas inteligentes de transporte sobre as redes veiculares centradas em interesses, estendendo os \textit{frameworks} disponíveis para o Omnet++. Neste trabalho futuro, também deve ser considerada uma maneira de trabalhar com mapas do Open Street Map;
        \item Investigar a possibilidade de utilização de comunicações centradas em interesses, utilizando comunicação D2D em redes celulares LTE;
        \item Avaliar o impacto de interesses de baixa popularidade na latência de comunicação entre nós, uma vez que tais interesses não ficam constantemente disponíveis em memória. Para tanto, devem ser utilizados dispositivos que possam ser embarcados em veículos e sinalizações semafóricas;
        \item Pesquisar e desenvolver estudos relacionados à segurança das redes centradas em interesses, utilizando nomes auto-certificáveis;
        
    \end{itemize}
    \item \textbf{Controle Inteligente de Tráfego Utilizando Sinalizações Semafóricas e Veículos Conectados:}

    \begin{itemize}
    	\item Melhorar o mecanismo de leitura de fluxos de tráfego, de modo que ele possa utilizar dados microscópicos de veículos, tais como a classe dos mesmos. Com isto, o controle de tráfego passará a trabalhar com quantidades equivalentes de veículos;
        \item Melhorar o mecanismo de cálculo de demandas, de modo que ele possa considerar diferentes tamanhos vias; 
        \item Desenvolver novos cenários para avaliações experimentais, utilizando mapas viários cujas vias apresentam mais de uma faixa;
        \item Desenvolver estudos, utilizando o sistema multiagente de controle de tráfego, em mapas viários do Open Street Map. Estes estudos devem utilizar dados realísticos de tráfego, de modo que seja possível avaliar o comportamento das estratégias de controle de tráfego em condições reais de tráfego;
        \item Estender o algoritmo SMER de controle de interseções isoladas, de modo que ele possa levar em consideração o fluxo de pedestres em interseções controladas por sinalizações semafóricas;
        \item Desenvolver cenários para avaliações experimentais, utilizando o controle de interseções por meio de veículos conectados, sem que exista um centro de controle de tráfego para manter dados de controle acerca de interseções. Neste cenário, os agentes Veículo devem ser capazes de negociar as passagens de seus veículos conectados pelas interseções, enquanto outros veículos esperam as liberações das mesmas, para, então, atravessá-las;
    \end{itemize}
    
    \item \textbf{Planejamento e Orientação Inteligente de Rotas Baseados em Interesses de Motoristas:}
    
    \begin{itemize}
    	\item Avaliar o impacto do número de sinalizações semafóricas no crescimento das agendas de intervalos de luzes verdes;
        \item Avaliar o impacto de diferentes classes de veículos, utilizando o algoritmo de planejamento e orientação de rotas orientado à ondas verdes;
        \item Avaliar o tempo de resposta do algoritmo de planejamento e orientação de rotas em mapas viários cujas vias possuem mais de uma faixa;
        \item Avaliar o tempo de atualização do estado global do sistema multiagente de planejamento e orientação de rotas em função da quantidade de veículos presentes no mapa viário;
        \item Utilizar os dados publicados por elementos urbanos, de modo que possam ser utilizados para ponderar dinamicamente as vias. Isto seria um recurso bastante útil, pois, em um ambiente urbano, existem áreas, tais como as proximidades de hospitais e escolas, onde o fluxo de veículos precisa ser gerenciado com maiores cuidados. Outra possibilidade seria o uso destes dados em eventos, tais como shows, jogos de futebol, entre outros;
        \item Desenvolver cenários para avaliações experimentais, envolvendo atividades de sensoriamento urbano. Com isto, dispositivos móveis utilizados por pedestres ou sensores instalados em dispositivos computacionais poderiam ser utilizados para alimentar o sistema. Com base nisto, novos critérios de para o cálculo de rotas ótimas poderiam ser explorados.
    \end{itemize}
\end{itemize}

  \backmatter
  \bibliographystyle{coppe-unsrt}
  \bibliography{thesis}
  
\appendix
  
    \chapter{Interesses do Controle de Tráfego}

\small\begin{longtable}{|p{4.7cm}|c|p{1.5cm}|p{6cm}|}
  \caption{Interesses de registrados por um agente Centro de Controle de Tráfego e as ações executadas pelo agente e ou pelo ambiente.}
  \label{tab:interesses_cct}
    \\ \hline
    
    \centering\textbf{Interesse} & \centering\textbf{NMS} & \centering\textbf{TAC} & \textbf{Ação Executada} \\ \hline
    \textit{roadway\_vehicle\_amount} & 1 & \centering LTE & Atualiza a quantidade de veículos de uma via no sistema supervisório e encaminha a mensagem para as sinalizações semafóricas.\\ \hline
    \textit{corridor\_vehicle\_amount} & 1 & \centering LTE & Atualiza a demanda de um sistema coordenado de sinalizações semafóricas no sistema supervisório e encaminha a mensagem para as sinalizações semafóricas.\\ \hline
    \textit{corridor\_reversibility\_change} & 1 & \centering LTE & Encaminha a mensagem para as sinalizações semafóricas. No entanto, se estiver controlando sistemas coordenados de sinalizações semafóricas, também age como um líder destes sistemas. \\ \hline
    \textit{corridor\_edge\_reversal} & 1 & \centering LTE & Encaminha a mensagem para as sinalizações semafóricas. No entanto, se estiver controlando sistemas coordenados de sinalizações semafóricas, também age como um líder destes sistemas. \\ \hline
    \textit{group\_member\_vehicle\_amount} & 1 & \centering LTE & Encaminha a mensagem para as sinalizações semafóricas. No entanto, se estiver controlando sistemas coordenados de sinalizações semafóricas, também age como um líder destes sistemas.  \\ \hline
    \textit{roadway\_segment\_vehicle\_amount} & 1 & \centering LTE & Encaminha a mensagem para as sinalizações semafóricas. No entanto, se estiver controlando sistemas coordenados de sinalizações semafóricas, também age como um líder destes sistemas.  \\ \hline
    \textit{traffic\_light\_coordination} & 1 & \centering LTE & Encaminha a mensagem para as sinalizações semafóricas. \\ \hline
    \textit{traffic\_light} & 1 & \centering LTE & O agente registra os identificadores dos agentes Sinalização Semafórica, bem como, os parâmetros de configuração de performance informados pelo engenheiro de tráfego. \\ \hline
    \textit{intersection\_control\_data\_request} & 1 & \centering LTE & O agente fornece os dados de controle de uma interseção, caso não exista um agente Veículo sendo responsável pelo controle da interseção. Se houver um agente com tal responsabilidade, a mensagem de interação é encaminhada para o mesmo. \\ \hline
    \textit{intersection\_control\_data} & 1 & \centering LTE & O agente atualiza os dados de controle de uma interseção e encaminha a mensagem para um agente Veículo que esteja sendo responsável pelo controle de uma interseção. \\ \hline
    
\end{longtable}

\begin{table}[H]
  \centering
  \caption{Interesses de registrados por um agente Veículo e as ações executadas pelo agente e ou pelo ambiente.}
  \label{tab:interesses_veiculo}
  \small\begin{tabular}{|p{4.7cm}|c|p{1.5cm}|p{6cm}|}
    \hline
    \centering\textbf{Interesse} & \textbf{NMS} & \centering\textbf{TAC} & \textbf{Ação Executada} \\ \hline
    \textit{roadway\_presence\_request} & 8 & \centering IEEE 802.11 & Ativa notificação da presença de veículo na via. \\  \hline
    \textit{roadway\_leader\_location}  & 8 & \centering IEEE 802.11 & Registra posição de um veículo da via e, em seguida, reexecuta a eleição de líder.                        \\ \hline
    \textit{vehicle\_on\_}$<$id. da via$>$ & 8 & \centering IEEE 802.11 & Se o veículo conectado for líder, registra presença de um veículo conectado em uma via. Caso contrário, o agente não executa qualquer ação, mas o ambiente somente coopera com a HRAdNet-VE.                         \\ \hline
    \textit{roadway\_presence\_confirmation} & 8 & \centering IEEE 802.11 & Se for o destino da mensagem de interação, interrompe notificação da presença do veículo na via. Caso contrário, o agente não executa qualquer ação, mas o ambiente somente coopera com a HRAdNet-VE. \\ \hline  
    \textit{vehicle\_out\_}$<$id. da via$>$ & 8 & \centering IEEE 802.11 & Se o veículo conectado for líder, remove presença de um veículo conectado em uma via. Caso contrário, o agente não executa qualquer ação, mas o ambiente somente coopera com a HRAdNet-VE. \\ \hline
    \textit{roadway\_ou\_confirmation} & 8 & \centering IEEE 802.11 & Se for o destino da mensagem de interação, interrompe notificação da presença do veículo na via. Caso contrário, o agente não executa qualquer ação, mas o ambiente somente coopera com a HRAdNet-VE. \\ \hline  
    \textit{intersection\_control\_data} & 1 & \centering LTE & Obtém os dados de controle de uma interseção, caso as sinalizações semafóricas desta apresentem ausência de funcionamento. \\ \hline  
    \textit{intersection\_control\_data\_request} & 1 & \centering LTE & Fornece dados de controle de uma interseção, caso o agente Veículo seja o líder de uma via. \\ \hline  
  \end{tabular}
\end{table}

\begin{table}[H]
  \centering
  \caption{Interesses comuns entre agentes Sinalização Semafórica e as ações executadas pelo agente e ou pelo ambiente.}
  \label{tab:interesses_sinalizacao}
  \small\begin{tabular}{|p{4.7cm}|c|p{1.5cm}|p{6cm}|}
    \hline
    \centering\textbf{Interesse} & \textbf{NMS} & \textbf{TAC} & \textbf{Ação Executada} \\ \hline
    \textit{vehicle\_on\_}$<$id. da via$>$ & 1 & \centering IEEE 802.11 & Registra presença de um veículo conectado em uma via. \\ \hline     \textit{vehicle\_out\_}$<$id. da via$>$ & 1 & \centering IEEE 802.11 & Remove presença de um veículo em uma via.  \\ \hline  	 \textit{reversibility\_change} & 1 & \centering IEEE 802.11 & Atualiza reversibilidade de um agente no multigrafo utilizado no controle de uma interseção isolada. \\ \hline
    \textit{edge\_reversal} & 1 & \centering IEEE 802.11 & Reverte arestas de um agente no multigrafo utilizado no controle de uma interseção isolada.  \\ \hline
    \textit{roadway\_vehicle\_amount} & 1 & \centering IEEE 802.11 & Atualiza a quantidade de veículos conectados obtida por um agente no multigrafo utilizado no controle de uma interseção isolada. \\ \hline
  \end{tabular}
\end{table}

\begin{table}[H]
  \centering
  \caption{Interesses registrados também em um agente Sinalização Semafórica participante de um sistema coordenado de sinalizações semafóricas e as ações executadas pelo agente e ou pelo ambiente.}
  \label{tab:interesses_participante}
  \small\begin{tabular}{|p{4.7cm}|c|p{1.5cm}|p{6cm}|}
    \hline
    \centering\textbf{Interesse} & \centering\textbf{NMS} & \centering\textbf{TAC} & \textbf{Ação Executada} \\ \hline
    \textit{traffic\_light\_coordination} & 1 & \centering LTE & Escalona ação para participação no sistema coordenado de sinalizações semafóricas. \\ \hline
    \textit{reversal\_of\_all\_edges} & 1 & \centering IEEE 802.11 & Reverte todas as arestas de um agente, fazendo o algoritmo SMER para controle de interseções isoladas se comportar como o algoritmo SER. \\ \hline
    \textit{corridor\_controller\_traffic\_light} & 1 & \centering LTE & O agente registra o identificador do líder de um sistema coordenado de sinalizações semafóricas. \\ \hline
  \end{tabular}
\end{table}

\begin{table}[H]
  \centering
  \caption{Interesses registrados também em um agente Sinalização Semafórica líder de um sistema coordenado de sinalizações semafóricas e as ações executadas pelo agente e ou pelo ambiente.}
  \label{tab:interesses_controlador}
  \small\begin{tabular}{|p{4.7cm}|c|p{1.5cm}|p{6cm}|}
    \hline
    \centering\textbf{Interesse} & \centering\textbf{NMS} & \centering\textbf{TAC} & \textbf{Ação Executada} \\ \hline
    \textit{corridor\_reversibility\_change} & 1 & \centering LTE & Atualiza a reversibilidade de um agente no multigrafo de controle de sistemas coordenados de sinalizações semafóricas. \\ \hline
    \textit{corridor\_edge\_reversal} & 1 & \centering LTE & Reverte arestas de um agente no multigrafo utilizado no controle de sistemas coordenados de sinalizações semafóricas. \\ \hline
    \textit{corridor\_vehicle\_amount} & 1 & \centering LTE & Atualiza a demanda de um agente no multigrafo utilizado no controle de sistemas coordenados de sinalizações semafóricas. \\ \hline
    \textit{group\_member\_vehicle\_amount} & 1 & \centering LTE & Registra a demanda de um agente em um agrupamento de sistemas coordenados de sinalizações semafóricas. \\ \hline
    \textit{roadway\_segment\_vehicle\_amount} & 1 & \centering LTE & Registra a demanda de um dos segmentos de via componentes de um corredor. \\ \hline
    \textit{reversal\_of\_all\_edges} & 1 & \centering IEEE 802.11 & Reverte todas as arestas de um agente, fazendo o algoritmo SMER para controle de interseções isoladas se comportar como o algoritmo SER. \\ \hline
    \textit{corridor\_participant\_traffic\_light} & 1 & \centering LTE & O agente registra o identificador de uma participante do sistema coordenado de sinalizações semafóricas, assim como, dados relativos ao seguimento de via deste participante. \\ \hline
    \textit{group\_member} & 1 & \centering LTE & O agente registra o identificador de um membro do mesmo grupo de sistemas coordenados de sinalizações semafóricas. \\ \hline
  \end{tabular}
\end{table}

\begin{table}[H]
  \centering
  \caption{Interesses registrados também em um agente Sinalização Semafórica participante de um sistema coordenado de sinalizações semafóricas e as ações executadas pelo agente e ou pelo ambiente.}
  \label{tab:interesses_participante}
  \small\begin{tabular}{|p{4.7cm}|c|p{1.5cm}|p{6cm}|}
    \hline
    \centering\textbf{Interesse} & \centering\textbf{NMS} & \centering\textbf{TAC} & \textbf{Ação Executada} \\ \hline
    \textit{traffic\_light\_coordination} & 1 & \centering LTE & Escalona ação para participação no sistema coordenado de sinalizações semafóricas. \\ \hline
    \textit{reversal\_of\_all\_edges} & 1 & \centering IEEE 802.11 & Reverte todas as arestas de um agente, fazendo o algoritmo SMER para controle de interseções isoladas se comportar como o algoritmo SER. \\ \hline
    \textit{corridor\_controller\_traffic\_light} & 1 & \centering LTE & O agente registra o identificador do líder de um sistema coordenado de sinalizações semafóricas. \\ \hline
  \end{tabular}
\end{table}

\begin{table}[!h]
  \centering
  \caption{Interesses registrados também em um agente Sinalização Semafórica participante permanente ou temporário de operações de coordenação de sinalizações semafóricas e as ações executadas pelo agente e ou pelo ambiente.}
  \label{tab:interesses_nao_participante}
  \small\begin{tabular}{|p{4.7cm}|c|p{1.5cm}|p{6cm}|}
    \hline
    \centering\textbf{Interesse} & \centering\textbf{NMS} & \centering\textbf{TAC} & \textbf{Ação Executada} \\ \hline
    \textit{participation\_in\_traffic\_light\_coordination} & 1 & \centering IEEE 802.11 & Prepara o agente para cooperar em interseções partipantes de um ou mais sistemas coordenados de sinalizações semafóricas.\\ \hline
    \textit{confirmation\_in\_traffic\_light\_coordination} & 1 & \centering IEEE 802.11 & Confirma a participação de um agente na coordenação de sinalizações semafóricas.\\ \hline
    \textit{reversal\_of\_all\_edges} & 1 & \centering IEEE 802.11 & Reverte todas as arestas de um agente, fazendo o algoritmo SMER para controle de interseções isoladas se comportar como o algoritmo SER. \\ \hline
  \end{tabular}
\end{table}
    \chapter{Algoritmos do Controle de Tráfego}

\begin{algorithm}[!h]
  \SetAlgoLined
  \scriptsize
  idVia := obterVia(posicaoGPS$_i$)\;
  \eSe{$idVia \neq idViaAtual_i$}{
  	presencaSinalizacao$_i$ := \textbf{falso}\;
    removerInteresseCamadaRede( ``vehicle\_out\_'' + idViaAnterior$_i$, ``IEEE 802.11'' )\;
    idViaAnterior$_i$ := idViaAtual$_i$\;
    idViaAtual$_i$ := idVia\;
    interesse := ``vehicle\_out\_'' + idViaAnterior\;
    Msg$_i$ := $\langle$ interesse, \textbf{nulo}, idViaAtual$_i$, 1, ``IEEE 802.11''  $\rangle$\;
    \textbf{enviar} Msg$_i$\;
    numTentNotifMudanca$_i$ = numTentNotifMudanca$_i$ + 1\;
    confirmacaoSaida$_i$ := \textbf{falso}\;
    removerViaCamadaRede(idViaAnterior$_i$)\;
    adicionarViaCamadaRede(idViaAtual$_i$)\;
    removerInteresseCamadaRede( ``vehicle\_on\_'' + idViaAnterior$_i$, ``IEEE 802.11'' )\;
    adicionarInteresseCamadaRede( ``vehicle\_on\_'' + idViaAtual$_i$, 8, ``IEEE 802.11'' )\;
    adicionarInteresseCamadaRede( ``vehicle\_out\_'' + idViaAtual$_i$, 8, ``IEEE 802.11'' )\;
    desfazerConfiguracaoAgenteVeiculoComoLider()\;
  }{
  	\Se{$confirmacaSaida_i$ = \textbf{falso}} {
    	\eSe{$numTentNotifMudanca_i < paramNumMaxTentNotifMudanca_i$}{
        	interesse := ``vehicle\_out\_'' + idViaAnterior\;
            \eSe{ctrlIntersecaoAtivo$_i$ = \textbf{falso}}{
    		
    			Msg$_i$ := $\langle$ interesse, \textbf{nulo}, idViaAnterior$_i$, 1, ``IEEE 802.11''  $\rangle$\;
    			\textbf{enviar} Msg$_i$\;
            }{
            	params := $\{\langle$ ``idViaAnterior'', idViaAnterior$_i$ $\rangle\}$\;
            	Msg$_i$ := $\langle$ interesse, \textbf{nulo}, \textbf{nulo}, 0,``LTE'', params  $\rangle$\;
    			\textbf{enviar} Msg$_i$\;
            }
            numTentNotifMudanca$_i$ = numTentNotifMudanca$_i$ + 1\;
        }{
        	\Se{ctrlIntersecaoAtivo$_i$ = \textbf{falso}}{
            	numTentNotifMudanca$_i$ := paramNumMaxTentNotifMudanca$_i$\;
            }
        }
    }
  }
  
  \caption{Monitoramento de mudança de faixa ou via.}
  \label{alg:monitoramento_faixas}
\end{algorithm}

\newpage

\begin{algorithm}[H]
  \SetAlgoLined
  	\scriptsize
    \Se{existeSinalizacaoSemaforica(idViaAtual) = \textbf{verdadeiro}} {
      \Se{estaNaAreaMotiramentoTrafego(idViaAtual$_i$) = \textbf{verdadeiro}}{
        \Se{presencaSinalizacao$_i$ = \textbf{falso}}{
            \eSe{numTentativasNotificacao$_i$ $<$ paramNumMaxTentativasNotificacao$_i$}{
            	interesse := ``vehicle\_on\_'' + idVia\;
                params := \{$\langle$ ``posicao'',posicaoGPS$_i\rangle$\}\;
                \eSe{ctrlIntersecaoAtivo$_i$ = \textbf{falso}}{
                  
                  Msg$_i := \langle$ interesse, \textbf{nulo}, idVia, 1, ``IEEE 802.11'', params$\rangle$\;
                  \textbf{enviar} Msg$_i$\;
                }{
                  	Msg$_i := \langle$ interesse, \textbf{nulo}, \textbf{nulo}, 1, ``LTE'', params$\rangle$\;
                    \textbf{enviar} Msg$_i$\;
                }
                numTentativasNotificacao$_i$ := numTentativasNotificacao$_i$ + 1\;
            }{
            	\Se{ctrlIntersecaoAtivo$_i$ = \textbf{falso}}{
                	iniciarControleIntersecaoComVeiculos()\;
                	numTentativasNotificacao$_i$ := paramNumMaxTentativasNotificacao$_i$ 
                }
            }
        }
      }
    }
    
  \caption{Notificação de presença de um veículo em uma via.}
  \label{alg:notificacao_presenca}
\end{algorithm}

\begin{algorithm}[H]
  \SetAlgoLined
  	\scriptsize
    \Entrada{Msg$_j$ := Msg$_i$, \textbf{se} Msg$_i$.interesse = \textit{vehicle\_on\_}$<$id. da via$>$}
    distancia := calcDistancia(posicaoGPS$_i$, Msg$_j$.parametros[``posicao''])\;
    areaMonitoramentoTrafego := obterAreaMonitoramentoTrafego(Msg$_j$.via)\;
    \Se{$Msg_j.direcao = 1 \wedge distancia \leq areaMonitoramentoTrafego_i$}{
    	\Se{$Msg_j.origem \notin veiculosVia_i $}{
        	\textbf{inserir} Msg$_j$.origem \textbf{em} $veiculosVia_i$\;
        }
        interesse := \textit{roadway\_presence\_confirmation}\;
        Msg$_i$ := $\langle$interesse,Msg$_j$.origem,Msg$_j$.via,-1, Msg$_j$.tac $\rangle$\;
        \textbf{enviar} Msg$_i$\;        
    }
    
  \caption{Tratamento do interesse \textit{vehicle\_on\_}$<$id. da via$>$.}
  \label{alg:tratamento_vehicle_on}
\end{algorithm}

\begin{algorithm}[H]
  \SetAlgoLined
  	\scriptsize
    \Entrada{Msg$_j$ := Msg$_i$, \textbf{se} Msg$_j$.interesse = \textit{roadway\_presence\_confirmation}}
    \Se{$Msg_j.destino = idAgente_i$} {
    	
    	\eSe{$Msg_j.tac$ = ``IEEE 802.11''}{
        	
            \eSe{$existeParametro(Msg_j.parametros, ``ctrIntersecao'' ) = \textbf{verdadeiro}$}{
                \Se{$Msg_j.parametros[ ``ctrIntersecao'' ] = \textbf{verdadeiro}$}{
                	iniciarControleIntersecaoComVeiculos()\;
                	presencaSinalizacao$_i$ := \textbf{falso}\;
                }
            }{
            	presencaSinalizacao$_i$ := \textbf{verdadeiro}\;
            }
        }{
        	\Se{$Msg_j.tac$ = ``LTE''}{
            	presencaSinalizacao$_i$ := \textbf{falso}\;
            }
        }
        
        numTentativaNotificacao$_i$ := 0\;
    }
    
  \caption{Tratamento do interesse \textit{roadway\_presence\_confirmation}.}
  \label{alg:tratamento_roadway_presence_confirmation}
\end{algorithm}

\newpage

\begin{algorithm}[H]
  \SetAlgoLined
  	\scriptsize
    \Entrada{Msg$_j$ := Msg$_i$, \textbf{se} Msg$_j$.interesse = \textit{vehicle\_out\_}$<$id. da via$>$}
    distancia := calcDistancia(posicaoGPS$_i$, Msg$_j$.parametros[``posicao''])\;
    areaMonitoramentoTrafego := obterAreaMonitoramentoTrafego(Msg$_j$.via)\;
    \Se{$Msg_j.direcao = 1 \wedge distancia \leq areaMonitoramentoTrafego_i$}{
    	\Se{$Msg_j.origem \in veiculosVia_i $}{
        	\textbf{remover} Msg$_j$.origem \textbf{de} $veiculosVia_i$\;
        }
        interesse := \textit{roadway\_left\_confirmation}\;
        Msg$_i$ := $\langle$interesse,Msg$_j$.origem,Msg$_j$.via,-1, Msg$_j$.tac $\rangle$\;
        \textbf{enviar} Msg$_i$\;        
    }
    
  \caption{Tratamento do interesse \textit{vehicle\_out\_}$<$id. da via$>$.}
  \label{alg:tratamento_vehicle_out}
\end{algorithm}

\begin{algorithm}[H]
  \SetAlgoLined
  	\scriptsize
    numObtQtdeVeiculos$_i$ := numObtQtdeVeiculos$_i$ + 1\;
    \Se{$|veiculosVia_i| > 0$}{
    	somaVeiculosVia$_i$ := somaVeiculosVia$_i$ + $|$veiculosVia$_i|$\;
        contSomasVeiculosVia$_i$ := contSomasVeiculosVia$_i$ + 1\;
    
    \Se{numObtQtdeVeiculos$_i$ = paramNumObtQtdeVeiculos$_i$}{
    	mediaQtdeVeiculos := somaVeiculosVia$_i$ / contSomasVeiculosVia$_i$\;
        params := $\{\langle$``mediaQtdeVeiculos'', mediaQtdeVeiculos$\rangle\}$\;
        interesse := \textit{roadway\_vehicle\_amount}\;
        Msg$_i^{intrsc}$ := $\langle$interesse, \textbf{nulo},idIntersecao$_i$, 0,``IEEE 802.11'',params$\rangle$\;
        \textbf{enviar} Msg$_i^{intrsc}$\;
         Msg$_i^{intrsc}$ := $\langle$interesse, \textbf{nulo},idIntersecao$_i$, 0,``LTE'',params$\rangle$\;
        \textbf{enviar} Msg$_i^{cct}$\;
        \Se{participaCorredor$_i$ = \textbf{verdadeiro}}{
            \Se{controlaCorredor$_i$ = \textbf{falso}}{
            	interesse := ``roadway\_segment\_vehicle\_amount''\;
              \ParaCada{$idCtrlCorredor \in listaIdCtrlCorredor_i$}{
                  params := $\{\langle$``mediaQtdeVeiculos'', mediaQtdeVeiculos$\rangle, \langle$``corredor'', idCtrlCorredor$\rangle\}$\;
                  Msg$_i^{idCtrlCorredor}$ := $\langle$interesse, \textbf{nulo},\textbf{nulo}, 0,``LTE'',params$\rangle$\;
                  \textbf{enviar} Msg$_i^{idCtrlCorredor}$\;
              }
            }
        }
        numObtQtdeVeiculos$_i$ := 0\;
    }  
    }
    
  \caption{Obtenção e compartilhamento da média da quantidade de veículos de uma via de entrada de uma interseção.}
  \label{alg:coleta_compartilhamento_demanda}
\end{algorithm}

\begin{algorithm}[H]
  \SetAlgoLined
  	\scriptsize
    interesse := ``roadway\_presence\_request''\;
    Msg$_i$ := $\langle$interesse, \textbf{nulo}, idVia$_i$, -1, ``IEEE 802.11''$\rangle$\;
    \textbf{enviar} Msg$_i$\;
    
  \caption{Requisição de notificação de presença na via.}
  \label{alg:requisicao_notificacao_presenca_via}
\end{algorithm}

\newpage

\begin{algorithm}[H]
  \SetAlgoLined
  	\scriptsize
    \Entrada{Msg$_j$ := Msg$_i$, \textbf{se} $Msg_j.interesse$ = \textit{roadway\_presence\_request}}
    \Se{$Msg_j.via = idViaAtual_i$}{
    	interesse := ``vehicle\_on\_'' + idViaAtual$_i$\;
        params := $\{\langle$ ``posicao'', posicaoGPS$_i$ $\rangle\}$\;
        \eSe{$Msg_j.tac$ = ``IEEE 802.11''} {
        	Msg$_i$ := $\langle$interesse, Msg$_j$.origem, Msg$_j$.via, 1, Msg$_j$.tac$, params\rangle$\;
            \textbf{enviar} Msg$_i$\;
        }{
        	\Se{$Msg_j.tac$ = ``LTE''} {
            	Msg$_i$ := $\langle$interesse, Msg$_j$.origem, \textbf{nulo}, 0, Msg$_j$.tac$, params\rangle$\;
            	\textbf{enviar} Msg$_i$\;
            }
        }
    }
    
    \caption{Tratamento do interesse \textit{roadway\_presence\_request}}
    \label{alg:tratamento_roadway_presence_request}
\end{algorithm}

\begin{algorithm}[H]
  \SetAlgoLined
  	\scriptsize
    \Entrada{Msg$_j$ := Msg$_i^{intrsc}$, \textbf{se} $Msg_j^{intrsc}.interesse$ = \textit{roadway\_vehicle\_amount}}
    \Se{$agenteEVizinho(idAgente_i,Msg_j.origem) = \textbf{verdadeiro}$}{
    	numMediasQtdeVeiculos$_i$ := numMediasQtdeVeiculos$_i$ + 1\;
        
        mediasQtdeVeiculosItrsc$_i$[$Msg_j.origem$] := Msg$_j$.parametros[ ``mediaQtdeVeiculos'' ]\;
        
        \Se{$numMediasQtdeVeiculos_i < paramNumeroSinSemIntrsc_i$}{
        	\Se{$numMediasQtdeVeiculos_i$ = paramNumeroSinSemIntrsc$_i$}{
        		
                somaMediasQtdeVeiculos := somarMediasQtdeVeiculos(mediasQtdeVeiculos$_i$)\;
                
                \ParaCada{$agente \in multigrafoSMERIntrsc_i$}{
                	demandaAgentes$_i$[agente] := 100 * ((mediasQtdeVeiculosItrsc$_i$[agente]/somaMediasQtdeVeiculos) + 0.5)\;
                }
                
                numMediasQtdeVeiculos$_i$ := 0\;
                valorMMCIntrsc$_i$ := calcularMMC(demandaAgentes$_i$)\;
                calculadaDemandaItrsc$_i$ := \textbf{verdadeiro}\;
        	}
        }
        
    }
    
    \caption{Tratamento do interesse \textit{roadway\_vehicle\_amount}}
    \label{alg:tratamento_roadway_vehicle_amount}
\end{algorithm}

\begin{algorithm}[H]
  \SetAlgoLined
  	\scriptsize
    acenderLuzVermelha()\;
    
    \eSe{coordenacaoAtiva$_i$ = \textbf{verdadeiro}}{
    	interesse := ``corridor\_edge\_reversal''\;
              	params := $\langle$``corredor'',idCorredor$_i\rangle$\;
              	Msg$_i^{cor}$ := $\langle$interesse, \textbf{nulo}, idIntersecao$_i$,0,``LTE'',params$\rangle$\; 
    }{
    	interesse := ``edge\_reversal''\;
    Msg$_i^{intrsc}$ := $\langle$interesse, \textbf{nulo}, idIntersecao$_i$,0,``IEEE 802.11''$\rangle$\; 
    \textbf{enviar} Msg$_i$\;
    }
    \caption{Acionamento da luz vermelha.}
    \label{alg:acionamento_luz_vermelha}
\end{algorithm}

\newpage

\begin{algorithm}[H]
  \SetAlgoLined
  	\scriptsize
    \eSe{$coordenacaoAtiva_i = \textbf{falso}$}{
    	reverterArestasItrsc(idAgente$_i$, multigrafoSMERIntrsc$_i$, revAgentesIntrsc$_i$[idAgente$_i$])\; 
    }{
    	reverterTodasArestas(idAgente$_i$, multigrafoSMERIntrsc$_i$)\;
        interesse := ``reversal\_of\_all\_edges''\;
        Msg$_i^{Intrsc}$ := $\langle$interesse, \textbf{nulo}, idIntersecao$_i$,0,``IEEE 802.11''$\rangle$\; 
        \textbf{enviar} Msg$_i^{Intrsc}$\;
        \eSe{controladorCorredor$_i$ = \textbf{verdadeiro}}{
        	\Se{numeroCiclos$_i$ = 0}{
              reverterArestasCor(idCorredor$_i$, multigrafoSMERCor$_i$, revAgentesCor$_i$[idCorredor$_i$])\;
              
              \Se{corMudaParaAmarelo(idCorredor$_i$, multigrafoSMERCor$_i$) = \textbf{falso}}{
              	numeroCiclos$_i$ := paramNumeroCiclos$_i$\;
                interesse := ``corredor\_edge\_reversal''\;
              	params := $\langle$``corredor'',idCorredor$_i\rangle$\;
              	Msg$_i^{cor}$ := $\langle$interesse, \textbf{nulo}, \textbf{nulo},0,``LTE'',params$\rangle$\; 
              \textbf{enviar} Msg$_i$\;
              }
            }
        }{
        	\Se{numeroCiclos$_i$ = 0}{
            	coordenacaoAtiva$_i$ := \textbf{falso}\;
            }
        }
    }
    
    \eSe{sinSemafMudaParaAmarelo(idAgente$_i$, multigrafoSMERIntrsc$_i$) = \textbf{verdadeiro}}{
    	acenderLuzAmarela()\;
        escalonarAcendimentoLuzVermelha(paramIntervaloIndLuzAmarela$_i$)\;
    }{
    	escalonarAcendimentoLuzAmarela(paramIntMinIndLuzVerde$_i$)\;
        interesse := ``edge\_reversal''\;
        Msg$_i^{intrsc}$ := $\langle$interesse, \textbf{nulo}, idIntersecao$_i$,0,``IEEE 802.11''$\rangle$\; 
        \textbf{enviar} Msg$_i$\;
    } 
    
    \caption{Acionamento da luz amarela.}
    \label{alg:acionamento_luz_amarela}
\end{algorithm}

\begin{algorithm}[H]
  \SetAlgoLined
  	\scriptsize
    \Entrada{Msg$_j$ := Msg$_i^{intrsc}$, \textbf{se} $I_{Msg_i}$ = \textit{edge\_reversal}}
    \Se{$agenteEVizinho(idAgente_i,Msg_j.origem) = \textbf{verdadeiro}$}{
    	reverterArestas($Msg_j.origem$, multigrafoSMERIntrsc$_i$, revAgentesIntrsc$_i$[$Msg_j.origem$])\;
        \Se{possuiArestasRevertidas(idAgente$_i$, multigrafoSMERIntrsc$_i$,revAgentesIntrsc$_i$[idAgente$_i$]) = \textbf{verdadeiro}}{
        	escalonarAcendimentoLuzVerde()\;
        }
    }
    
    \caption{Tratamento do interesse \textit{edge\_reversal}.}
    \label{alg:tratamento_edge_reversal}
\end{algorithm}

\newpage

\begin{algorithm}[!t]
  \SetAlgoLined
  	\scriptsize
    acenderLuzVerde()\;
    
    escalonarAcendimentoLuzAmarela(paramIntMinIndLuzVerde$_i$)\;
    
    \eSe{$coordenacaoAtiva_i = \textbf{verdadeiro}$}{
    	numeroCiclos$_i$ := numeroCiclos$_i$ - 1\;
    }{
    	\Se{calculadaDemandaIntrsc$_i$ = \textbf{verdadeiro}}{
        	revAgente := 1\;
            \eSe{$demandaAgentes_i[idAgente_i] = 0$}{
            	revAgente := valorMMCIntrsc$_i$\;
            }{
            	\Se{$demandaAgentes[idAgente_i] > 0$}{
                    revAgente := valorMMCIntrs$_i$/demandaAgentes[idAgente$_i$]\;
                }
            }
            \Se{$revAgente \neq revAgentesIntrsc_i[idAgente_i]$}{
                	revAgentesIntrsc$_i$[idAgente$_i$] := revAgente\;
                    \ParaCada{$agente \in multigrafoSMERIntrsc_i$}{
                    	\Se{$agente \neq idAgente_i \wedge \wedge agenteEVizinho(idagente_i,agente)$}{
                        	valMDC := calcularMDC(revAgente,revAgentesIntrsc$_i$[agente])\;
                            numArestas := revAgente + revAgentesItrsc$_i$[agente] - valMDC\;
                            multigrafoSMERIntrsc[idAgente$_i$][agente] := numArestas\;
                            multigrafoSMERIntrsc[agente][idAgente$_i$] := 0\;
                        }
                    }
                    
                    interesse := ``reversibility\_change''\;
                    Msg$_i^{intrsc}$ := $\langle$interesse,\textbf{nulo},idVia$_i$,0,``IEEE 802.11''$\rangle$\;
                    \textbf{enviar} Msg$_i^{Intrsc}$\;
                    Msg$_i^{cct}$ := $\langle$interesse,\textbf{nulo},idVia$_i$,0,``LTE''$\rangle$\;
                    \textbf{enviar} Msg$_i^{cct}$\;
                }
           calculadaDemandaIntrsc$_i$ := \textbf{falso}\;
        }
    }
    \caption{Acendimento da luz verde.}
    \label{alg:acionamento_luz_verde}
\end{algorithm}

\begin{algorithm}[H]
  \SetAlgoLined
  	\scriptsize
    \Entrada{Msg$_j$ := Msg$_i^{intrsc}$, \textbf{se} $I_{Msg_i^{Intrsc}}$ = \textit{reversibility\_change}}
    
    \Se{$agenteEVizinho(idAgente_i,Msg_j.origem) = \textbf{verdadeiro}$}{
		revAgente := 1\;
        \eSe{$demandaAgentes_i[Msg_j.origem] = 0$}{
        	revAgente := valorMMCIntrsc\;
        }{
          	\Se{$demandaAgentes[Msg_j.origem] > 0$}{
                revAgente := valorMMCIntrs$_i$/demandaAgentes[$Msg_j.origem$]\;
            }
        }
        \Se{$revAgente \neq revAgentesIntrsc_i[Msg_j.origem]$}{
           	revAgentesIntrsc$_i$[$Msg_j.origem$] := revAgente\;
            \ParaCada{$agente \in multigrafoSMERIntrsc_i$}{
               	\Se{$agente \neq Msg_j.origem \wedge agenteEVizinho(Msg_j.origem,agente)$}{
                   	valMDC := calcularMDC(revAgente,revAgentesIntrsc$_i$[agente])\;
                    numArestas := revAgente + revAgentesItrsc$_i$[agente] - valMDC\;
                    multigrafoSMERIntrsc[$Msg_j.origem$][agente] := numArestas\;
                    multigrafoSMERIntrsc[agente][$Msg_j.origem$] := 0\;
                 }
            }
        }    	        
    }
    
    \caption{Tratamento do interesse \textit{reversibility\_change}}
    \label{alg:tratamento_reversibility_change}
\end{algorithm}

\newpage

\begin{algorithm}[!t]
  \SetAlgoLined
  	\scriptsize
    
    coordenacaoAtiva$_i$ := \textbf{verdadeiro}\;
    
    cpMultigrafoSMERCor = criaCopiaMultigrafo(multigrafoSMERCor$_i$)\;
    
    mudarParaAmarelo := \textbf{falso}\;
    
    numeroCiclos$_i$ := paramNumeroCiclos$_i$\;
    somaOffSets := paramIntervaloIndLuzAmarela$_i$\;
    instanteEscAcao := tempoAtual()\;
    instanteAtual := 0\;
    
    interesse := ``participation\_in\_traffic\_light\_coordination''\;
    params := $\{\langle$ ``numeroCiclos'', numeroCiclos$_i$ $\rangle\}$\;
    Msg$_i^{Intrsc}$ := $\langle$interesse,\textbf{nulo},\textbf{nulo},0, ``IEEE 802.11''$\rangle$\;
    \textbf{enviar} Msg$_i^{Intrsc}$\;
    numeroVizinhosParticipantes$_i$ := 0\;
    
        	\Se{calculadaDemandaCor$_i$ = \textbf{verdadeiro}}{
            	revCorredor := 1\;
                \eSe{demandaCorredores[idCorredor$_i$] = 0}{
                	revCorredor := valorMMCCor$_i$\;
                }{
                	\Se{$demandaCorredores[idCorredor_i] > 0$}{
                    	revCorredor := valorMMCCor$_i$/demandaCorredores[idCorredor$_i$]\;
                    }
                }
                \Se{$revCorredor \neq revAgentesCor_i[idCorredor_i]$}{
                	revAgentesCor$_i$[idCorredor$_i$] := revCorredor\;
                    \ParaCada{$vertice \in multigrafoSMERCor_i$}{
                    	\Se{$vertice \neq idCorredor_i \wedge corredorEVizinho(idCorredor,vertice)$}{
                        	valMDC := calcularMDC(revCorredor,revAgentesCor$_i$[vertice])\;
                            numArestas := revCorredor + revAgentesCor$_i$[vertice] - valMDC\;
                            multigrafoSMERCor[idCorredor$_i$][vertice] := numArestas\;
                            multigrafoSMERCor[vertice][idCorredor$_i$] := 0\;
                        }
                    }
                    
                    interesse := ``corridor\_reversibility\_change''\;
                    params := $\{\langle$``corridor'',idCorridor$_i\rangle$\;
                    Msg$_i^{cor}$ := $\langle$interesse,\textbf{nulo},idVia$_i$,0,``LTE'',params $\rangle$\;
                    \textbf{enviar} Msg$_i^{cor}$\;
                }
                calculadaDemandaCor$_i$ := \textbf{falso}\;
            }

    \Enqto{mudarParaAmarelo = \textbf{falso}}{
    	reverterArestasCor(idCorredor$_i$, cpMultigrafoSMERCor, revAgentesCor$_i$[idCorredor$_i$])\;
        \ParaCada{$idAgente \in sinSemafsCorredor_i$}{
        	idSegmentoVia := segViaCorredor[idAgente]\;
            tamSegVia := obterTamanhoVia(idSegmentoVia)\;
            velSegVia := obterVelPermitidaVia(idSegmentoVia)\;
            offSet := tamSegVia / velSegVia\;
            somaOffSets := somaOffSets + offSet\;
            instanteEscAcao := somaOffSets\;
            \Se{$offSet - paramIntervaloIndLuzAmarela_i > 0$}{
            	instanteEscAcao := instanteEscAcao - paramIntervaloIndLuzAmarela$_i$\;
            }
            interesse := ``traffic\_light\_coordination''\;
            params := $\{\langle$``instanteEscAcao'', instanteEscAcao + instanteAtual$\rangle$,$\langle$ ``numeroCiclos'', numeroCiclos$_i$ * 2 $\rangle$,$\langle$``corredor'',idCorredor$_i$ $\rangle\}$\;
            Msg$_i^{agente}$ = $\langle$interesse, idAgente, \textbf{nulo}, 1, "LTE", params $\rangle$\;
            \textbf{enviar} Msg$_i$\;
        }
        instanteAtual := instanteAtual + (tamanhoCicloMinIntrsc$_i$ * paramNumeroCiclos$_i$)\;
        \Se{corMudaParaAmarelo(idCorredor,cpMultigrafoSMERCor) = \textbf{verdadeiro}}{
        	mudarParaAmarelo := \textbf{verdadeiro}\;
        }
    }
    reverterTodasArestas(idAgente$_i$,multigrafoSMERIntrsc)\;
    \caption{Ativa sistema coordenado de sinalizações semafóricas.}
    \label{alg:inicia_coordenacao}
\end{algorithm}

\begin{algorithm}[H]
  \SetAlgoLined
  	\scriptsize
    \Entrada{Msg$_j$ := Msg$_i^{intrsc}$, \textbf{se} $I_{Msg_i^{Intrsc}}$ = \textit{participation\_in\_traffic\_light\_coordination}}
    
    \Se{$agenteEVizinho(idAgente_i,Msg_j.origem) = \textbf{verdadeiro}$}{
		coordenacaoAtiva$_i$ := \textbf{verdadeiro}\;
        reverterTodasArestas($Msg_j.origem$, multigrafoSMERIntrsc$_i$)\;
        numeroCiclos$_i$ := Msg$_j$.parametros[``numeroCiclos'']\; 
        \eSe{indicacaoSinSemaforica$_i$ = ``verde''}{
        	acenderLuzAmarela()\;
            cancelarAcendimentoLuzAmarela()\;
            escalonarAcendimentoLuzVermelha(paramIntervaloIndLuzAmarela$_i$)\;
        }{
        	\Se{indicacaoSinSemaforica$_i$ = ``amarelo''}{
            	cancelarAcendimentoLuzVermelha()\;
            	escalonarAcendimentoLuzVermelha(paramIntervaloIndLuzAmarela$_i$)\;
            }
        }
        interesse := ``confirmation\_in\_traffic\_light\_coordination''\;
        Msg$_i$ := $\langle$interesse,$Msg_j.origem$,\textbf{nulo},0,Msg$_j$.tac $\rangle$ \;
        \textbf{enviar} Msg$_i$\;
    }
    
    \caption{Tratamento do interesse \textit{participation\_in\_traffic\_light\_coordination}}
    \label{alg:tratamento_participation_in_traffic_light_coordination}
\end{algorithm}

\begin{algorithm}[H]
  \SetAlgoLined
  	\scriptsize
    \Entrada{Msg$_j$ := Msg$_i^{intrsc}$, \textbf{se} $I_{Msg_i^{Intrsc}}$ = \textit{confirmation\_in\_traffic\_light\_coordination}}
    
    \Se{$agenteEVizinho(idAgente_i,Msg_j.origem) = \textbf{verdadeiro}$}{
		numeroVizinhosParticipantes$_i$ := numeroVizinhosParticipantes$_i$ + 1\;
        \Se{$numeroVizinhosParticipantes_i = |multigrafoSMERIntrsc[idAgente_i]|$}{
        	cancelarAcendimentoLuzAmarela()\;
            cancelarAcendimentoLuzVermelha()\;
            cancelarAcendimentoLuzVerde()\;
            acenderLuzVerde()\;
        	escalonarAcendimentoLuzAmarela(paramIntervaloIndLuzVerde$_i$)\;
        }
    }
    
    \caption{Tratamento do interesse \textit{confirmation\_in\_traffic\_light\_coordination}}
    \label{alg:tratamento_confirmation_in_traffic_light_coordination}
\end{algorithm}

\begin{algorithm}[H]
  \SetAlgoLined
  	\scriptsize
    \Entrada{Msg$_j$ := Msg$_i^{agente}$, \textbf{se} $I_{Msg_i^{agente}}$ = \textit{traffic\_light\_coordination}}
    
    \Se{$Msg_j.parametros[``corredor''] \in idCorredores_i \wedge Msg_j.origem \in idControladores_i$}{
		escalonarCoordenacaoCorredor(Msg$_j$.parametros[``instanteEscAcao''],Msg$_j$.parametros[``numeroCiclos''])\;
    }
    
    \caption{Tratamento do interesse \textit{traffic\_light\_coordination}}
    \label{alg:traffic_light_coordination}
\end{algorithm}

\begin{algorithm}[H]
  \SetAlgoLined
  	\scriptsize
	\Entrada{numeroCiclos}    
    coordenacaoAtiva$_i$ := \textbf{verdadeiro}\;
    numeroCiclos$_i$ := numeroCiclos\;
    numeroVizinhosParticipantes$_i$ := 0\;
    reverterTodasArestas(idAgente$_i$, multigrafoSMERItrsc$_i$)\;
    interesse := ``participation\_in\_traffic\_light\_coordination''\;
    params := $\{\langle$ ``numeroCiclos'', numeroCiclos$_i$ $\rangle\}$\;
    Msg$_i$ := $\langle$interesse,\textbf{nulo},\textbf{nulo},0, ``IEEE 802.11''$\rangle$\;
    \textbf{enviar} Msg$_i$\;

    \caption{Início da participação de uma sinalização semafórica na coordenação de sinalizações semafóricas.}
    \label{alg:inicio_participacao_coordenacao}
\end{algorithm}

\newpage

\begin{algorithm}[!t]
  \SetAlgoLined
  	\scriptsize
    \Entrada{Msg$_j$ := Msg$_i^{intrsc}$, \textbf{se} $I_{Msg_i}$ = \textit{reversal\_of\_all\_edges}}
    \Se{$agenteEVizinho(idAgente_i,Msg_j.origem) = \textbf{verdadeiro}$}{
    	reverterTodasArestas($Msg_j.origem$, multigrafoSMERIntrsc$_i$)\;
        \Se{possuiArestasRevertidas(idAgente$_i$, multigrafoSMERIntrsc$_i$,revAgentesIntrsc$_i$[idAgente$_i$]) = \textbf{verdadeiro}}{
        	escalonarAcendimentoLuzVerde()\;
        }
    }
    
    \caption{Tratamento do interesse \textit{reversal\_of\_all\_edges}.}
    \label{alg:tratamento_reversal_of_all_edges}
\end{algorithm}

\begin{algorithm}[H]
  \SetAlgoLined
  	\scriptsize
    \Entrada{Msg$_j$ := Msg$_i^{intrsc}$, \textbf{se} $I_{Msg_i}$ = \textit{corridor\_edge\_reversal}}
    \Se{$corredorEVizinho(idCorredor_i,Msg_j.parametros[``corredor'']) = \textbf{verdadeiro}$}{
    	reverterArestasCor(Msg$_j$.parametros[``corredor''], multigrafoSMERCor$_i$, revAgentesCor$_i$[Msg$_j$.parametros[``corredor'']])\;
        \Se{possuiArestasRevertidas(idCorredor$_i$, multigrafoSMERCor$_i$,revAgentesCor$_i$[idCorredor$_i$]) = \textbf{verdadeiro}}{
        	escalonarInicioCoordenacao()\;
        }
    }
    
    \caption{Tratamento do interesse \textit{corridor\_edge\_reversal}.}
    \label{alg:tratamento_corridor_edge_reversal}
\end{algorithm}

\begin{algorithm}[H]
  \SetAlgoLined
  	\scriptsize
    \Entrada{Msg$_j$ := Msg$_i^{idCtrlCorredor}$, \textbf{se} $I_{Msg_i^{idCtrlCorredor}}$ = \textit{roadway\_segment\_vehicle\_amount}}
    \Se{$Msg_j.origem \in sinSemafsCorredor_i \wedge idCorredor_i = Msg_j.parametros[``corredor'']$}{
    	mediasQtdeVeiculosCor$_i$[$Msg_j.origem$] := Msg$_j$.parametros[``mediaQtdeVeiculos'']\; 
    }
    
    \caption{Tratamento do interesse \textit{roadway\_segment\_vehicle\_amount}.}
    \label{alg:tratamento_roadway_segment_vehicle_amount}
\end{algorithm}

\begin{algorithm}[H]
  \SetAlgoLined
  	\scriptsize
    \Se{controladorCorredor$_i$ = \textbf{verdadeiro}}{
        maiorMedia := 0\;
        \ParaCada{$media \in mediasQtdeVeiculosCor_i$}{
            \Se{$media > maiorMedia$}{
                maiorMedia := media\;
            }
        }
        mediasQtdeVeiculosGrupo$_i$[idCorredor$_i$] := maiorMedia\;
        interesse := ``group\_member\_vehicle\_amount''\;
        params := $\{\langle$ ``grupo'', idGrupoCors$_i$ $\rangle$, $\langle$ ``corredor'', idCorredor$_i$ $\rangle$ ,$\langle$ ``mediaQtdeVeiculos'', maiorMedia $\rangle$\}\;
        Msg$_i$ := $\langle$interesse, \textbf{nulo}, \textbf{nulo}, 0, ``LTE'', params$\rangle$\;
        \textbf{enviar} Msg$_i$.
    }
    
    \caption{Seleção e compartilhamento da maior média de quantidades de veículos dos segmentos de via pertencentes a um sistema coordenado de sinalizações semafóricas.}
    \label{alg:selecao_maior_media_sistema}
\end{algorithm}

\begin{algorithm}[H]
  \SetAlgoLined
  	\scriptsize
    \Entrada{Msg$_j$ := Msg$_i$, \textbf{se} $I_{Msg_i}$ = \textit{group\_member\_vehicle\_amount}}
    \Se{$Msg_j.parametros[``grupo''] = idGrupoCors_i$}{
    	mediasQtdeVeiculosGrupo$_i$[Msg$_j$.parametros[``corredor'']] := Msg$_j$.parametros[``mediaQtdeVeiculos'']\;
    }
    
    \caption{Tratamento do interesse \textit{group\_member\_vehicle\_amount}.}
    \label{alg:tratamento_group_member_vehicle_amount}
\end{algorithm}

\newpage

\begin{algorithm}[H]
  \SetAlgoLined
  	\scriptsize
    \Se{controladorCorredor$_i$ = \textbf{verdadeiro}}{
        maiorMedia := 0\;
        \ParaCada{$media \in mediasQtdeGrupoCors_i$}{
            \Se{$media > maiorMedia$}{
                maiorMedia := media\;
            }
        }
        \ParaCada{$idCorredor \in mediasQtdeVeiculosCor_i$}{
        	mediasQtdeVeiculosCor$_i$[idCorredor] := maiorMedia\;
        }
        interesse := ``corridor\_vehicle\_amount''\;
    	params := $\{\langle$ ``mediaQtdeVeiculos'', maiorMedia $\rangle$, $\langle$ ``corredor'', idCorredor$_i$ $\rangle\}$\;
    	Msg$_i$ := $\langle$interesse, \textbf{nulo}, \textbf{nulo}, 0, ``LTE'',params$\rangle$\;
    	\textbf{enviar} Msg$_i$\;
    }
    
    \caption{Seleção e compartilhamento da maior média de quantidades de veículos de um mesmo grupo de sistemas coordenados de sinalizações semafóricas.}
    \label{alg:selecao_maior_media_grupo}
\end{algorithm}

\begin{algorithm}[H]
  \SetAlgoLined
  	\scriptsize
    \Entrada{Msg$_j$ := Msg$_i$, \textbf{se} $I_{Msg_i}$ = \textit{corridor\_vehicle\_amount}}
    \Se{$corredorEVizinho(idCorredor_i,Msg_j.parametros[``corredor'']) = \textbf{verdadeiro}$}{\textbf{}
    	mediasQtdeVeiculosCor$_i$[Msg$_j$.parametros[``corredor'']] := Msg$_j$.parametros[``mediaQtdeVeiculos'']\;
    }
    
    \caption{Tratamento do interesse \textit{corridor\_vehicle\_amount}.}
    \label{alg:tratamento_corridor_vehicle_amount}
\end{algorithm}

\begin{algorithm}[H]
  \SetAlgoLined
  	\scriptsize
    \Se{controladorCorredor$_i$ = \textbf{verdadeiro}} {
    	somaQtdeVeiculosCor := somarMediasQtdeVeiculos(mediasQtdeVeiculosCor$_i$)\;
        \ParaCada{$corredor \in multigrafoSMERCor_i$}{
        	demandasCorredores$_i$[corredor] := 100 * ((mediasQtdeVeiculosCor$_i$[corredor]/somaQtdeVeiculosCor) + 0.5)\;
        }
        valorMMCCor$_i$ := calcularMMC(demandaCorredores$_i$)\;
        calculadaDemandaCor$_i$ := \textbf{verdadeiro}\;
    }
    
    \caption{Atualização das demandas dos corredores pertencentes aos sistemas coordenados de sinalizações semafóricas.}
    \label{alg:atualizacao_demanda_corredores}
\end{algorithm}

\begin{algorithm}[H]
  \SetAlgoLined
  	\scriptsize
    idIntersecao := obterIdIntersecao(idViaAtual$_i$)\;
    interesse := ``intersection\_control\_data\_request''\;
    params := $\{\langle$ ``idIntersecao'', idIntersecao$\rangle\}$\;
    Msg$_i$ := $\langle$interesse, \textbf{nulo}, \textbf{nulo}, 0, ``LTE'', params$\rangle$\;
    \textbf{enviar} Msg$_i$\;
    
    \caption{Início do controle de interseções utilizando veículos conectados.}
    \label{alg:inicio_controle_intersecao_veiculos}
\end{algorithm}

\newpage

\begin{algorithm}[!t]
  \SetAlgoLined
  	\scriptsize
    \Entrada{Msg$_j$ := Msg$_i^{cor}$, \textbf{se} $I_{Msg_i^{cor}}$ = \textit{corridor\_reversibility\_change}}
    \Se{$corredorEVizinho(idCorredor_i,Msg_j.parametros[``corredor'']) = \textbf{verdadeiro}$}{
    	revCorredor := 1\;
        \eSe{demandaCorredores[Msg$_j$.parametros[``corredor'']] = 0}{
          	revCorredor := valorMMCCor$_i$\;
        }{
           	\Se{$demandaCorredores[idCorredor_i] > 0$}{
               	revCorredor := valorMMCCor$_i$/demandaCorredores[Msg$_j$.parametros[``corredor'']]\;
            }
        }
        \Se{$revCorredor \neq revAgentesCor_i[idCorredor_i]$}{
                	revAgentesCor$_i$[Msg$_j$.parametros[``corredor'']] := revCorredor\;
                    \ParaCada{$vertice \in multigrafoSMERCor_i$}{
                    	\Se{$vertice \neq Msg_j.parametros[``corredor''] \wedge corredorEVizinho(Msg_j.parametros[``corredor''],vertice)$}{
                        	valMDC := calcularMDC(revCorredor,revAgentesCor$_i$[vertice])\;
                            numArestas := revCorredor + revAgentesCor$_i$[vertice] - valMDC\;
                            multigrafoSMERCor[Msg$_j$.parametros[``corredor'']][vertice] := numArestas\;
                            multigrafoSMERCor[vertice][Msg$_j$.parametros[``corredor'']] := 0\;
                        }
                    }
                    
                }
    }
    
    \caption{Tratamento do interesse \textit{corridor\_reversibility\_change}.}
    \label{alg:tratamento_corridor_reversibility_change}
\end{algorithm}

\begin{algorithm}[!t]
  \SetAlgoLined
  	\scriptsize
    \Entrada{Msg$_j$ := Msg$_i$, \textbf{se} Msg$_i$.interesse = \textit{intersection\_control\_data\_request}}
    \Se{$Msg_j.parametros[``idIntersecao''] \in intersecoes_i$} {
        dadosControleIntrsc := obterDadosControleItrsc(Msg$_j$.parametros[``idIntersecao''])\;
        interesse := ``intersection\_control\_data\_request''\;
        \eSe{dadosControleIntrsc.possuiResponsavel = \textbf{falso}}{
          params := $\{\langle$ ``idIntersecao'', Msg$_j$.parametros[``idIntersecao''] $\rangle,\langle$   ``dadosControleIntrsc'', dadosControleIntrsc $\rangle\}$\;
          Msg$_i$ := $\langle$interesse, Msg$_j$.origem, \textbf{nulo}, 0, ``LTE'', params $\rangle$\;
        }{
        	Msg$_i$ := $\langle$interesse, Msg$_j$.origem, dadosControleIntrsc.idResponsavel, 0, ``LTE'' $\rangle$\;
        }
        \textbf{enviar} Msg$_i$\;
    }
    
    \caption{Tratamento do interesse \textit{intersection\_control\_data\_request}.}
    \label{alg:tratamento_intersection_control_data_request}
\end{algorithm}

\begin{algorithm}[!t]
  \SetAlgoLined
  	\scriptsize
    \Entrada{Msg$_j$ := Msg$_i$, \textbf{se} Msg$_i$.interesse = \textit{intersection\_control\_data}}
    idIntersecao := obterIdIntersecao(idViaAtual$_i$)\;
    \Se{$Msg_j.parametros[``idIntersecao''] = idIntersecao$} {
		dadosControleIntrsc = Msg$_j$.parametros[``dadosControleIntrsc'']\;
		
        \Se{sinSemaforica$_i$ = \textbf{nulo}} {
        
          sinSemaforica$_i$ := criarSinalizacaoSemaforicaVirtual(dadosControleIntrsc)\;
          iniciarOperacaoSinSemaforicaVirtual(sinSemaforica)\;
          ctrlIntersecaoAtivo$_i$ := \textbf{verdadeiro}\;
        }
        \Se{$sinSemaforica.indicacao \neq ``verde''$ } {
        	iniciarIdentificacaoLider()\;
        }
	}
    
    \caption{Tratamento do interesse \textit{intersection\_control\_data}.}
    \label{alg:tratamento_intersection_control_data}
\end{algorithm}

\begin{algorithm}[!t]
  \SetAlgoLined
  	\scriptsize
    \Se{$sinSemaforica_i.indicacao \neq ``verde'' $}{
    	interesse := ``vehicle\_position''\;
        params := $\{\langle$``posicaoGPS'', posicaoGPS$_i$ $\rangle\}$\;
        Msg$_i$ := $\langle$interesse, \textbf{nulo}, idViaAtual$_i$, 0, ``IEEE 802.11'', params $\rangle\}$\;
        \textbf{enviar} Msg$_i$\;
    }
    
    \caption{Envio da posição do veículo para eleição de líder.}
    \label{alg:envio_posicao_veiculo}
\end{algorithm}

\begin{algorithm}[H]
  \SetAlgoLined
  	\scriptsize
    \Entrada{Msg$_j$ := Msg$_i$, \textbf{se} Msg$_i$.interesse = \textit{vehicle\_position}}
    \Se{$sinSemaforica_i.indicacao \neq ``verde'' $}{
    	posicoesVeiculosVia$_i$[Msg$_j$.origem] = Msg$_j$.parametros[``posicaoGPS'']\;
        idLider := idAgente\;
        posicaoFimVia := obterPosicaoFimVia(idViaAtual$_i$)\;
        distFaixaRetencao := calcularDistancia(posicaoGPS$_i$, posicaoFimVia)\;
        \ParaCada{$idAgente \in posicoesVeiculosVia_i$} {
        	distancia := calcularDistancia(posicoesVeiculosVia$_i$[idAgente], posicaoFimVia)\;
            \Se{$distancia < distFaixaRetencao$} {
            	idLider := idAgente\;
                distFaixaRetencao := distancia\;
            }
        }
        
        \eSe{idLider = $idAgente_i$} {
        	configurarAgenteVeiculoComoLider()\;
        }{
        	desfazerConfiguracaoAgenteVeiculoComoLider()\;
        }
    }
    
    \caption{Tratamento do interesse \textit{vehicle\_position}.}
    \label{alg:tratamento_vehicle_position}
\end{algorithm}
    \chapter{Interesses do Planejamento e Orientação de Rotas}

\begin{table}[H]
  \centering
  \caption{Interesses registrados pelo agente Centro de Controle de Tráfego e as ações executadas pelo agente e ou pelo ambiente.}
  \label{tab:interesses_cct_2}
  \small\begin{tabular}{|p{4.7cm}|c|p{1.5cm}|p{6cm}|}
    \hline
    \centering\textbf{Interesse} & \textbf{NMS} & \centering\textbf{TAC} & \textbf{Ação Executada} \\ \hline
    \textit{urban\_element\_data} & \centering 1 & \centering LTE & Encaminha a mensagem para os agentes responsáveis pelos controles de interseções. \\ \hline
    \textit{new\_traffic\_light\_schedule} & \centering 1 & \centering LTE & Gera uma nova agenda de intervalos de indicações de luzes verdes de uma interseção. Em seguida, encaminha a mensagem para os agentes responsáveis pelos controles de interseções. \\ \hline
    \textit{roadway\_space\_allocation} & \centering 1 & \centering LTE & Atualiza alocações de espaços nas vias. Após isto, encaminha a mensagem para os agentes responsáveis pelos controles de interseções. \\ \hline
  \end{tabular}
\end{table}

\newpage

\begin{table}[H]
  \centering
  \caption{Interesses registrados pelo agente Centro de Controle de Tráfego e as ações executadas pelo agente e ou pelo ambiente.}
  \label{tab:interesses_cct_2}
  \small\begin{tabular}{|p{4.7cm}|c|p{1.5cm}|p{6cm}|}
    \hline
    \centering\textbf{Interesse} & \textbf{NMS} & \centering\textbf{TAC} & \textbf{Ação Executada} \\ \hline
    \textit{urban\_element\_data} & \centering 1 & \centering LTE & Encaminha a mensagem para os agentes responsáveis pelos controles de interseções. \\ \hline
    \textit{new\_traffic\_light\_schedule} & \centering 1 & \centering LTE & Gera uma nova agenda de intervalos de indicações de luzes verdes de uma interseção. Em seguida, encaminha a mensagem para os agentes responsáveis pelos controles de interseções. \\ \hline
    \textit{roadway\_space\_allocation} & \centering 1 & \centering LTE & Atualiza alocações de espaços nas vias. Após isto, encaminha a mensagem para os agentes responsáveis pelos controles de interseções. \\ \hline
  \end{tabular}
\end{table}

\begin{table}[H]
  \centering
  \caption{Interesses registrados por um agente Sinalização Semafórica e as ações executadas pelo agente e ou pelo ambiente.}
  \label{tab:interesses_sinalizacao_2}
  \small\begin{tabular}{|p{4.7cm}|c|p{1.5cm}|p{6cm}|}
    \hline
    \centering\textbf{Interesse} & \textbf{NMS} & \centering\textbf{TAC} & \textbf{Ação Executada} \\ \hline
    \textit{urban\_element\_data} & \centering 1 & \centering LTE & Extrai e registra interesses relativos a um elemento urbano. Em seguida, encaminha a mensagem de interação para os veículos conectados da via, utilizando uma interface de acesso à comunicação baseada no padrão IEEE 802.11p.\\ \hline
    \textit{new\_traffic\_light\_schedule} & \centering 1 & \centering LTE & Gera uma nova agenda de intervalos de indicações de luzes verdes de uma interseção. Em seguida, encaminha a mensagem de interação para os veículos conectados da via, utilizando uma interface de acesso à comunicação baseada no padrão IEEE 802.11p. \\ \hline
    \textit{roadway\_space\_allocation} & \centering 1 & \centering LTE & Atualiza alocações de espaços nas vias. \\ \hline
    \textit{green\_wave\_request} & \centering 1 & \centering IEEE 802.11 & Fornece dados acerca da onda verde da via. \\ \hline
  \end{tabular}
\end{table}

\begin{table}[H]
  \centering
  \caption{Interesses registrados por um agente Veículo e as ações executadas pelo agente e ou pelo ambiente.}
  \label{tab:interesses_veiculo_2}
  \small\begin{tabular}{|p{4.7cm}|c|p{1.5cm}|p{6cm}|}
    \hline
    \centering\textbf{Interesse} & \textbf{NMS} & \centering\textbf{TAC} & \textbf{Ação Executada} \\ \hline
    \textit{urban\_element\_data} & \centering 4 & \centering IEEE 802.11p & Extrai e registra interesses relativos a um elemento urbano. Além disto, o ambiente coopera com a HRAdNet-VE.\\ \hline
    \textit{new\_traffic\_light\_schedule} & \centering 4 & \centering IEEE 802.11p & Envia uma mensagem de interação, requisitando um novo cálculo de rota ótima. Além disto, o ambiente coopera com a HRAdNet-VE. \\ \hline
    \textit{green\_wave\_request} & \centering 8 & \centering IEEE 802.11 & O agente não executa qualquer ação, mas o ambiente somente coopera com a HAdNet-VE. \\ \hline
    \textit{green\_wave} & \centering 8 & \centering IEEE 802.11 & Se o agente for o destino da mensagem de interação, ele toma ciência da onda verde. Caso contrário, o agente não executa qualquer ação, mas o ambiente somente coopera com a HAdNet-VE. \\ \hline
    \textit{hello} & \centering 1 & \centering IEEE 802.11 & Registra a presença de um veículo próximo ao veículo conectado do agente. \\ \hline
  \end{tabular}
\end{table}

\begin{table}[H]
  \centering
  \caption{Interesses registrados por um agente Sinalização Semafórica e as ações executadas pelo agente e ou pelo ambiente.}
  \label{tab:interesses_sinalizacao_2_1}
  \small\begin{tabular}{|p{4.7cm}|c|p{1.5cm}|p{6cm}|}
    \hline
    \centering\textbf{Interesse} & \textbf{NMS} & \centering\textbf{TAC} & \textbf{Ação Executada} \\ \hline
    \textit{route\_to\_$<$id. da via$>$} & \centering 1 & \centering IEEE 802.11 & Calcula uma rota ótima até a via de um elemento urbano. \\ \hline
    \textit{route\_to\_$<$id. da via$>$} & \centering 1 & \centering IEEE 802.11p & Calcula uma rota ótima até a via de um elemento urbano. \\ \hline
    \textit{route\_to\_$<$id. da via$>$} & \centering 1 & \centering LTE & Calcula uma rota ótima até a via de um elemento urbano. \\ \hline
  \end{tabular}
\end{table}

\begin{table}[H]
  \centering
  \caption{Interesses registrados por um agente Veículo e as ações executadas pelo agente e ou pelo ambiente.}
  \label{tab:interesses_veiculo_3}
  \small\begin{tabular}{|p{4.7cm}|c|p{1.5cm}|p{6cm}|}
    \hline
    \centering\textbf{Interesse} & \textbf{NMS} & \centering\textbf{TAC} & \textbf{Ação Executada} \\ \hline
    \textit{route\_to\_$<$id. da via$>$} & \centering 8 & \centering IEEE 802.11 & O agente realiza não qualquer ação, mas o ambiente coopera com a HAdNet-VE.  \\ \hline
    \textit{route\_to\_$<$id. da via$>$} & \centering 4 & \centering IEEE 802.11p & O agente realiza não qualquer ação, mas o ambiente coopera com a HAdNet-VE. \\ \hline
    \textit{calculated\_route\_to\_$<$id. da via$>$} & \centering 8 & \centering IEEE 802.11 & Caso o agente seja o destino da mensagem, ele informa a rota ao motorista. Caso contrário, o agente não qualquer ação, mas o ambiente somente coopera com a HAdNet-VE. \\ \hline
    \textit{calculated\_route\_to\_$<$id. da via$>$} & \centering 4 & \centering IEEE 802.11p & Caso o agente seja o destino da mensagem, ele informa a rota ao motorista. Caso contrário, o agente não qualquer ação, mas o ambiente somente coopera com a HAdNet-VE. \\ \hline
    \textit{calculated\_route\_to\_$<$id. da via$>$} & \centering 1 & \centering LTE & O agente informa a rota ao motorista. Caso contrário, o agente não qualquer ação, mas o ambiente somente coopera com a HAdNet-VE. \\ \hline
    
  \end{tabular}
\end{table}
    \chapter{Algoritmos do Planejamento e Orientação de Rotas}

\begin{algorithm}[!h]
  \SetAlgoLined
  \footnotesize
  interesse := ``urban\_element\_data''\;
  params := $\{\langle$ ``interessesUsuario'', listaIntUsuarios$_i\rangle, \langle$ ``idVia'', idVia$_i\rangle\}$\;
  Msg$_i$ := $\langle$ interesse, \textbf{nulo}, \textbf{nulo}, 0, ``LTE'', params $\rangle$\;  
  \textbf{enviar} Msg$_i$\;
  \caption{Enviando mensagens de interação com o interesse \textit{urban\_element\_data}.}
  \label{alg:envio_urban_element_data}
\end{algorithm}

\newpage

\begin{algorithm}[!h]
  \SetAlgoLined
  	\scriptsize
    \Entrada{multiplicador}
    
    periodObtQtdeVeic := dadosControleIntrsc.paramPerObtQtdeVeiculos\;
    numObtQtdeVeic := dadosControleIntrsc.paramNumObtQtdeVeiculos\;
    
    tempo := (periodObtQtdeVeic * numObtQtdeVeic) * multiplicador\;
    
    tempoInicial := tempoAtualGPS$_i$\;
    acumuladorTempo := 0\;
    fase := 0;
    
    tamanhosIntervalos$_i$ := $\emptyset$\;
    limparAlocEspacosVias(alocacoesEspacosVias$_i$, viasEntradaIntrsc$_i$)\; 
    temposIniciais := $\emptyset$\;
    tamIntVertices := $\emptyset$\;
    verticesOp := $\emptyset$\;
    verticesBlc := $\emptyset$\;
    iniciosIntervalos$_i$ := $\emptyset$\;
    finsIntervalos$_i$ := $\emptyset$\;
    
    vertices := obterVertices(multigrafoSMER$_i$)\;
    
    multigrafo := multigrafoSMER$_i$\;
    
    reversibilidades := revAgentes$_i$\;
    
    intMinIndVerdes := intMinIndVerdes$_i$\;
    
    intIndAmarelos := intIndAmarelos$_i$\;
    
    mapVerticesVias := mapVerticesVias$_i$\;
    
    \Enqto{$tempo > acumuladorTempo$}{
    	
        \Se{$|verticesOp = 0|$}{
        	\ParaCada{$vertice \in vertices$}{
            	\Se{possuiArestasRevertidas(vertice, multigrafo, revertibilidades[vertice]) = \textbf{verdadeiro}}{
                	inserir(verticesOp, vertice)\;
                    temposIniciais[vertice] := tempoInicial\;
                    tamIntVertices[vertice] := 0\;
                }
            }
        }
        
        \ParaCada{$vertice \in verticesOp$}{
        	\eSe(dadosControleIntrsc.coordenacaoAtiva = \textbf{falso}) {
        		reverterArestas(vertice, multigrafo, reversibilidades[vertice])\;
            }{
            	reverterTodasArestas(vertice, multigrafo)\;
            }
            
            \eSe{possuiArestasRevertidas(vertice, multigrafo, reversibilidades[vertice]) = \textbf{falso}}{
            
            	inserir(verticesBlc, vertice)\;
                inserir(iniciosIntervalos, indice, $\langle$mapVerticesVias[vertice], tempoInicial$\rangle$)\;
                fimIntervalo := tamIntVertices[vertice] + intMinIndVerdes[vertice] + intIndAmarelos[vertice]\;
                inserir(finsIntervalos, fase, $\langle$mapVerticesVias[vertice], fimIntervalo$\rangle$)\;
                alocacoesEspacosVias[mapVerticesVias[vertice]] := criarLista()\;
                inserir(alocacoesEspacosVias[mapVerticesVias[vertice]], fase, $\emptyset$)\;
                tamIntVertices[vertice] := tamIntVertices[vertice] + intMinIndVerdes[vertice] + intIndAmarelos[vertice]\;
            
            }{
            	
                tamIntVertices[vertice] := tamIntVertices[vertice] + intMinIndVerdes[vertice]\;
                
            }
        }
        
        \ParaCada{$vertice \in verticesBlc$}{
        	remover(verticesOp, vertice)\;
        }
        
        \Se{$|verticesOp| = 0$}{
        	maiorTamIntervalo := 0\;
            intervaloAmarelo := 0\;
            \ParaCada{$vertice, tamIntervalo \in tamIntVertices$}{
            	\Se{$tamIntervalo > maiorTamIntervalo$} {
                	maiorTamIntervalo := tamIntervalo\;
                    intervaloAmarelo := intIndAmarelos[vertice]\;
                }
            }
            
            tempo := tempo + maiorTamIntervalo + intervaloAmarelo\;
            tempoInicial := tempoInicial + maiorTamIntervalo + intervaloAmarelo\;
            
            inserir(tamanhosIntervalos$_i$, fase, tamIntVertices)\;
            
            fase := fase + 1\;
        }
    }

    \caption{Geração de agendas de intervalos de indicação de luzes verde.}
    \label{alg:geracao_agenda_intersecao_isolada}
\end{algorithm}

\newpage

\begin{algorithm}[!h]
  \SetAlgoLined
  	\scriptsize
    \Entrada{Msg$_j$ := Msg$_i$, \textbf{se} ${Msg_i.interesse}$ = \textit{new\_traffic\_light\_schedule}}
    
    dadosControleIntrsc := Msg$_j$.parametros[``dadosControleIntrsc'']\;
    
    periodObtQtdeVeic := dadosControleIntrsc.paramPerObtQtdeVeiculos\;
    numObtQtdeVeic := dadosControleIntrsc.paramNumObtQtdeVeiculos\;
    
    tempo := (periodObtQtdeVeic * numObtQtdeVeic) * 1000\;
    
    tempoInicial := tempoAtualGPS$_i$\;
    acumuladorTempo := 0\;
    fase := 0;
    
    vias := obterViasEntrada(dadosControleIntsrc)\;
    
    limparAlocEspacosVias(alocacoesEspacosVias$_i$, vias)\; 
    temposIniciais := $\emptyset$\;
    tamIntVertices := $\emptyset$\;
    verticesOp := $\emptyset$\;
    verticesBlc := $\emptyset$\;
    iniciosIntervalos$_i$ := $\emptyset$\;
    finsIntervalos$_i$ := $\emptyset$\;
    
    vertices := obterVertices(dadosControleIntrsc.multigrafoSMER)\;
    
    multigrafo := dadosControleIntrsc.multigrafoSMER\;
    
    reversibilidades := dadosControleIntrsc.revAgentes\;
    
    intMinIndVerdes := dadosControleIntrsc.intMinIndVerdes\;
    
    intIndAmarelos := dadosControleIntrsc.intIndAmarelos\;
    
    mapVerticesVias := dadosControleIntrsc.mapVerticesVias\;
    
    \Enqto{$tempo > acumuladorTempo$}{
    	
        \Se{$|verticesOp = 0|$}{
        	\ParaCada{$vertice \in vertices$}{
            	\Se{possuiArestasRevertidas(vertice, multigrafo, revertibilidades[vertice]) = \textbf{verdadeiro}}{
                	inserir(verticesOp, vertice)\;
                    temposIniciais[vertice] := tempoInicial\;
                    tamIntVertices[vertice] := 0\;
                }
            }
        }
        
        \ParaCada{$vertice \in verticesOp$}{
        	\eSe(dadosControleIntrsc.coordenacaoAtiva = \textbf{falso}) {
        		reverterArestas(vertice, multigrafo, reversibilidades[vertice])\;
            }{
            	reverterTodasArestas(vertice, multigrafo)\;
            }
            
            \eSe{possuiArestasRevertidas(vertice, multigrafo, reversibilidades[vertice]) = \textbf{falso}}{
            
            	inserir(verticesBlc, vertice)\;
                inserir(iniciosIntervalos, indice, $\langle$mapVerticesVias[vertice], tempoInicial$\rangle$)\;
                fimIntervalo := tamIntVertices[vertice] + intMinIndVerdes[vertice] + intIndAmarelos[vertice]\;
                inserir(finsIntervalos, fase, $\langle$mapVerticesVias[vertice], fimIntervalo$\rangle$)\;
                alocacoesEspacosVias[mapVerticesVias[vertice]] := criarLista()\;
                inserir(alocacoesEspacosVias[mapVerticesVias[vertice]], fase, $\emptyset$)\;
                tamIntVertices[vertice] := tamIntVertices[vertice] + intMinIndVerdes[vertice] + intIndAmarelos[vertice]\;
            
            }{
            	
                tamIntVertices[vertice] := tamIntVertices[vertice] + intMinIndVerdes[vertice]\;
                
            }
        }
        
        \ParaCada{$vertice \in verticesBlc$}{
        	remover(verticesOp, vertice)\;
        }
        
        \Se{$|verticesOp| = 0$}{
        	maiorTamIntervalo := 0\;
            intervaloAmarelo := 0\;
            \ParaCada{$vertice, tamIntervalo \in tamIntVertices$}{
            	\Se{$tamIntervalo > maiorTamIntervalo$} {
                	maiorTamIntervalo := tamIntervalo\;
                    intervaloAmarelo := intIndAmarelos[vertice]\;
                }
            }
            
            tempo := tempo + maiorTamIntervalo + intervaloAmarelo\;
            tempoInicial := tempoInicial + maiorTamIntervalo + intervaloAmarelo\;
            
            fase := fase + 1\;
        }
    }
   	
    Msg$_i$ := $\langle$Msg$_j$.interesse, \textbf{nulo}, idVia$_i$, -1, ``IEEE 802.11p''$\rangle$\;
    \textbf{enviar} Msg$_i$\;
    
    \caption{Geração de agendas de intervalos de indicações de luzes verde baseada nos dados de controle de outras interseções.}
    \label{alg:geracao_agenda_dados_outras_intersecoes}
\end{algorithm}

\newpage

\begin{algorithm}[!h]
  \SetAlgoLined
  	\scriptsize
    \Entrada{destino}
	
    \eSe{possuiRotaCalculada$_i$ = \textbf{falso}}{
      \eSe{$numTentativasCalcRotas_i < paramNumTentativasCalcRotas_i$ }{

          interesse := ``route\_to\_'' + idVia\;
          params := $\{\langle$ ``destino'', destino$\rangle,\langle$ ``posicaoGPS'', posicaoGPS $\rangle,\langle$ ``tamanho'', tamanho$_i \rangle\}$\;

          \eSe{$tacCalcRotas_i = ``IEEE 802.11'' \vee tacCalcRotas_i = ``IEEE 802.11p''$}{
              Msg$_i$ := $\langle$ interesse, \textbf{nulo}, idViaAtual$_i$, 1, tacCalcRotas$_i$, params $\rangle$\;
              \textbf{enviar} Msg$_i$\;
          }{
          	  params := params $\cup$ $\{\langle$ ``idViaAtual'', idViaAtual$_i\rangle\}$\;
              Msg$_i$ := $\langle$ interesse, \textbf{nulo}, \textbf{nulo}, 0, tacCalcRotas$_i$, params $\rangle$\;
              \textbf{enviar} Msg$_i$\;
          }

          numTentativasCalcRotas$_i$ := numTentativasCalcRotas$_i$ + 1\;
      }{
          \eSe{tacCalcRotas$_i$ = ``IEEE 802.11''}{
              tacCalcRotas$_i$ := ``IEEE 802.11p''\;
              numTentativasCalcRotas$_i$ := paramNumTentativasCalcRotas\;
          }{
              \Se{tacCalcRotas$_i$ = ``IEEE 802.11''}{
                  tacCalcRotas$_i$ := ``LTE''\;
              }
          }
      }
    }{
    	numTentativasCalcRotas$_i$ := 0\;
        tacCalcRotas$_i$ := ``IEEE 802.11''\;
    }
    
    \caption{Solicitação de cálculo de rotas ótimas.}
    \label{alg:solicitacao_calculo_rotas}
\end{algorithm}

\begin{algorithm}[!h]
  \SetAlgoLined
  	\scriptsize
    \Entrada{Msg$_j$ := Msg$_i$, \textbf{se} $I_{Msg_i}$ = \textit{route\_to\_$<$id. da via$>$}}
    
    idVia := idVia$_i$\;
    \Se{existeParametro(Msg$_j$.parametros, ``idViaAtual'' ) = \textbf{verdadeiro}}{
    	idVia := Msg$_j$.parametros[``idViaAtual'']\;
    }
    
    \eSe{existeParametro(Msg$_j$.parametros, ``rota'' ) = \textbf{verdadeiro}}{
    	instantes := Msg$_j$.parametros[ ``instantes'' ]\;
        rota := Msg$_j$.parametros[ ``rota'' ]\;
        \eSe{rotaEIgual(rota, instantes, paramToleranciaRota$_i$) = \textbf{verdadeiro}}{
        	devolverRotaOtima(Msg$_j$)\;
        }{
        	calcularRotaOtima(Msg$_j$, idVia)\;
        }
    }{
    	calcularRotaOtima(Msg$_j$, idVia)\;
    }
    
    \caption{Tratamento do interesse \textit{route\_to\_$<$id. da via$>$}}
    \label{alg:tratamento_route_to}
\end{algorithm}

\newpage

\begin{algorithm}[!t]
  \SetAlgoLined
  	\scriptsize
    \Entrada{Msg$_j$, idViaAtual}
    
    destino := Msg$_j$.parametros[``destino'']\;
    posicao := Msg$_j$.parametros[``posicaoGPS'']\;
    rota := $\emptyset$\;
    instantes := $\emptyset$\;
    
    distancia := obterDistancia(posicaoGPS, posicaoGPS$_i$, idVia)\;
    velocidadeMedia := obterVelocidadeMediaVia(idVia)\;
    custoInicialMedio := distancia / velocidadeMedia\;
    
    faixas := obterFaixasVia(idVia);
    
    viasControladas := obterViasControladas()\;
    
    tempoViagem := $\infty$\;
        
    \ParaCada{$faixa \in faixas$}{
        viasVisitadas := $\emptyset$\;
        caminho := $\emptyset$\;
        novosInstantes := $\emptyset$
    	
        \ParaCada{$via \in viasControladas$}{
        	viasVisitadas[via] := $\infty$\;
            novosInstantes[via] := $\infty$\;
            caminho[via] := ``''\;
        }
        
        viasVisitadas[faixa] := tempoAtualGPS + custoInicial\;
        novosIntantes := tempoAtualGPS + custoInicial\;
        
        \Enqto{$|viasVisitadas| > 0$}{
        	menorCusto := $\infty$\;
            viaAtual := ``''\;
        	
            \ParaCada{$via, custo \in viasVisitadas$}{
            	\Se{custo $<$ menorCusto}{
                	menorCusto := custo\;
                    viaAtual := via\;
                }
            }
            
            \Se{$viaAtual \neq $ ``''}{
            	remover(viasVisitadas, viaAtual)\;
                
                proximasVias := obterProximasVias(viaAtual)\;
                
                \ParaCada{$proximaVia \in proximasVias$}{
                	tamanhoProxVia := obterTamanhoVia(proximaVia)\;
                    velocidadeProxVia := obterVelocidadeVia(proximaVia)\;
                	instante := novosInstantes[proximaVia] + (tamanhoProxVia / velocidadeProvia)\;
                    
                    fins := finsIntervalos$_i$[proximaVia]\;
                    inicios := iniciosIntervalos$_i$[proximaVia]\;
                    
                    \Para{i := 1 \Ate $|fins|$}{
                    	\Se{instante $<$ fins[i]}{
                        	
                            atraso := 0\;
                            
                            \Se{inicios[i] $>$ instante}{
                            	atraso := inicios[i] - instante\;
                            }
                            atraso := atraso + obterVeiculosFrente(proximaVia, instante) * paramPenPorVeic$_i$\;
                            custo := menorCusto + atraso + (tamanhoProxVia / velocidadeProvia)\;
                            tamFila := calcTamFila(alocacoesEspacosVias[proximaVia][i])\;
                            \Se{tamFila $\leq$ tamanhoProxVia}{
                            	\Se{custo $<$ novosInstantes[proximaVia]}{
                                	viasVisitadas[proximaVia] := custo\;
                                    novosInstantes[proximaVia] := custo\;
                                    caminhos[proximaVia] := viaAtual\;
                                }
                            }
                            \textbf{parar}\;
                        }
                    }
                }
            }
        }
        atualizarMenorRota(rota, instantes, novosInstantes, tempoViagem)\;
    }
    
	alocarEspacosVias(Msg$_j$.origem, rota, instantes, paramTempoAlocParcial$_i$)\;
    enviarMensagemComRotaCalculada(rota, instantes, Msg$_j$)\;    
    
    \caption{Cálculo de rota ótima.}
    \label{alg:calculo_rota_otima}
\end{algorithm}

\newpage

\begin{algorithm}[!h]
  \SetAlgoLined
  	\scriptsize
    \Entrada{rota, instantes, Msg$_j$}
    
    destino := Msg$_j$.parametros[``destino'']\;
    interesse := ``calculated\_route\_to'' + destino\;
    params := $\{\langle$ ``rota'', rota $\rangle,\langle$ ``instantes'', instantes $\rangle\}$\;

    \eSe{Msg$_j$.tac = ``IEEE 802.11'' $\vee$ Msg$_j$.tac = ``IEEE 802.11p''}{
    	Msg$_i$ := $\langle$interesse, \textbf{nulo}, Msg$_j$.origem, -1, Msg$_j$.tac, params$\rangle$\;
        \textbf{enviar} Msg$_i$\;
    }{
    	params := params $\cup \{\langle$ ``origemMsg'', Msg$_i$.origem $\rangle\}$\; 
    	Msg$_i$ := $\langle$interesse, \textbf{nulo}, \textbf{nulo}, 0, Msg$_j$.tac, params$\rangle$\;
        \textbf{enviar} Msg$_i$\;
    }
    
    \caption{Envio mensagem cujo interesse é \textit{calculated\_route\_to\_$<$id. da via$>$}}
    \label{alg:envio_calculated_route_to}
\end{algorithm}

\begin{algorithm}[!h]
  \SetAlgoLined
  	\scriptsize
	
    viasEntrada := viasEntrada$_i$\;
    
    \ParaCada{$via \in viasEntrada$}{
    	posInicioOndaVerde := posInicioOndasVerdes$_i$[via]\;
        sinalX := sinaisX$_i$[via]\;
        sinalY := sinaisY$_i$[via]\;
        anguloVia := angulosVias$_i$[via]\;
        
        tamanhoVia := obterTamanhoVia(via)\;
        
        velocidadeOndasVerdes$_i$[via] := tamanhoVia / tamanhosIntervalos$_i$[fase$_i$][via]\;
        avanco := velocidadeOndasVerdes$_i$[via] / 1000\;
        
        posInicioOndaVerde.x := posInicioOndaVerde.x + ((sinalX * avanco) * cosseno(anguloVia))\;
        posInicioAreaMont.y := posInicioAreaMont.y + ((sinalY * avanco) * seno(anguloVia))\;
        posInicioOndasVerdes$_i$[via] := posInicioOndaVerde\;

        tempoParaProxIntVerde$_i$[via] := tempoParaProxIntVerde$_i$[via] - 0.001\;
        tempoDurOndaVerde$_i$[via] := tempoDurOndaVerde$_i$[via] - 0.001;
    }
    
    \caption{Atualização dos posicionamentos das ondas verdes de uma interseção.}
    \label{alg:calculo_avancos_ondas_verdes}
\end{algorithm}

    \chapter{Parâmetros de Rede para a HRAdNet-VE}

\begin{table}[!h]
    \centering
    \caption{Configurações da camada física relativas as interfaces de acesso à comunicação sem fio baseadas no padrão IEEE 802.11n.}
    \label{tab:config_IEEE_80211n}
    \small\begin{tabular}{|l|l|}
      \hline
      \textbf{Parâmetro} & \textbf{Valor} \\
      \hline
      Potência de transmissão & 158.48mW\\ \hline
      Limite de atenuação de sinal & -90dBm\\ \hline
      Frequência & 5.0 GHz\\ \hline
      Ruído térmico & -160dbm\\ \hline
      Sensibilidade da antena & -87dBm\\ \hline
      Tamanho do cabeçalho & 128 bits \\ \hline
      Alcance de comunicação & 200m \\ \hline
    \end{tabular}
\end{table}

\begin{table}[H]
    \centering
    \caption{Configurações do CSMA/CA na camada de enlace das interfaces de acesso à comunicação sem fio baseadas no padrão IEEE 802.11n.}
    \label{tab:config_MAC_IEEE_80211n}
    \small\begin{tabular}{|l|l|}
      \hline
      \textbf{Parâmetro} & \textbf{Valor} \\
      \hline
      Tamanho da fila & 100\\ \hline
      Duração do \textit{slot} & 0.0005s\\ \hline
      Difs & 0.00011s\\ \hline
      Número de tentativas de retransmissão & 14\\ \hline
      Taxa de transmissão & 11.35 Mbps \\ \hline
      Tamanho da janela de contenção & 20\\ \hline
      Tamanho do cabeçalho MAC & 256 bits \\ \hline
    \end{tabular}
\end{table}

\begin{table}[H]
    \centering
    \caption{Configurações da camada física relativas as interfaces de acesso à comunicação sem fio baseadas no padrão IEEE 802.11p.}
    \label{tab:config_IEEE_80211p}
    \small \begin{tabular}{|l|c|}
      \hline
      \textbf{Parâmetro} & \textbf{Valor} \\
      \hline
      Potência de transmissão & 20mW\\ \hline
      Limite da atenuação de sinal  & -89dBm\\ \hline
      Frequência & 5.89 GHz\\ \hline
      Ruído térmico & -110dbm\\ \hline
      Sensibilidade da camada física & -89dBm\\ \hline
      Tamanho do cabeçalho PHY & 46 bits \\ \hline
      Alcance do rádio & 500m \\ \hline
    \end{tabular}
\end{table}

\begin{table}[H]
    \centering
    \caption{Configurações do CSMA/CA na camada de enlace das interfaces de acesso à comunicação sem fio baseadas no padrão IEEE 802.11p.}
    \label{tab:config_MAC_IEEE_80211p}
    \small \begin{tabular}{|l|c|}
      \hline
      \textbf{Parâmetro} & \textbf{Valor} \\ \hline
      Duração do \textit{slot} & 0.00013s\\ \hline
      Difs & 0.00032s\\ \hline
      Taxa de transmissão & 18 Mbps \\ \hline
      Tamanho da janela de contenção & 15s\\ \hline
      Tamanho do cabeçalho MAC &  256 bits \\ \hline
    \end{tabular}
\end{table}

\begin{table}[H]
    \centering
    \caption{Configurações das interfaces de acesso à comunicação sem fio baseadas no padrão LTE usadas pelos nós UE.}
    \label{tab:LTEUEsettings}
    \small \begin{tabular}{|l|c|}
      \hline
      \textbf{Parâmetro} & \textbf{Valor} \\
      \hline
      Potência de transmissão & 251,18mW \\ \hline
      Modelo de macrocélula & RURAL\_MACROCELL \cite{3GPP:2017} \\ \hline
      Frequência & 2.1 GHz \\ \hline
      Largura de banda & 10 MHz \\ \hline
      BLER (Block Error Rate)  & 0.001 \\ \hline
      Redução HARQ & 0.2 \\ \hline
      Ruído térmico & -104.5 dBm \\ \hline
      Figura de ruído & 7 dB\\ \hline
      Sensibilidade da antena & -107.5 dBm \\ \hline
      Alcance de comunicação & 5000m \\ \hline
    \end{tabular}
\end{table}

\begin{table}[H]
    \centering
    \caption{Configurações das interfaces de acesso à comunicação sem fio baseadas no padrão LTE usadas pelo nó eNodeB.}
    \label{tab:LTEeNodeBsettings}
    \small \begin{tabular}{|l|c|}
      \hline
      \textbf{Parâmetro} & \textbf{Valor} \\
      \hline
      Potência de transmissão & 10W \\ \hline
      Modelo de macrocélula & RURAL\_MACROCELL \cite{3GPP:2017} \\ \hline
      Frequência & 2.1 GHz \\ \hline
      Largura de banda & 10 MHz \\ \hline
      BLER (Block Error Rate)  & 0.001 \\ \hline
      Redução HARQ & 0.2 \\ \hline
      Ruído térmico & -118.4 dBm \\ \hline
      Figura de ruído & 2 dB\\ \hline
      Sensibilidade da antena & -123.4 dBm \\ \hline
      Ganho da antena & 18 dBi \\ \hline
      Perda do cabo & 2 dB \\ \hline
      Alcance de Comunicação & 5000m \\ \hline
      Altura & 170m \\ \hline
    \end{tabular}
\end{table}
    \chapter{Configurações de Interesses Adotadas em Experimentos}

\begin{table}[H]
    \centering
    \caption{Identificação dos serviços de dados fornecidos pelas aplicações em experimentos com CCNs.}
    \label{tab:nomes_dados_CCNs}
    \small \begin{tabular}{|p{2cm}|c|p{5.5cm}|}
      \hline
      \textbf{Aplicação} & \textbf{Nome do Serviço de Dados} & \textbf{Descrição}\\ \hline
      \multirow{6}{*}{\centering AD} & \multirow{3}{*}{\textit{vehicle\_entering\_in\_$<$id. da Via$>$}} & Fornece dados acerca da entrada de veículos em uma via de entrada de uma interseção. \\ \cline{2-3}
      						& \multirow{3}{*}{\textit{vehicle\_leaving\_$<$id. da Via$>$}} & Fornece dados acerca de quais veículos estão deixando uma via de entrada de uma interseção. \\ \hline
      \multirow{8}{*}{\centering CSS} & \multirow{4}{*}{\textit{ack\_vehicle\_entering\_in\_$<$id. da Via$>$}} & Fornece o reconhecimento do recebimento dos dados acerca da entrada de veículos em uma via de entrada de uma interseção.\\ \cline{2-3}
        & \multirow{4}{*}{\textit{ack\_vehicle\_leaving\_$<$id. da Via$>$}} & Fornece o reconhecimento do recebimento dos dados relativos aos veículos que deixam uma via de entrada de uma interseção.\\ \hline
      \multirow{3}{*}{\centering NO} & \multirow{3}{*}{\textit{obstacle}} & Fornece dados relativos ao obstáculo no segmento de estrada. \\ \hline
      \multirow{6}{*}{\centering CACC} & \multirow{2}{*}{\textit{presence}} & Fornece dados sobre a presença de veículos em uma vizinhança.\\ \cline{2-3}
       & \multirow{4}{*}{\textit{state}} & Fornece dados relativos a veículos conectados posicionados imediatamente à frente em uma via. \\ \hline 
    \end{tabular}
\end{table}

\begin{table}[H]
    \centering
    \caption{Definição dos interesses registrados pelas aplicações em experimentos com a RAdNet-VE e RAdNet.}
    \label{tab:nomes_dados_interesses}
    \small \begin{tabular}{|l|l|}
      \hline
      \textbf{Nome do Serviço de Dados} & \textbf{Interesse} \\ \hline
      \multirow{2}{*}{\textit{vehicle\_entering\_in\_$<$id. da Via$>$}} & \textit{vehicle\_entering\_in\_$<$id. da Via$>$\_req} \\ \cline{2-2}
       												   & \textit{vehicle\_entering\_in\_$<$id. da Via$>$\_data} \\ \hline
      \multirow{2}{*}{\textit{vehicle\_leaving\_$<$id. da Via$>$}} & \textit{vehicle\_leaving\_$<$id. da Via$>$\_req} \\ \cline{2-2}
      											  & \textit{vehicle\_leaving\_$<$id. da Via$>$\_data} \\ \hline
      \multirow{2}{*}{\textit{ack\_vehicle\_entering\_in\_$<$id. da Via$>$}} & \textit{ack\_vehicle\_entering\_in\_$<$id. da Via$>$\_req} \\ \cline{2-2}
       												   & \textit{ack\_vehicle\_entering\_in\_$<$id. da Via$>$\_data} \\ \hline
      \multirow{2}{*}{\textit{ack\_vehicle\_leaving\_$<$id. da Via$>$}} & \textit{ack\_vehicle\_leaving\_$<$id. da Via$>$\_req} \\ \cline{2-2}
      											  & \textit{ack\_vehicle\_leaving\_$<$id. da Via$>$\_data} \\ \hline                                                  
      \multirow{2}{*}{\textit{obstacle}} & \textit{obstacle\_req} \\ \cline{2-2}
       												   & \textit{obstacle\_data} \\ \hline
      \multirow{2}{*}{\textit{presence}} & \textit{presence\_req} \\ \cline{2-2}
      											  & \textit{presence\_data} \\ \hline              
	  \multirow{2}{*}{\textit{state}} & \textit{state\_req} \\ \cline{2-2}
      											  & \textit{state\_data} \\ \hline        
    \end{tabular}
\end{table}

\begin{table}[H]
    \centering
    \caption{Definições de números máximos de saltos e direções de propagação de mensagens de rede para experimentos com RAdNet-VE.}
    \label{tab:interesses_saltos_direcoes}
    \small \begin{tabular}{|l|c|c|}
      \hline
      \textbf{Interesse} & \textbf{Número Máximo de Saltos} & \textbf{Direção} \\ \hline
      \textit{vehicle\_entering\_in\_$<$id. da Via$>$\_req} & \centering 5 & -1 \\ \hline
      \textit{vehicle\_entering\_in\_$<$id. da Via$>$\_data} & \centering 5 & 1 \\ \hline
      \textit{vehicle\_leaving\_$<$id. da Via$>$\_req} & 5 & 1 \\ \hline
      \textit{vehicle\_leaving\_$<$id. da Via$>$\_data} & 5 & -1 \\ \hline
      \textit{ack\_vehicle\_entering\_in\_$<$id. da Via$>$\_req} & \centering 5 & 1 \\ \hline
      \textit{ack\_vehicle\_entering\_in\_$<$id. da Via$>$\_data} & \centering 5 & -1 \\ \hline
      \textit{ack\_vehicle\_leaving\_$<$id. da Via$>$\_req} & 5 & -1 \\ \hline
      \textit{ack\_vehicle\_leaving\_$<$id. da Via$>$\_data} & 5 & 1 \\ \hline
      \textit{obstacle\_req} & \centering 50 & 1 \\ \hline
      \textit{obstacle\_data} & \centering 50 & -1 \\ \hline
      \textit{presence\_req} & 1 & 0 \\ \hline
      \textit{presence\_data} & 1 & 0 \\ \hline
      \textit{state\_req} & 1 & 1 \\ \hline
      \textit{state\_data} & 1 & -1 \\ \hline
    \end{tabular}
\end{table}

\begin{table}[H]
    \centering
    \caption{Configurações dos interesses registrados pelos agentes Veículo durante os experimentos, contendo tecnologia de acesso a comunicação (TAC), número máximo de saltos (NMS), direção de propagação da mensagem de rede (DIR) e o tipo do interesse (C - Controle ou D - Dados)}
    \label{tab:interessesVeiculo}
    \small \begin{tabular}{|l|c|c|c|c|}
      \hline
      \textbf{Interesse} & \textbf{TAC} &\textbf{NMS} & \textbf{DIR} & \textbf{Tipo} \\ \hline
      \textit{hello} & IEEE 802.11n & 1 & 0 & D \\ \hline
      \textit{hello} & IEEE 802.11p & 1 & 0 & D \\ \hline
      \textit{vehicle\_on\_$<$id. da via$>$} & IEEE 802.11n & 8 & 1 & D \\ \hline
      \textit{vehicle\_out\_$<$id. da via$>$} & IEEE 802.11n & 8 & -1 & D \\ \hline
      \textit{roadway\_presence\_request} & IEEE 802.11n & 8 & 1 e -1 & D \\ \hline
      \textit{roadway\_presence\_confirmation} & IEEE 802.11n & 8 & 1 & D \\ \hline
      \textit{roadway\_left\_confirmation} & IEEE 802.11n & 8 & -1 & D \\ \hline
      \textit{urban\_element\_data} & IEEE 802.11n & 8 & -1 & C \\ \hline
      \textit{urban\_element\_data} & IEEE 802.11p & 4 & -1 & C \\ \hline
      \textit{route\_to\_$<$id. da via$>$} & IEEE 802.11n & 8 & 1 & C \\ \hline
      \textit{route\_to\_$<$id. da via$>$} & IEEE 802.11p & 4 & 1 & C \\ \hline
      \textit{calculated\_route\_to\_$<$id. da via$>$} & IEEE 802.11n & 8 & -1 & C \\ \hline
      \textit{calculated\_route\_to\_$<$id. da via$>$} & IEEE 802.11p & 4 & -1 & C \\ \hline
      \textit{calculated\_route\_to\_$<$id. da via$>$} & LTE & 1 & 0 & C \\ \hline
      \textit{new\_traffic\_light\_schedule} & IEEE 802.11n & 8 & -1 & C \\ \hline
      \textit{new\_traffic\_light\_schedule} & IEEE 802.11p & 4 & -1 & C \\ \hline
      \textit{green\_wave\_request} & IEEE 802.11n & 8 & 1 & C \\ \hline
      \textit{green\_wave} & IEEE 802.11n & 8 & -1 & C \\ \hline
      \textit{intersection\_control\_data} & LTE & 1 & 0 & C \\ \hline
      \textit{vehicle\_position} & IEEE 802.11n & 8 & 0 & D \\ \hline
      \textit{intersection\_control\_data\_request} & LTE & 1 & 0 & C \\ \hline
    \end{tabular}
\end{table}

\begin{table}[H]
    \centering
    \caption{Configurações dos interesses registrados por todos os agentes Sinalização Semafórica, contendo tecnologia de acesso a comunicação (TAC), número máximo de saltos (NMS), direção de propagação da mensagem de rede (DIR) e o tipo do interesse (C - Controle ou D - Dados)}
    \label{tab:interessesSinalizacaoSemaforica}
    \small \begin{tabular}{|l|c|c|c|c|}
      \hline
      \textbf{Interesse} & \textbf{TAC} &\textbf{NMS} & \textbf{DIR} & \textbf{Tipo} \\ \hline
      \textit{vehicle\_on\_$<$id. da via$>$} & IEEE 802.11n & 1 & 1 & D \\ \hline
      \textit{vehicle\_out\_$<$id. da via$>$} & IEEE 802.11n & 1 & -1 & D \\ \hline
      \textit{reversibility\_change} & IEEE 802.11n & 1 & 0 & C \\ \hline
      \textit{edge\_reversal} & IEEE 802.11n & 1 & 0 & C \\ \hline
      \textit{roadway\_vehicle\_amount} & IEEE 802.11n & 1 & 0 & D \\ \hline
      \textit{urban\_element\_data} & LTE & 1 & 0 & C \\ \hline
      \textit{route\_to\_$<$id. da via$>$} & IEEE 802.11n & 1 & 1 & C \\ \hline
      \textit{route\_to\_$<$id. da via$>$} & IEEE 802.11p & 1 & 1 & C \\ \hline
      \textit{route\_to\_$<$id. da via$>$} & LTE & 1 & 0 & C \\ \hline
      \textit{new\_traffic\_light\_schedule} & LTE & 1 & 0 & C \\ \hline
      \textit{roadway\_space\_allocation} & LTE & 1 & 0 & C \\ \hline
    \end{tabular}
\end{table}

\begin{table}[H]
    \centering
    \caption{Configurações dos interesses registrados somente pelos agentes Sinalização Semafórica controladores de um sistema coordenado de sinalizações semafóricas, contendo tecnologia de acesso a comunicação (TAC), número máximo de saltos (NMS), direção de propagação da mensagem de rede (DIR) e o tipo do interesse (C - Controle ou D - Dados)}
    \label{tab:interessesSinalizacaoSemaforicaCoord}
    \small \begin{tabular}{|l|c|c|c|c|}
      \hline
      \textbf{Interesse} & \textbf{TAC} &\textbf{NMS} & \textbf{DIR} & \textbf{Tipo} \\ \hline
      \textit{corridor\_reversibility\_change} & LTE & 1 & 0 & C \\ \hline
      \textit{corridor\_edge\_reversal} & LTE & 1 & 0 & C \\ \hline
      \textit{corridor\_vehicle\_amount} & LTE & 1 & 0 & C \\ \hline
      \textit{group\_member\_vehicle\_amount} & LTE & 1 & 0 & D \\ \hline
      \textit{roadway\_segment\_vehicle\_amount} & LTE & 1 & 0 & D \\ \hline
      \textit{group\_member} & LTE & 1 & 0 & D \\ \hline
      \textit{reversal\_of\_all\_edges} & IEEE 802.11n & 1 & 0 & C \\ \hline
    \end{tabular}
\end{table}

\begin{table}[H]
    \centering
    \caption{Configurações dos interesses registrados somente pelos agentes Sinalização Semafórica participantes de um sistema coordenado de sinalizações semafóricas, contendo tecnologia de acesso a comunicação (TAC), número máximo de saltos (NMS), direção de propagação da mensagem de rede (DIR) e o tipo do interesse (C - Controle ou D - Dados)}
    \label{tab:interessesSinalizacaoSemaforicaPart}
    \small \begin{tabular}{|l|c|c|c|c|}
      \hline
      \textbf{Interesse} & \textbf{TAC} &\textbf{NMS} & \textbf{DIR} & \textbf{Tipo} \\ \hline
      \textit{traffic\_light\_coordination} & LTE & 1 & 0 & C \\ \hline
      \textit{participation\_in\_traffic\_light\_coordination} & IEEE 802.11n & 1 & 0 & D \\ \hline
      \textit{confirmation\_in\_traffic\_light\_coordination} & IEEE 802.11n & 1 & 0 & D \\ \hline
      \textit{corridor\_controller\_traffic\_light} & LTE & 1 & 0 & D \\ \hline
      \textit{reversal\_of\_all\_edges} & IEEE 802.11n & 1 & 0 & C \\ \hline
    \end{tabular}
\end{table}

\begin{table}[H]
    \centering
    \caption{Configurações de interesses registrados pelo agente Centro de Controle de Tráfego, contendo tecnologia de acesso a comunicação (TAC), número máximo de saltos (NMS), direção de propagação da mensagem de rede (DIR) e o tipo do interesse (C - Controle ou D - Dados)}
    \label{tab:interessesCentroControleTrafego}
    \small \begin{tabular}{|l|c|c|c|c|}
      \hline
      \textbf{Interesse} & \textbf{TAC} &\textbf{NMS} & \textbf{DIR} & \textbf{Tipo} \\ \hline
      \textit{urban\_element\_data} & LTE & 1 & 0 & C \\ \hline
      \textit{corridor\_reversibility\_change} & LTE & 1 & 0 & C \\ \hline
      \textit{corridor\_edge\_reversal} & LTE & 1 & 0 & C \\ \hline
      \textit{corridor\_vehicle\_amount} & LTE & 1 & 0 & C \\ \hline
      \textit{group\_member\_vehicle\_amount} & LTE & 1 & 0 & D \\ \hline
      \textit{roadway\_segment\_vehicle\_amount} & LTE & 1 & 0 & D \\ \hline
      \textit{group\_member} & LTE & 1 & 0 & D \\ \hline
      \textit{traffic\_light\_coordination} & LTE & 1 & 0 & C \\ \hline
      \textit{corridor\_controller\_traffic\_light} & LTE & 1 & 0 & D \\ \hline
      \textit{intersection\_control\_data\_request} & LTE & 1 & 0 & C \\ \hline
      \textit{intersection\_control\_data} & LTE & 1 & 0 & C \\ \hline
      \textit{calculated\_route\_to\_$<$id. da via$>$} & LTE & 1 & 0 & C \\ \hline
      \textit{route\_to\_$<$id. da via$>$} & LTE & 1 & 0 & C \\ \hline
      \textit{new\_traffic\_light\_schedule} & LTE & 1 & 0 & C \\ \hline
      \textit{roadway\_space\_allocation} & LTE & 1 & 0 & C \\ \hline
      \textit{traffic\_litht} & LTE & 1 & 0 & C \\ \hline
    \end{tabular}
\end{table}
  
\end{document}